\documentclass[useAMS,usegraphicx]{mn2e}

\usepackage{times}

\title[Molecular gas in LIRGs]{The molecular gas in Luminous Infrared
Galaxies I. CO lines,\\
  extreme physical conditions, and their drivers}

\author[Papadopoulos, Van der Werf, Xilouris, Isaak, Gao, \& M\"uhle]{Padelis P.\ Papadopoulos,$^1$\thanks{Email: padelis@mpifr-bonn.mpg.de} Paul P.\ van der Werf,$^2$
  E.M.\ Xilouris,$^3$ K.G.\ Isaak,$^4$ Yu Gao,$^5$ \and S.\ M\"uhle,$^6$\\
$^1$Max Planck Institute f\"ur Radioastronomie, Auf dem H\"ugel 69, D-53121 Bonn, Germany\\
$^2$Leiden Observatory, Leiden University, P.O.~Box 9513, NL-2300 RA Leiden, The Netherlands\\
$^3$Institute of Astronomy and Astrophysics, National Observatory of Athens,
I.Metaxa \& Vas.Pavlou str., GR-15236, Athens, Greece\\
$^4$Research and Scientific Support Department, European Space Agency, Keplerlaan 1, 2200~AG, Noordwijk, The Netherlands\\
$^5$Purple Mountain Observatory, Chinese Academy of Sciences, Nanjing, Jiangsu 210008, China\\
$^6$Argelander-Institut f\"ur Astronomie, Auf dem H\"ugel 71, D-53121 Bonn, Germany}

\begin{document}

\date{Accepted ... Received ...; in original form ...}

\pagerange{\pageref{firstpage}--\pageref{lastpage}} \pubyear{...}

\maketitle

\label{firstpage}

\begin{abstract}
We  report results  from a  large  molecular line  survey of  Luminous
Infrared Galaxies  (LIRGs: $\rm L_{IR}$$\ga  $10$^{11}$\,L$_{\odot }$)
in the  local Universe (z$\leq$0.1), conducted during  the last decade
with  the James  Clerk  Maxwell  Telescope (JCMT)  and  the IRAM  30-m
telescope.  This work presents the CO  and $ ^{13}$CO line data for 36
galaxies,  further augmented  by  multi-J total  CO line  luminosities
available  for other  IR-bright  galaxies from  the literature.   This
yields a combined sample of N=70 galaxies with the star-formation (SF)
powered  fraction  of  their  IR luminosities  spanning  $\rm  L^{(*)}
_{IR}$$\sim  $(10$^{10}$-2$\times $10$^{12}$)\,$\rm L_{\odot}$,  and a
wide range of morphologies.   Simple comparisons of their available CO
Spectral Line Energy Distributions (SLEDs) with local ones, as well as
radiative transfer models discern a surprisingly wide range of average
ISM   conditions,  with   most   of  the   surprises   found  in   the
high-excitation  regime.   These take  the  form  of  global CO  SLEDs
dominated by a very warm ($\rm T_{kin}$$\ga $100\,K) and dense (n$\geq
$10$^4$\,cm$^{-3}$)   gas   phase,   involving   galaxy-sized   ($\sim
$(few)$\times   $10$^9$\,M$_{\odot}$)   gas   mass  reservoirs   under
conditions that  are typically found  only for $\sim $(1-3)\%  of mass
per typical SF molecular cloud in the Galaxy.  Furthermore some of the
highest  excitation CO  SLEDs  are found  in  Ultra Luminous  Infrared
Galaxies  (ULIRGs,  $\rm  L_{IR}$$\geq  $10$^{12}$\,L$_{\odot}$),  and
surpass  even those found  solely in  compact (star-formation)-powered
´hot´-spots   in  Galactic   molecular   clouds.   Strong   supersonic
turbulence  and high  cosmic  ray (CR)  energy  densities rather  than
far-UV/optical photons or SNR-induced  shocks from individual SF sites
can  globally warm  the  large amounts  of  dense gas  found in  these
merger-driven starbursts and easily  power their extraordinary CO line
excitation.   This  exciting  possibility  can now  be  systematically
investigated with  Herschel and ALMA.  As expected  for an IR-selected
(and  thus (SF  rate)-selected) galaxy  sample, only  few  ``cold'' CO
SLEDs are found,  and for fewer still a  cold low/moderate-density and
gravitationally bound state (i.e.   Galactic-type) emerges as the most
likely   one.   The   rest   remain  compatible   with   a  warm   and
gravitationally unbound low-density phase often found in ULIRGs.  Such
degeneracies,  prominent when  only  the low-J  SLED segment  (J=1--0,
2--1, 3--2) is available, advise  against using its CO line ratios and
the  so-called  $\rm   X_{co}$=M(H$_2$)/$\rm  L_{co}$(1-0)  factor  as
star-formation  mode indicators,  a  practice that  may  have lead  to
misclasification of the ISM environments of IR-selected gas-rich disks
in the distant Universe.  Finally we expect that the wide range of ISM
conditions  found among LIRGs  will strongly  impact the  $\rm X_{co}$
factor, an  issue we examine in detail in paper II  (Papadopoulos et
al. 2012).

\end{abstract}


\section{Introduction}

The population  of luminous  infrared galaxies (LIRGs),  discovered by
the {\it  Infrared Astronomical  Satellite (IRAS)} to  have bolometric
luminosities dominated  by the infrared part of  their Spectral Energy
Distributions (SEDs) (e.g.  Soifer et al.  1987), contains some of the
most  extreme star-forming  systems in  the local  Universe.   At $\rm
L_{IR}\ga   10^{11}\,   L_{\odot}$   these  deeply   dust   enshrouded
star-forming  systems dominate  the luminosity  function of  the local
Universe,  and at  $\rm L_{IR}\geq  10^{12}\, L_{\odot}$  surpass even
optically selected QSOs (Soifer  \& Neugebauer 1991; Sanders \& Ishida
2004 and references therein).  Their large reservoirs of molecular gas
mass ($\sim  $10$^9$-10$^{10}$\,M$_{\odot}$) discovered via  CO J=1--0
observations (Tinney et al.   1990; Sanders, Scoville, \& Soifer 1991;
Solomon et  al.~1997), along with  clear evidence of  strong dynamical
interactions and mergers in many LIRGs (e.g.  Sanders \& Ishida 2004),
make  these systems  unique local  examples of  dust-enshrouded galaxy
formation in the distant Universe (e.g.  Smail, Ivison, \& Blain 1997;
Hughes et al.  1998).

 Interferometric  imaging of  the  CO 1-0  line  (and occasionally  of
 J=2--1),  revealed  gas  disks   with  D$\la  $0.5\,kpc  and  surface
 densities  $\rm \Sigma  (H_2)$$\sim  $($10^2$--$10^4$)\,$\rm M_{\odot
 }\, pc^{-2} $, often decoupled from the stellar components of merging
 galaxies (e.g.  Sanders et al.   1988a; Wang et al. 1991; Planesas et
 al. 1991;  Bryant \&  Scoville 1996, 1999;  Downes \&  Solomon 1998).
 Naturally   the   CO(2--1)/(1--0)  ratio   was   the   first  to   be
 systematically  measured (Kr\"ugel  et  al.  1990;  Braine \&  Combes
 1992; Horellou et al 1995;  Aalto et a.  1995; Albrecht, Kr\"ugel, \&
 Chini 2007)  even if it is  insensitive to the presence  of dense and
 warm gas typical near SF  sites.  The advent of submm interferometric
 CO(3--2) imaging revealed the distributions  of such gas in LIRGs via
 the  distribution  of CO  (3--2)/(1--0),  (3--2)/(2--1) ratios  (e.g.
 Sakamoto et al.  2008; Wilson et al.  2009; Iono et al.  2007, 2009),
 but its utility  remains limited by dissimilar u-v  coverage and lack
 of zero-spacing  information (Iono et  al.  2004).  Thus  single dish
 measurements of  total molecular  line luminosities and  their ratios
 remain  a  primary  tool   for  probing  the  average  molecular  gas
 properties in  LIRGs (and a prerequisite  for interferometric images
 that contain all spatial information).

 Ideally a combination of low to mid-J rotational lines of heavy rotor
 molecules with high critical densities such as HCN ($\rm n_{cr}$$\sim
 $(2-40)$\times  10^{5}$\,$\rm  cm^{-3}$  for  J=1--0, 3--2),  and  CO
 J+1$\rightarrow $J lines from J=0, 1,  up to at least J$\geq $2 ($\rm
 E_{3}/k_B$$\sim  $33\,K, $\rm n_{crit}$$\sim  $$10^4$\,cm$^{-3}$) are
 necessary to probe the large range of physical properties within GMCs
 ($\rm            T_{kin}$$\sim           $(15-100)\,K,           $\rm
 n(H_2)$$\sim$(few)$\times$($10^2$-$10^6$)\,cm$^{-3}$).     Sensitivity
 limitations  and/or  lack   of  multi-beam  receivers  confined  such
 measurements  to few  nearby LIRGs  and mostly  towards  their nuclei
 (e.g.  Devereux et al.  1994; Dumke et al.  2001; Meier et al.  2001;
 Zhu et al.   2003), while heavy rotor molecular  lines are faint with
 HCN(1--0)$\sim$(1/5-1/10)$\times $CO(1--0) even in ULIRGs, and as low
 as  $\sim$ 1/40$\times$CO(1--0)  in typical  spirals (Solomon  et al.
 1992).  Multi-J observations of the luminous CO line emission are not
 limited  by sensitivity rather  by the  lack of  multi-beam receivers
 and/or  beam-matched observations  in  widely different  frequencies.
 Thus, while  several studies use CO(3--2)/(1--0) as  a warm/dense gas
 tracer in substantial LIRG  samples (e.g.  Mauersberger et al.  1999;
 Yao  et al.   2003;  Narayanan 2005;  Mao  et al.   2011), these  are
 typically confined solely within their nuclear regions.  CO J=4--3 or
 higher-J  line  observations  are  even more  sporadic,  hindered  by
 increasing  atmospheric absorptions  at $\nu  $$\ga$  460\,GHz, while
 also confined to the nuclear  regions of nearby LIRGs (e.g.  White et
 al.  1994; G\"usten et al.   1996; Petitpas \& Wilson 1998; Nieten et
 al.   1999;  Mao  et  al.   2000;  Bayet  et  al.   2006).   This  is
 unfortunate since, while the difficulty of measuring reliable CO line
 ratios increases  with wider J-level separations (for  the same dish:
 $\rm      \Omega^{(beam)}     _{J+1\rightarrow     J}/\Omega^{(beam)}
 _{1\rightarrow 0}$=$\rm (J+1)^{-2}$), so does their diagnostic power.
 Nevertheless  molecular Spectral  Line  Energy Distributions  (SLEDs)
 remain the  key tool for probing  the average state  of the molecular
 gas in galaxies, and  for estimating total and star-forming molecular
 gas masses.  Finally local CO SLEDs provide a necessary benchmark for
 the usually more sparsely sampled ones for galaxies at high redshifts
 (e.g.  Weiss  et al  2007), and  have been recently  used as  SF mode
 (merger  versus disk-driven star  formation) indicators  for gas-rich
 near-IR selected  disks in the  distant Universe (Dannerbauer  et al.
 2009; Daddi et al. 2010).

We  used the  James Clerk  Maxwell Telescope  (JCMT\footnote{The James
  Clerk Maxwell  Telescope is operated  by the Joint  Astronomy Centre
  (JAC)  on behalf of  the Science  and Technology  Facilities Council
  (STFC)  of  the United  Kingdom,  the  Netherlands Organisation  for
  Scientific Research,  and the National Research  Council of Canada})
on Mauna Kea in Hawaii (USA),  and the IRAM 30-meter telescope at Pico
Veleta  (Spain) to  conduct  a multi-J  CO,  HCN line  survey of  such
systems  and study  the  molecular ISM  in  some of  the most  extreme
star-forming systems  in the local Universe.   First results regarding
the CO J=6--5 lines and dust emission SEDs have already been published
(Papadopoulos et  al.  2010a,b, hereafter  P10a,b).  In this  paper we
present the  entire (JCMT)+(IRAM) CO, $^{13}$CO  line dataset, further
augmented  with  all reliable  measurements  of  {\it  total} CO  line
luminosities of LIRGs from  the literature, which yields currently the
largest such database assembled  for local star-forming galaxies.  For
a few  LIRGs in our sample  their CO SLEDs have  been extended towards
higher-J levels (from J=5--4 up to J=13--12) with the SPIRE/FTS aboard
the Herschel Space Observatory.

 The layout of  this work is as follows: in Sections  2, 3 we describe
 the  sample  and  the  observations,  Section  4  contains  the  data
 reduction, the  literature search,  and line flux  rectifications for
 any  discrepant  values  found.   In  Section 5  we  investigate  the
 molecular  ISM  exitation  in  LIRGs  using  comparisons  with  local
 environments,  and propose  a  new two-phase  ISM  model for  extreme
 starbursts,  motivated  by   recent  views  on  SF-powered  radiative
 feedback onto  the ISM. In the  same saction we  also briefly discuss
 the role of  AGN in driving the global ISM  excitation.  In section 6
 one-phase radiative transfer  modeling of the CO line  ratios is used
 to  systematically extract the  average densities,  temperatures, and
 dynamical states  of the average ISM environments  encountered in our
 sample,  and assemble  a comprehensive  picture of  their  range.  In
 section  7 we  present the  CO  SLED excitation  range possible  from
 J=1--0 up to J=7--6, and  discuss their power sources.  There we also
 discuss the  (CO SLED)$\leftrightarrow$(ISM state)  degeneracies, and
 their impact on the classification of SF modes in galaxies, using the
 CO line ratios of recently discovered near-IR selected gas-rich disks
 at  high redshifts  as an  example.   We present  our conclusions  in
 Section 8, where we also list a few important questions regarding the
 expected effects  of ISM excitation conditions on  molecular gas mass
 estimates  in LIRGs  and  ULIRGs,  the subject  of  our second  paper
 (Papadopoulos et  al. 2012).   Throughout this work  we adopt  a flat
 $\Lambda        $-dominated         cosmology        with        $\rm
 H_0$=71\,km\,s$^{-1}$\,Mpc$^{-1}$ and $\Omega_{\rm m}$=0.27.

\section{The sample}

 The sample was  drawn from two CO J=1--0 surveys  of LIRGs by Sanders
 et al.   1991, and  Solomon et al.   1997 (themselves drawn  from the
 {\it   IRAS}   BGS   flux-limited   sample   with   $\rm   f_{60\,\mu
   m}$$>$5.24\,Jy; Soifer  et al.  1987, 1989; Sanders  et al.  2003),
 so that all  galaxies in our sample have at  least one measurement of
 this basic  H$_2$-tracing line.  We imposed  two additional criteria,
 namely: a) $\rm  z\leq 0.1$ (the maximum redshift  for which the JCMT
 B-band receivers can  tune to CO J=3--2), and  b) compact CO-emitting
 regions  (sizes  from  CO  interferometric  images)  so  that  single
 telescope pointings or  small maps can record total  line fluxes.  If
 CO images were unavailable then radio continuum (Condon et al.  1990,
 1996; Crawford et al.  1996), sub-mm dust emission (Lisenfeld, Isaak,
 \& Hills  2000; Mortier  et al.  2011)  or near-IR images  (Zenner \&
 Lenzen 1993; Murphy et al.  1996; Scoville et al.  2000) helped place
 an upper limit on the  size of their CO-bright emission unaffected by
 extinction.  This  is because sub-mm emission, sensitive  to both the
 warm dust associated with molecular gas and the cold dust concomitant
 with  the more extended  HI (Thomas  et al.   2001) sets  $\rm \theta
 _{sub-mm}$$>$$\theta _{CO}$,  while the radio continuum,  tied to the
 far-IR  emission through  a well-known  correlation, traces  the star
 forming  regions of LIRGs  (Condon et  al.  1990)  where most  of the
 molecular gas~lies.

\begin{table*}
 \centering
 \begin{minipage}{111mm}
  \caption{JCMT: observing periods, CO lines, typical system temperatures}
  \begin{tabular}{@{}lccc@{}}
\hline
Year & periods & spectral lines & $\rm T_{sys}$ (K)$^{a}$\\
\hline
1999    & 06/07--20/07 & CO 3--2, 4--3 & 450-700 (B\,3), 1130--2380 (W/C)\\
2001    & 10/12--28/12 & CO 3--2       & 380-550 (B\,3)\\
2002    & 24/02--25/02 & CO 3--2       & 630-690 (B\,3)\\
``      & 17/04--18/04 & CO 3--2       & 760, 1100-1700 (B\,3)\\
``      & 17/06--30/06 & CO 2--1, 3--2 & 550 (A\,3), 430-650, 900-1200 (B\,3)\\
``      & 20/11--23/11 & CO 4--3       & 2600-4800, 6000-9300 (W/C)$^{b}$\\
2003    & 06/11        & CO 3--2       & 4000-4400 (B\,3)\\
2004    & 20/01        & CO 3--2       & 530-620 (A\,3)\\
``      & 02/04--29/05 & $ ^{12,13}$CO 2--1, CO 3--2 &300-450 (A\,3), 470-2400 (B\,3)$^{b}$\\ 
``      & 13/07--25/08 & $ ^{12,13}$CO 2--1, CO 3--2 & 320-510 (A\,3), 800-1000 (B\,3)\\
``      & 28/09--10/11 & $ ^{12,13}$CO 2--1, CO 3--2 & 300-600 (A\,3), 530-1350 (B\,3)\\
``      & 18/11        & CO 4--3       & 2450-2800 (W/C)\\
2005    & 20/02        & CO 6--5       & 4500-5200 (W/D)\\
``      & 17/04-23/04  & CO 3--2, 4--3, 6--5 & 720-1100 (B\,3), 1600 (W/C), 3700-5500 (W/D)\\
``      & 22/08        & $ ^{13}$CO 2--1 & 415-420 (A\,3)\\
``      & 10/10--28/10 & $ ^{12,13}$CO 2--1 & 230-420 (A\,3)\\
``      & 15/12--31/12 & $ ^{12,13}$CO 2--1, CO 3--2 & 330-490 (A\,3), 750-980 (B\,3)\\
2006    & 15/12--18/12 & $ ^{13}$CO 2--1 & 250-280 (A\,3/ACSIS)\\
2007    & 16/12--21/12 & $ ^{13}$CO 2--1 & 230-280 (A\,3/ACSIS)\\
``      & 22/02--24/02 & $ ^{13}$CO 2--1 & 310-320 (A\,3/ACSIS)\\
2008    & 09/05        &  CO 3--2        & 1630 (HARP-B/ACSIS) \\
2009    & 06/01--07/01 &  CO 6--5        & 2600-3200, 7500 (W/D, ACSIS)$^{b}$\\
``      & 22/01--25/01 &  CO 6--5        & 1700-2600, 9000 (W/D, ACSIS)$^{b}$\\ 
``      & 01/02        &  CO 6--5        & 2200-3500 (W/D, ACSIS)\\
``      & 27/01        &  CO 6--5        & 2000-2900 (W/D, ACSIS)\\
``      & 02/03        &  CO 6--5        & 1300-1500 (W/D, ACSIS)\\
``      & 13/03--15/03 &  CO 6--5        & 1900-3100 (W/D, ACSIS)\\
2010    & 11/09        &  CO 6--5        & 1400-1800 (W/D, ACSIS)\\
\hline
\end{tabular}
$^{a}$$\rm T_{sys}$ values include atmospheric absorption. High values of
 $\rm T_{sys}(B\,3)$$\ga $600\,K   and $\rm T_{sys}(A\,3)$$\ga $350\,K) are measured in a few cases
 close to tuning range limits ($\rm \nu _{sky}$$<$320\,GHz for B\,3, $\rm \nu _{sky}$$\leq $215\,GHz 
for A\,3), or when $\rm |\nu _{sky}-325\,GHz|$$<$5\,GHz, i.e. close to a strong atmospheric
absorption feature at $\rm 325\,GHz$.\\
$^{b}$The high $\rm T_{sys}$ values found only for VIIZw\,31, a circumpolar source
 at the JCMT latitude, observed at elevation of $\sim 30^{\circ}$
\end{minipage}
\end{table*}

\begin{figure}
\includegraphics[width=\columnwidth]{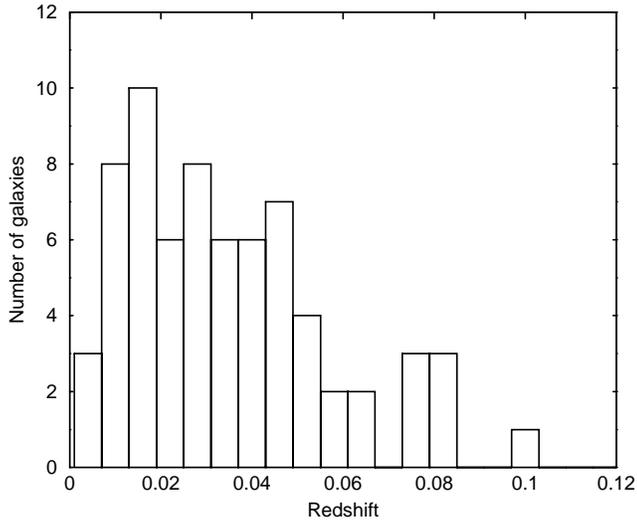}
\caption{The redshift distribution of the combined sample (see section 2).}
\end{figure}

 We have also extensively searched  the literature for all local LIRGs
 whose {\it total} CO J+1$\rightarrow $J fluxes from J+1=1 up to J+1=3
 are  measured (for  higher-J  lines such  measurements are  currently
 non-existent in  the local Universe).  Unfortunately  this forced the
 omission of many, often beam-matched, observations (e.g.  Devereux et
 al.  1994;  Mauersberger et al.  1999;  Mao et al.  2011)  as in many
 such cases  CO line emission  extends well beyond the  nuclear region
 where such  observations were conducted.   The combined sample  of 70
 LIRGs (Table  3) is currently  the largest for which  their molecular
 gas can be studied using CO lines which can probe physical conditions
 from  the  quiescent  to   the  star  forming  phase.   Its  redshift
 distribution  is shown  in Figure  1,  while its  IR luminosity  (the
 SF-powered    part)    range     is    $\rm    L^{(*)}    _{IR}$$\sim
 $(10$^{10}$--2$\times    $10$^{12}$)\,L$_{\odot}$    (computed   over
 $\lambda$=(8-1000)\,$\mu $m).   The sample  is sparce towards  low IR
 luminosities     since     galaxies     with    $\rm     L_{IR}$$\sim
 $(1-5)$\times$10$^{10}$\,L$_{\odot  }$  are  usually  extended  (e.g.
 Tinney et al.  1990) and thus rarely have total CO J+1$\rightarrow $J
 fluxes available for J+1$\geq $3.  We tried to alleviate this bias by
 including the  few low-L$_{\rm IR}$  galaxies whose total  CO J=1--0,
 3--2  line luminosities  have been  recently measured  (Leech  et al.
 2010).

\begin{table*}
 \centering
 \begin{minipage}{132mm}
  \caption{Point source $\rm S_{\nu }/T$ (Jy/K) conversion factors and HPBW values used for the observations and literature data}
  \begin{tabular}{@{}lccccc@{}}
\hline
Telescope$^{a}$ & 110--115\,GHz$^{b}$ & 210--230\,GHz
& 315--345\,GHz$^{b}$ & 430--461\,GHz$^{b}$ & 620--710\,GHz$^{b}$\\
\hline
IRAM 30-m ($\rm S_{\nu}/T^{*} _A$)  &  6.3$^{c}$\,\,($22''$) & 7.9, 8.7($11'')$  &   &   & \\
JCMT      ($\rm S_{\nu}/T^{*} _A$)  &     & 25-28($22''$)  & 28-38($14''$)& 50-74($11''$) & 49-62($8''$)\\
NRAO  12-m ($\rm S_{\nu}/T^{*} _R$) &  35$^{d}$\,\,($55''$) & 55($32''$) &   &   &  \\
FCRAO 14-m ($\rm S_{\nu}/T^{*} _A$) &  42($45''$)$^{e}$  &     &   &   &  \\
Onsala 20-m ($\rm S_{\nu}/T^{*} _A$)&  31($33''$)                   &   &   &   &  \\
SEST        ($\rm S_{\nu}/T_{mb}$)  &  27($45''$)$^{f}$  & 41($22''$)$^{f}$ &  
&   &    \\
HHSMT ($\rm S_{\nu}/T_{mb}$) &   &   & 50$^{g}$\,\,($23''$)&   &  \\ 
NRO 45-m   ($\rm S_{\nu}/T_{mb}$) & 2.4$^{h}$\,\,($14.5''$) &   &   &   &  \\
CSO        ($\rm S_{\nu}/T_{mb}$) &   &   & 43$^{i}$\,\,($21''$) &   &   \\
\hline
\end{tabular}
$^{a}$The telescope and temperature scale type (see Kutner \& Ulich 1981 for definitions).\\
$^{b}$The frequency range per receiver.\\
$^{c}$For $\rm T^*_A$=$\rm (B_{eff}/F_{eff})\times T_{mb}$=$\rm 0.789\times T_{mb}$:
                  $\rm S_{\nu }/T_{mb}$=4.95\,Jy/K, used to obtain the CO 1-0 fluxes from 
                  all the IRAM 30-m spectra in the literature that are reported in the $\rm T_{mb}$ scale
                   (unless a different $\rm S_{\nu}/T_{mb}$ is mentioned).\\
$^{d}$Measured at 110\,GHz (section 2.3.2), also in the NRAO
                 12-m User's Manual 1990 Edition, Figure~14.\\
$^{e}$The $\rm T^*_R$ scale is sometimes used to report data from the 
FCRAO 14-m (e.g. Sanders et al. 1986), for which  $\rm S_{\nu }/T^*_R = 
\eta_{fss} \left(S_{\nu }/T^* _A\right) = 31.5(Jy/K)$  (for $\rm \eta _{fss}=0.75$) is adopted.\\
$^{f}$http://www.ls.eso.org/lasilla/Telescopes/SEST/html/telescope\-instruments/telescope/index.html.\\
$^{g}$The Heinrich Hertz Submillimeter Telescope (Arizona, USA) (from Narayanan et al. 2005).\\
$^{h}$For the NRO 45-m telescope in Nobeyama (Japan) at 115\,GHz:
 $\rm \Gamma =(8 k_B/\pi D^2) (\eta_{mb}/\eta_{a})$ and adopting $\eta_a=0.32$, $\eta_{mb}=0.44$ 
(from the NRO website).\\
$^{i}$Caltech Submillimeter Observatory (CSO):
 $\rm \Gamma = (8 k_B/\pi D^2) (\eta_{mb}/\eta_{a})$, where $\rm D=10.4\,m$, $\eta_{mb}=0.746$ and
  $\rm \theta_{1/2}=1.22(\lambda /D)$ yielding $\eta_{a}/\eta_{mb}=0.76$
(using  $\Omega_A A_{e}=\lambda^2$ and $\rm \eta_a=A_e/A_g$, $\eta_{mb}=\Omega_{mb}/\Omega_A$.
\end{minipage}
\end{table*}

\section{The  observations}

We  used  the  A\,3  receiver  (211-276\,GHz, DSB  operation)  on  the
15-meter JCMT on  Mauna Kea in Hawaii (USA), to  observe the CO J=2--1
(230.538\,GHz), and  $ ^{13}$CO  J=2--1 (220.398\,GHz) lines,  and its
B\,3 (315-370\,GHz) and  W/C (430-510\,GHz) receivers operating single
sideband (SSB) for CO  J=3--2 (345.796\,GHz) and J=4--3 (461.041\,GHz)
line  observations in  our  sample.  The  observations were  conducted
during several periods from 1999 up  to 2010 (see Table 1 for specific
periods  and  typical system  temperatures).   CO J=3--2  observations
beyond   2008   utilized  the   new   16-beam   HARP-B  SSB   receiver
(325-375\,GHz).  The  decomissioning of  the W/C JCMT  receiver before
completion  of the  survey  as well  as  the several  CO J=4--3  lines
redshifted into  the deep  450\,GHz atmospheric absorption  band meant
that  such measurements  could be  conducted for  only 10  out  of the
original sample of 30~LIRGs.

The Digital  Autocorrelation Spectrometer (DAS)  was used in  all JCMT
observations until 2006, while the new spectrometer ACSIS was employed
aftewards.  At 345\,GHz we used its 920\,MHz ($\sim$800\,km\,$s^{-1}$)
or 1.8\,GHz  ($\sim $1565\,km\,s$^{-1}$) bandwidth  mode, depending on
the expected  line width  and the need  for maximum  sensitivity (i.e.
when 920\,MHz bandwidth was sufficient to cover the line, dual-channel
operation   was  possible  with   B\,3,  and   was  used   for  better
sensitivity).   For the  high  frequency W/C  observations the  widest
1.8\,GHz bandwidth was used throughout whose $\sim $1170\,km\,s$^{-1}$
velocity coverage adequately covered the FWZI of all the CO 4--3 lines
observed.  For  the CO, $  ^{13}$CO J=2--1 lines both  bandwidth modes
were   used,  yielding   $\sim   $(1200--2345)\,km\,s$^{-1}$  velocity
coverage.     Beam-switching   with    frequencies    of   $\rm    \nu
_{chop}$=(1-2)\,Hz, at throws  of 120$''$-180$''$ (in azimuth) ensured
flat    baselines.     The    beam    sizes    were:    $\rm    \Theta
_{HPBW}(230\,GHz)$=22$''$,  $\rm \Theta  _{HPBW}(345\,GHz)$=14$''$ and
$\rm  \Theta_{HPBW}(461\,GHz)$=11$''$.   We  checked and  updated  the
pointing model  offsets every hour  using continuum and  spectral line
observations  of strong  sources, with  average residual  pointing rms
scatter  $\rm  \sigma_r$=$\sqrt{\sigma^2  _{el}+\sigma^2 _{az}}$$\la
$2.5$''$.

\subsection{The CO J=6--5 observations}

The  first measurements  of  the CO  J=6--5  line (691.473\,GHz)  were
conducted for the luminous ULIRG/QSO Mrk\,231 and the LIRG Arp\,193 in
our  sample  using  the  old  JCMT W/D  band  (620-710\,GHz)  receiver
(operating  in SSB  mode)  on 20  of  February and  22  of April  2005
respectively,   under    excellent,   dry   conditions    ($\rm   \tau
_{220\,GHz}$$\la $0.035).   The typical system  temperatures were $\rm
T_{sys}$$\sim  $(3700--5500)\,K  (including  atmospheric  absorption).
The DAS  spectrometer was  used in its  widest mode of  1.8\,GHz ($\rm
\sim 780\,km\,s^{-1}$ at 690\,GHz),  and beam switching at frequencies
of $\rm \nu _{chop}$=2\,Hz with azimuthal throws of $60''$ resulted in
flat  baselines.    The  beam  size   at  691\,GHz  was   $\rm  \Theta
_{HPBW}$=$8''$.  Good  pointing with such narrow beams  is crucial and
was checked  every (45-60)\,mins using differential  pointing with the
B3  receiver (350\,GHz).   This allows  access to  many  more suitable
compact sources in the sky  than direct pointing with the W/D receiver
at 690\,GHz,  and was  found accurate to  within $\rm  \sigma _r$$\sim
$2.6$''$ (rms) during that observing period.

The other CO J=6--5 measurements  were conducted during 2009, with the
upgraded W/D  receiver equipped with  new SIS mixers  (effectively the
same type  installed at  the ALMA telescopes  in this  waveband) which
dramatically  enhanced its  performance.  The  resulting  low receiver
temperatures  ($\rm  T_{rx}$$\sim   $550\,K)  allowed  very  sensitive
observations   with   typical   $\rm  T_{sys}$$\sim   $(1500--3000)\,K
(including  atmospheric  absorption)  for  $\rm  \tau_{220\,GHz}$$\sim
$0.035-0.06.   Dual  channel  operation  (after the  two  polarization
channels were aligned to within  $\la $$1''$) further enhanced the W/D
band  observing capabilities  at the  JCMT (see  Table 1).   The ACSIS
spectrometer at  is widest mode of  1.8\,GHz was used, while  in a few
cases two separate tunings were  used to create an effective bandwidth
of $\sim $3.2\,GHz ($\sim  $1390\,km\,s$^{-1}$ at 690\,GHz) so that it
adequately     covers    (U)LIRG     CO    lines     with    FWZI$\sim
$(800--950)\,km\,s$^{-1}$.     Rapid    beam    switching   at    $\rm
\nu_{chop}$=4\,Hz  (continuum  mode)  and  azimuthal throw  of  $30''$
yielded very  flat baselines  under most circumstances.   The pointing
model  was updated  every  45-60\,mins using  observations of  compact
sources with the  W/D receiver, as well as  differential pointing with
the A\,3 receiver,  yielding rms residual error radius  of $\rm \sigma
_r$$\sim $2.2$''$.  The final CO J=6--5 observations were conducted in
2011 during which  I\,Zw\,1 was observed, with only  one W(D) receiver
channel functioning,  under dry conditions  ($\rm \tau_{220\,GHz}$$\la
$0.05)  that yielded  $\rm T_{sys}$$\sim  $(1400--1800)\,K.   The same
beam-switching scheme was used,  while two separate tunings yielded an
effective bandwidth  of $\sim $3.2\,GHz  covering the wide CO  line of
this  ULIRG/QSO   (e.g.   Barvainis  et  al.    1989).   The  pointing
uncertainty    remained   within    the   range    of    previous   CO
J=6--5~observations.  Nevertheless we wish to note that isolated cases
of large pointing offsets reducing  the observed CO J=6--5 line fluxes
have been found  (e.g. for Arp\,220, see P10a and  P10b), and may have
affected a few of these highly demanding CO line measurements.

\subsection{The IRAM 30-m observations}

Observations  of  CO, $  ^{13}$CO  J=1--0,  2--1  with the  IRAM  30-m
telescope were conducted during two sessions in 2006 namely, June from
20 to 25,  and November from 26 to 28.  In  both periods the A100/B100
(3\,mm) and  A230/B230 (1\,mm) receivers  were used, connected  to the
1\,MHz   (A100/B100,   512\,MHz)   and  4\,MHz   (A230/B230,   1\,GHz)
filterbanks.   During the  first period  the A230/B230  receivers were
used to observe the $ ^{12}$CO  J=2--1 line.  If the latter was strong
and detected  in about a hour  or less, the 1\,mm  receivers were then
re-tuned to $^{13}$CO  J=2--1.  For sources with very  weak $ ^{12}$CO
lines  (e.g.    08030+5243,  08572+3915),  no   attempt  of  observing
$^{13}$CO  J=2--1 was  made.   Data were  acquired  under New  Control
System (NCS) in  series of four-minute scans, each  comprised of eight
30-sec   subscans.   The   typical   system  temperatures   (including
atmospheric  absorption)  for  the  CO  2--1  observations  were  $\rm
T_{sys}$(210-230\,GHz)$\sim  $(220--500)\,K,  with  the lowest  mostly
during  the $  ^{13}$CO observations  (though for  occasional tunings
towards  the edge  of  the  band and/or  bad  weather conditions  $\rm
T_{sys}$$\sim $(700-900)\,K).  For most  sources data were acquired in
two  or more  different days  to  ensure a  line detection,  and as  a
consistency check.  Pointing and  focus were checked frequently during
the   observations   with  residual   pointing   errors  $\rm   \sigma
_r$$\sim$3$''$~(rms).

  During the November period  receivers A100/B100 were used to observe
  $ ^{13}$CO J=1--0 line simultaneously  to the $ ^{12}$CO J=2--1 line
  observed with  A230/B230 (tuned  to the same  line each  time).  The
  pointing error  stayed $\la $3$''$ (rms), except  during November 26
  when it went  up to $\sim $6$''$ (corresponding  data were omitted).
  The  typical system  temperatures were  $\rm T_{sys}(110\,GHz)$$\sim
  $110-160\,K,  $\rm T_{sys}(115\,GHz)$$\sim  $(200-380)\,K,  and $\rm
  T_{sys}$(210--230\,GHz)$\sim  $(330-425)\,K.  Finally,  in  order to
  maintain  very  flat  baselines,  the  wobbler  switching  (nutating
  subreflector)  observing mode with  a frequency  of 0.5~Hz  and beam
  throws  of  180$''$--240$´´$  was  employed  during  both  observing
  sessions.      The     beam      sizes     were:     $\rm     \Theta
  _{HPBW}(110\,GHz)$=22$''$          and          $\rm          \Theta
  _{HPBW}$(210-230\,GHz)=11$''$,      with      corresponding     beam
  efficiencies\footnote{http://www.iram.fr/IRAMES/index.htm}  of: $\rm
  B_{eff}(110\,GHz)$=0.75,   $\rm  B_{eff}(230\,GHz)$=0.52   and  $\rm
  B_{eff}(210\,GHz)$=0.57,  and  forward  beam  efficiencies  of  $\rm
  F_{eff}(3\,mm)$=0.95  and $\rm  F_{eff}(1\,mm)$=0.91.  We  also note
  that in most  cases we had redudant CO  J=2--1 measurements with the
  JCMT, and then: a) adopted  the average when JCMT/IRAM values agreed
  to within 20\% (most cases),  or b) adopted the JCMT measurement (as
  its wider  beam is less prone  to flux loss due  to pointing offsets
  and/or beam-throw/flux-loss  uncertainties) if a  discrepancy larger
  than the aforementioned was found.

\begin{figure*}
\centering
\includegraphics[width=0.325\textwidth]{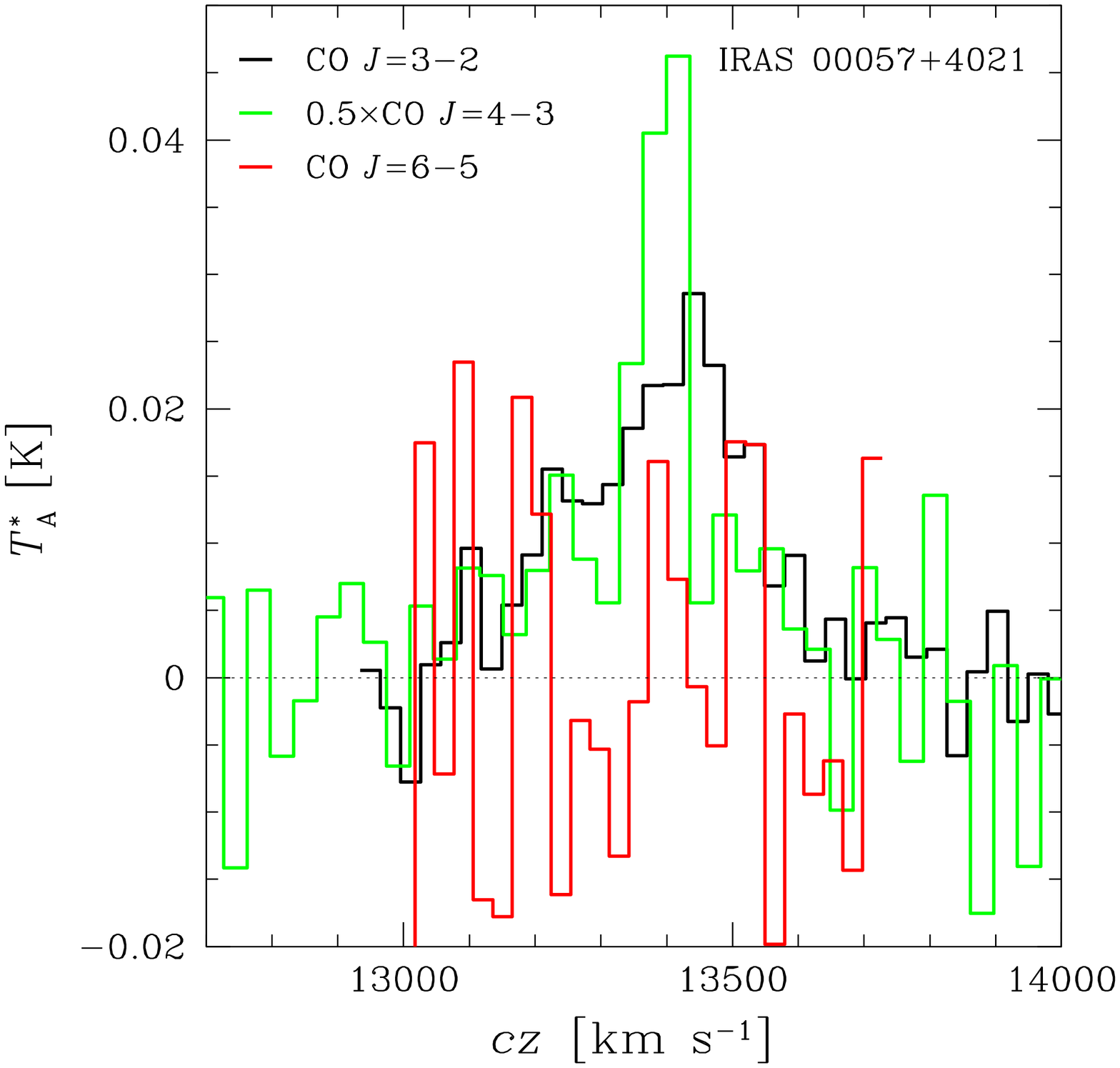}
\includegraphics[width=0.325\textwidth]{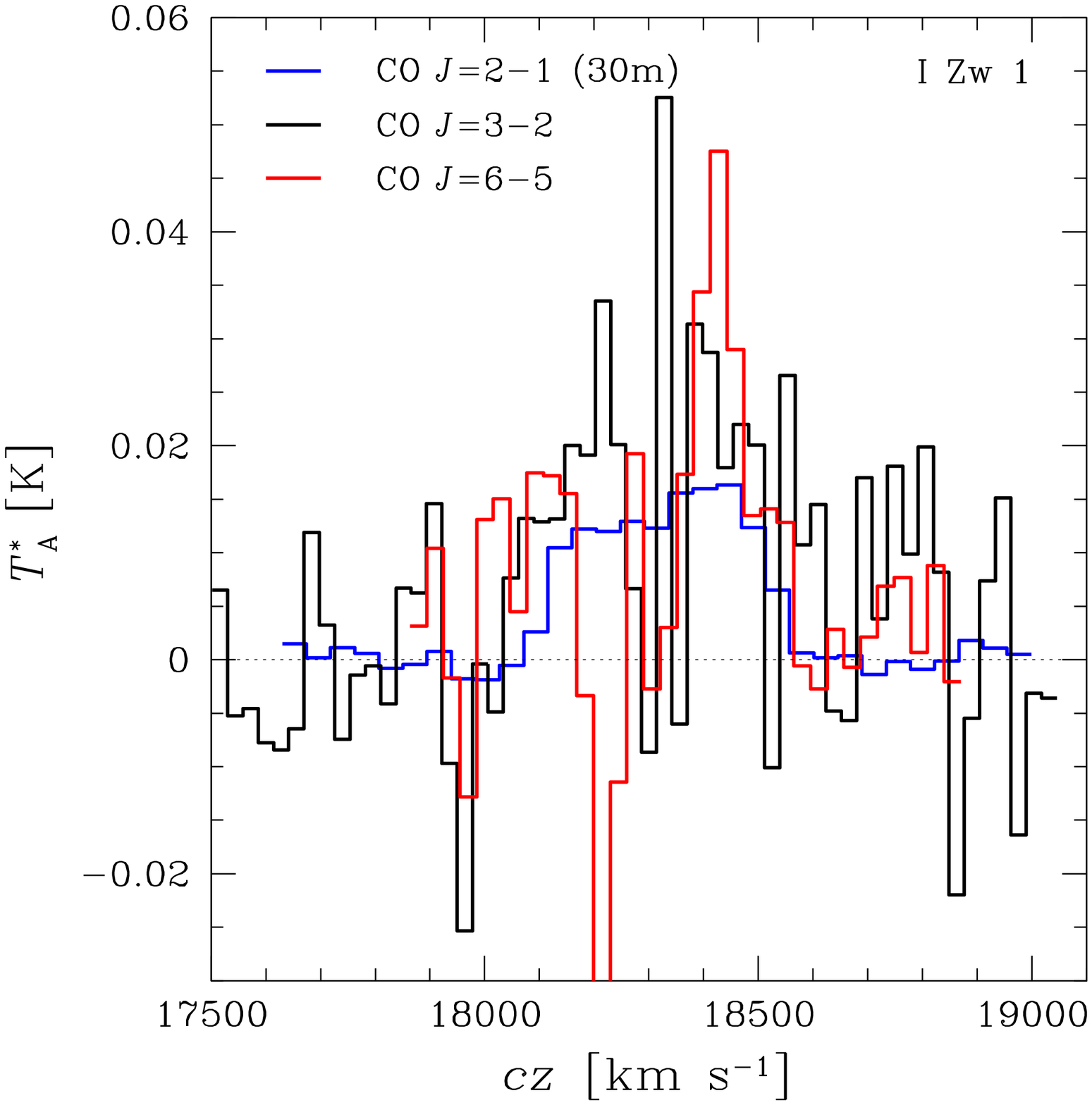}
\includegraphics[width=0.325\textwidth]{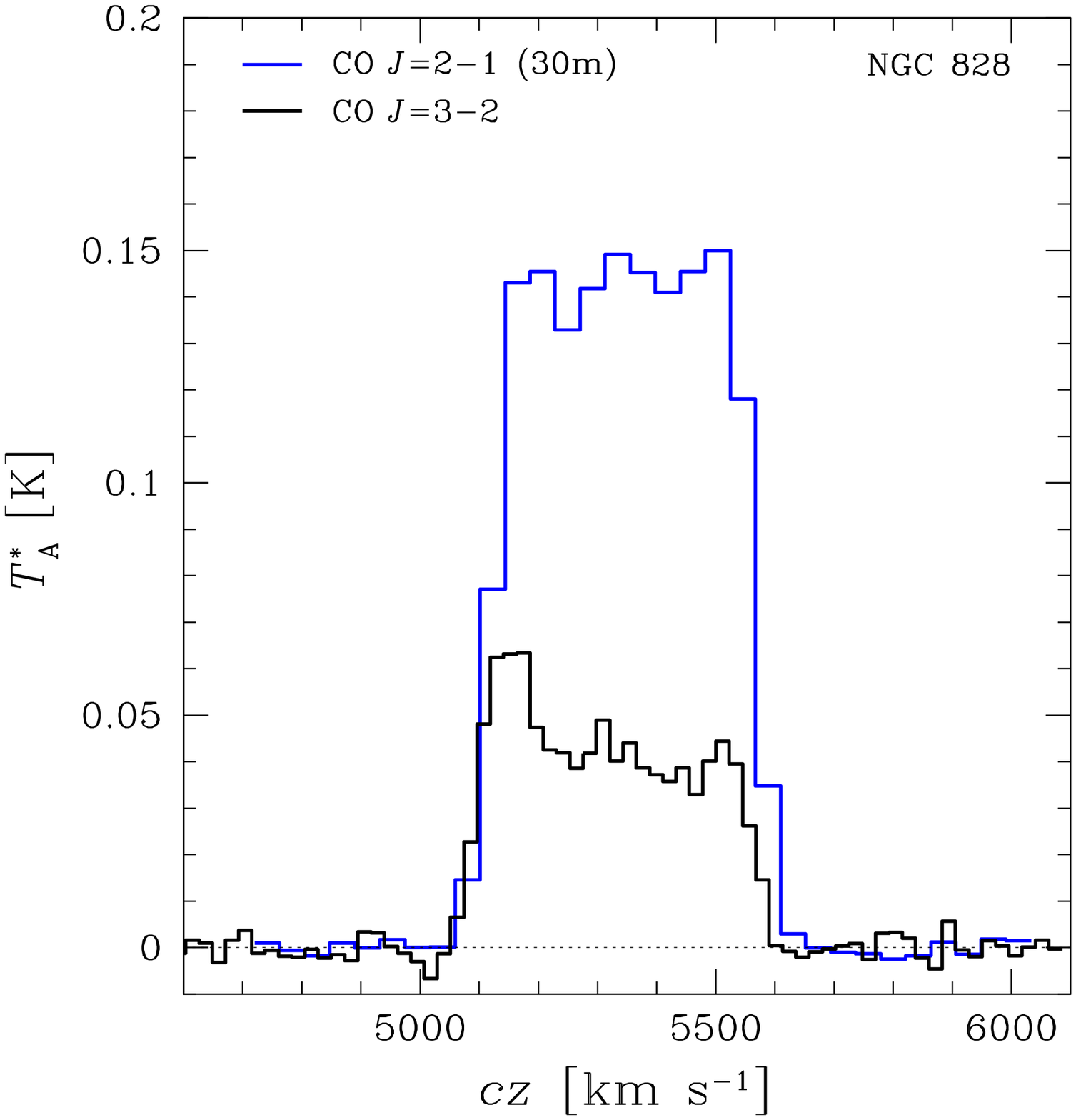}

\includegraphics[width=0.325\textwidth]{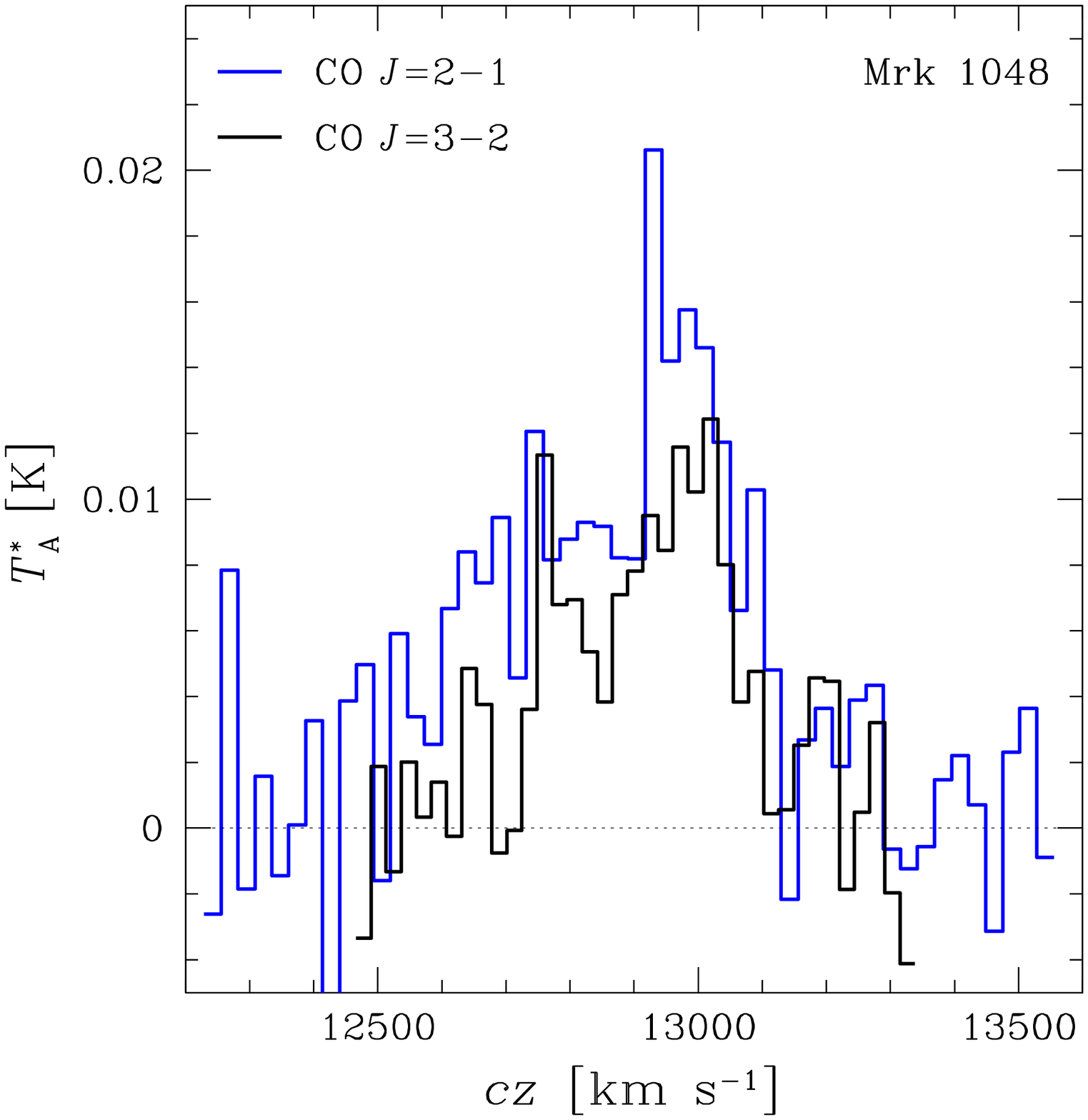}
\includegraphics[width=0.325\textwidth]{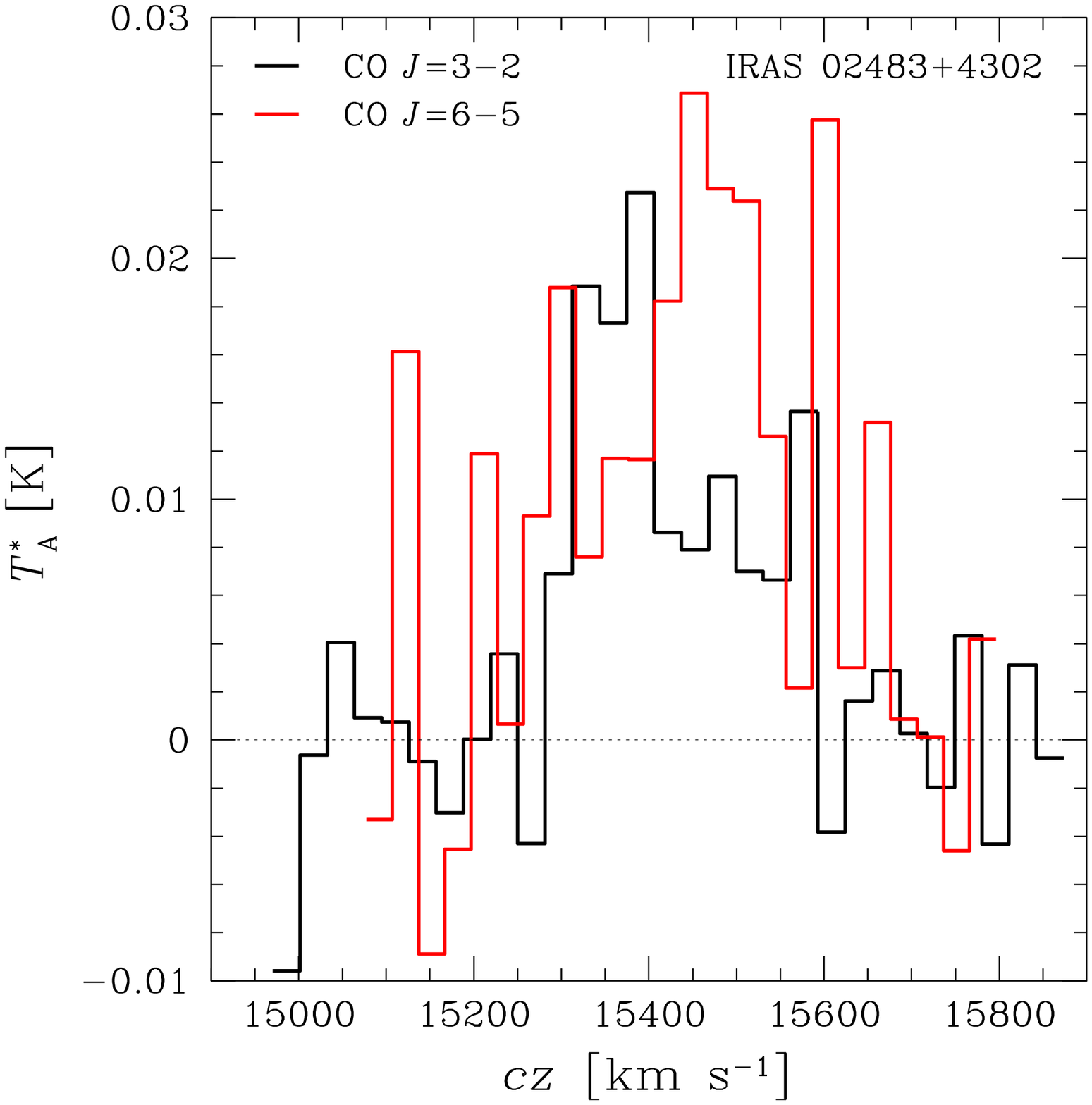}
\includegraphics[width=0.325\textwidth]{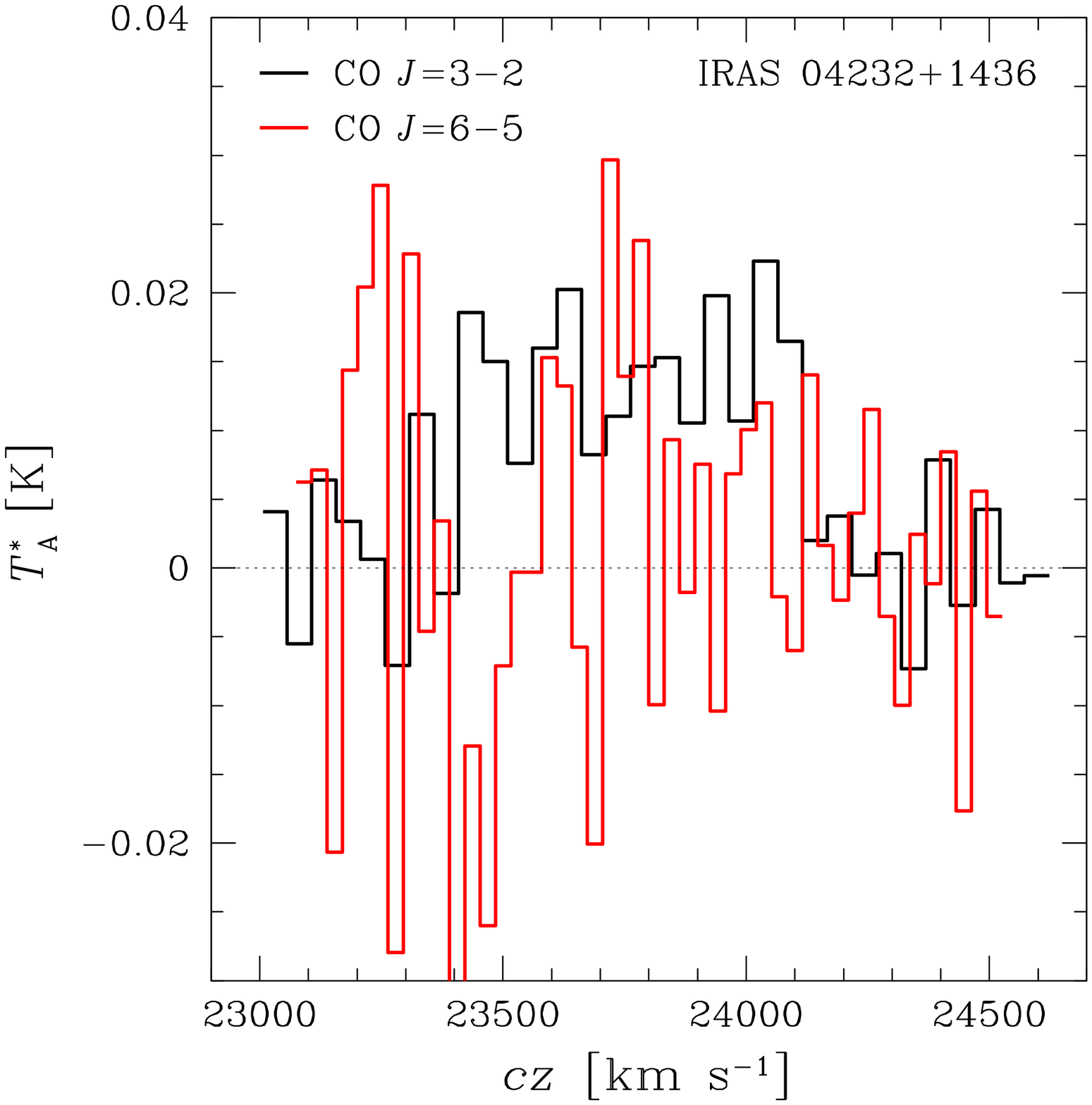} 

\includegraphics[width=0.325\textwidth]{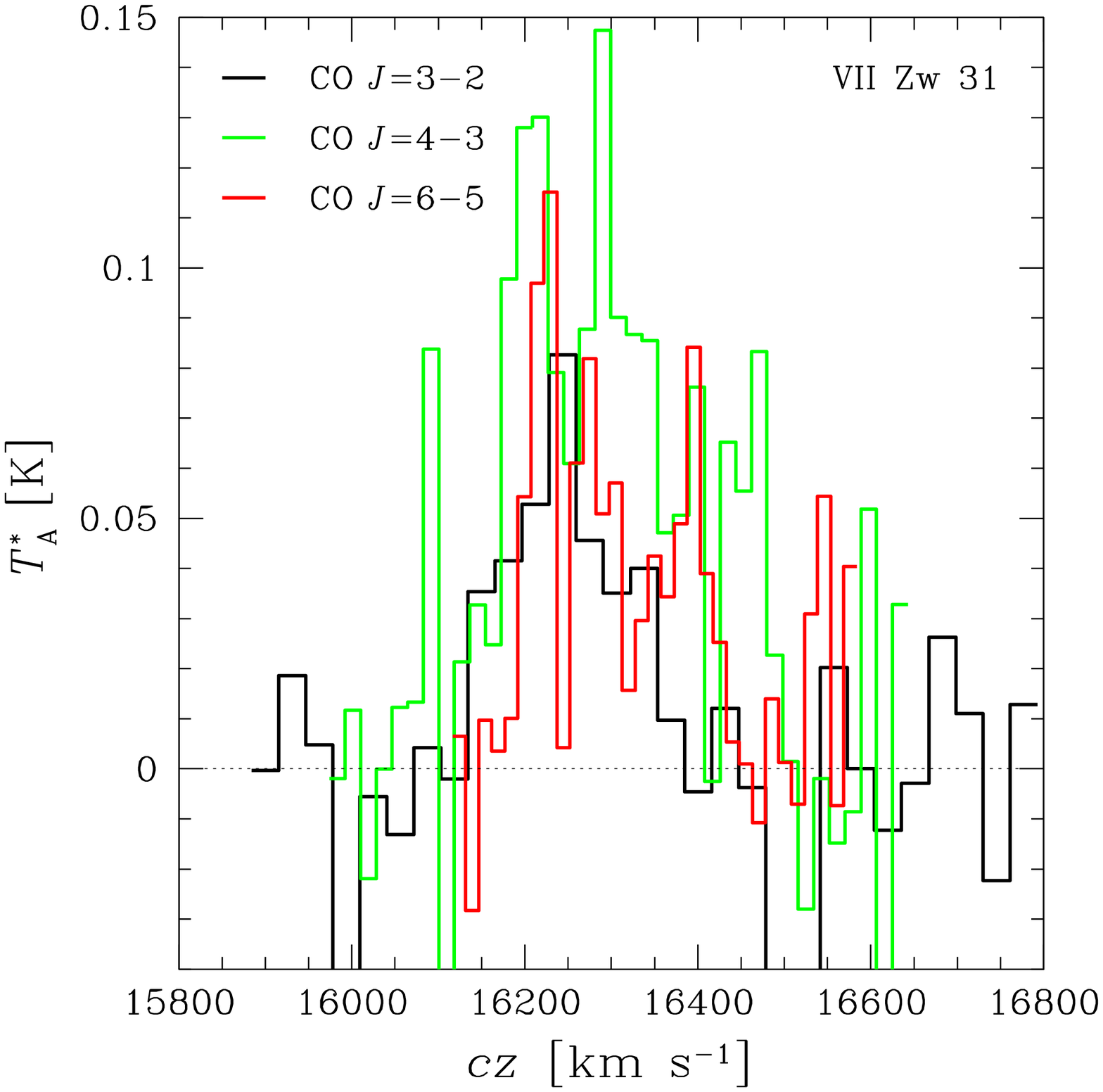}
\includegraphics[width=0.325\textwidth]{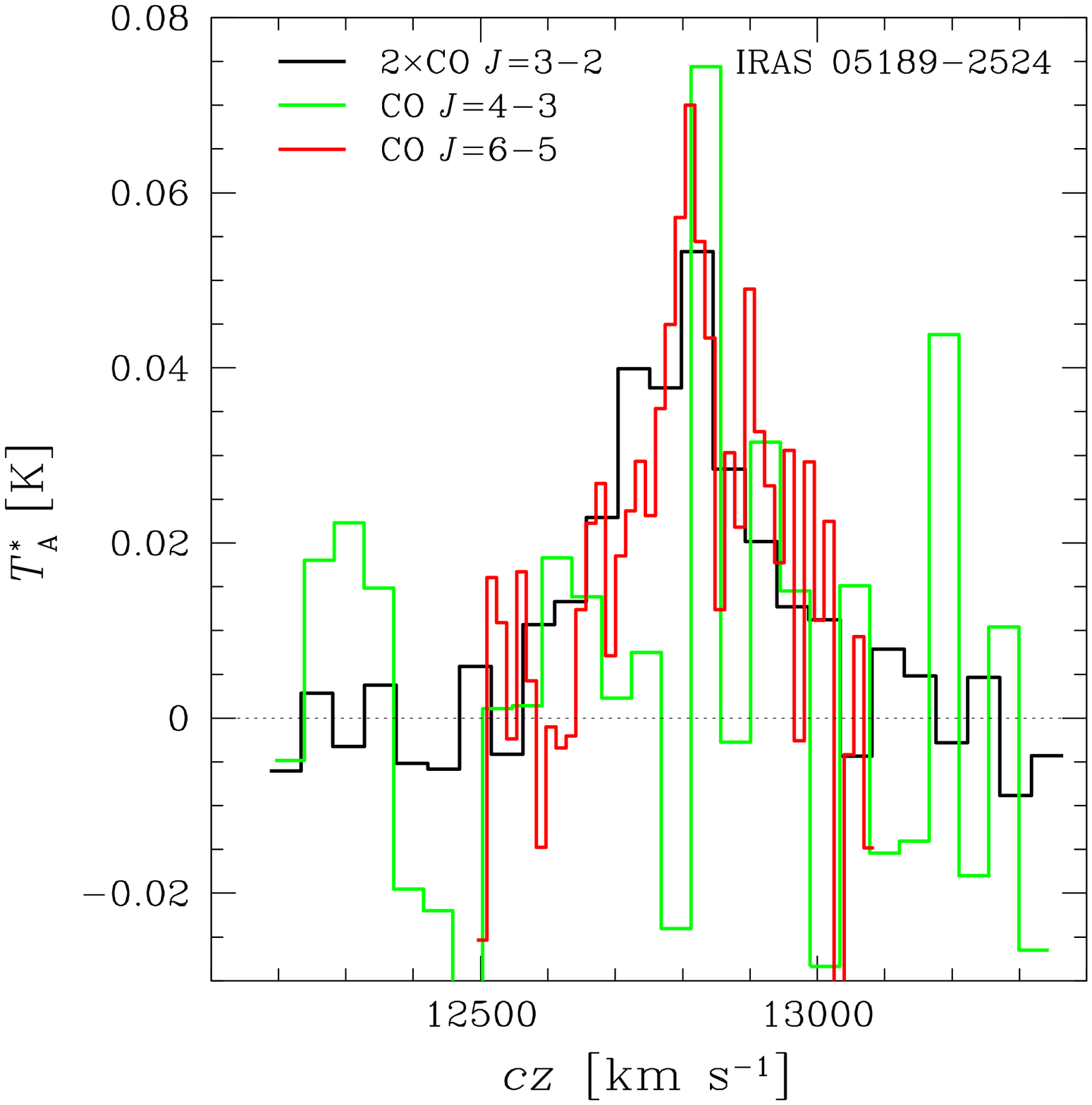}
\includegraphics[width=0.325\textwidth]{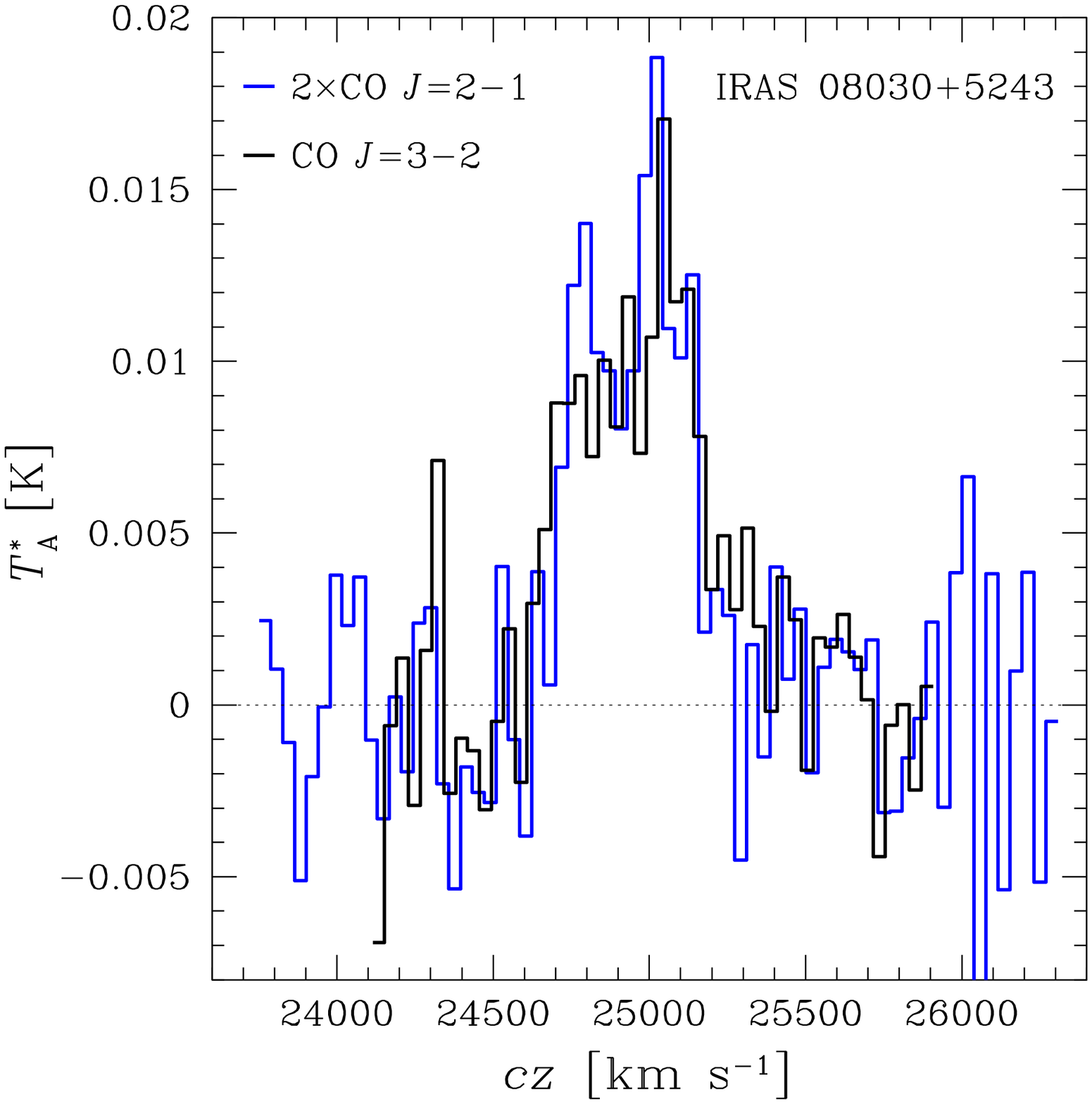}

\caption{The high-J  CO J+1$\rightarrow  $J, J+1$\geq $3  spectra.  In
  the few  cases where all  three CO J=3--2,  4--3 and 6--5  lines are
  available  we omit  the  overlay of  CO  J=2--1 in  order to  reduce
  confusion  (the  J=2--1 lines  are  all  shown  in Figure  3).   The
  velocities  are  with respect  to  $\rm V_{opt}$=$\rm  cz_{co}$(LSR)
  (Table  3), and  with typical  resolutions $\rm  \Delta V_{ch}$$\sim
  $(10--50)\,km\,s$^{-1}$.  A  common color designated  per transition
  is used in all frames. }

\end{figure*}

\begin{figure*}
\centering

\includegraphics[width=0.325\textwidth]{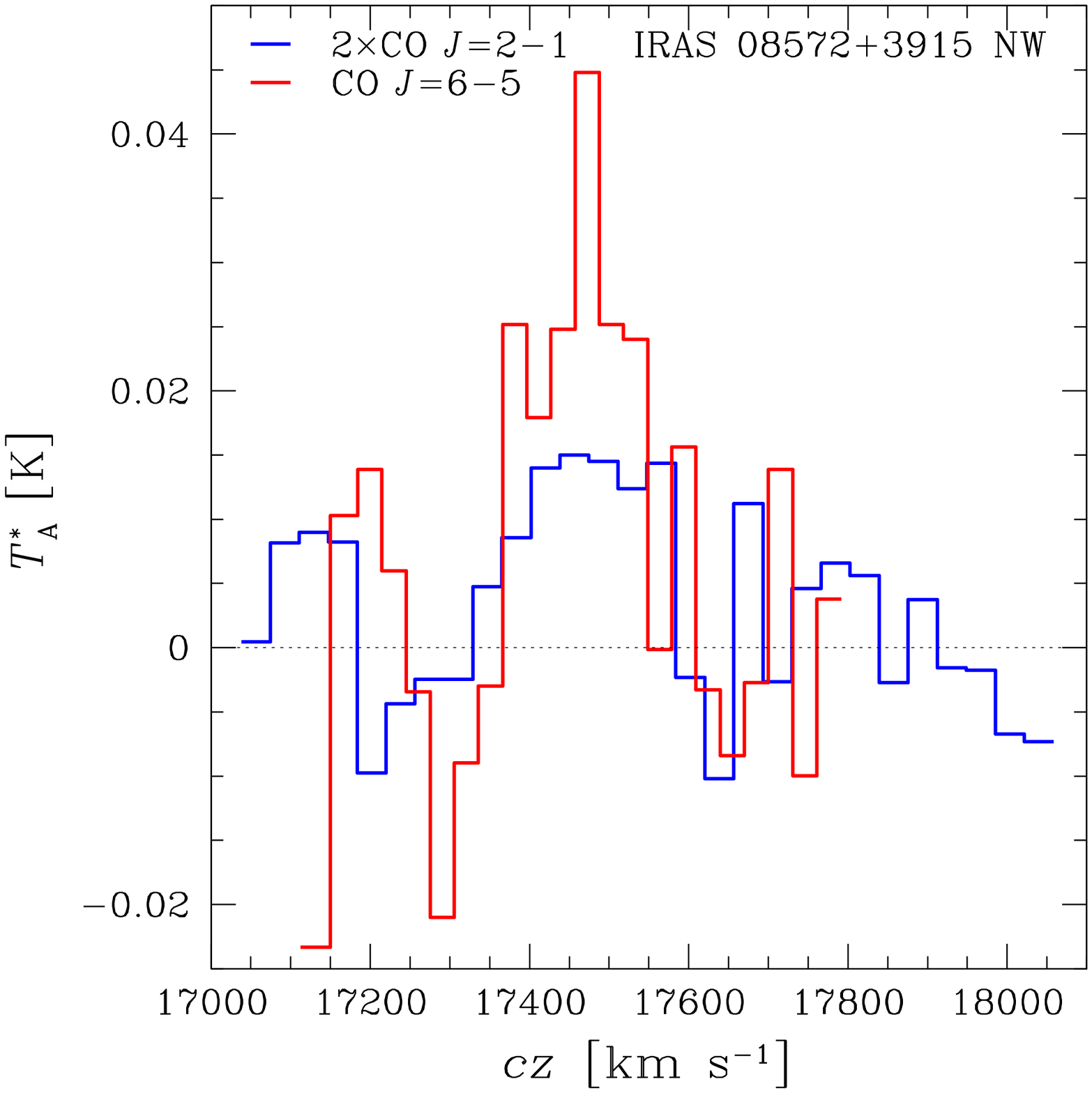}
\includegraphics[width=0.325\textwidth]{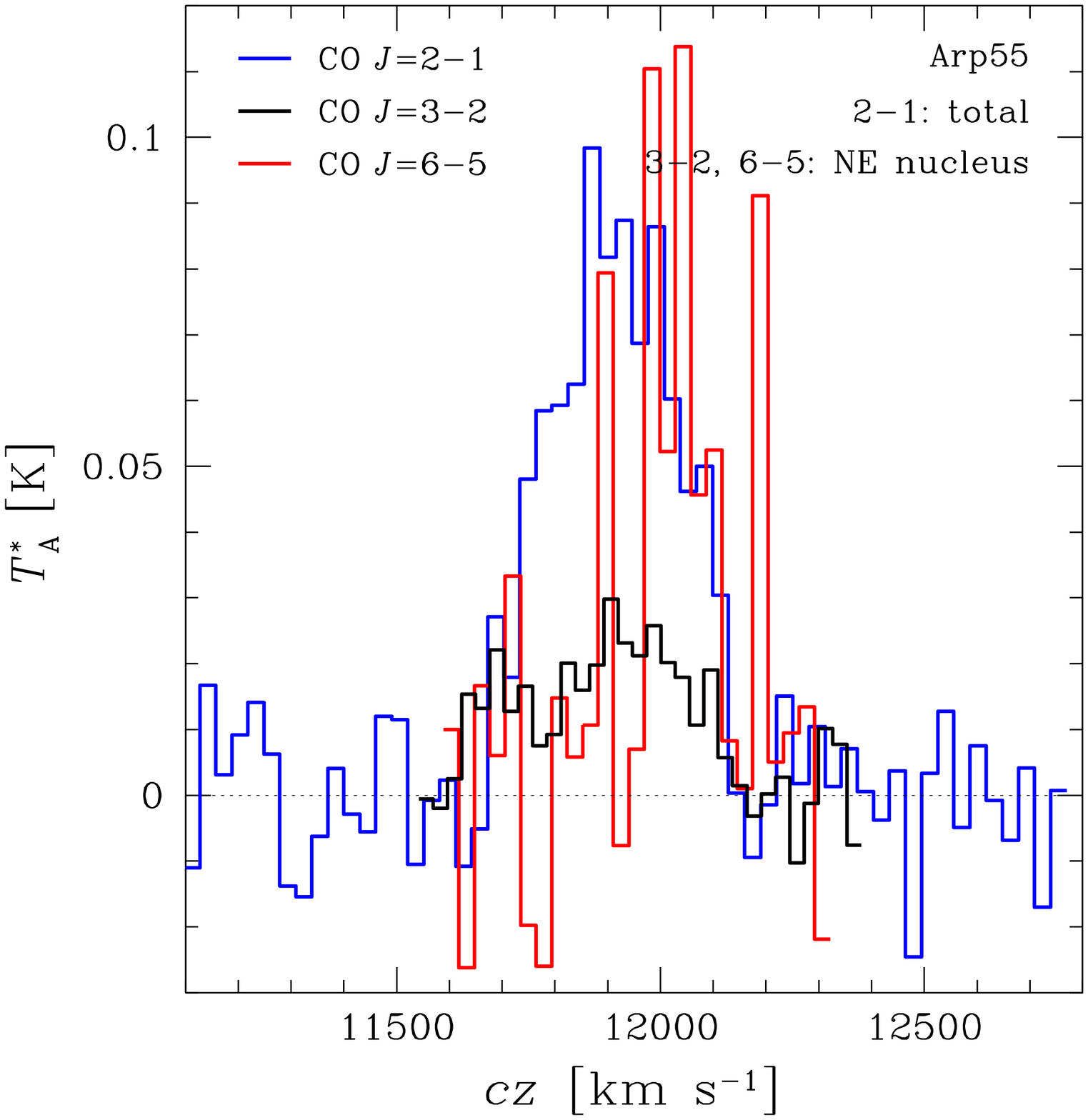} 
\includegraphics[width=0.325\textwidth]{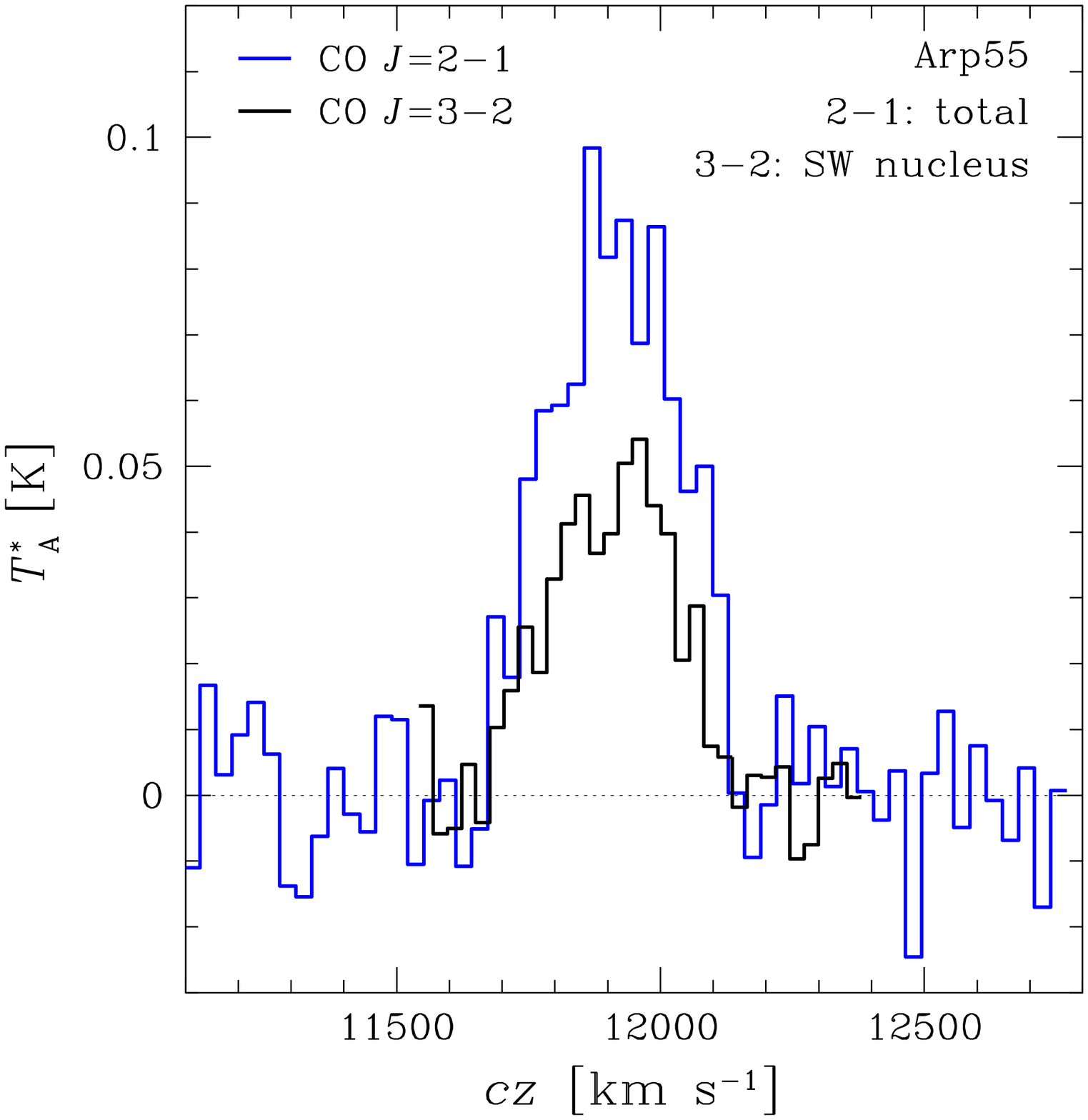}

\includegraphics[width=0.325\textwidth]{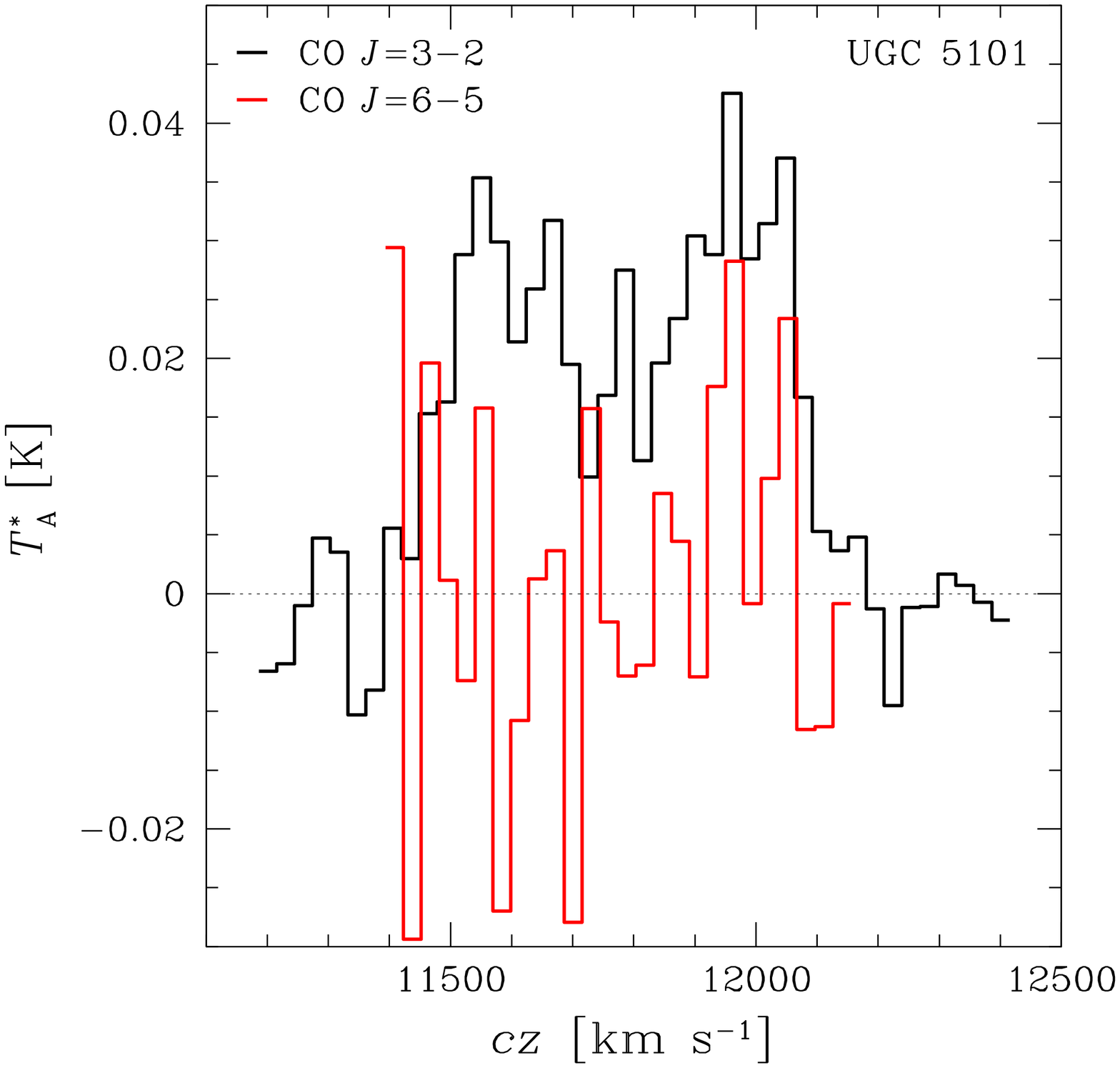} 
\includegraphics[width=0.325\textwidth]{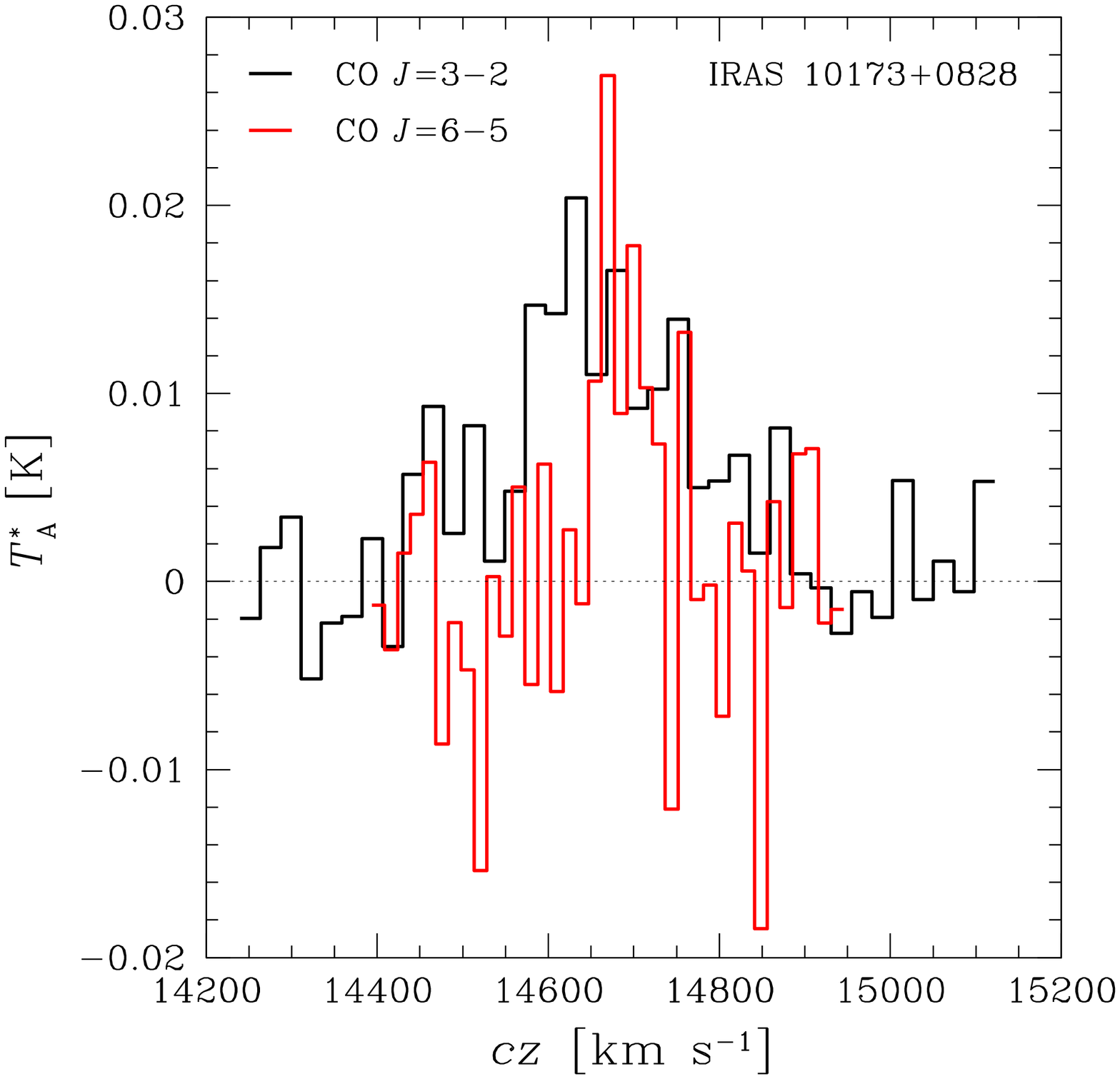}
\includegraphics[width=0.325\textwidth]{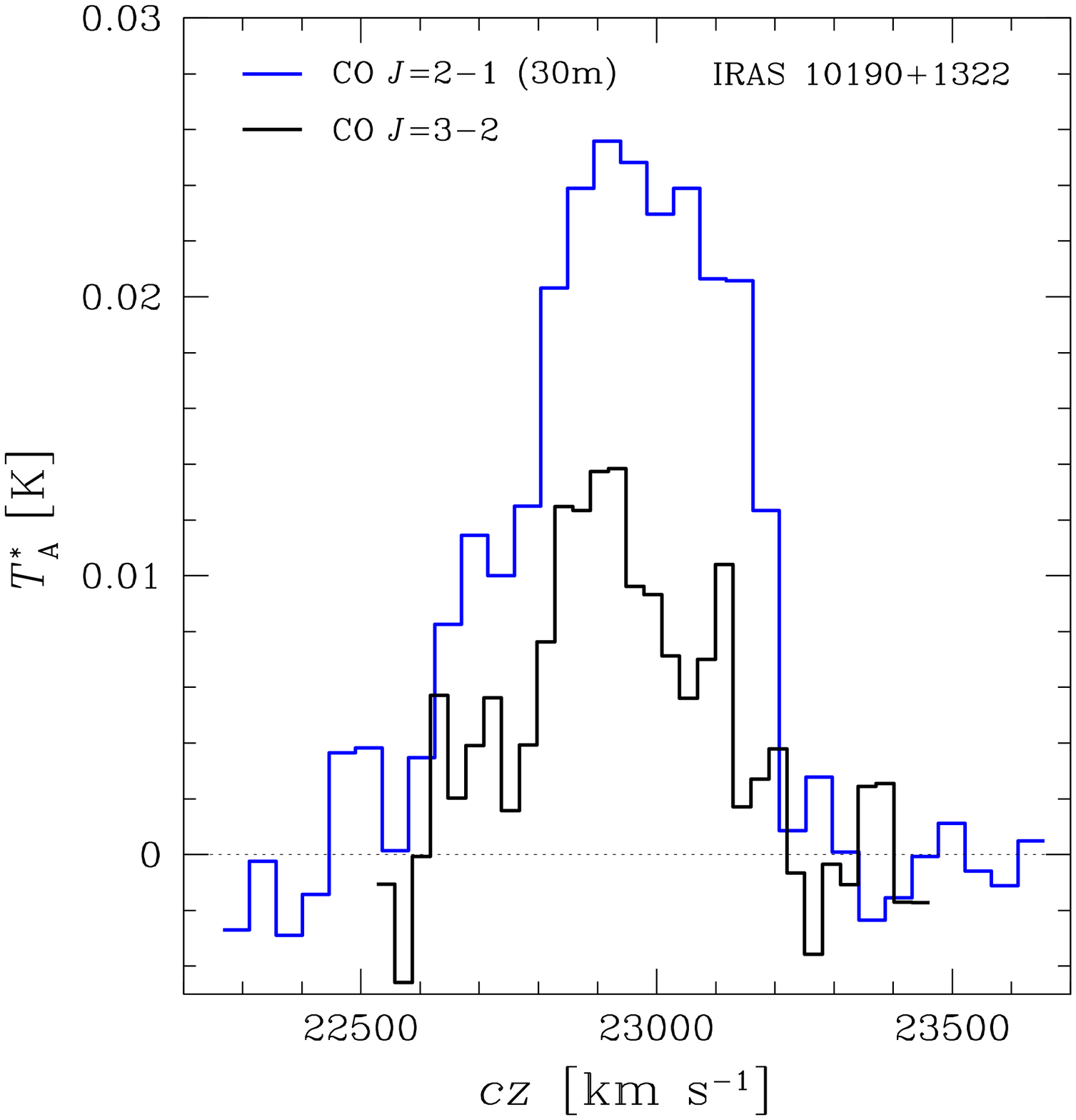}

\includegraphics[width=0.325\textwidth]{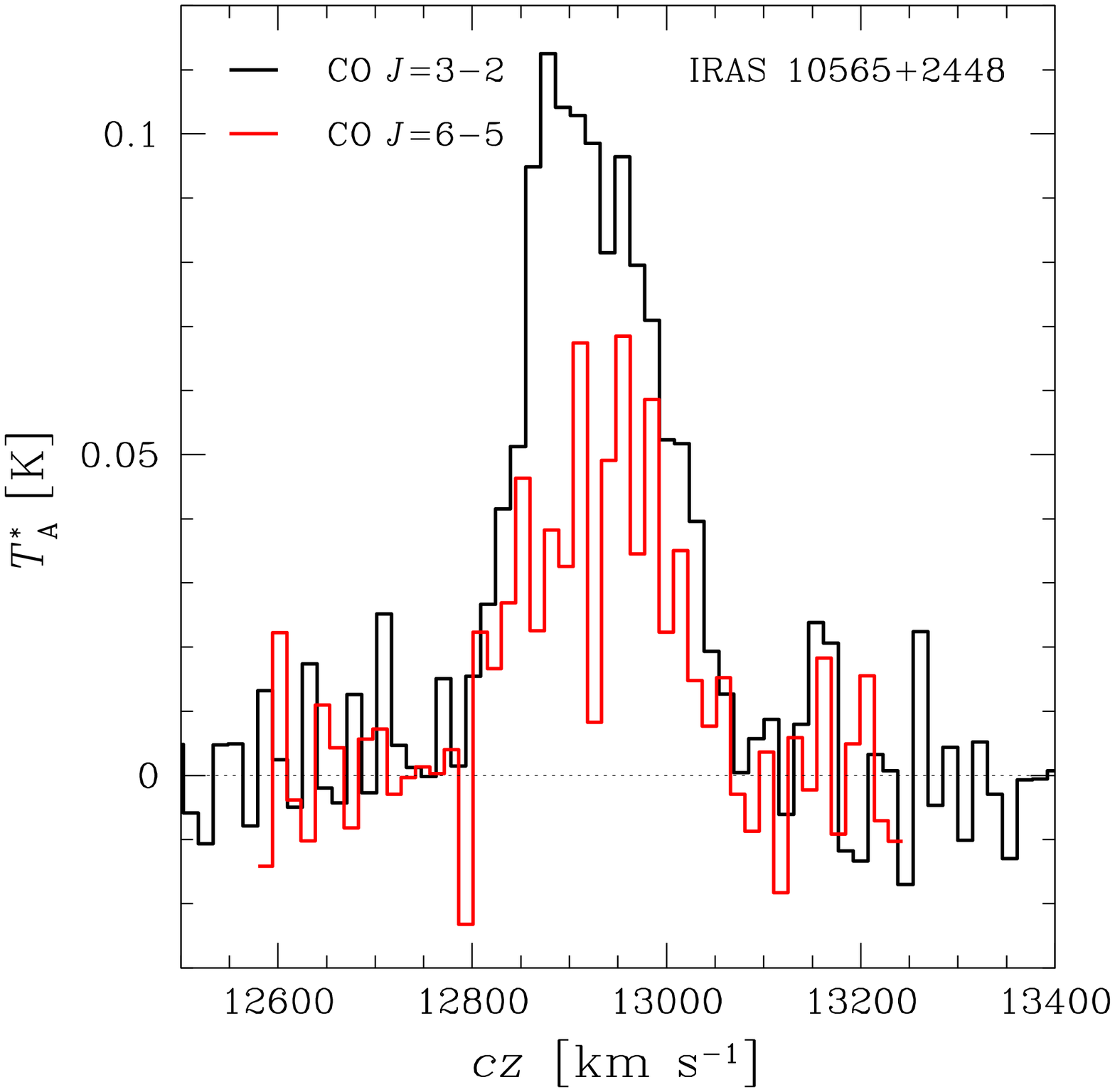}
\includegraphics[width=0.325\textwidth]{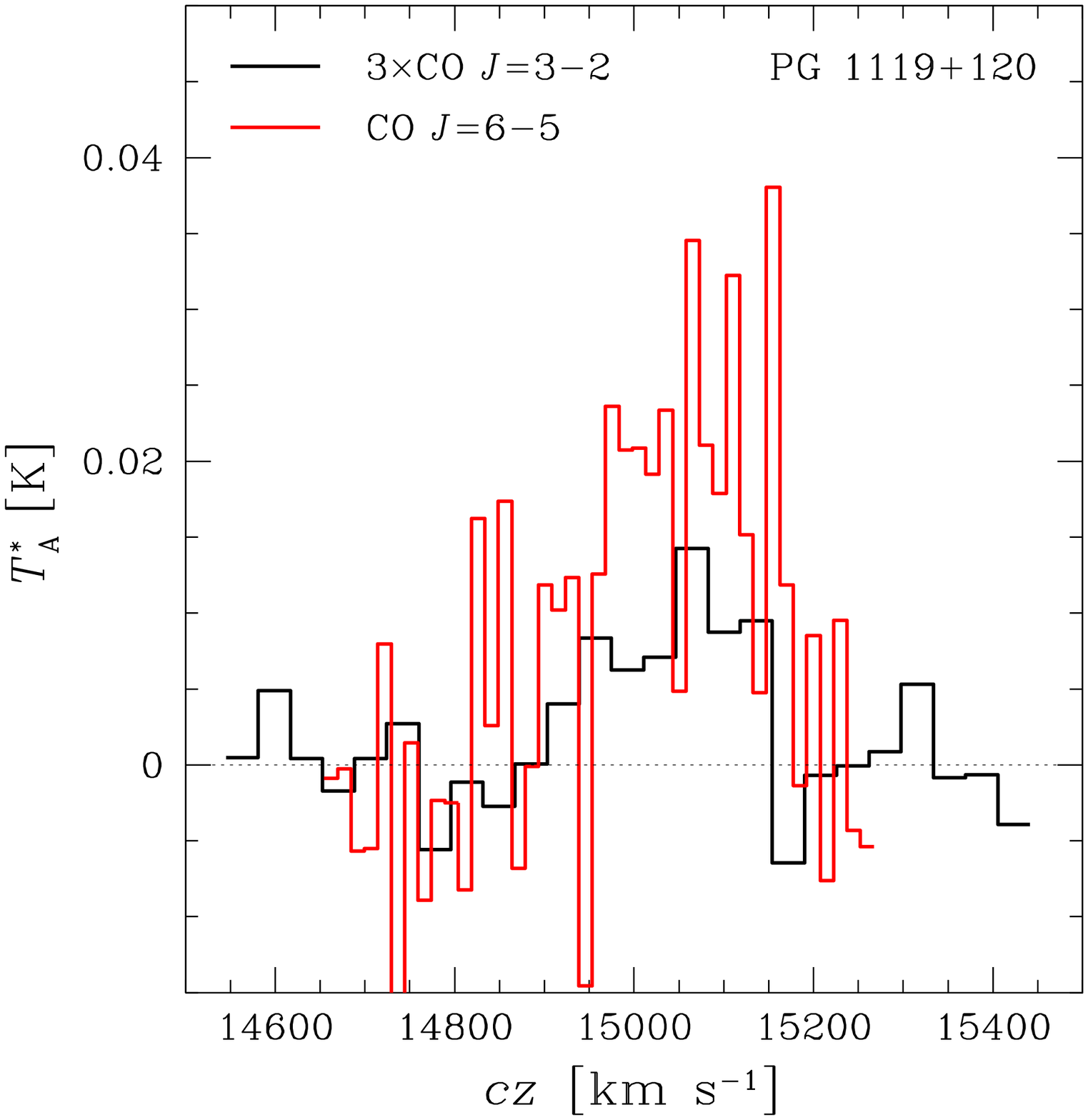}
\includegraphics[width=0.325\textwidth]{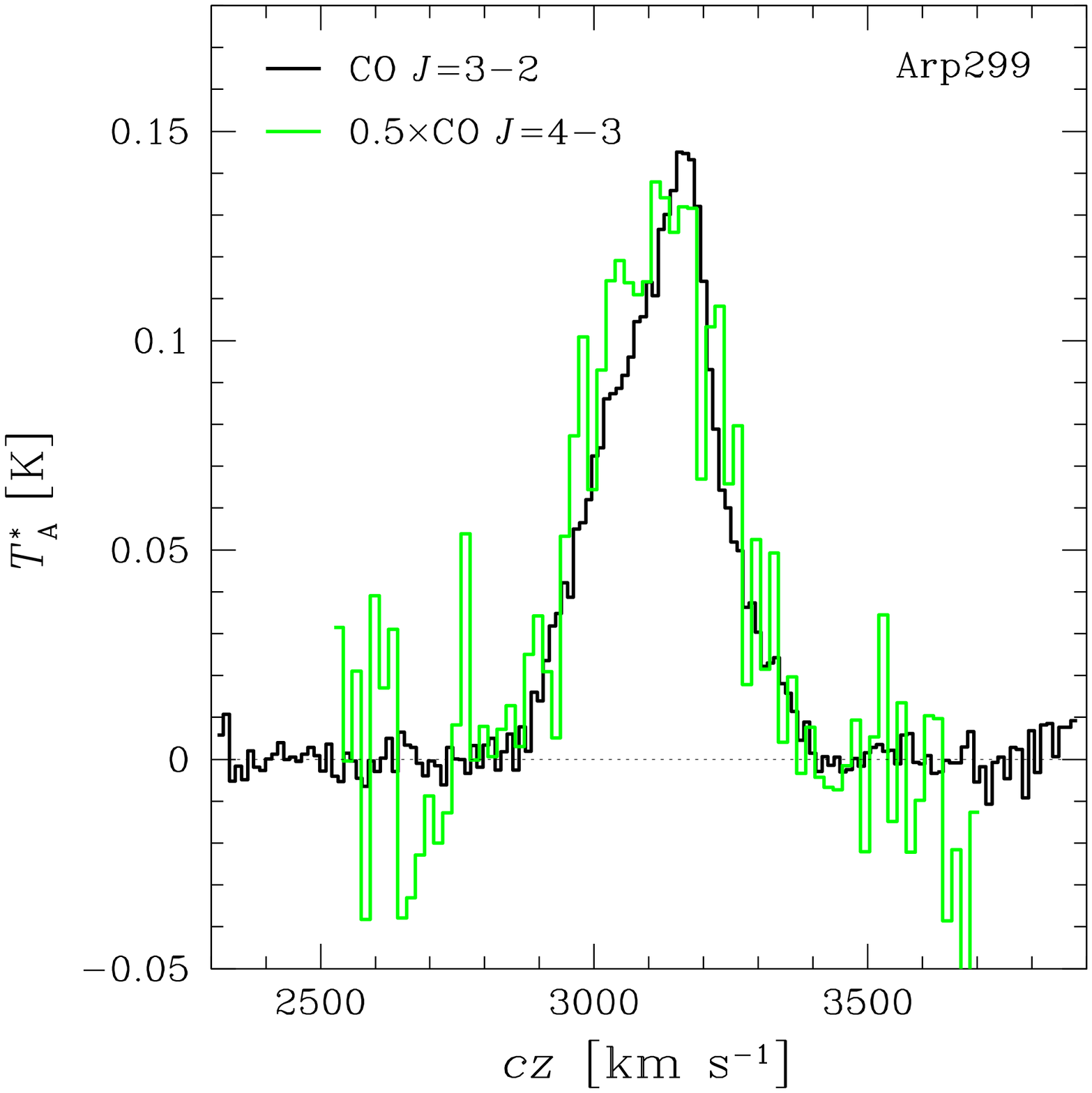}

\contcaption{The  high-J CO J+1$\rightarrow  $J, J+1$\geq  $3 spectra.
  In the few cases where all  three CO J=3--2, 4--3 and 6--5 lines are
  available  we omit  the  overlay of  CO  J=2--1 in  order to  reduce
  confusion  (the  J=2--1 lines  are  all  shown  in Figure  3).   The
  velocities  are  with respect  to  $\rm V_{opt}$=$\rm  cz_{co}$(LSR)
  (Table  3), and  with typical  resolutions $\rm  \Delta V_{ch}$$\sim
  $(10--50)\,km\,s$^{-1}$.  A  common color designated  per transition
  is used in all frames. }
\end{figure*}

\begin{figure*}
\centering

\includegraphics[width=0.325\textwidth]{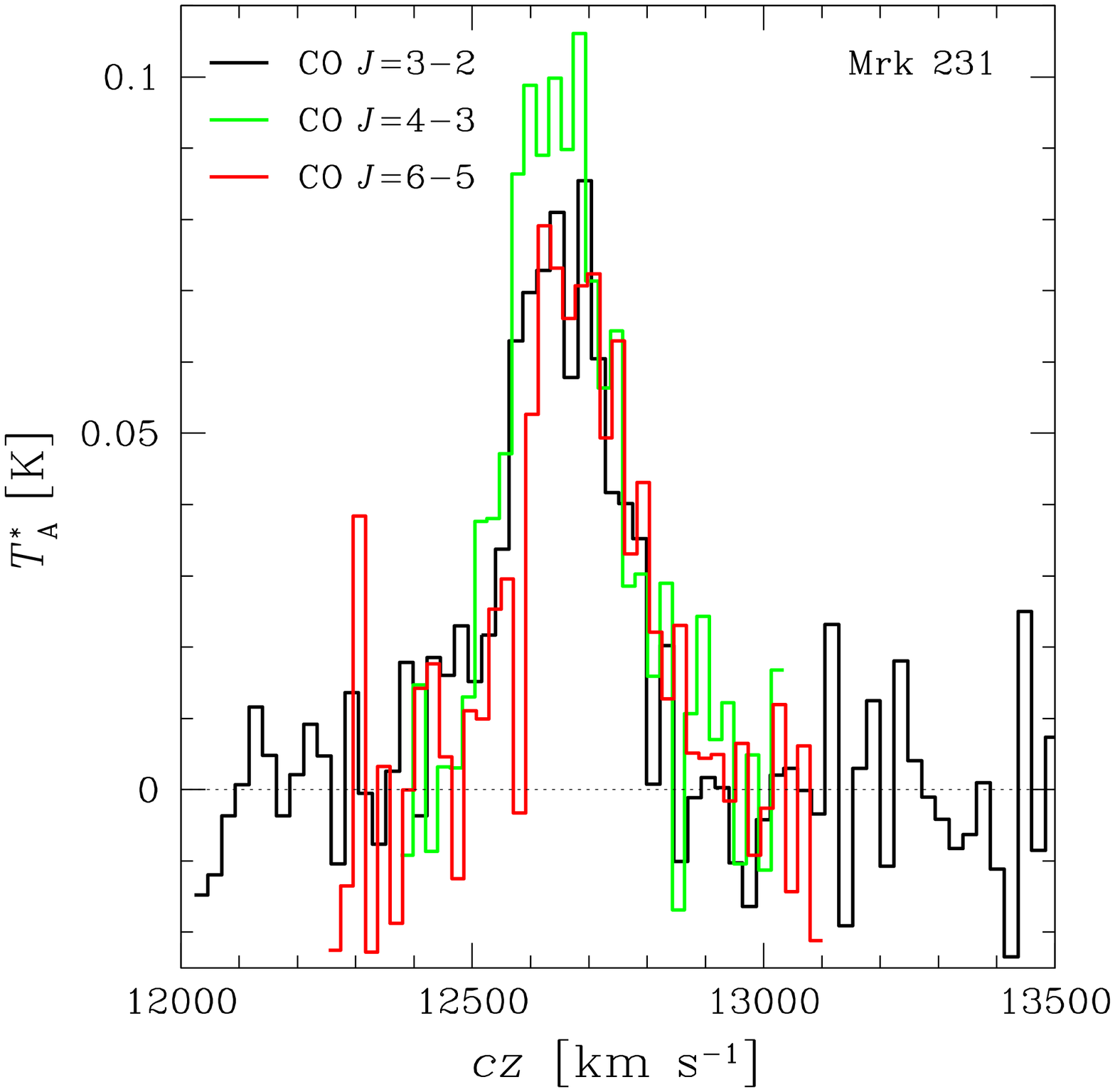}
\includegraphics[width=0.325\textwidth]{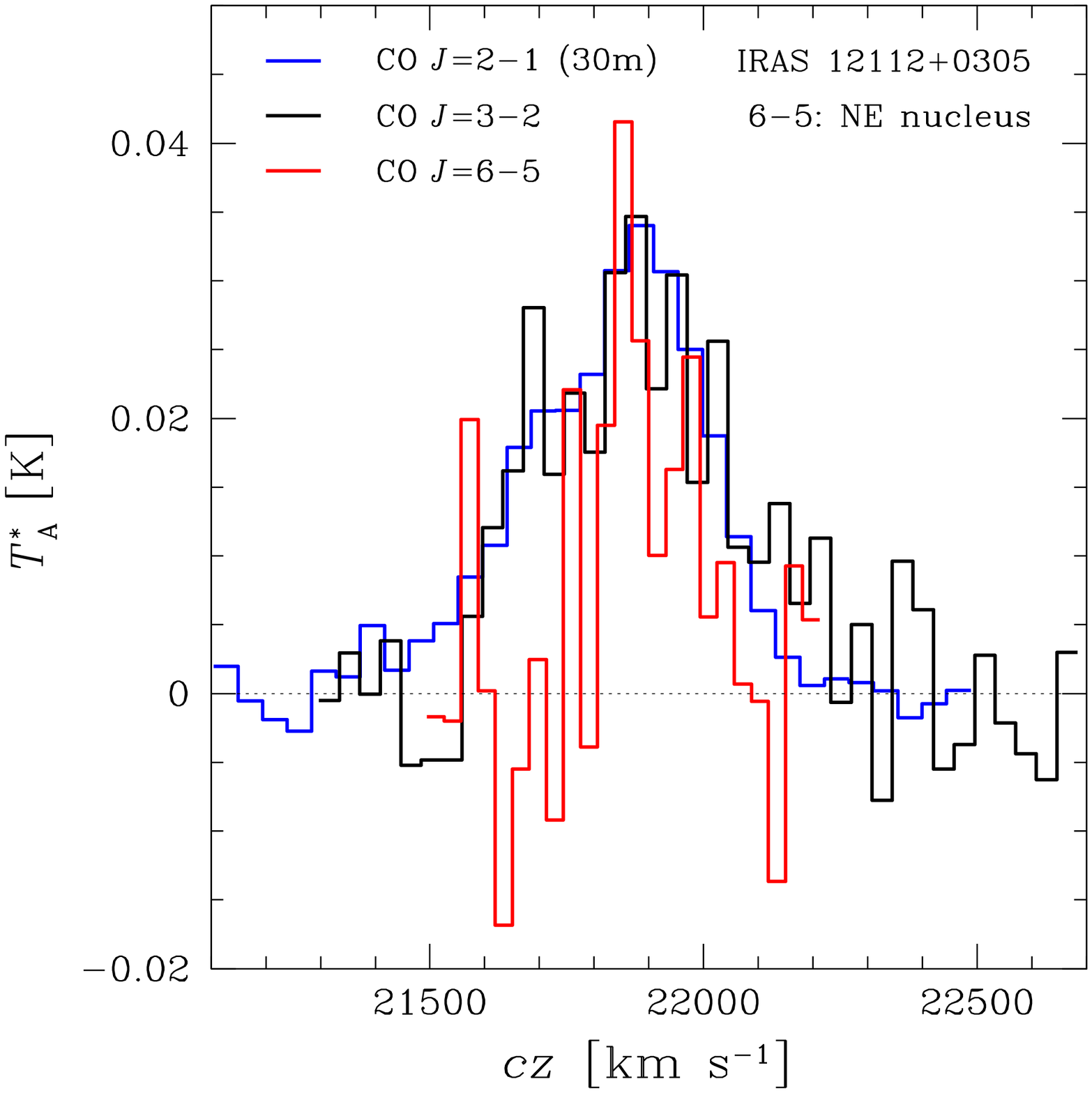}
\includegraphics[width=0.325\textwidth]{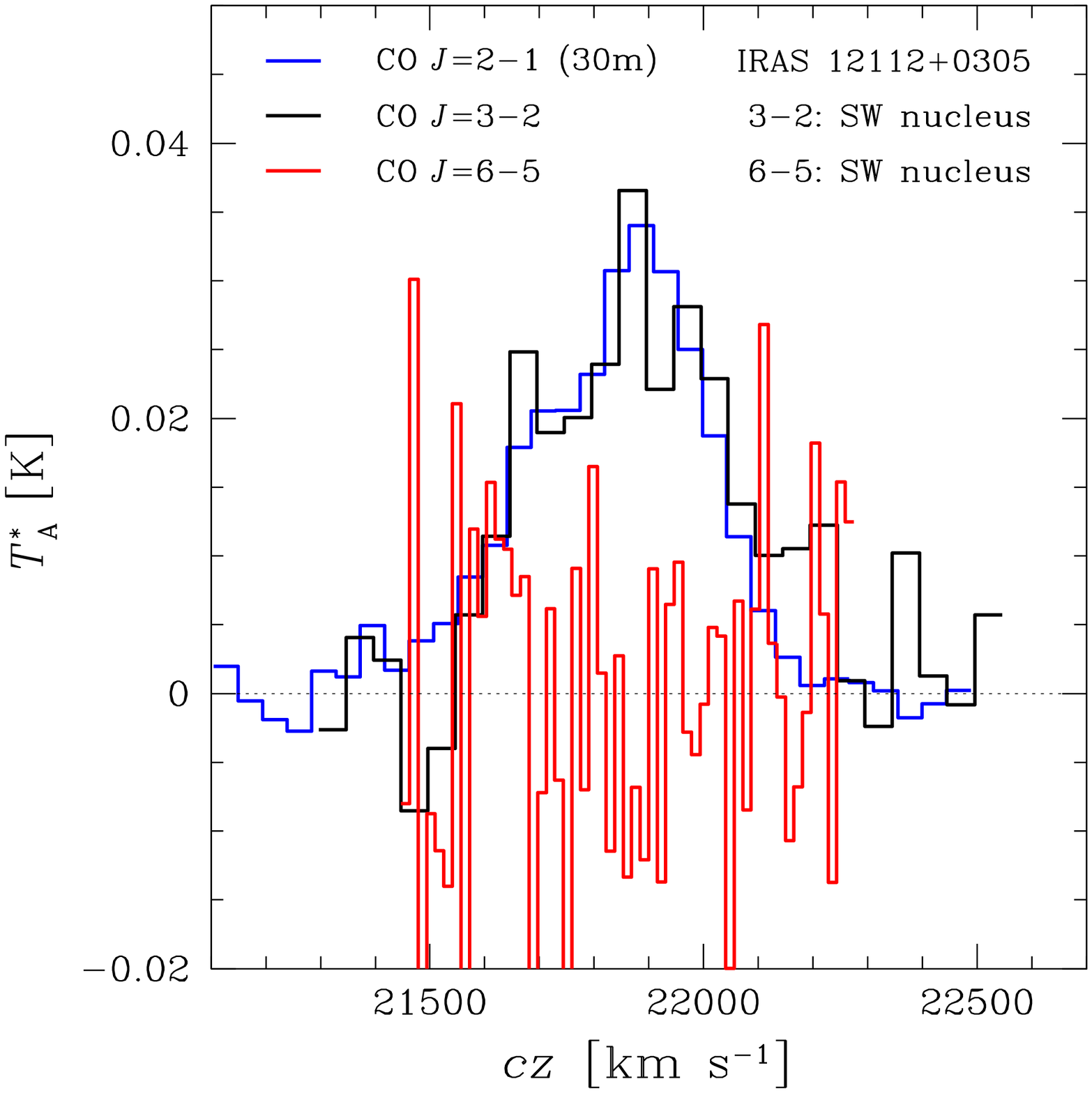}

\includegraphics[width=0.325\textwidth]{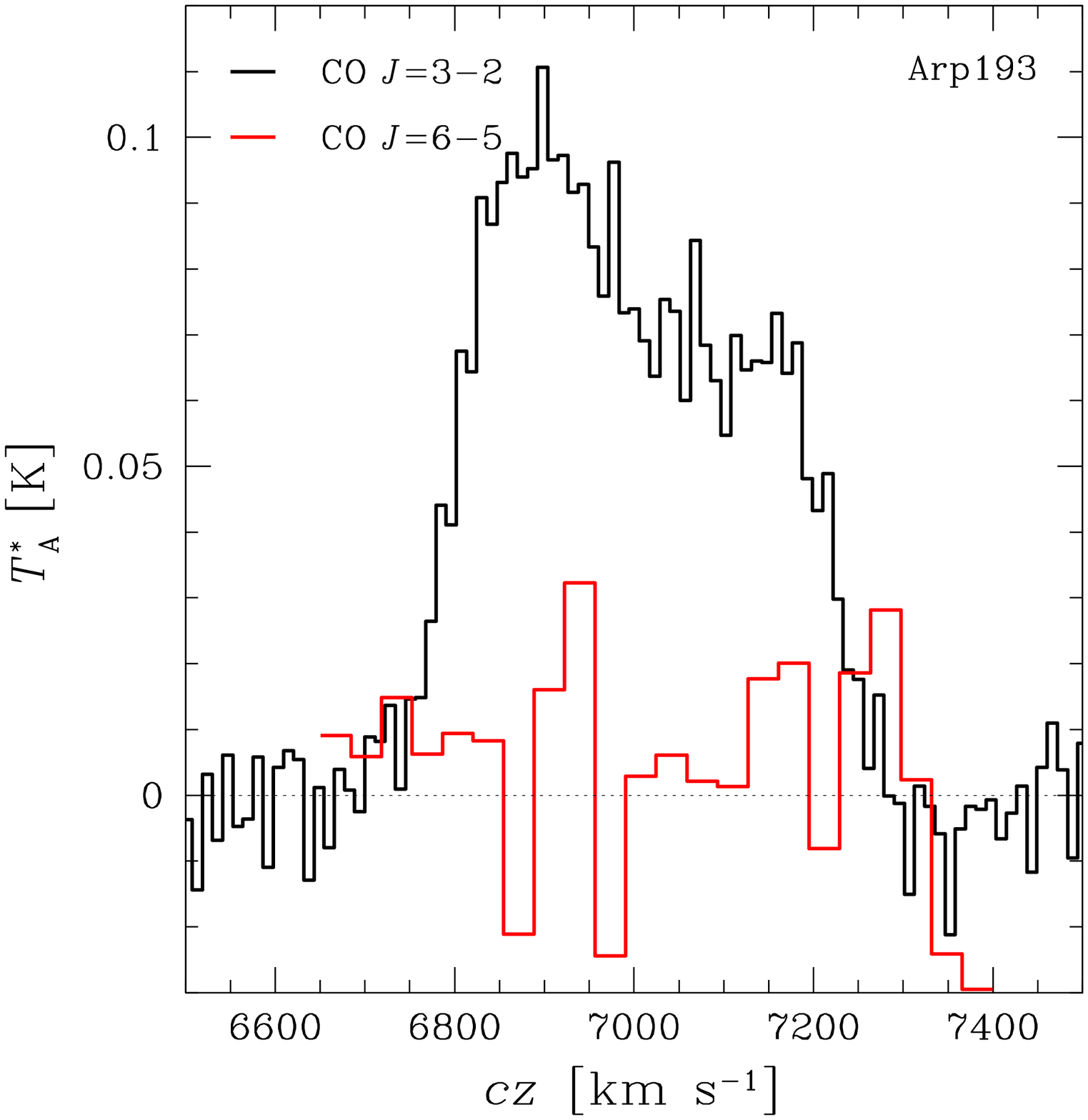}
\includegraphics[width=0.325\textwidth]{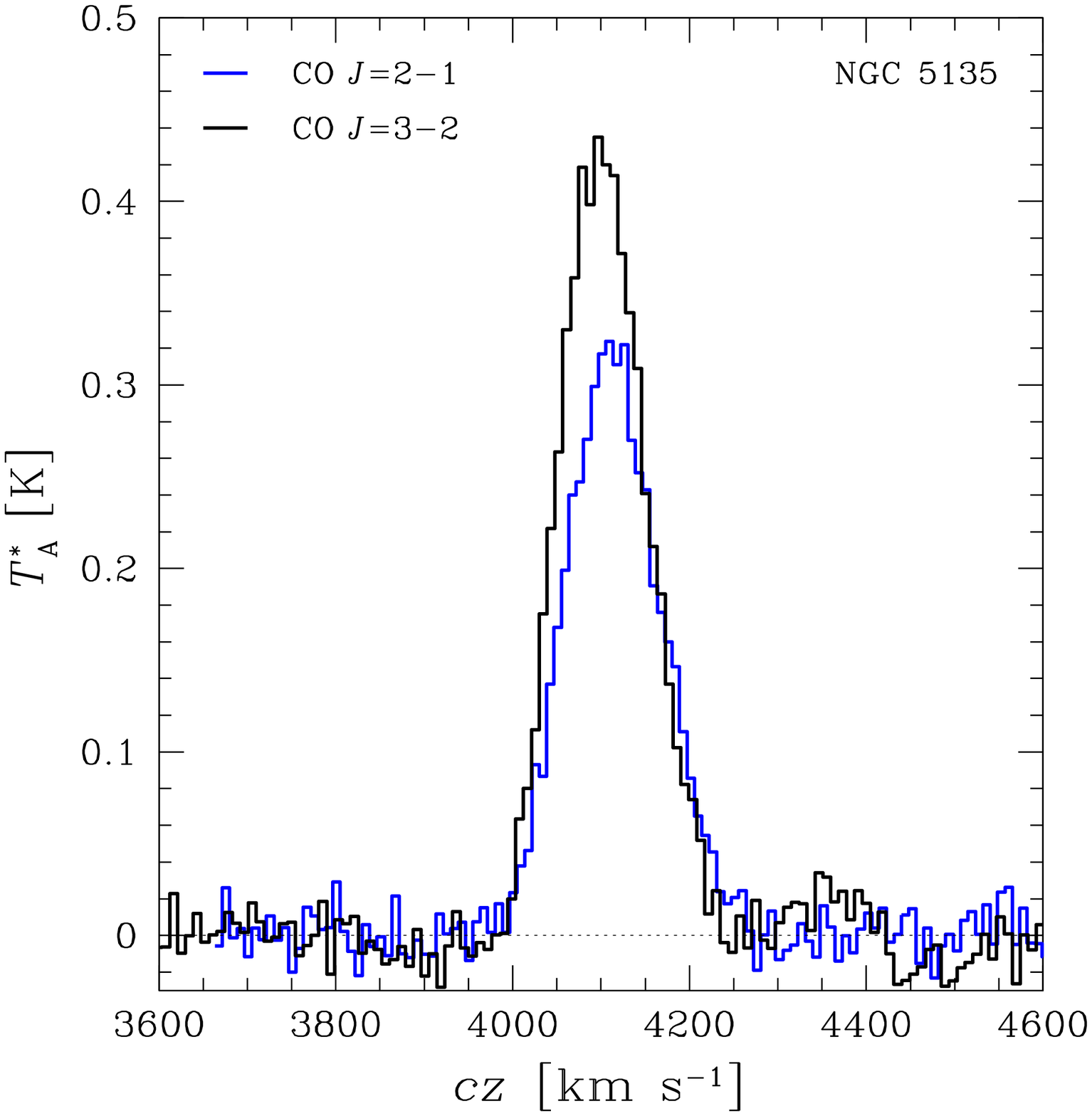} 
\includegraphics[width=0.325\textwidth]{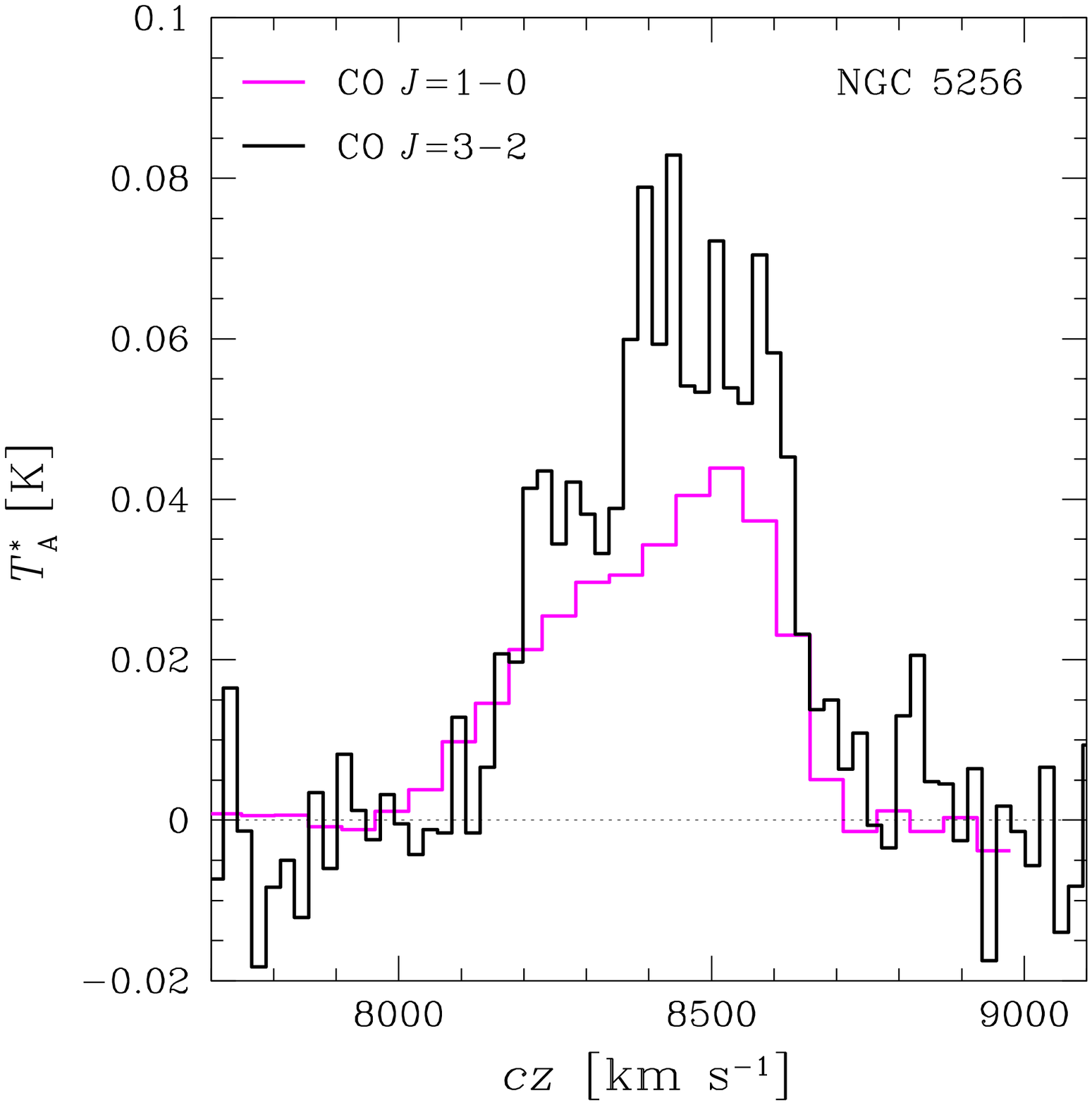} 

\includegraphics[width=0.325\textwidth]{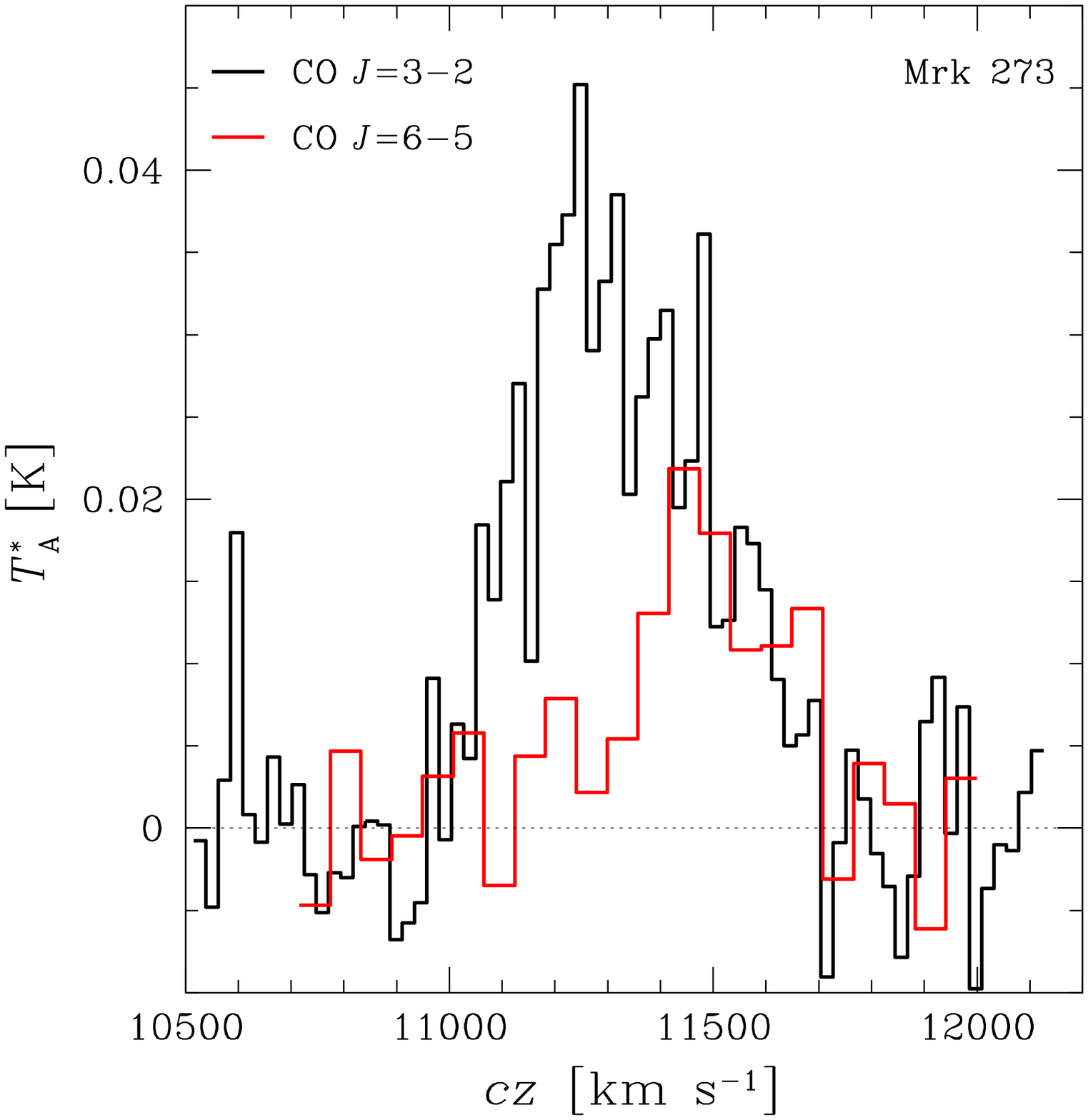}
\includegraphics[width=0.325\textwidth]{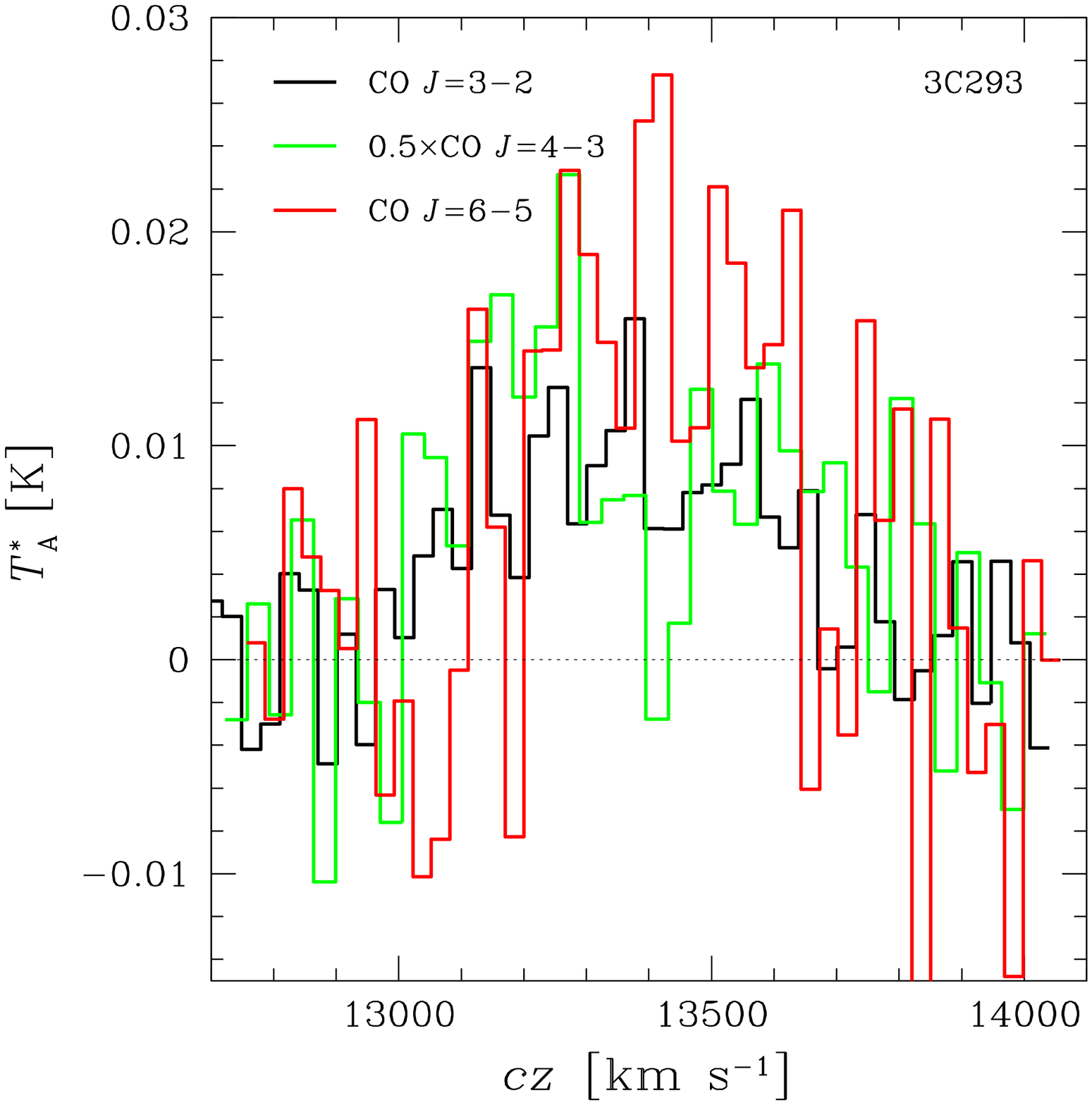} 
\includegraphics[width=0.325\textwidth]{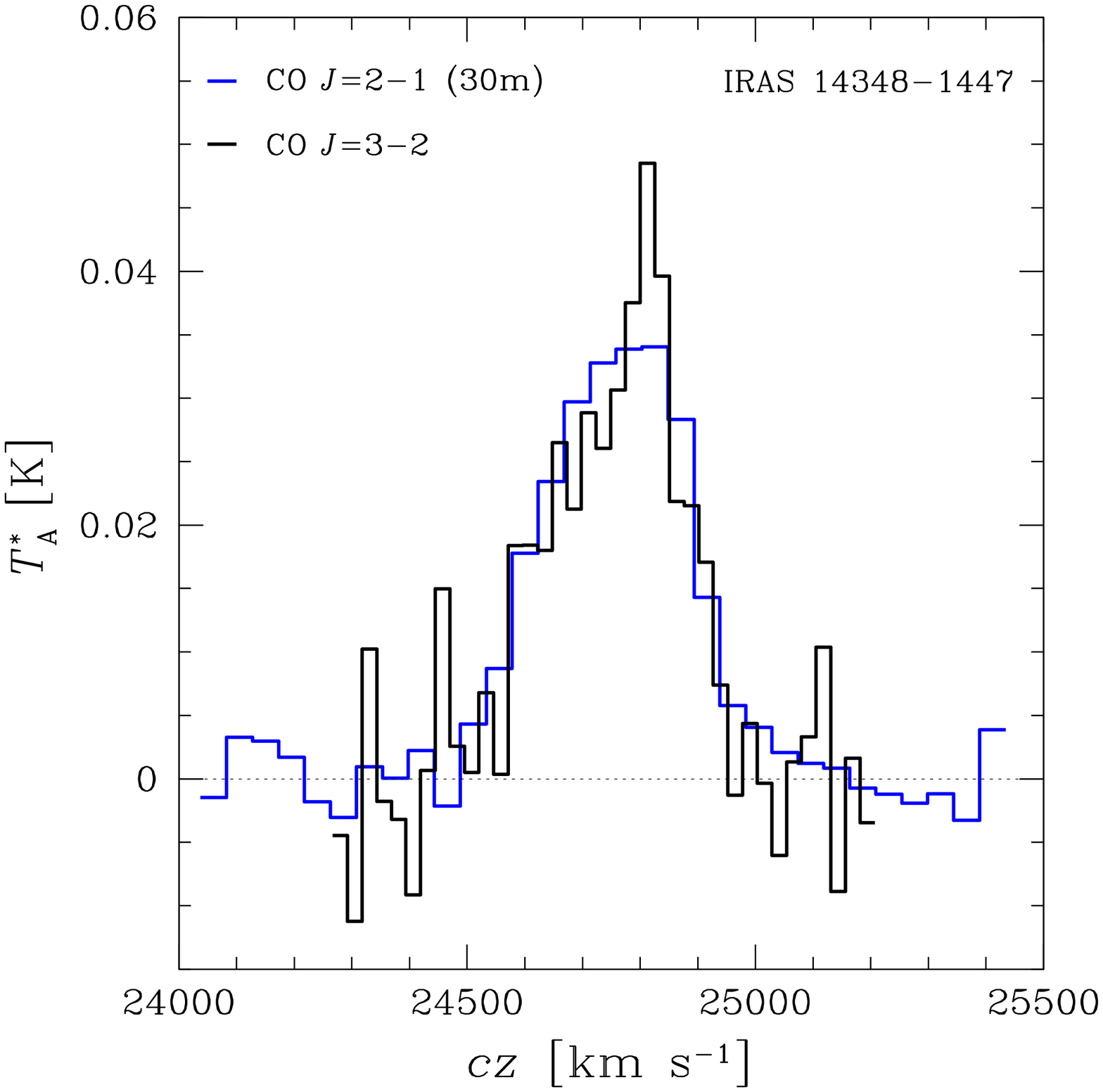}

 \contcaption{The high-J CO  J+1$\rightarrow $J, J+1$\geq $3 spectra.
   In the few cases where all three CO J=3--2, 4--3 and 6--5 lines are
   available  we omit  the overlay  of CO  J=2--1 in  order  to reduce
   confusion (the  J=2--1 lines  are all shown  in Figure 3).   In the
   case of  NGC\,5256 only  the CO J=1--0,  3--2 lines  are available.
   The velocities are with respect to $\rm V_{opt}$=$\rm cz_{co}$(LSR)
   (Table 3),  and with  typical resolutions $\rm  \Delta V_{ch}$$\sim
   $(10--50)\,km\,s$^{-1}$.  A common  color designated per transition
   is used in all frames. }
\end{figure*}

\begin{figure*}
\centering

\includegraphics[width=0.325\textwidth]{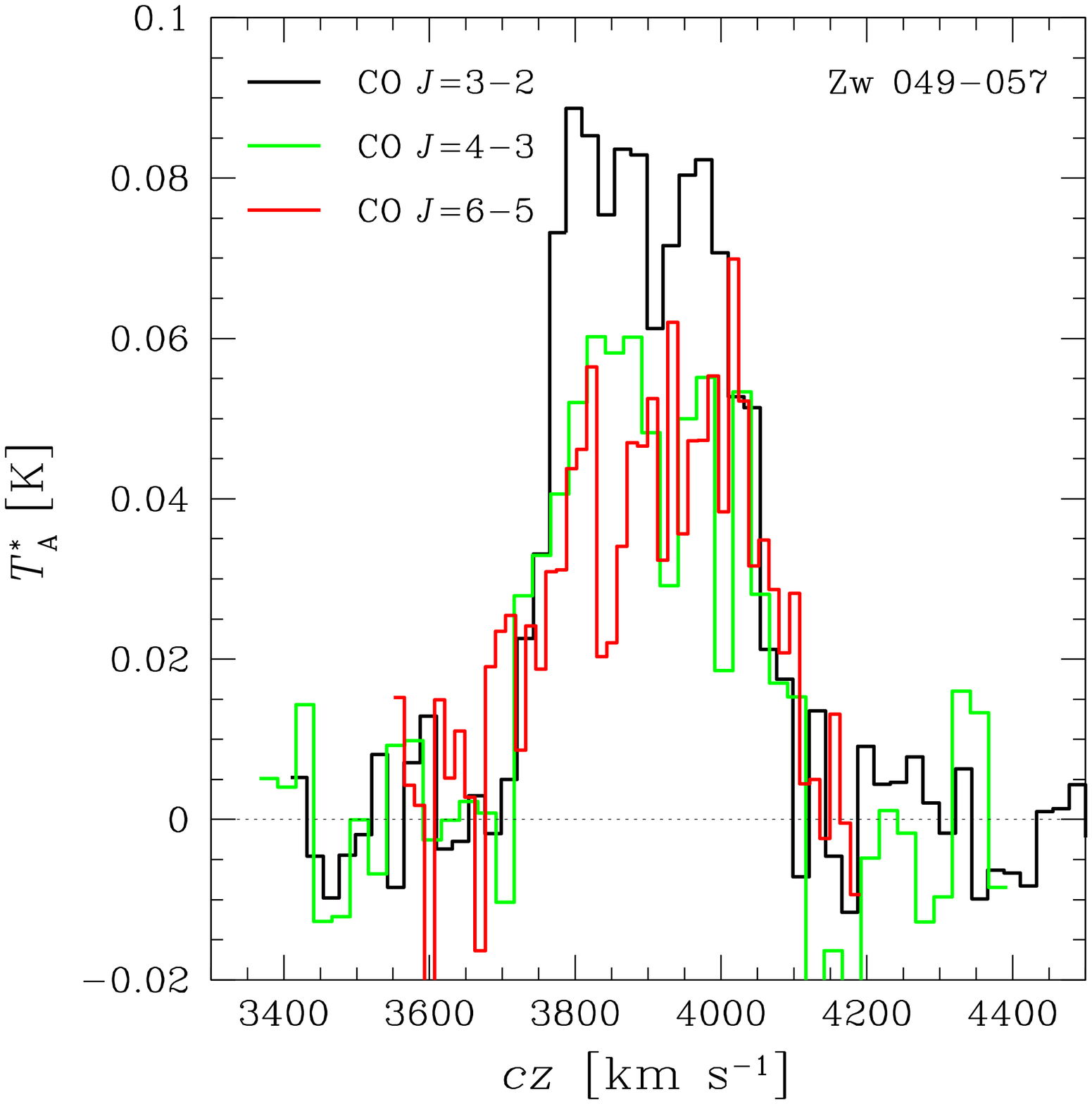} 
\includegraphics[width=0.325\textwidth]{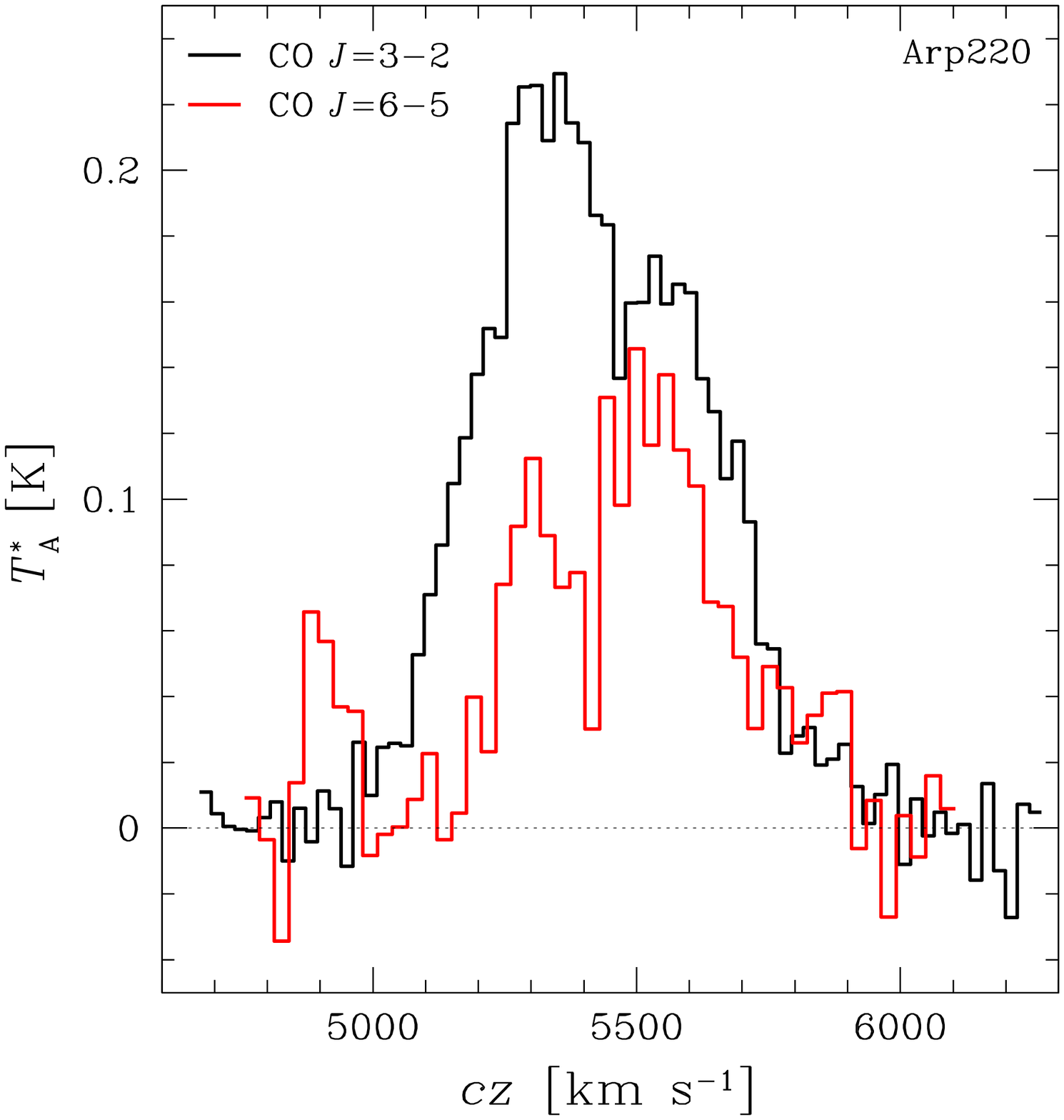}
\includegraphics[width=0.325\textwidth]{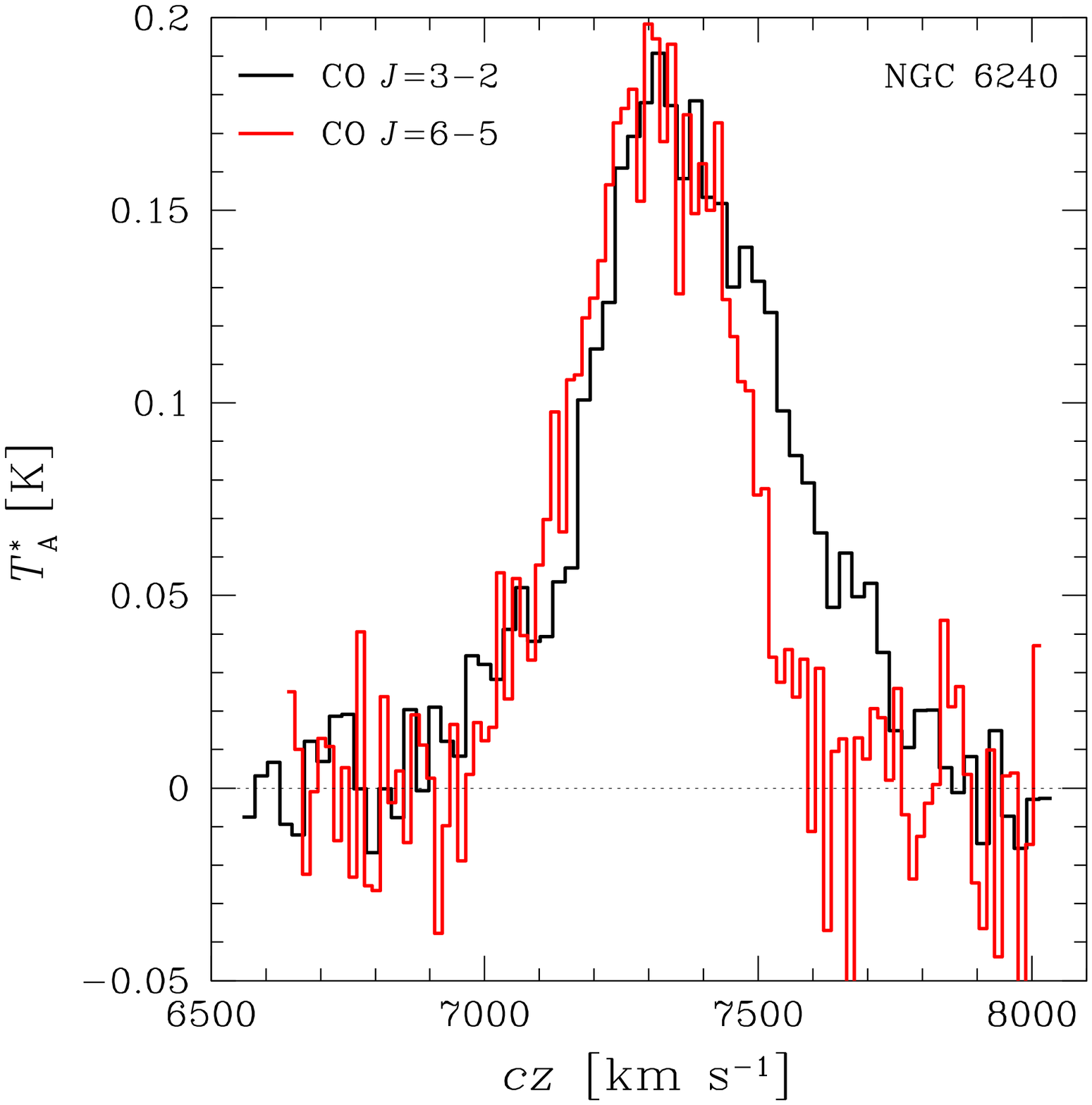}

\includegraphics[width=0.325\textwidth]{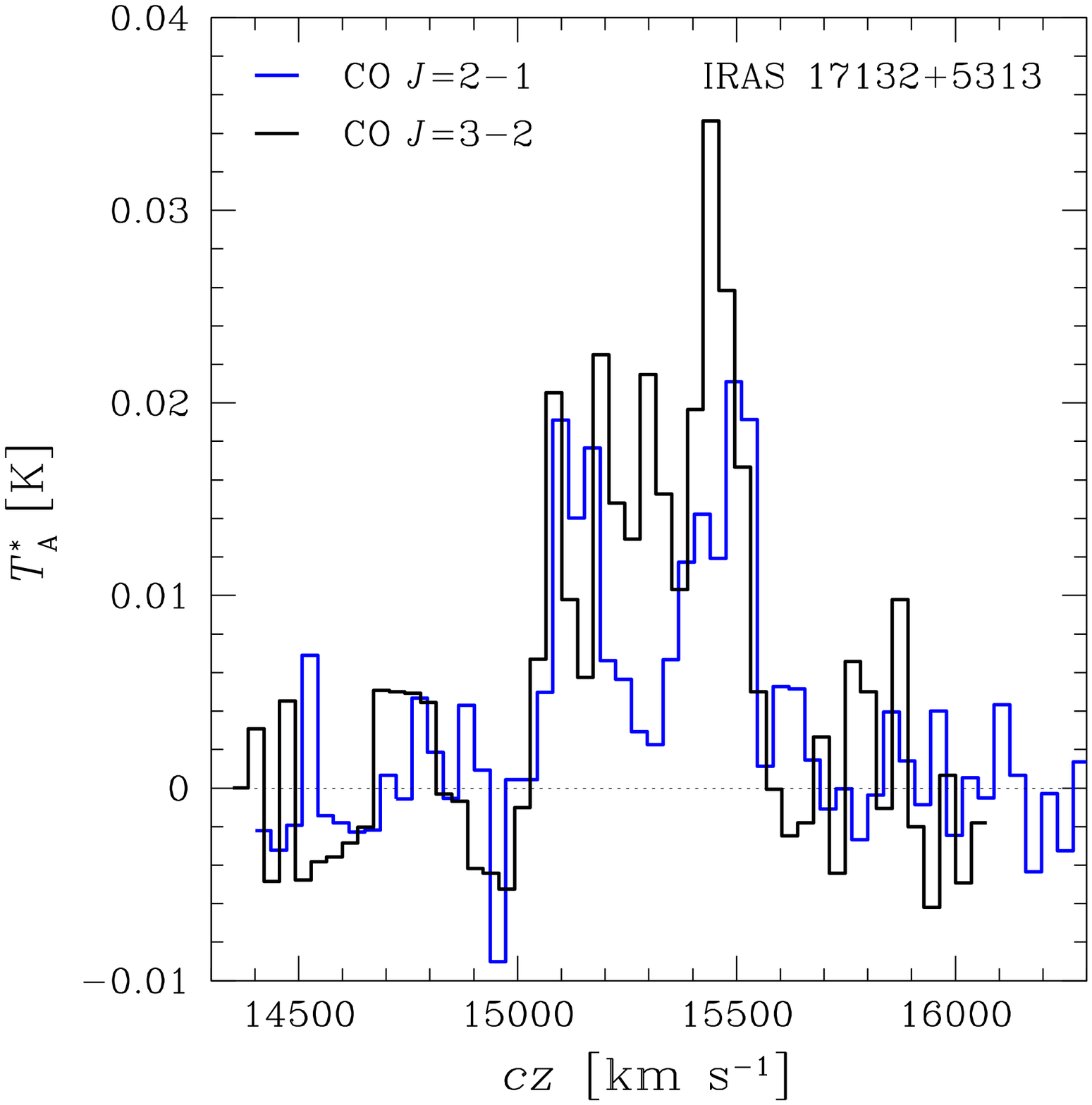}
\includegraphics[width=0.325\textwidth]{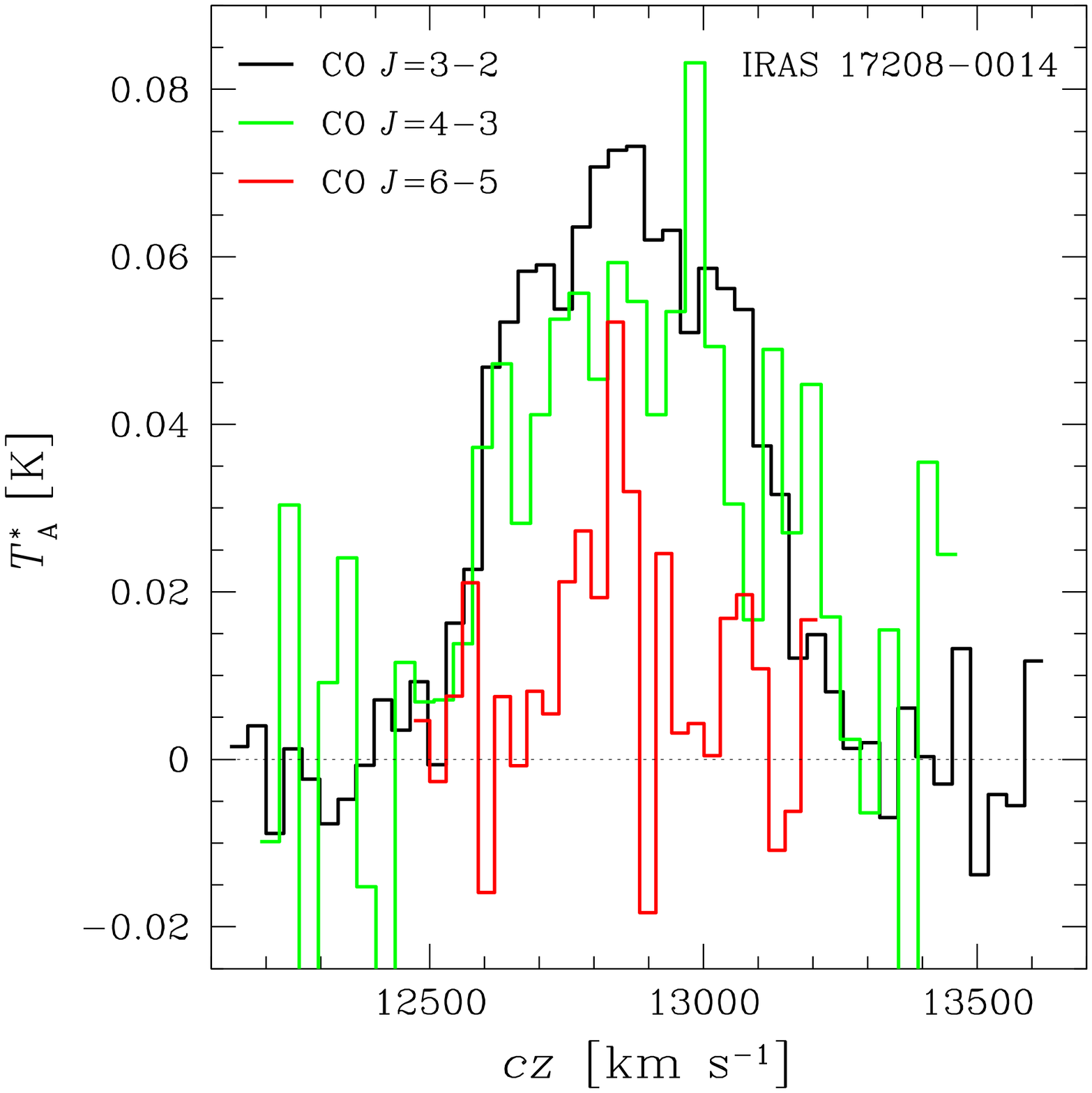}
\includegraphics[width=0.325\textwidth]{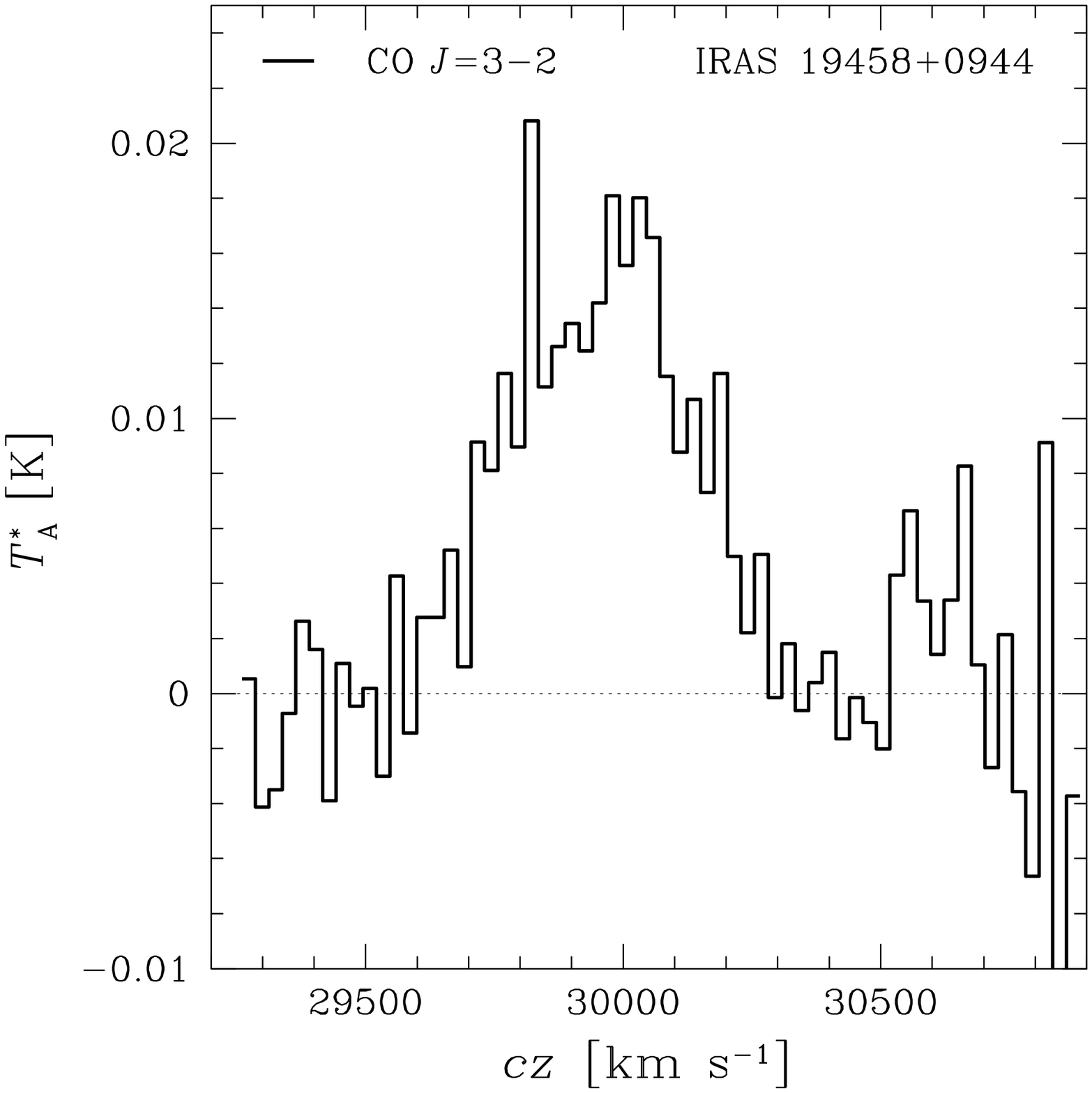}

\includegraphics[width=0.325\textwidth]{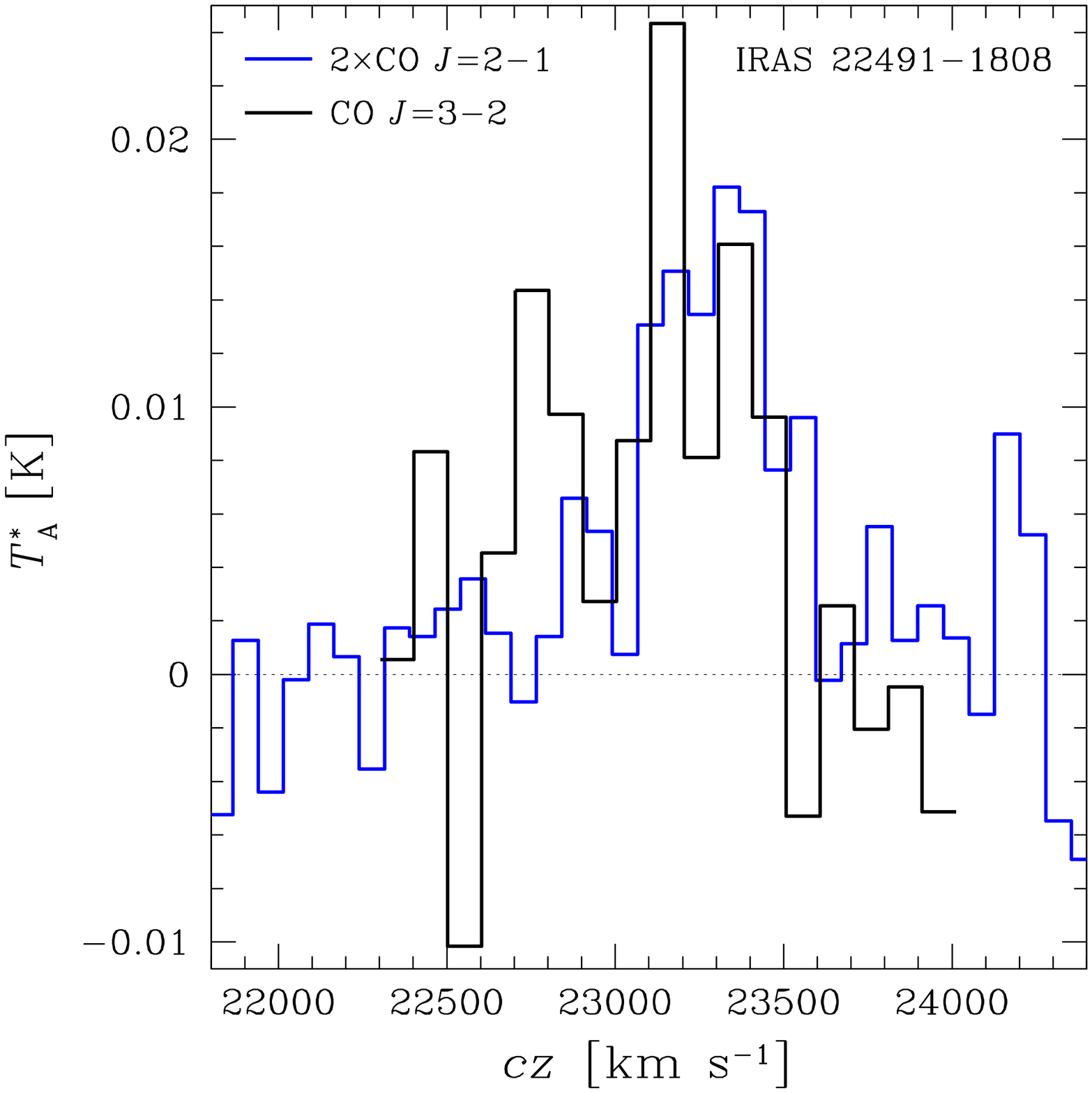}
\includegraphics[width=0.325\textwidth]{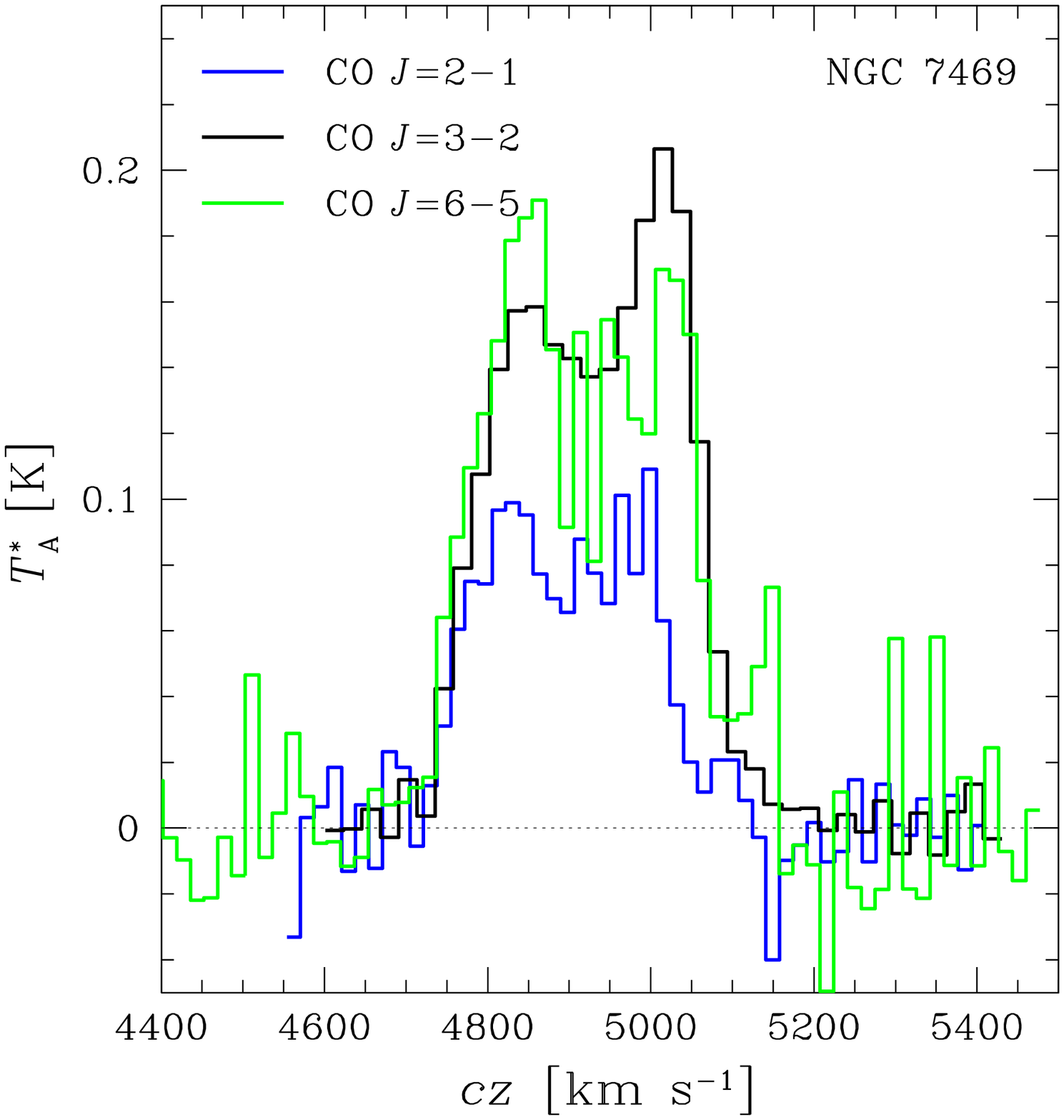}
\includegraphics[width=0.325\textwidth]{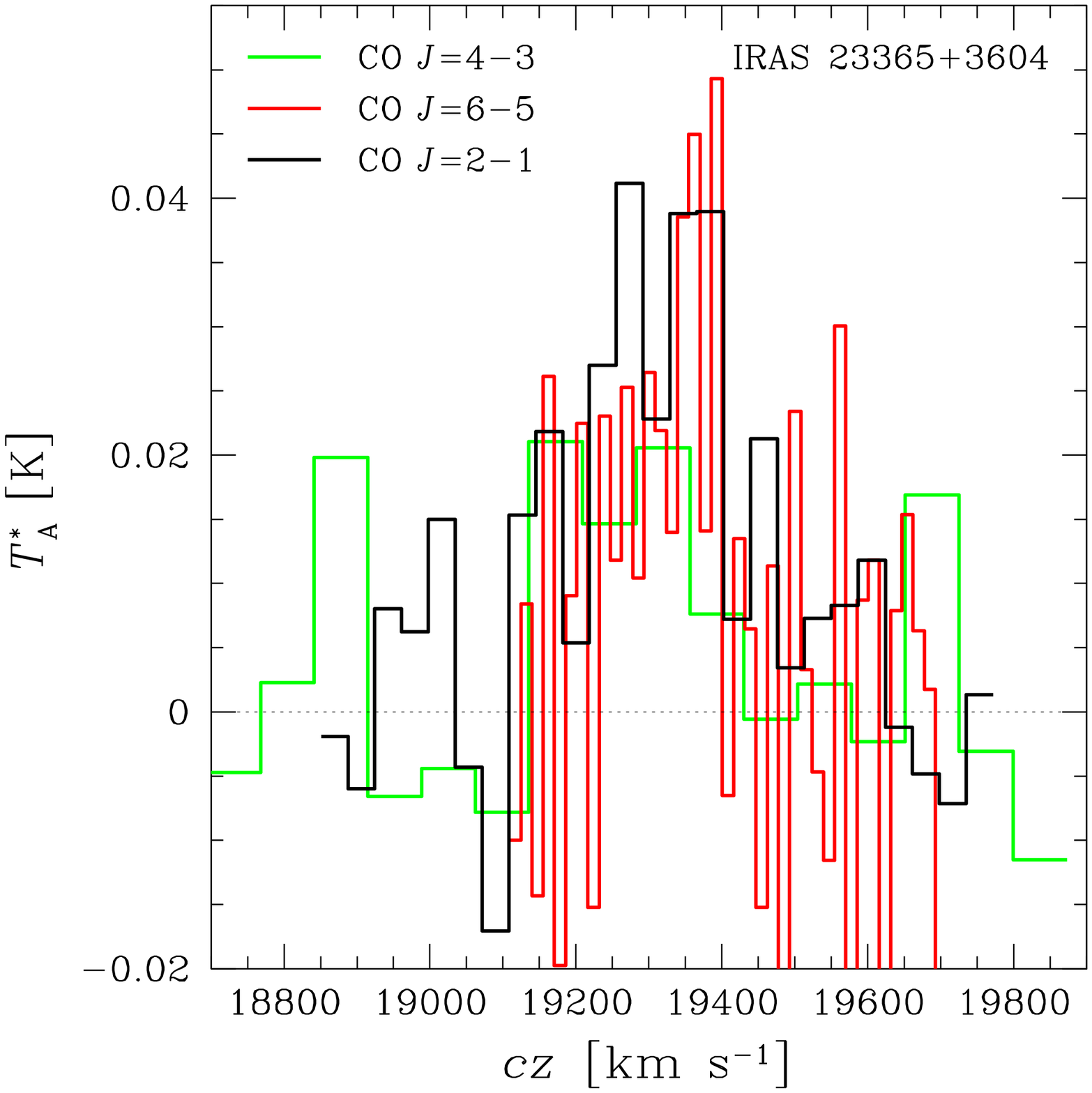}

\contcaption{The high-J CO  J+1$\rightarrow $J, J+1$\geq $3 spectra.
   In the few cases where all three CO J=3--2, 4--3 and 6--5 lines are
   available  we omit  the overlay  of CO  J=2--1 in  order  to reduce
   confusion (the  J=2--1 lines  are all shown  in Figure 3).
  The   velocities   are  with   respect  to   $\rm
V_{opt}$=$\rm  cz_{co}$(LSR) (Table 3),  and with  typical resolutions
$\rm  \Delta  V_{ch}$$\sim  $(10--50)\,km\,s$^{-1}$.  A  common  color
designated per transition is used in all frames. }
\end{figure*}

\begin{figure*}
\centering

\includegraphics[width=0.325\textwidth]{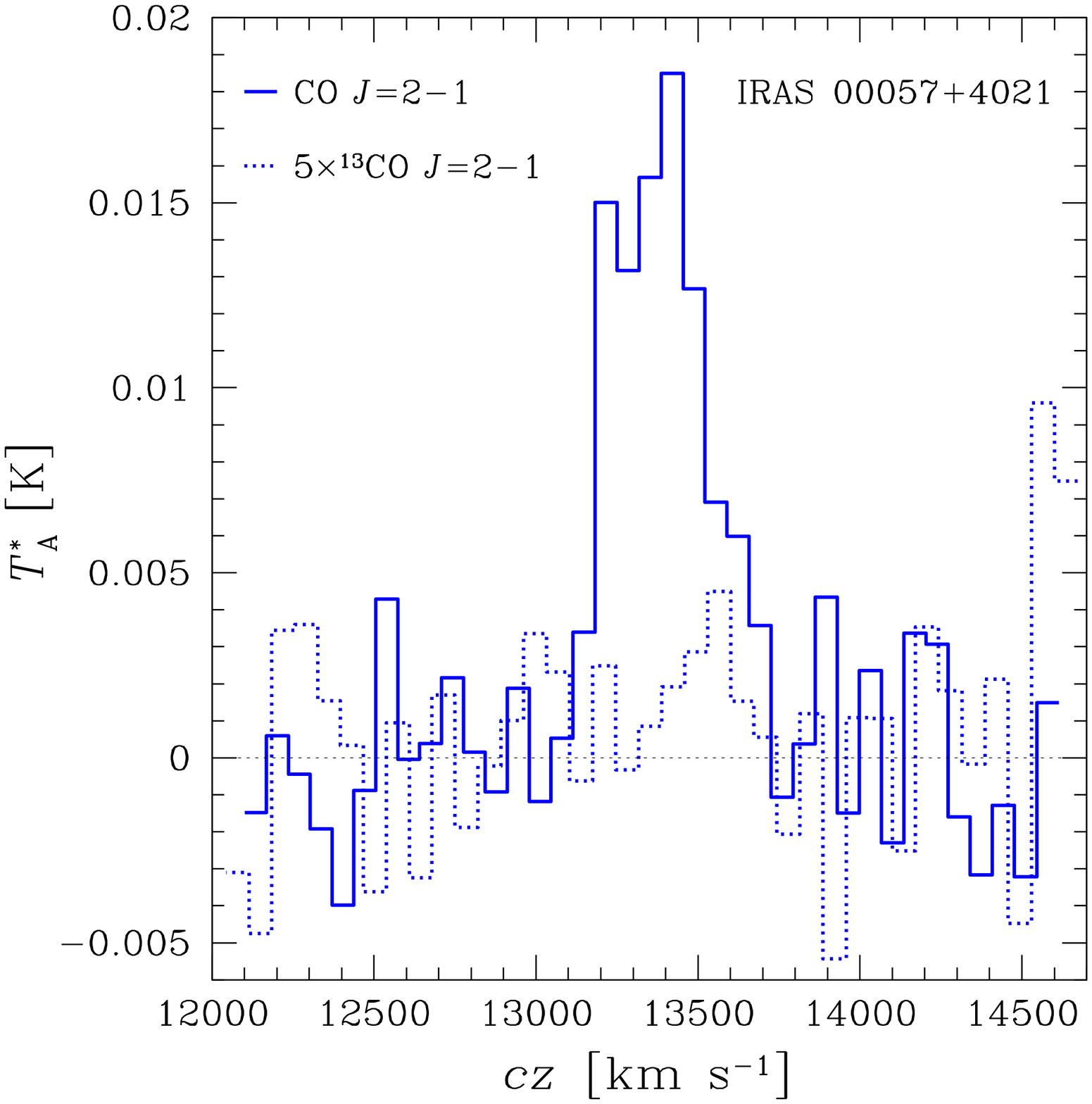}
\includegraphics[width=0.325\textwidth]{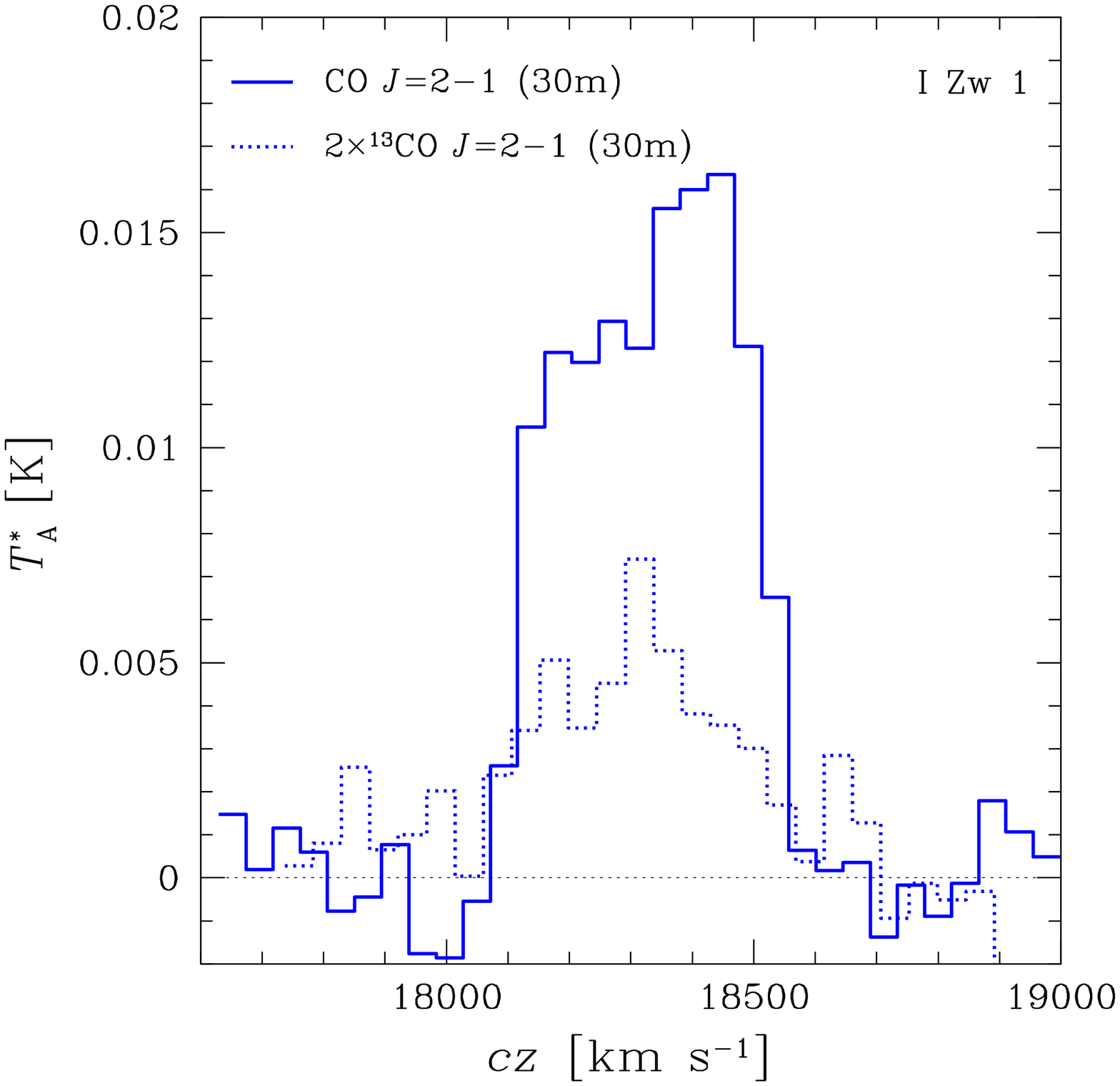} 
\includegraphics[width=0.325\textwidth]{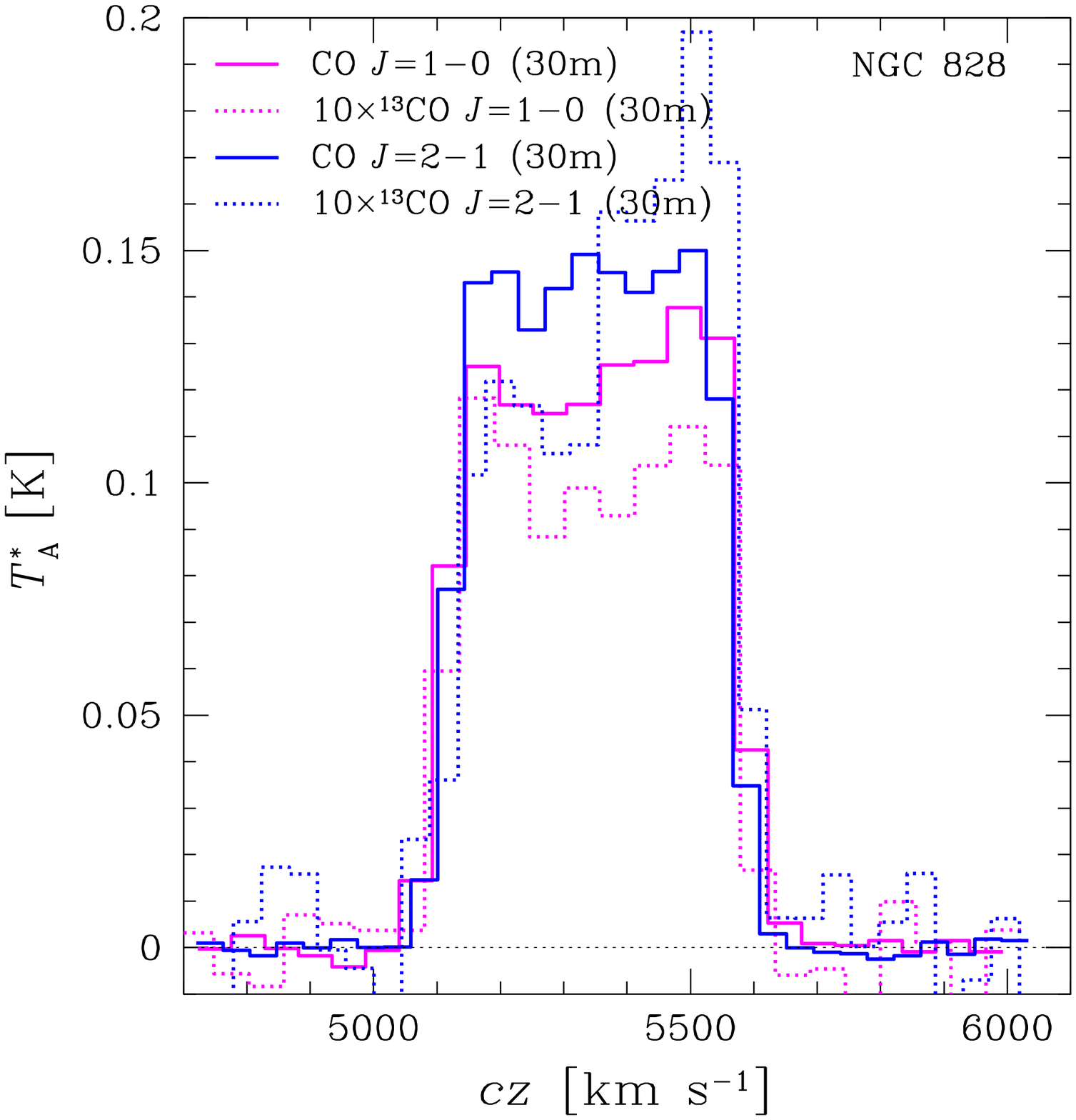}

\includegraphics[width=0.325\textwidth]{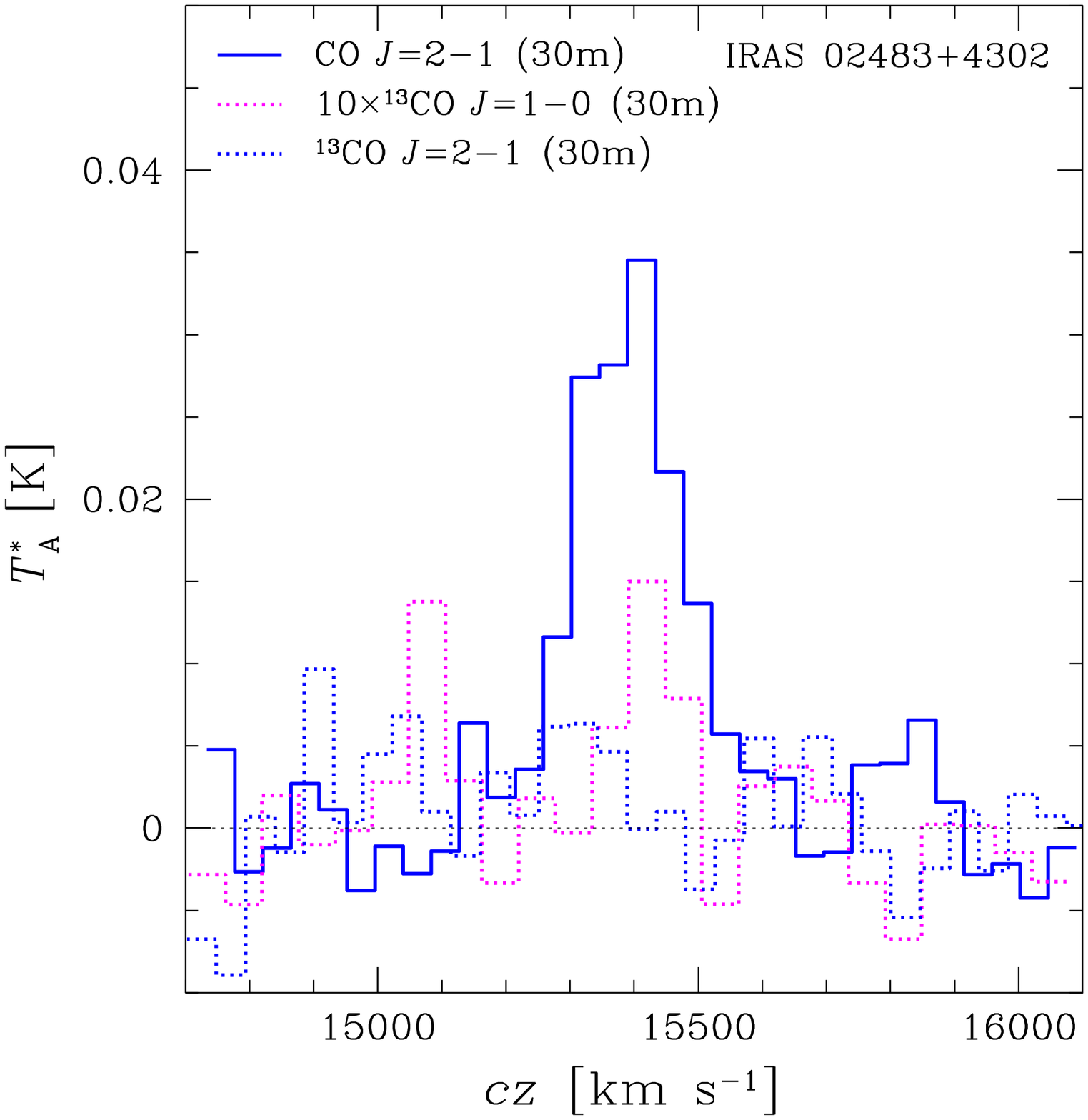}
\includegraphics[width=0.325\textwidth]{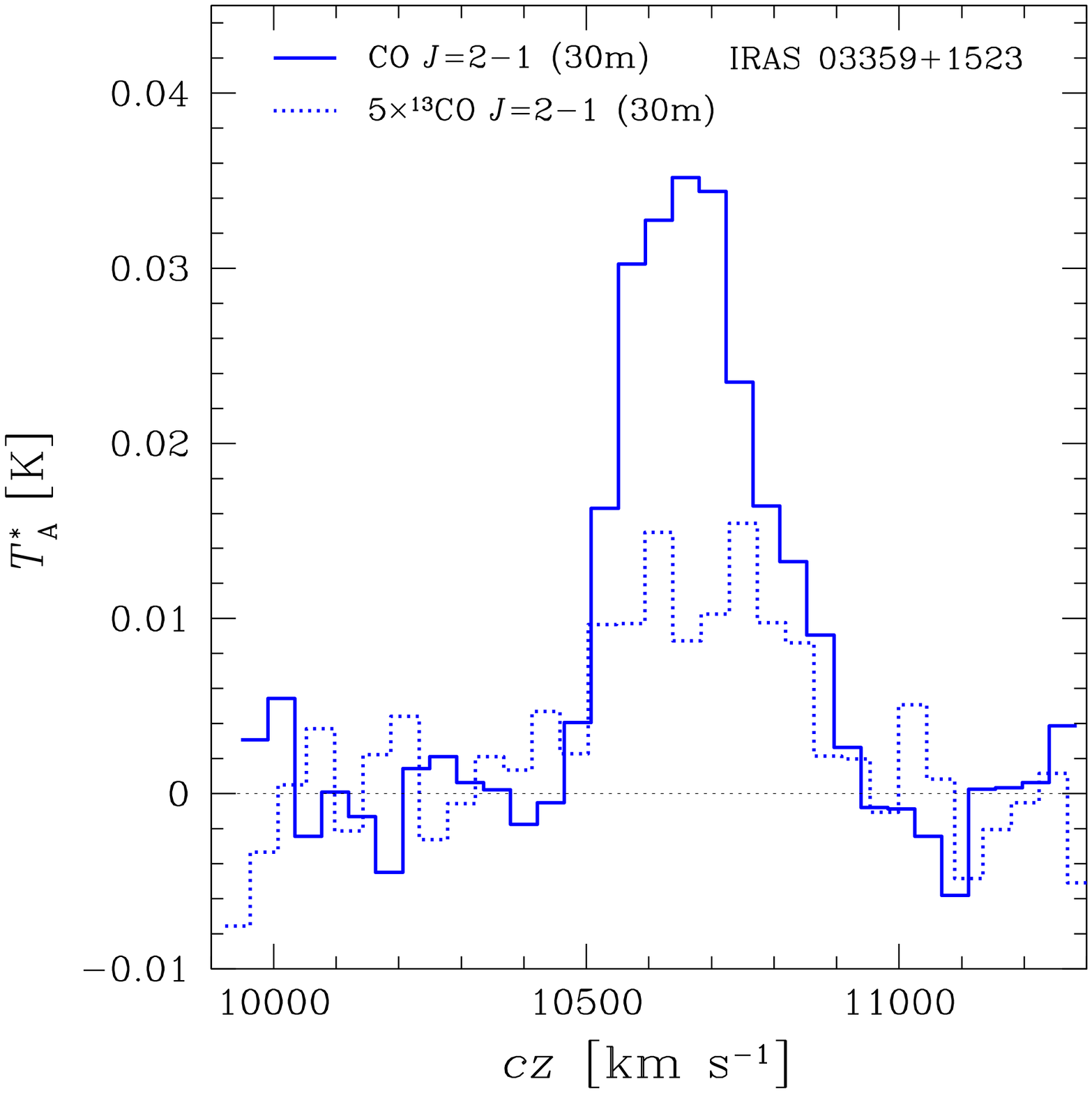}
\includegraphics[width=0.325\textwidth]{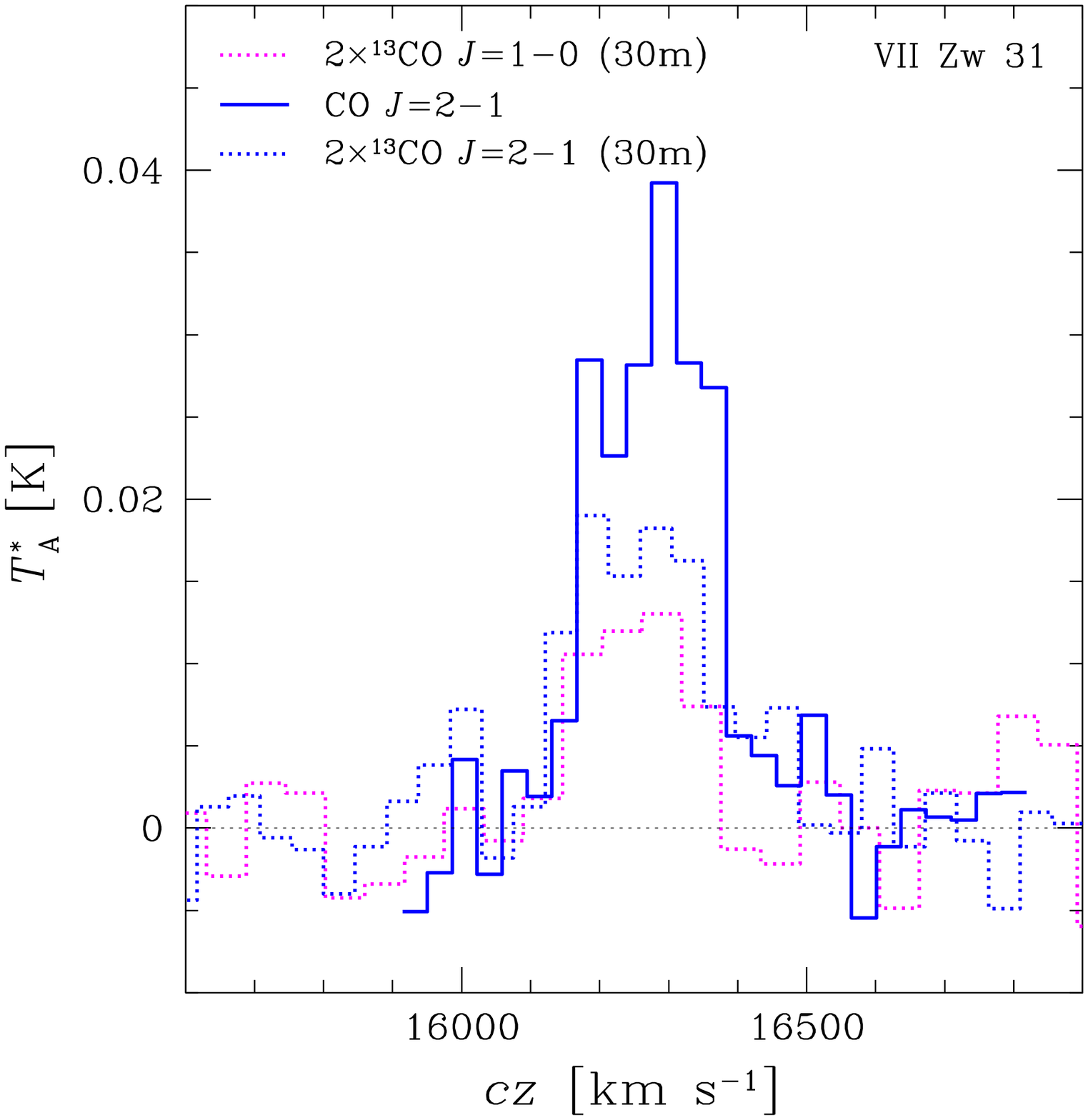}

\includegraphics[width=0.325\textwidth]{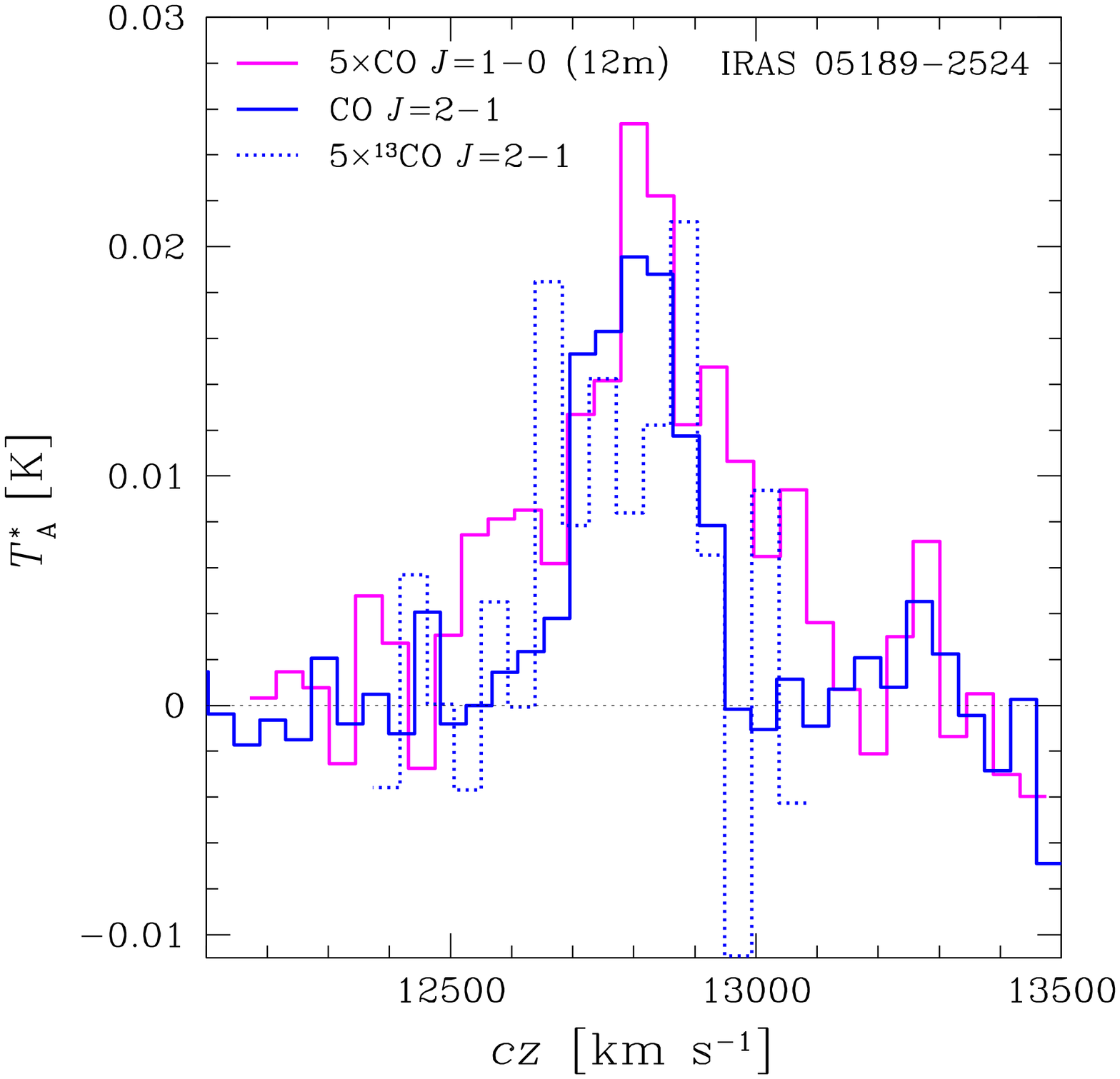}
\includegraphics[width=0.325\textwidth]{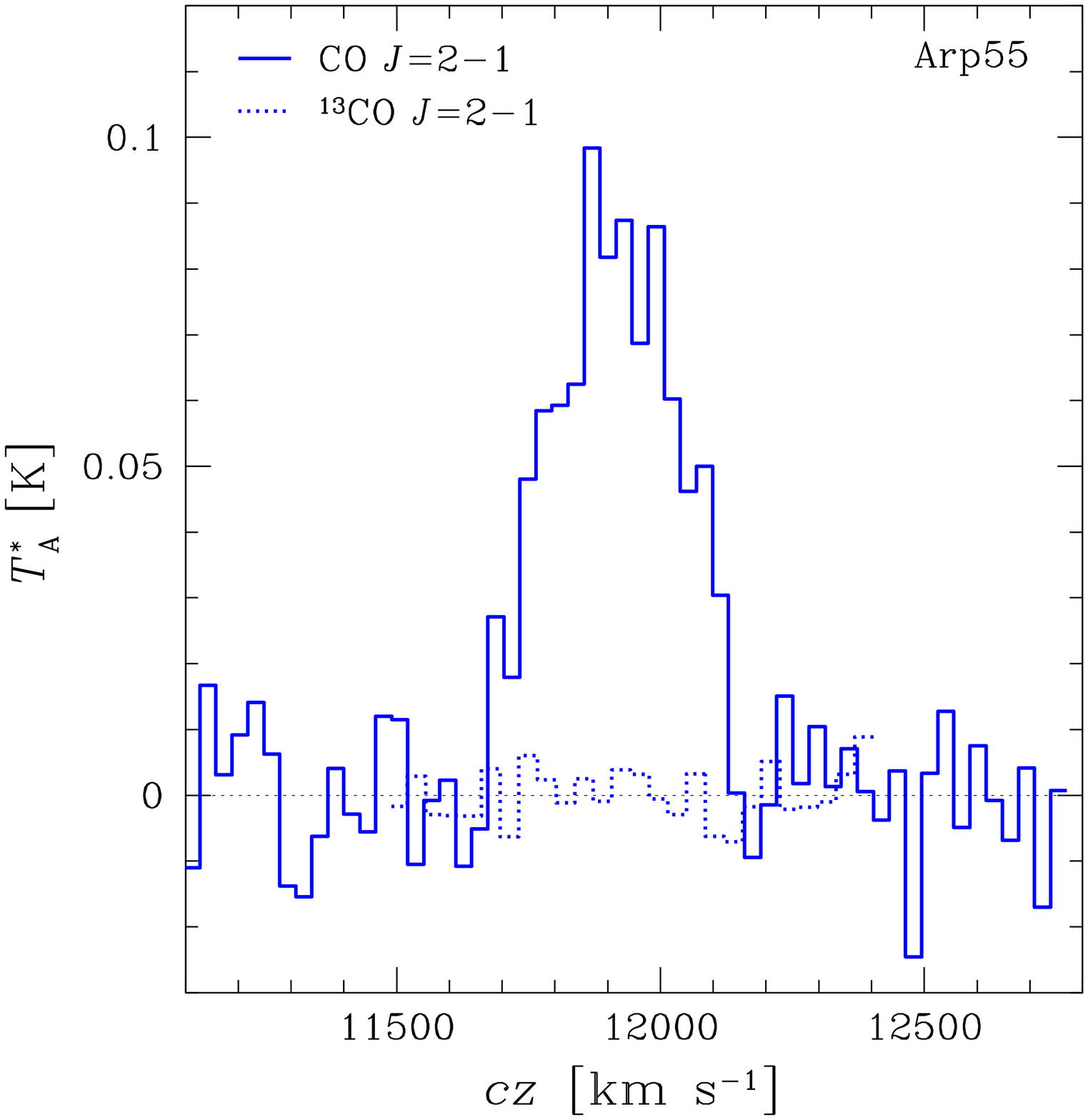}
\includegraphics[width=0.325\textwidth]{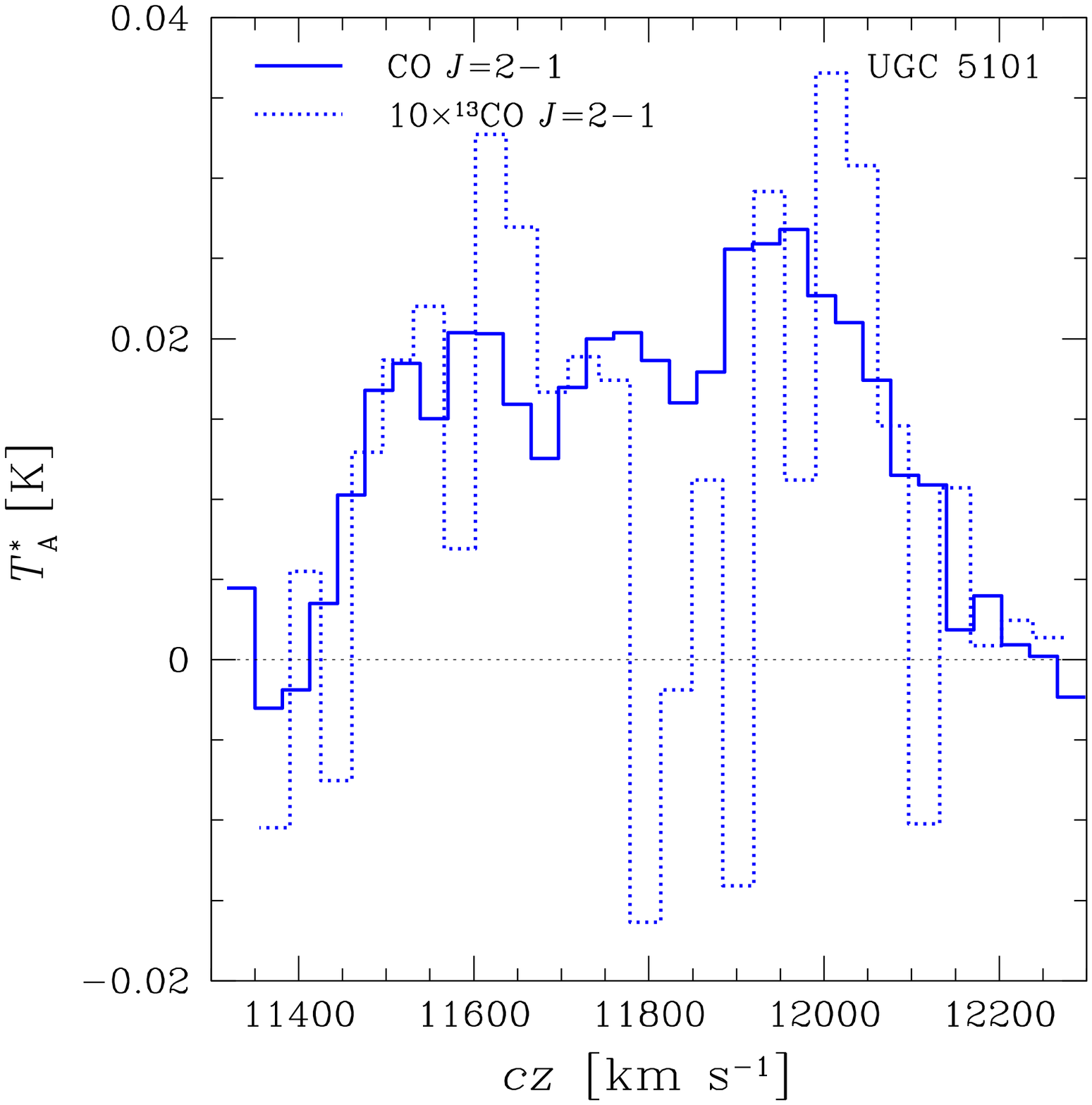} 

\caption{The CO, $^{13}$CO line data.  The velocities are with respect
to  $\rm  V_{opt}$=$\rm  cz_{co}$(LSR)  (Table 3),  and  with  typical
resolutions  $\rm   \Delta  V_{ch}$$\sim  $(35--90)\,km\,s$^{-1}$.   A
common color designated per transition is used in all frames. }
\end{figure*}

\begin{figure*}
\centering
 
\includegraphics[width=0.325\textwidth]{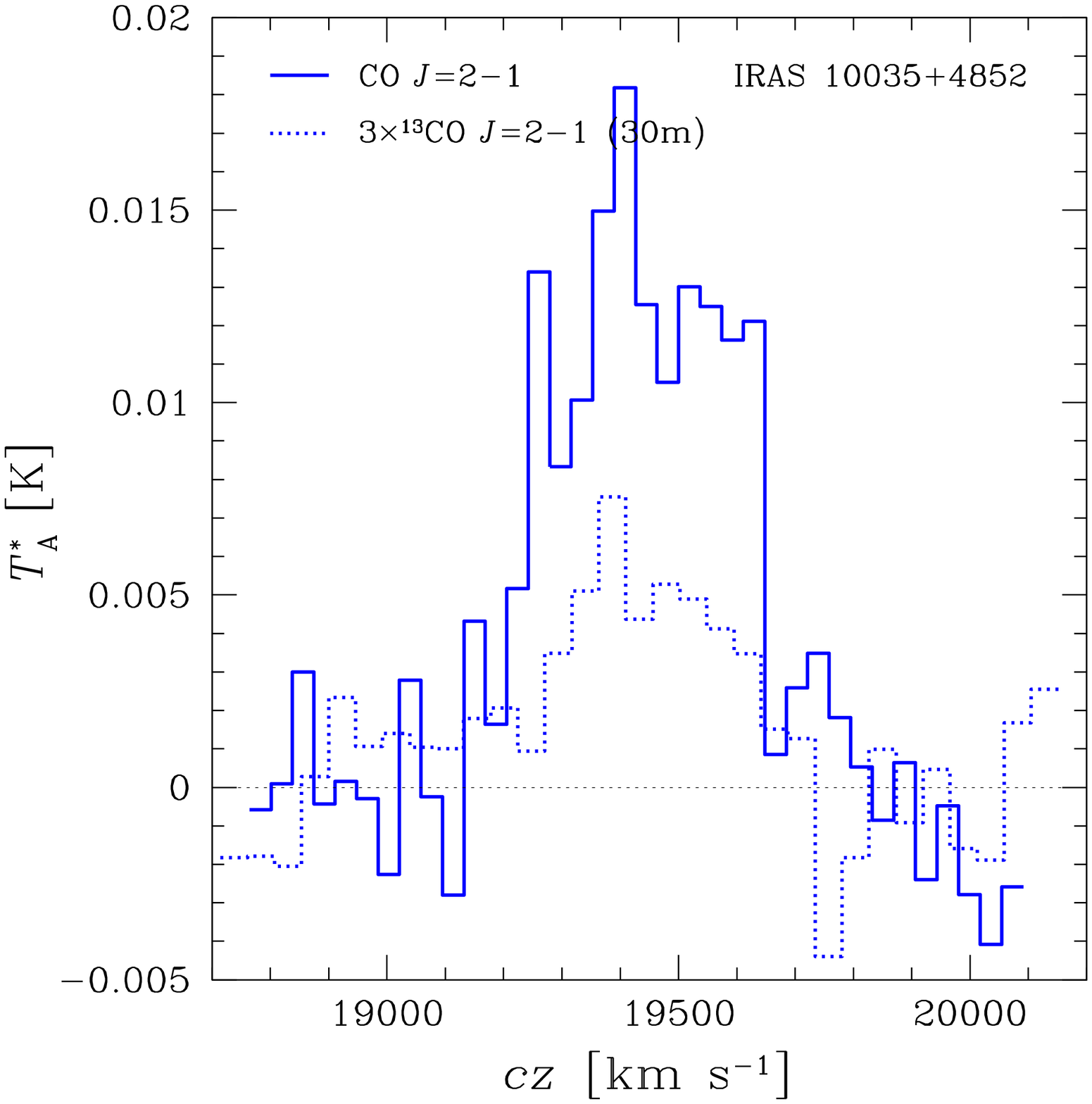} 
\includegraphics[width=0.325\textwidth]{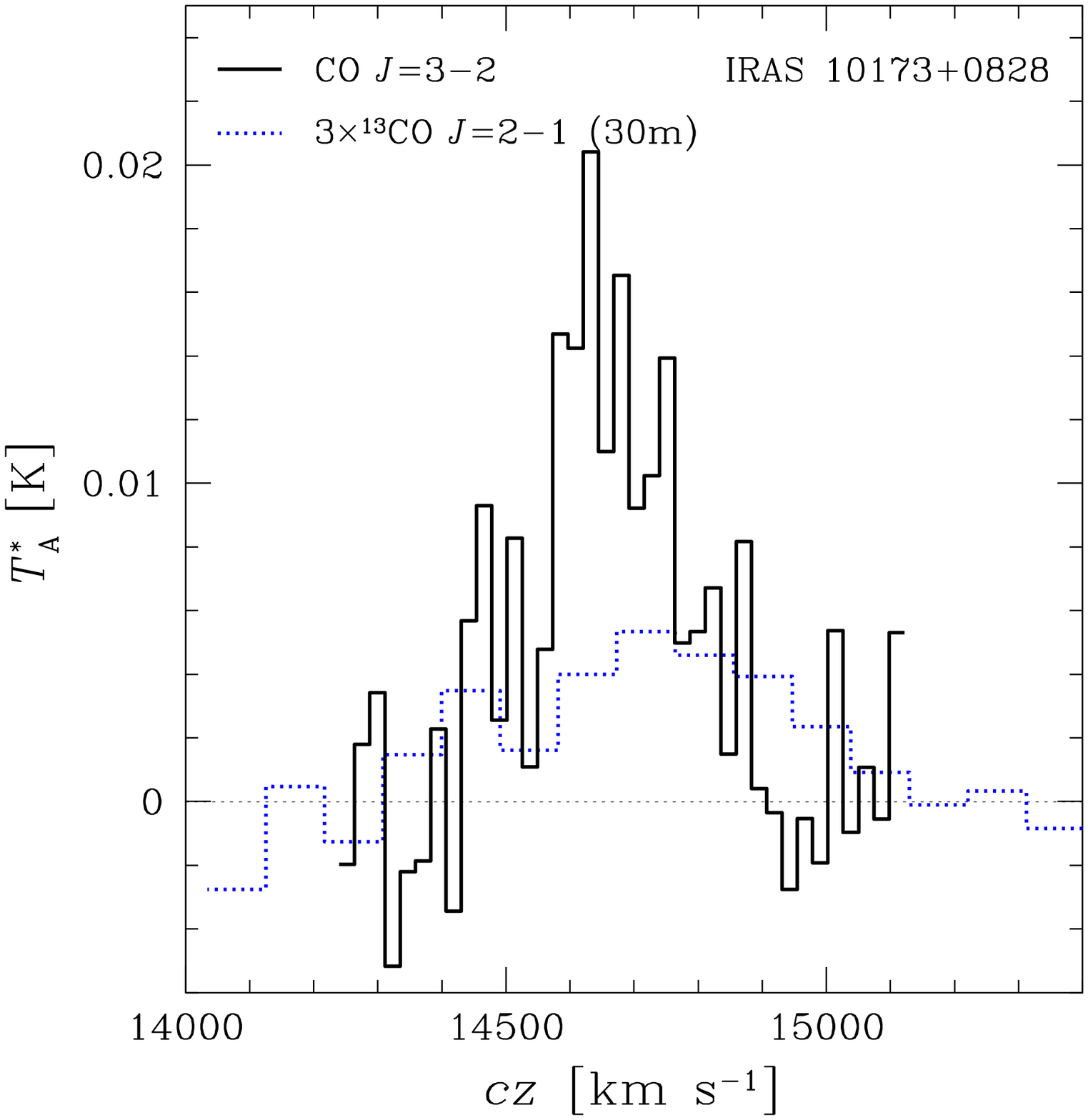} 
\includegraphics[width=0.325\textwidth]{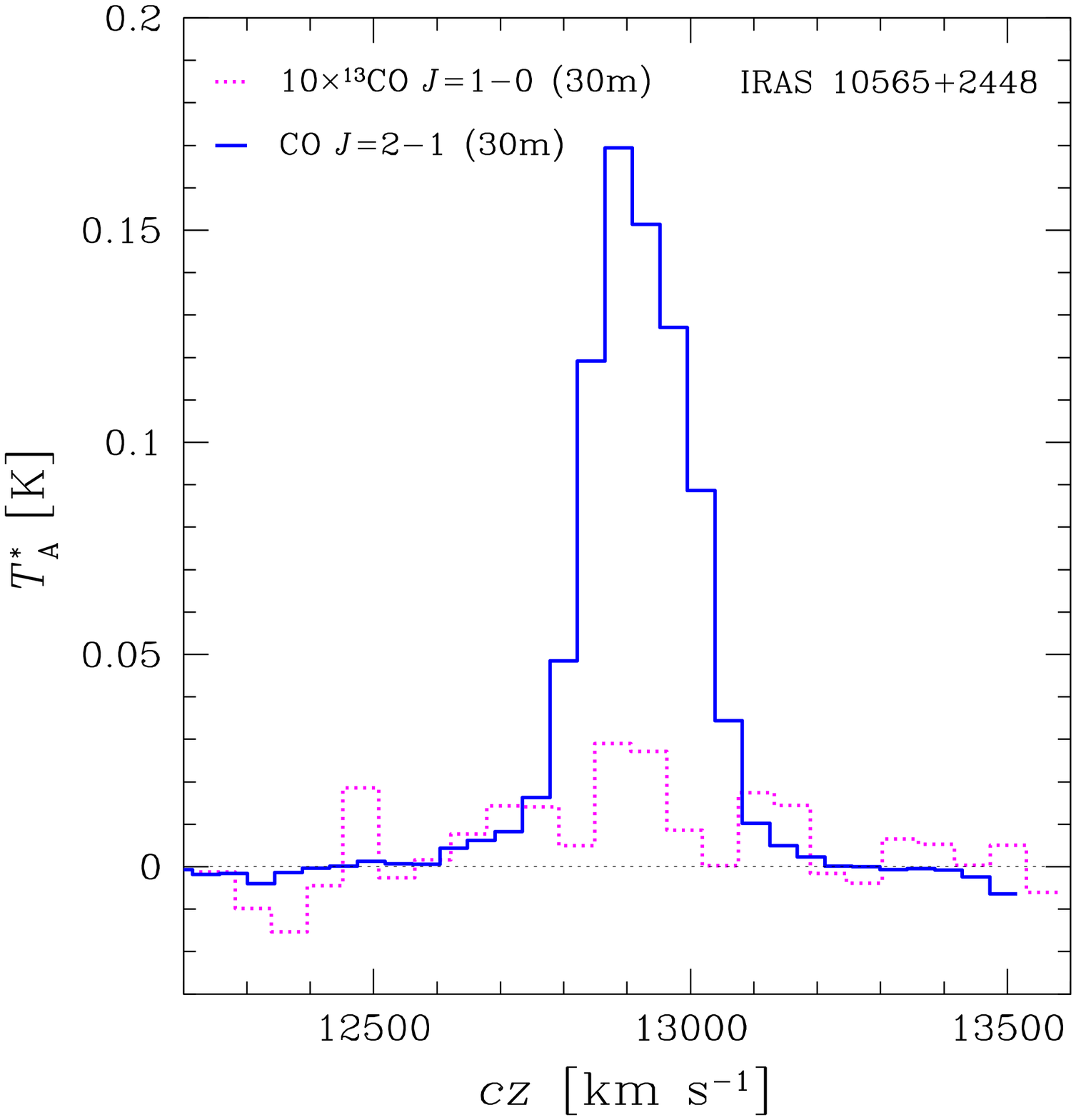} 

\includegraphics[width=0.325\textwidth]{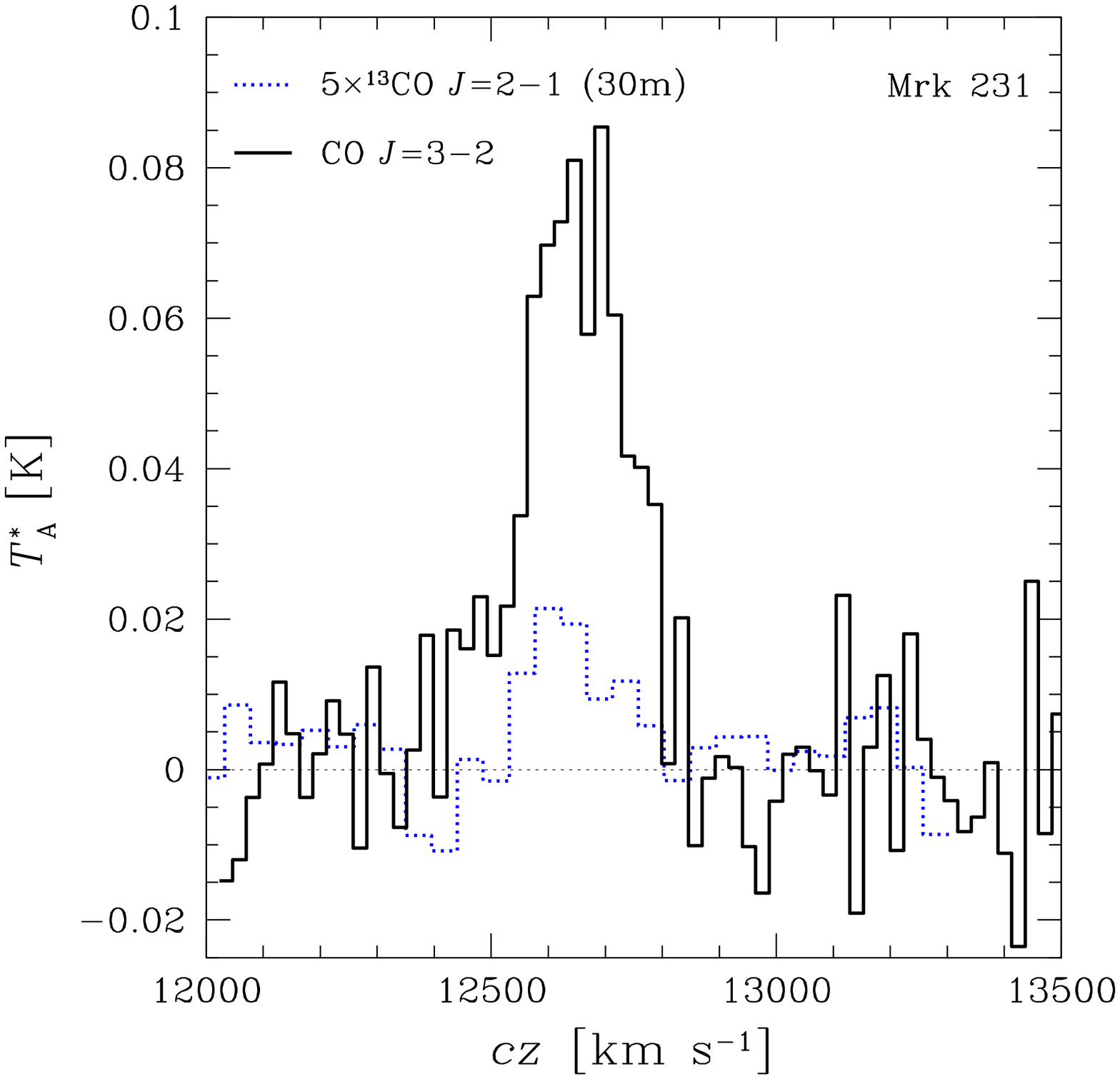} 
\includegraphics[width=0.325\textwidth]{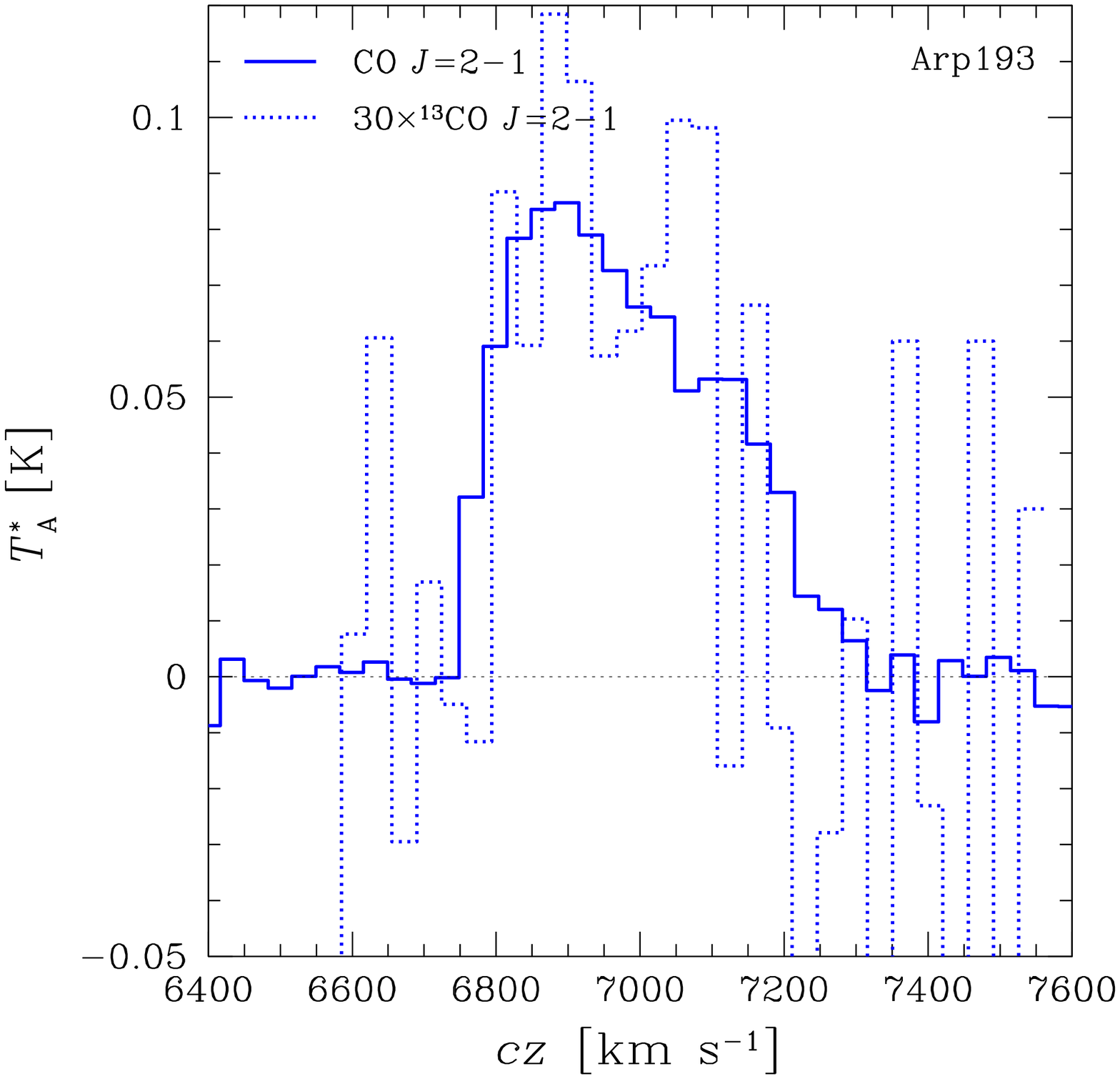} 
\includegraphics[width=0.325\textwidth]{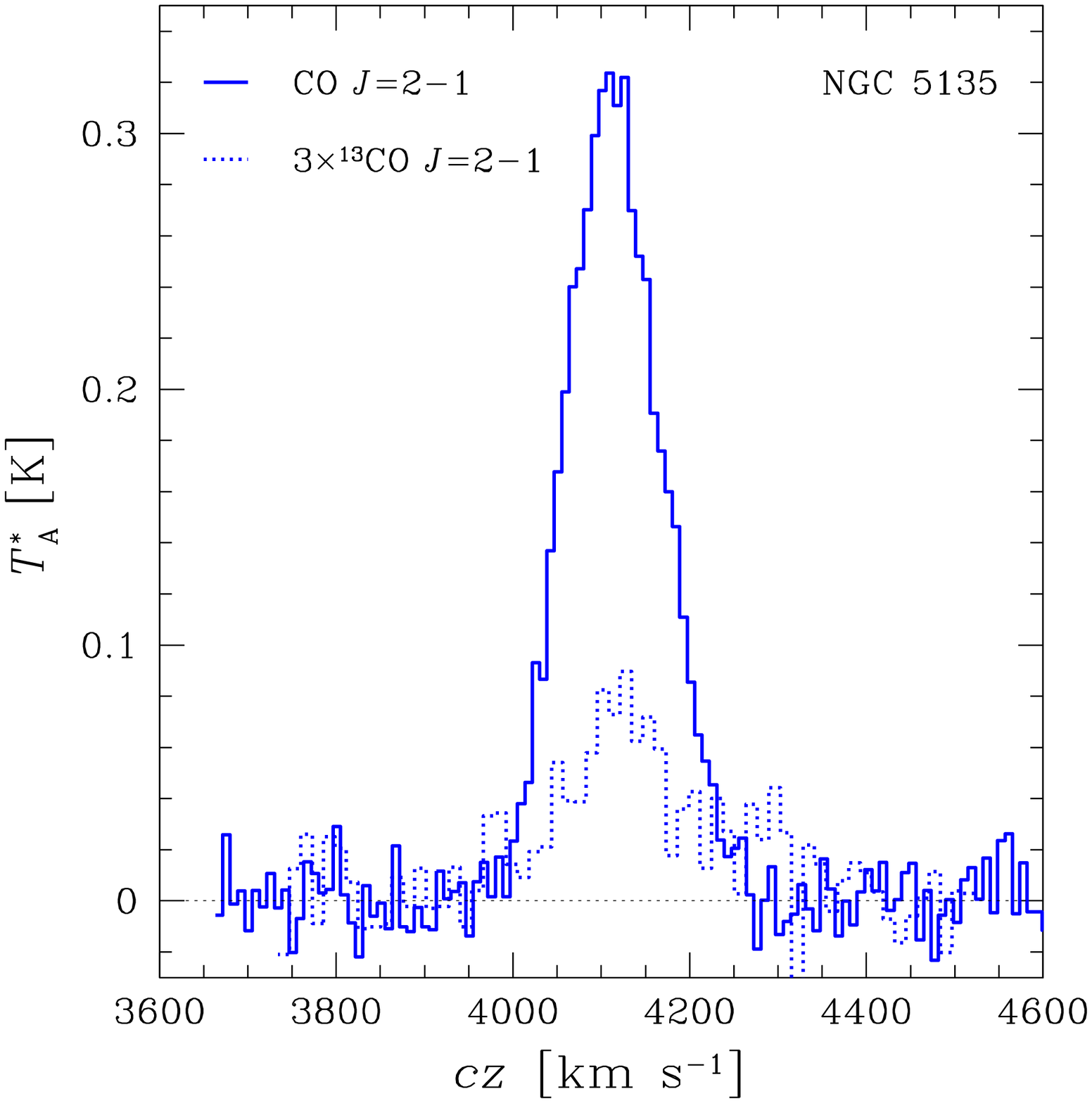} 

\includegraphics[width=0.325\textwidth]{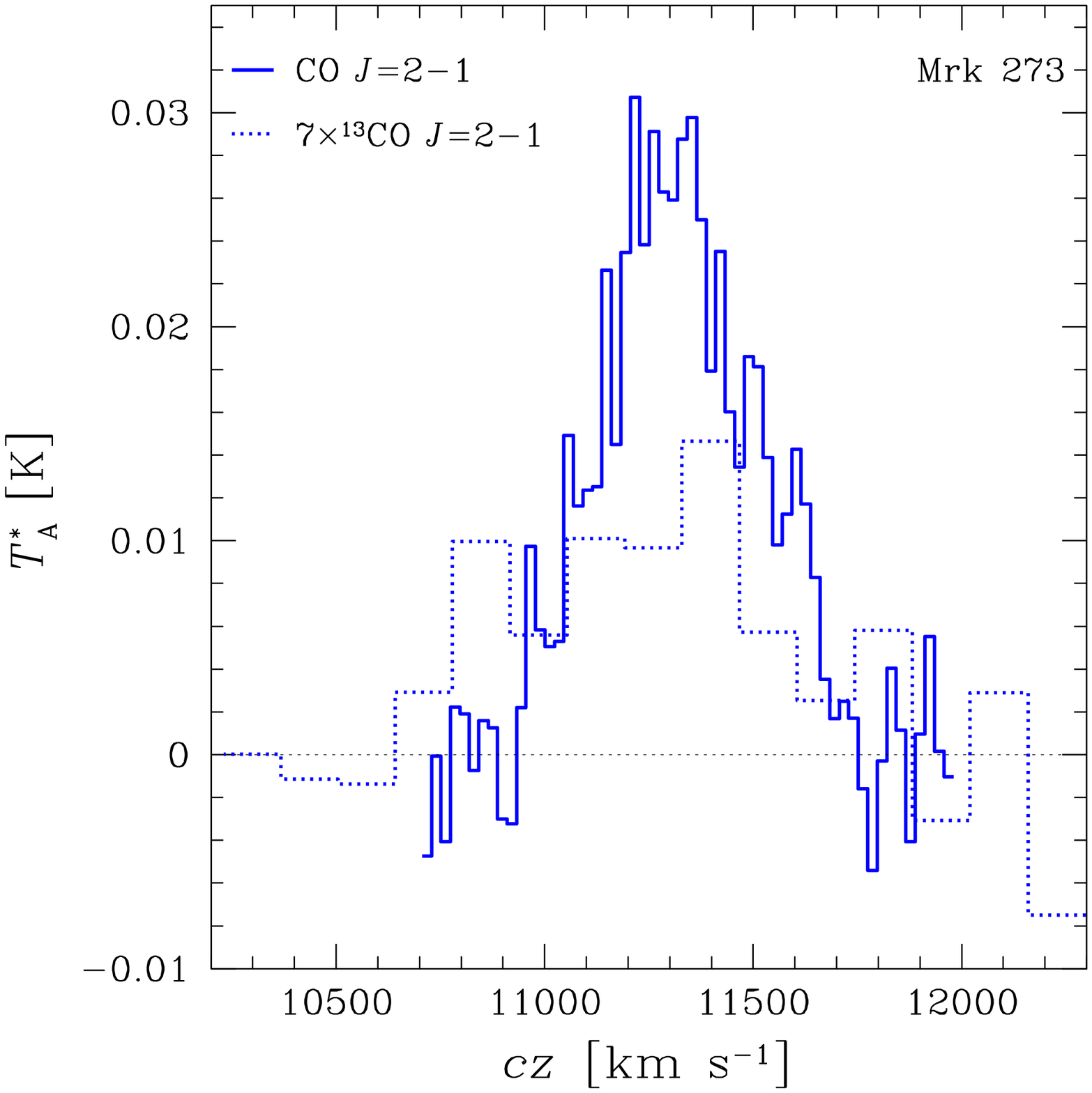}
\includegraphics[width=0.325\textwidth]{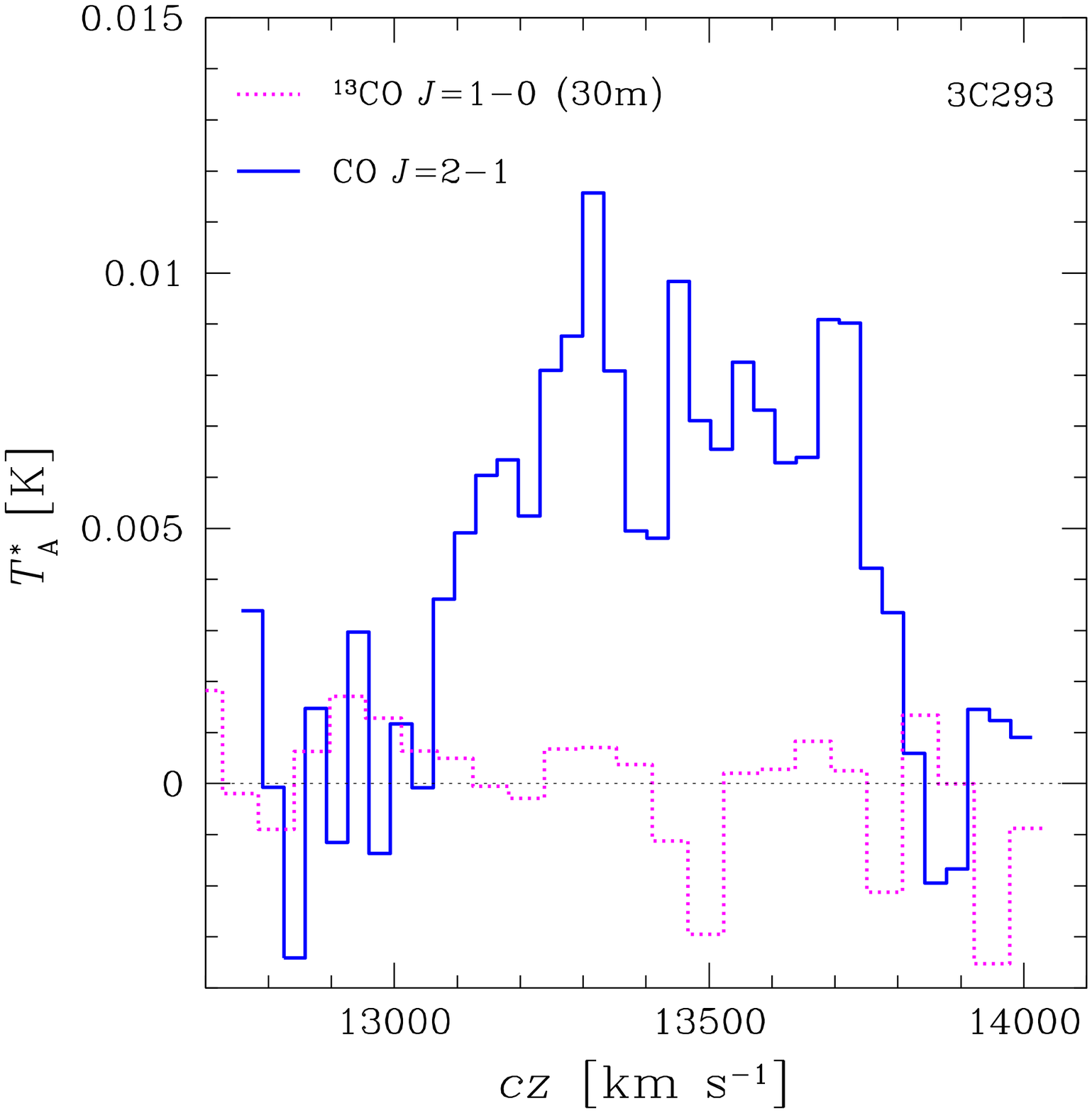} 
\includegraphics[width=0.325\textwidth]{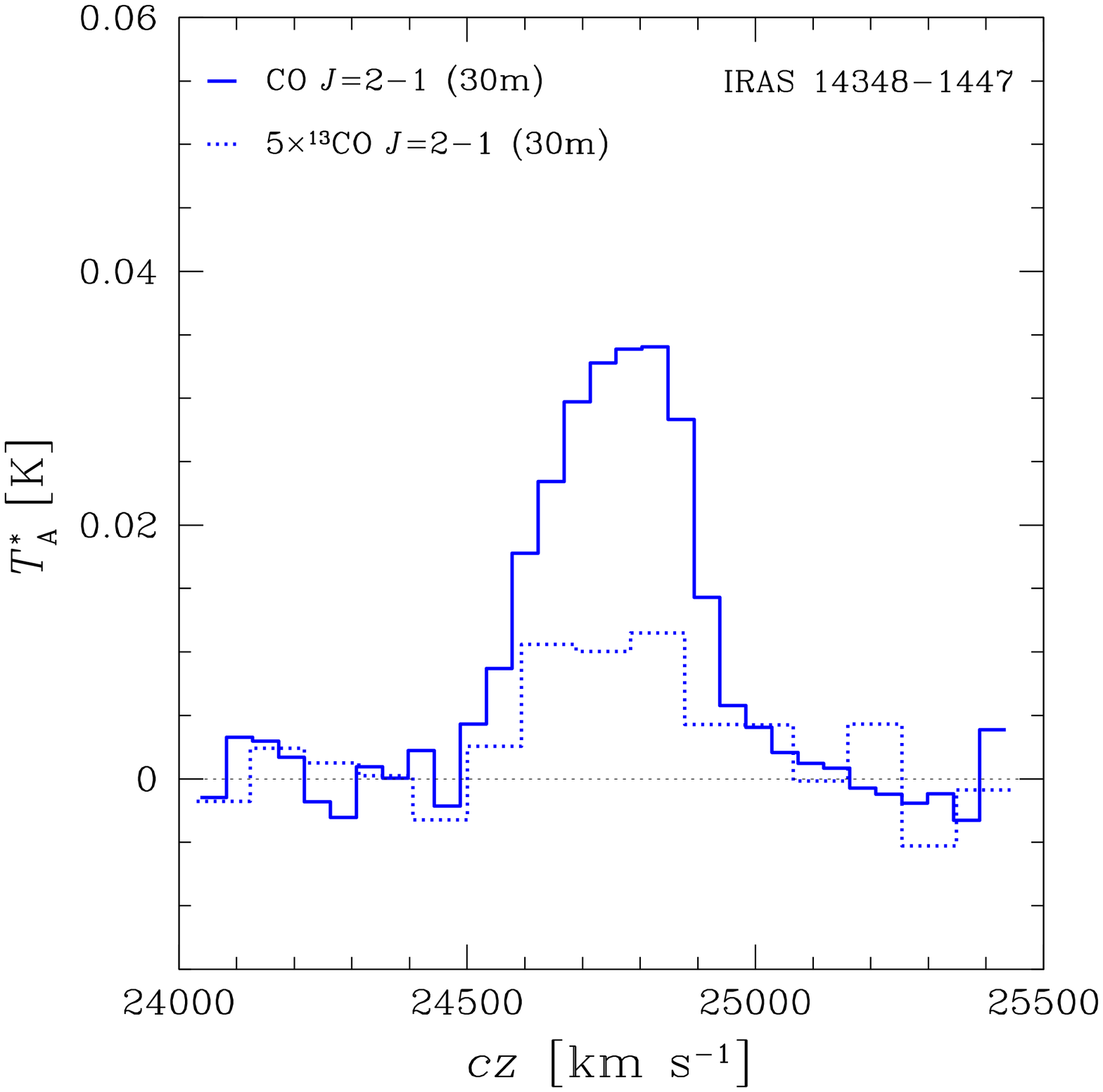}

\contcaption{The CO, $^{13}$CO  line data.  The velocities are with
respect  to  $\rm  V_{opt}$=$\rm  cz_{co}$(LSR) (Table  3),  and  with
typical resolutions  $\rm \Delta V_{ch}$$\sim $(35--90)\,km\,s$^{-1}$.
A common color designated per transition is used in all frames. }
\end{figure*}

\begin{figure*}
\centering
 
\includegraphics[width=0.325\textwidth]{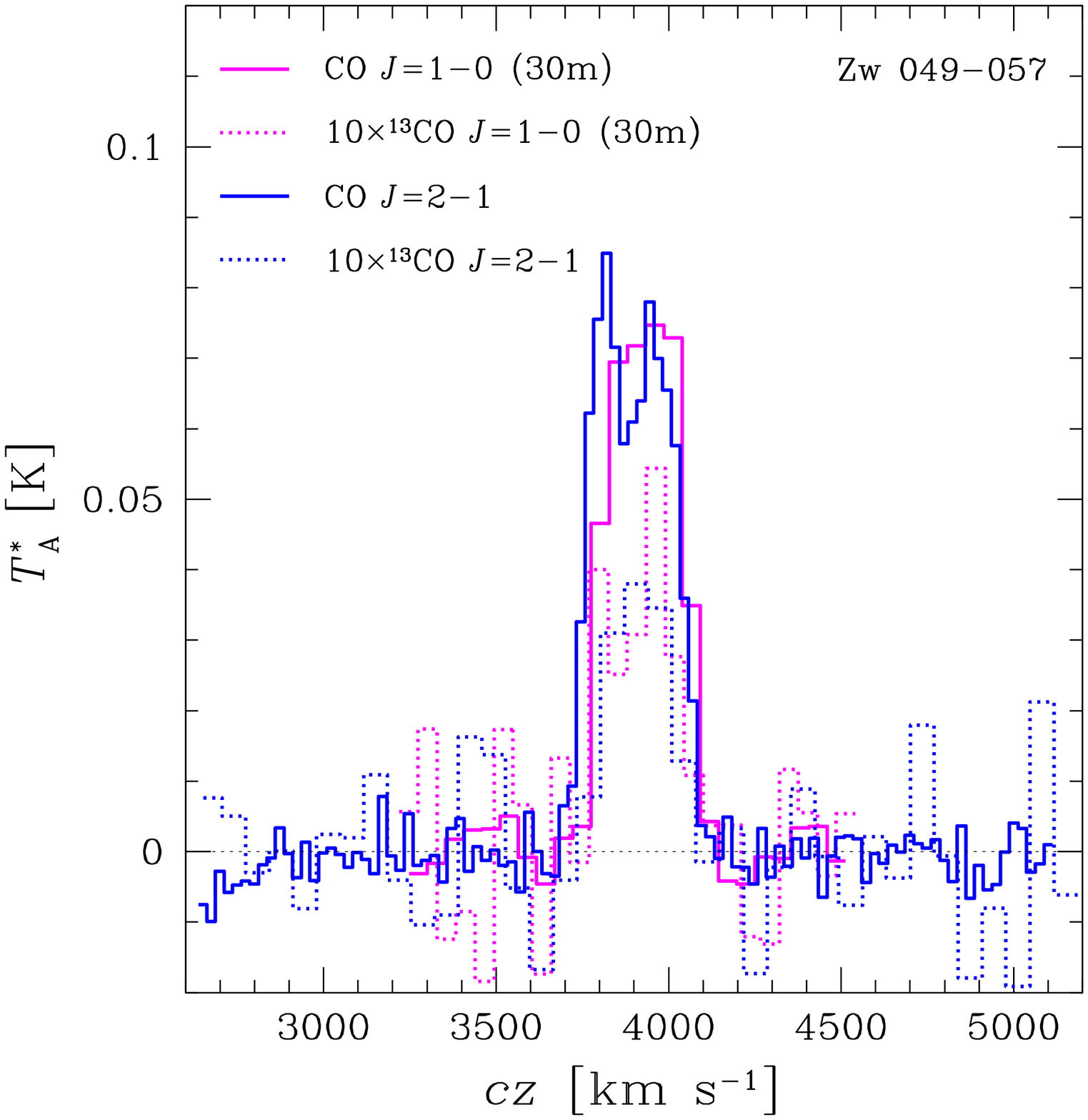} 
\includegraphics[width=0.325\textwidth]{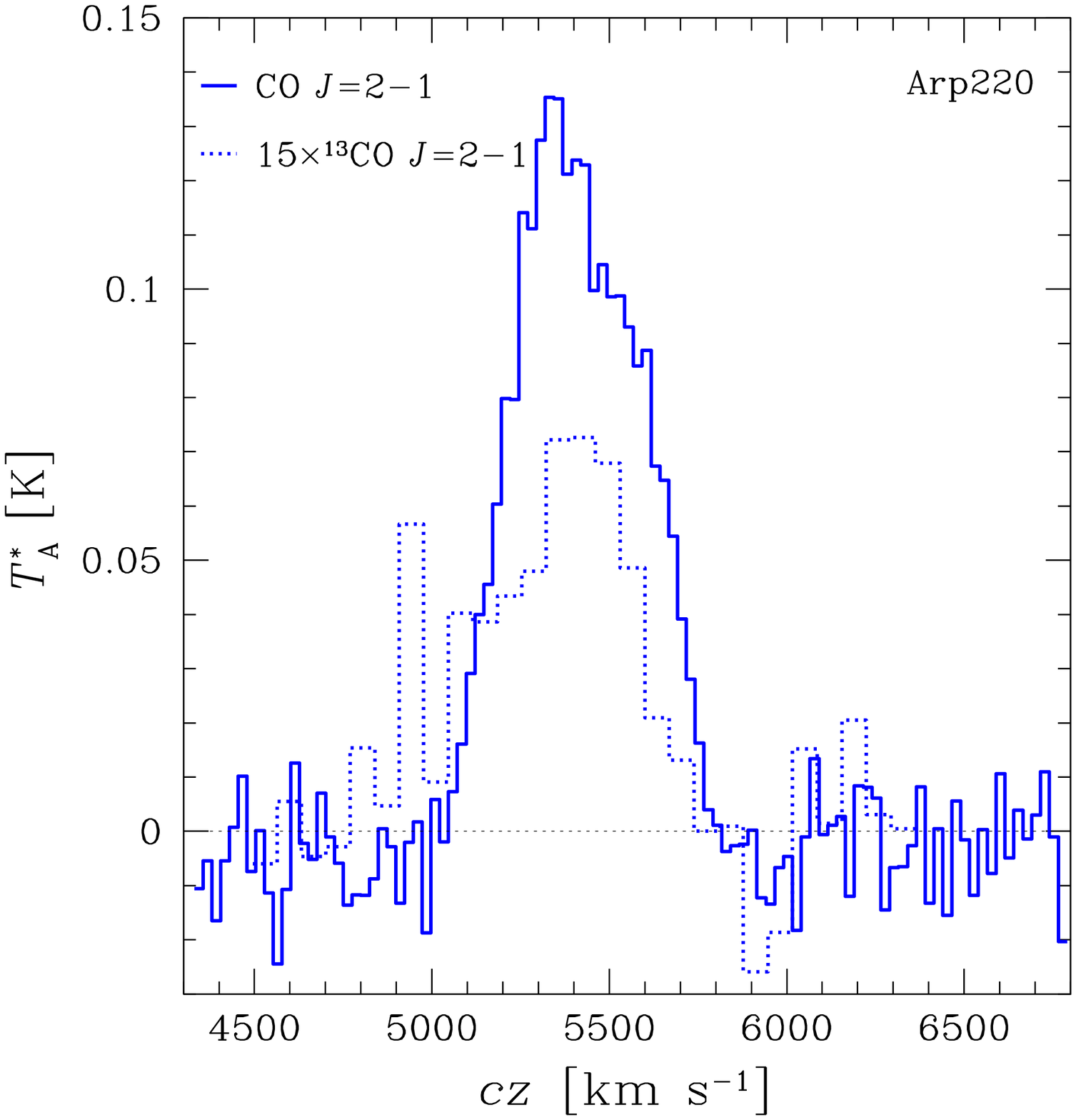} 
\includegraphics[width=0.325\textwidth]{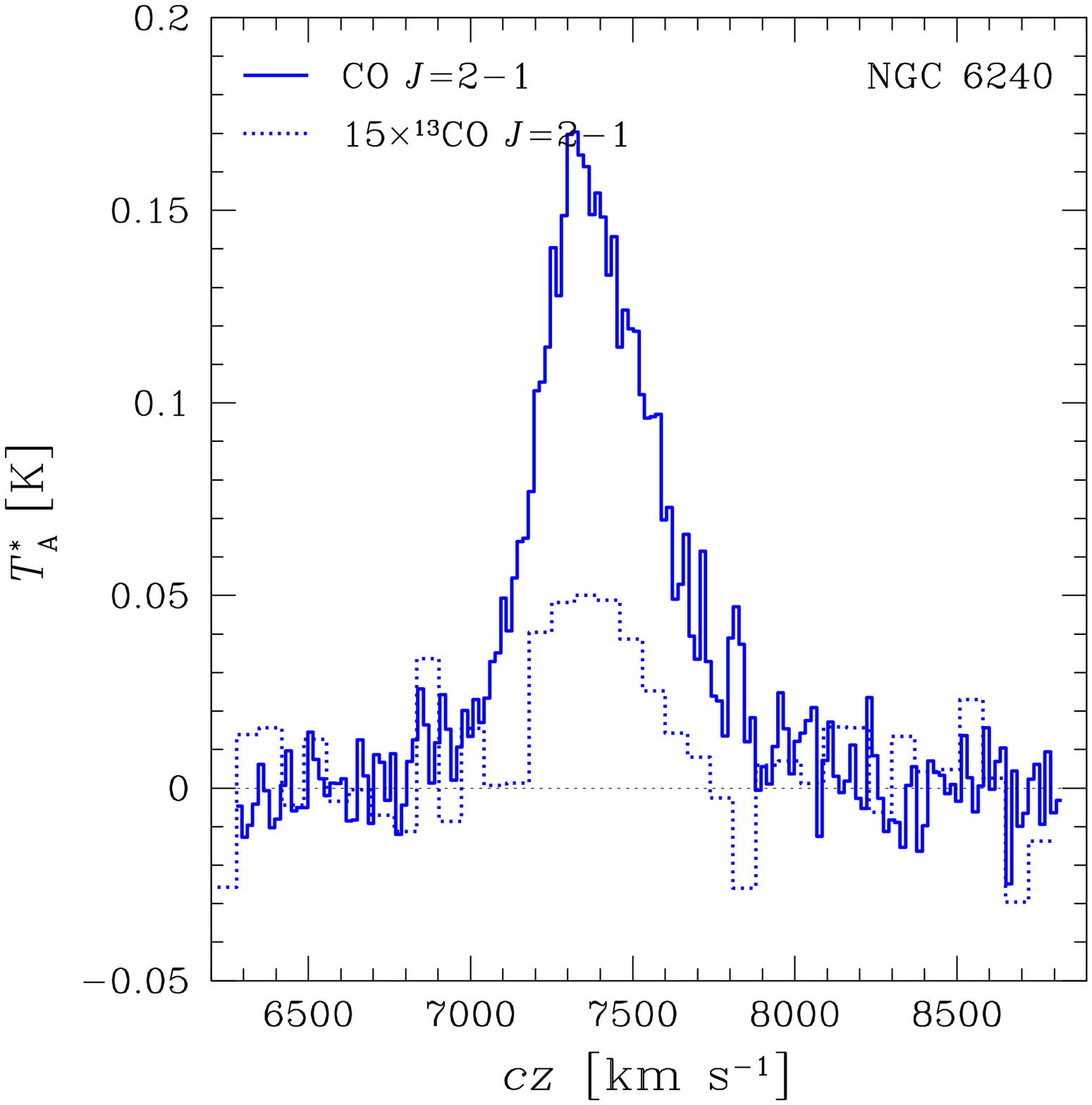}

\includegraphics[width=0.325\textwidth]{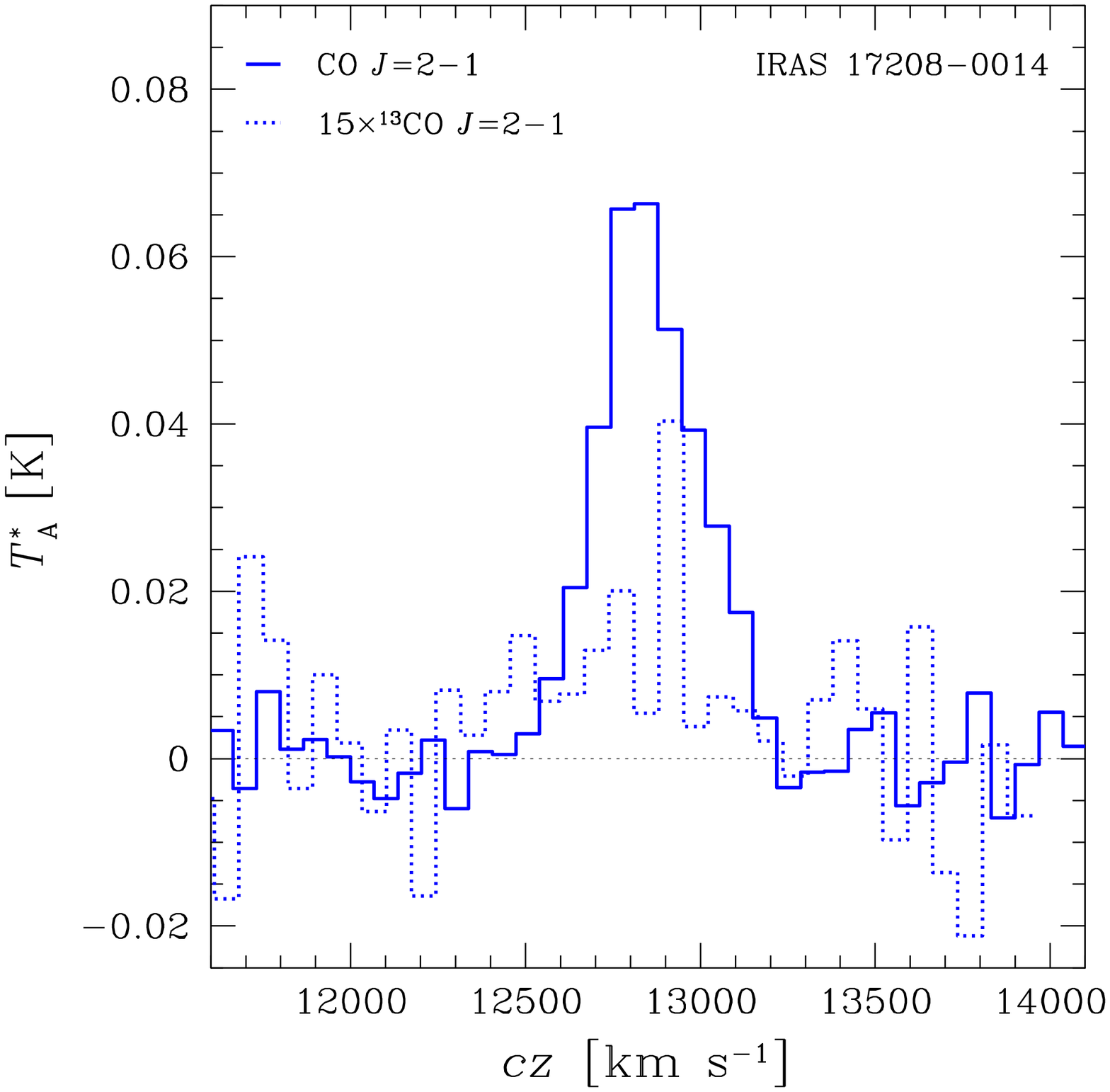} 
\includegraphics[width=0.325\textwidth]{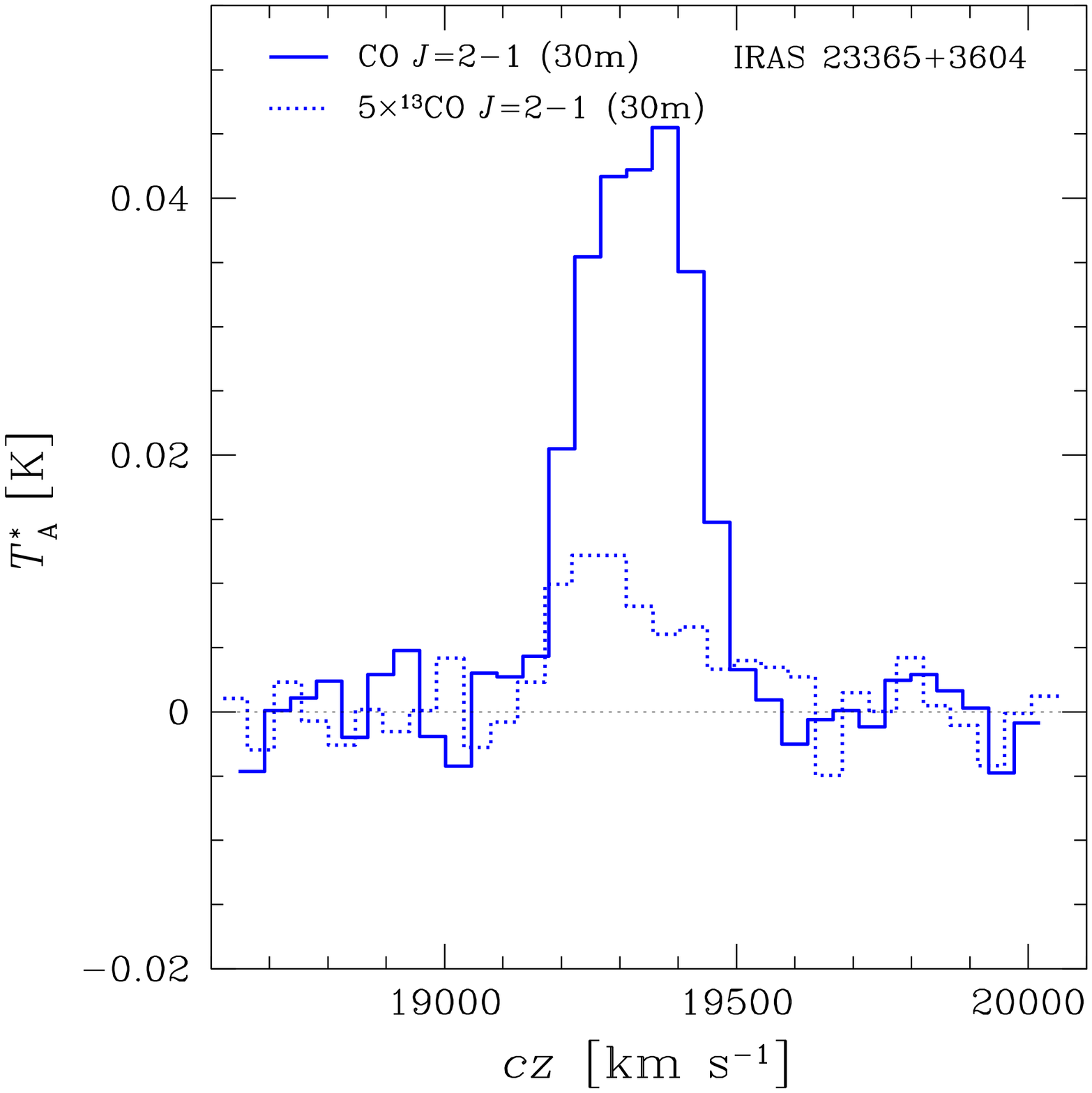} 
\includegraphics[width=0.325\textwidth]{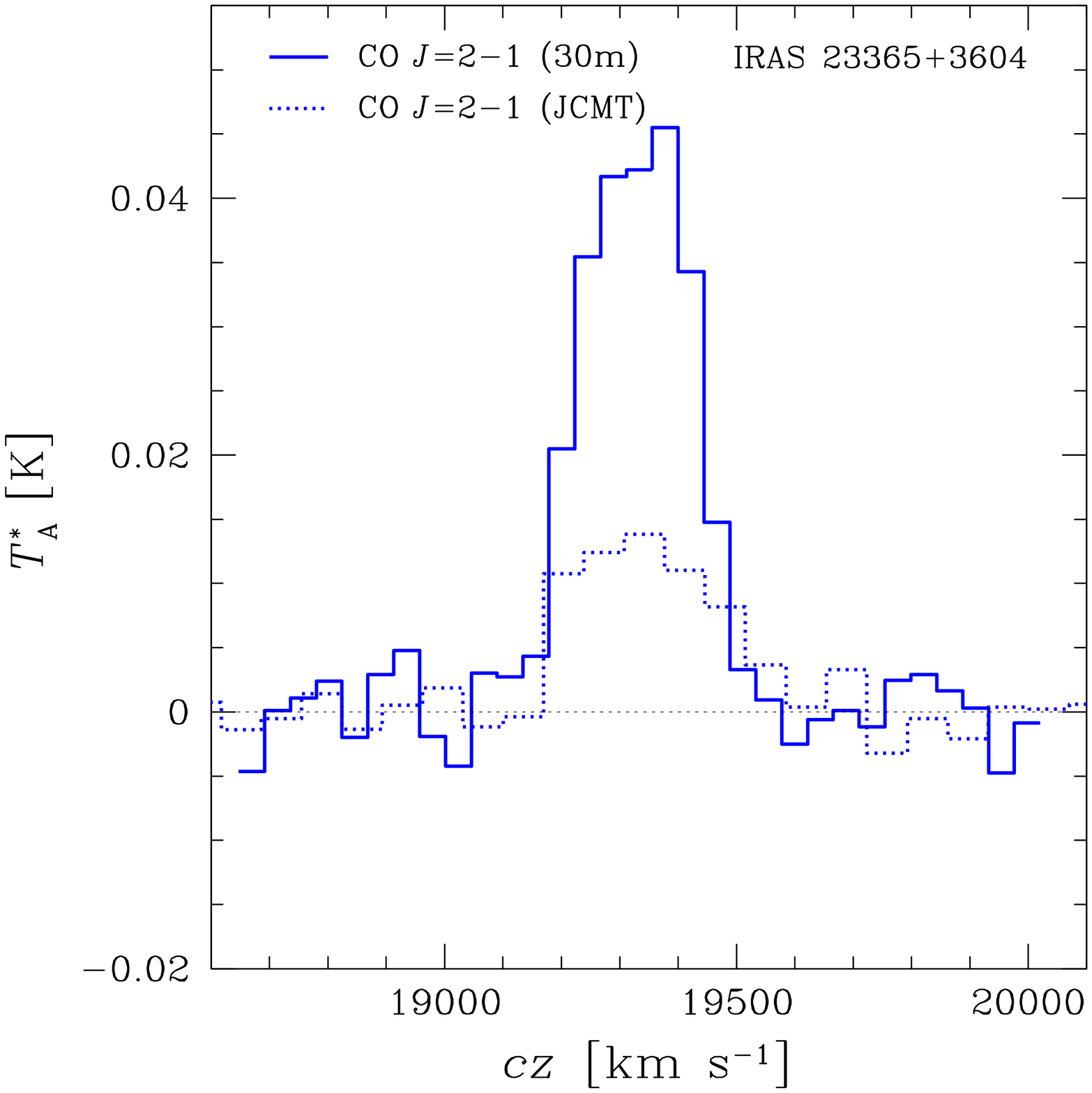}
                                                                
\contcaption{The CO, $^{13}$CO  line data.  The velocities are with
respect  to  $\rm  V_{opt}$=$\rm  cz_{co}$(LSR) (Table  3),  and  with
typical resolutions  $\rm \Delta V_{ch}$$\sim $(35--90)\,km\,s$^{-1}$.
A common color designated per transition is used in all frames. }
\end{figure*}

\section{Data reduction, line intensity estimates, literature data}

In both telescopes the output spectra  are in the $\rm T^{*} _A$ scale
(see   Kutner,  \&   Ulich   1981).   We   inspected  all   individual
10min/(4-6)min JCMT/IRAM spectra for  baseline ripples and to clip any
intensity ``spikes'' in individual  channels.  The edited spectra were
then  co-added  using  a  $1/\sigma ^2$-weighting  scheme  and  linear
baselines were  subtracted from  each final co-added  spectrum.  These
spectra  are shown in  Figures 2  and 3  and were  used to  derive the
velocity-integrated molecular line flux densities from

\begin{eqnarray}
\rm \rm S_{line} &=& \int _{\Delta V} S_{\nu } dV = \frac{8 k_B}{\eta ^{*}
_a  \pi  D^2}  K_c  (x)\int   _{\Delta  V}  T^*  _A  dV \nonumber \\
 &=&  \frac{\Gamma
(Jy/K)}{\eta ^{*} _a} K_c(x) \int _{\Delta V} T^* _A dV,
\end{eqnarray}

\noindent
where $\rm \Gamma  _{JCMT}$=15.62, $\rm \Gamma_{IRAM}$=3.905 and $\eta
^{*} _a$ is the aperture efficiency defined against the $\rm T^{*} _A$
scale  ($\eta^{*}  _a$=$\eta_a/\eta_{rss}$,   where  $\eta_a$  is  the
aperture efficiency measured against the  $\rm T'_A$ scale, as is more
typical, and $\rm \eta_{rss}$ is the rearward spillover and scattering
efficiency, Rohlfs \& Wilson (1996), Eqs 8.16, 8.17).  The factor $\rm
K_c (x) =x^2/(1-e^{-x^2})$, with x=$\theta _s/(1.2\theta _{HPBW})$ and
$\theta _s$=source  diameter, accounts  for the geometric  coupling of
the beam (its gaussian part) to a disk-like source, when a CO emission
size was available, and  $2\sigma _r$$\la $$\rm \theta _{s}$$\la $$\rm
\theta _{HPBW}$  where $\sigma _r$  is the pointing error  radius (see
4.1).   The total  point-source  conversion factors  $\rm S_{\nu  }/T$
adopted  for the  JCMT,  the IRAM  30-m  telescope, and  all the  data
gleaned  from the  literature  (for the  corresponding output  antenna
temperature scales) are comprehensively tabulated in Table~2.

\subsection{Aperture efficiencies, line intensity uncertainties, biases}

 Aperture efficiencies of  large high-frequency sub-mm telescopes such
as  the  JCMT  can  change  significantly  (especially  for  $\rm  \nu
$$\ga$460\,GHz) depending  on a  variety of factors  (e.g.,  elevation,
thermal relaxation of  the dish or its re-shaping  after an holography
session).   In order  to track  them  over an  decade of  observations
(during which the JCMT dish has been re-adjusted quite a few times) we
conducted frequent aperture  efficiency measurements using planets and
adopted the  average $\eta^*_a$ obtained  per observing  period for
deriving the  line fluxes  of all the  sources observed during  it. In
many cases, as a cross-check,  we distributed the measurements of very
CO-luminous LIRGs over several  widely separated periods, during which
very  different aperture  efficiencies (sometimes  up to  a  factor of
$\sim $2) were often measured. In  all such cases Equation 1, with the
appropriate  $\eta^*_a$  values,  yielded  velocity-integrated  line
fluxes   in   excellent   agreement.    Indicatively   most   aperture
efficiencies  measured  for  the  JCMT  lay  within  $\eta^*_a$$\sim
$0.41--0.56 (B-band,  315-350\,GHz), and $\eta^*  _a$$\sim $0.21--0.31
(C-band, 430-461\,GHz).   For the three periods of  the more demanding
W/D  band  observations we  derived  $\eta^*_{a}$=0.25 (2005),  0.32
(2009),  0.27 (2010) from  planetary measurements.   The uncertainties
for the reported velocity-integrated line flux densities have been
computed from

\begin{equation}
\rm \frac{\Delta S_{line}}{S_{line}}=
\left[\left(\frac{\delta T_{\Delta V}}{T_{\Delta V}}\right)^2 _{th}+
\left(\frac{\delta T_{\Delta V}}{T_{\Delta V}}\right)^2 _{cal}+
\left(\frac{\delta \eta}{\eta}\right)^2\right]^{1/2},
\end{equation}

\noindent
where  $\rm \delta  T  _{\Delta V}$  is  the stochastic  error of  the
average line  intensity $\rm  T_{\Delta V} $  (averaged over  the line
FWZI  $\rm \Delta V  $), $\eta  $ is  the telescope  efficiency factor
(used to derive the integrated line flux from the temperature scale of
the output  spectrum, e.g., $\eta^*_a$  for $\rm T^*_A$)\footnote{More
  precisely the  3nd term  in Equation 2  embodies the  uncertainty of
  total efficiency  factor $\eta_l  \eta_{fss}$ needed to  convert the
  $\rm T'_A$  scale (corrected only  for atmospheric extinction)  to a
  final  fully  corrected   (apart  from  source-beam  coupling)  $\rm
  T^*_{R}$  (or  $\rm  T_{mb}$)   scale.},  and  $\delta  \eta  $  its
uncertainty.  The  first term is  estimated from the spectra  shown in
Figures 2, 3 using

\begin{equation}
\rm \left(\frac{\delta T_{\Delta V}}{T_{\Delta V}}\right)_{th} =
\frac{\delta T_{chan}}{T_{\Delta V}}\left(\frac{N_{\Delta V}+
N_{bas}}{N_{\Delta V}N_{bas}}\right)^{1/2}, 
\end{equation}

\noindent
where $\rm  \delta T_{chan}$  is the stochastic  intensity dispersion,
estimated  from  the line-free  part  of  the  spectrum (for  a  given
velocity channel  width $\rm  \Delta V_{chan}$), while  $\rm N_{\Delta
  V}$=$\rm  \Delta V/\Delta V_{chan}$  and $\rm  N_{bas}$=$\rm 2\Delta
V_{bas}/\Delta V_{chan}$  are the number  of channels within  the line
FWZI  and  the line-free  baseline  (with  $\rm \Delta  V_{bas}/\Delta
V_{chan}$ channels  symmetrically around the  line) respectively.  The
second term in Equation 2 accounts for line calibration errors (due to
a  host of  factors such  as  imprecise knowledge  of the  calibration
loads,  uncertainties  in  the   atmospheric  model  and  the  derived
extinction  etc).   Observations  of  numerous  strong  spectral  line
standards and  planets during each observing  period yielded intensity
dispersions  of  $\sim  $15\%  ($\rm 230\,GHz$  and  $\rm  345\,GHz$),
$\sim$20\% ($\rm  460\,GHz$), and $\sim $25\%  ($\rm 690\,GHz$), which
we  adopt as  the combined  calibration (cal)  and  $\delta \eta/\eta$
uncertainties  per  observing band  at  the  JCMT.   For the  30-m  we
consider these to be $\sim $15\% for both 3\,mm and 1\,mm~bands.

Finally, even  with the accurate  tracking and pointing  achievable by
enclosed telescopes such as the JCMT, the residual rms pointing errors
and the narrow beams of large mm/sub-mm telescopes at high frequencies
can lead to a substantial  and systematic reduction of measured fluxes
of compact sources.  We try to  account for this as described in P10a,
by  applying a  $\rm \langle  G \rangle$  scaling factor  to  the line
fluxes of all point-like sources  with CO emission region diameters of
$\rm   \theta   _{co}$$\leq    $$\rm   2\sigma_{r}   $   (where   $\rm
\sigma_{r}(JCMT)$=2.5$''$  and  $\rm  \sigma_{r}(IRAM)$=3$''$ are  the
pointing error radii).  This factor is (see P10a):

\begin{equation}
\rm G(\sigma_r)=1+8\,ln2\left(\frac{\sigma_r}{\sqrt{2}\,
\theta _{1/2}}\right)^2,
\end{equation}

\noindent
where $\theta_{1/2}$ is the  beam HPBW, $\sigma_r/\sqrt{2}$ is the rms
pointing error per pointing coordinate.  For $\rm \theta_{co}$$ \leq
$$\rm  2\sigma_r$, $\rm  K_c(x)$ is  replaced  in Equation  1 by  $\rm
G(\sigma_r)$ as  the beam-source  coupling correction is  overtaken by
the pointing error correction.  At  345\,GHz and 460\,GHz for the JCMT
we  obtain  $\rm  \langle  G_{345}  \rangle$=1.087  and  $\rm  \langle
G_{460}\rangle$=1.15,  (with neglible  correction at  230\,GHz), while
for  the  30-m  $\rm  \langle G_{230}\rangle  $=1.20  (and  negligible
correction at  115\,GHz).  The  CO J=6--5 observations  with HPBW$\sim
$8$''$  are those  most  succeptible  to this  bias  and $\rm  \langle
G_{690} \rangle  $ has been estimated  from the pointing  rms {\it per
  observing    session},    yielding     a    range    $\rm    \langle
G_{690}\rangle$=1.17--1.37.   For  sources  with  CO (or  sub-mm  dust
emission)   sizes  of  $\rm   2\sigma_{r}$$\la  $$\rm   \theta  _{s}$$
\la  $$\theta _{1/2} $  the $\rm  K_c$ factor  is used  in Equation~1.
Finally  in the  cases  where  large offsets  were  found between  the
presumed CO source center and the observed positions in the literature
(see discussion in 4.2) we applied a ´beam-shift´ correction factor of
$\rm              K_{sh}$=$\rm             exp\left[4\,ln2\left(\Delta
  \theta/\Theta_{HPBW}\right)^2\right]$  where $\Delta \theta$  is the
(beam center)-source offset (see Table 4).

\subsection{Incorporating data from the literature, the final dataset}

A  detailed literature  search for  all  total CO  and $^{13}$CO  line
fluxes available  for LIRGs enlarged  our sample (see Table  3), while
allowing also a consistency check using the duplicate measurements per
object (especially for the J=1--0  transition).  In most cases we find
good  agreement among  the various  CO J=1--0  fluxes reported  in the
literature, and  between the (much  fewer) reported J=2--1,  3--2 line
fluxes     and    our     measurements,     within    the     expected
uncertainties\footnote{In   the   literature   the  uncertainties   of
  mm/sub-mm line measurements are  often underestimated, with only the
  thermal  rms error  reported (the  first term  in the  expression in
  Equation 2).  In all such cases we assumed a 15\% of calibration and
  $\eta  _{a}$ uncertainty in  addition to  the one  reported.}.  Most
cases  of serious discrepancies  among CO  line fluxes  were rectified
after accounting for the  different source positions often used.  This
typically occured because of  the positional uncertainties inherent in
the optical identification of heavily dust-obscured sources within the
large IRAS position error ellipse  (e.g.  Solomon et al.  1997) and/or
the  better   CO  positions  obtained   for  some  LIRGs   (e.g.   via
interferometry)  and used  for  subsequent CO  observations after  the
original J=1--0  detections with usually  large ($\sim $45$''$-50$''$)
beams  were made (e.g.   Sanders et  al.  1991;  Young et  al.  1995).
Moreover,  even when a  particular LIRG  is optically  identified, the
multi-component/interacting nature  of many such  systems can confound
the choice of a pointing  center for CO single dish observations (e.g.
Leech et al.  2010).  In  the cases where we found different positions
used   for   CO   observations   reported  in   the   literature,   we
``shift/scale'' the  corresponding line  fluxes to a  common position,
the interferometrically-derived  CO emission peak of the  LIRG (or its
near-IR  and/or radio continuum  peak if  the former  was unavailable)
assuming a source much smaller than  the beams used (the case for most
LIRGs).  This  rectified many of  the discrepant CO line  fluxes, with
only a  few inconsistent ones  remaining, mostly between new  data and
much older  CO observations (mainly  those reported in the  Sanders et
al.  1991 CO J=1--0 survey).   These are likely the result of improved
calibration  techniques  (e.g.   good  sideband  rejection),  pointing
accuracy, and faster beam-switching available for mm/sub-mm telescopes
now than in the~past.

\subsubsection{Resolving a large discrepancy: new CO J=1--0 observations of IRAS\,05189--2524}

The  largest  remaining CO  line  flux  discrepancy  is for  the  LIRG
IRAS\,05189--2524  with  $\rm  S_{10}$=$\rm  (91\pm  18)\,  Jy\,  km\,
s^{-1}$ obtained for  CO J=1--0 with the NRAO  12-m telescope (Sanders
et al.   1991) and $\rm S_{10}$=$\rm  (52\pm 10)\,Jy\,km\,s^{-1}$ with
SEST (Strong et  al.  2004).  The larger value  would place this ULIRG
to the  lowest end of gas  excitation of the entire  sample, with $\rm
r_{32}$=0.24  and  $\rm r_{21}$=0.32,  typical  for  the coldest  most
quiescent GMCs  found in the Galaxy  and M\,31 (Loinard  et al.  1995;
Allen  et  al.  1995;  Fixsen  et  al.~1999).   We used  the  12-meter
telescope\footnote{The  Kitt  Peak  12\,m  telescope is  operated  the
  Arizona Radio Observatory  (ARO), Steward Observatory, University of
  Arizona},  to re-observe the  CO J=1--0  line in  this ULIRG  on the
nights of  12, 13 and  24 March, 2008.   The final co-added  CO J=1--0
spectrum (Figure  4) was used  to estimate a  velocity-integrated line
flux  of   $\rm  S_{10}$=$\rm  (43\pm   8)\,Jy\,km\,s^{-1}$,  in  good
agreement with the value reported by  Strong et al. 2004 but less than
half that  reported by Sanders et  al.~1991. We report  the average of
ours and the Strong et al.  2004 value (Table 4).

\begin{figure}
\includegraphics[width=\columnwidth]{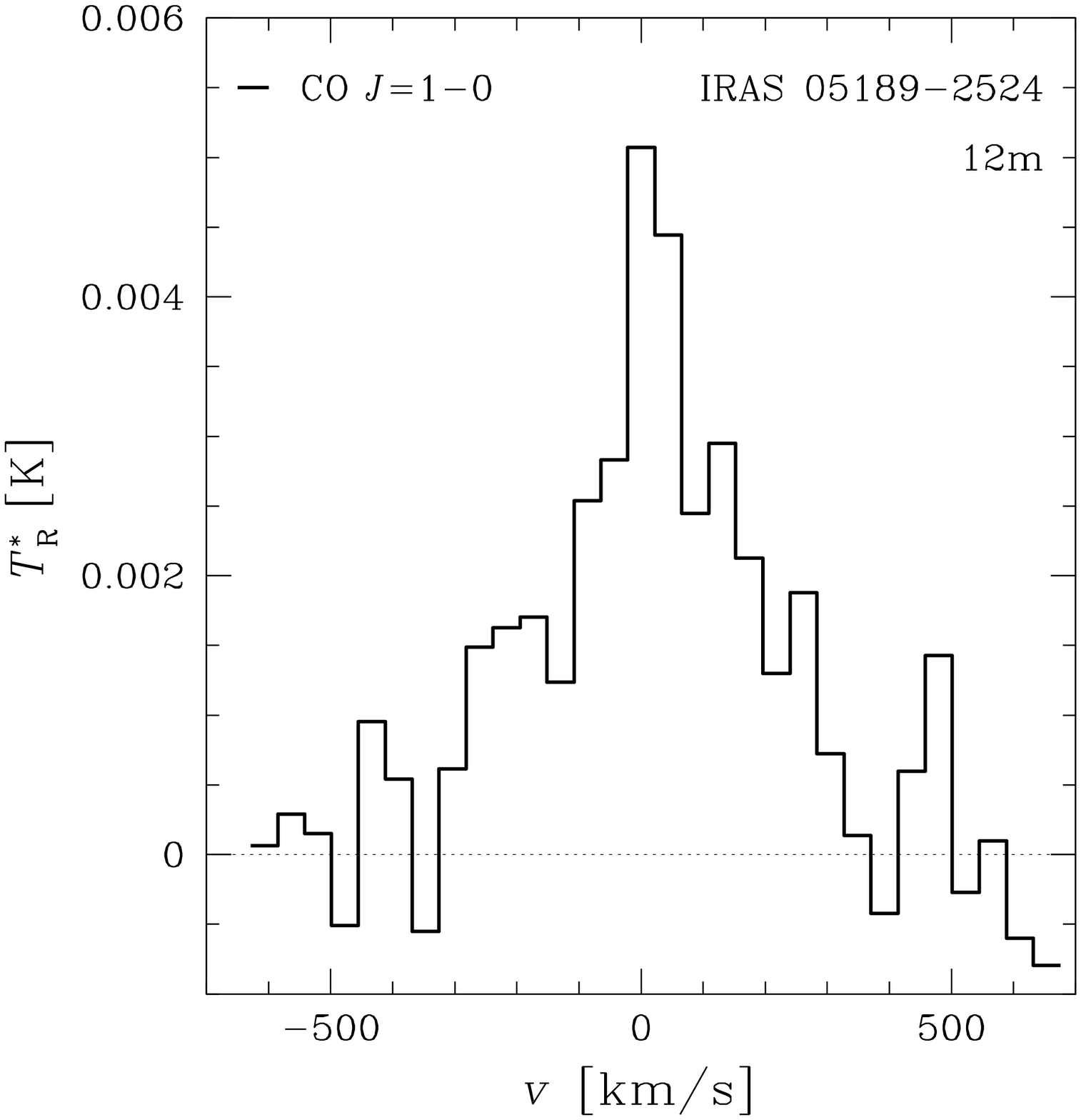}
\caption{The CO J=1--0 spectrum of IRAS\,05189-2524, centered at cz=12800\,km\,s$^{-1}$, obtained
with the 12\,m telescope (see section 4.2.1).}
\end{figure}

\newpage

\begin{table*}
\centering
\begin{minipage}{152mm}
\caption{Observational parameters of the combined LIRG sample}
\begin{tabular}{@{}lcclcll@{}}
\hline
Name$^{a}$ & RA (J2000)$^{b}$ & Dec (J2000)$^{b}$ & 
$z$ ($\rm D_L$)$^{c}$ & $(\Delta \theta_{\alpha}, \Delta \theta_{\delta})^{d}$ &
$\rm \langle \theta_{s}\rangle^{e}$ & Refs$^{f}$ \\  
\hline
00057+4021                          & 00 08 20.58 & $+$40 37 55.5  & 0.0445 (194.5) &
$(0'',0'')$ & $0.81''$(co)          & 1$\rm ^{p,m}$\\
00322--0840$^*$ (NGC\,157)          & 00 34 46.48 & $-$08 23 47.8  & 0.0055 (23.3)  &      
$(0'',0'')$ & $80''$(co,sm,x)       & 2$\rm ^{p,m}$\\
00509+1225 (I\,Zw\,1, PG\,0050+124) & 00 53 34.92 & $+$12 41 35.5  & 0.0611 (270.3) &
$(0'',0'')$ & $\la 12''$(co)        & 3$^{\rm m}$,4$^{\rm p}$\\
01053--1746$^*$ (Arp\,236)          & 01 07 47.00 & $-$17 30 24.0  & 0.0200 (85.8)  & 
$(0'',0'')$ & $30''$(sm,x)          & 5$^{\rm p}$,6$^{\rm m}$\\
01077--1707$^*$                     & 01 10 08.20 & $-$16 51 11.0  & 0.0351 (152.3) &
$(0'',0'')$ & $\la 15''$(sm)        & 5$^{\rm p}$,6$^{\rm m}$\\
01418+1651$^*$ (III\,Zw\,35)        & 01 44 30.50 & $+$17 06 08.0  & 0.0274 (118.2) & 
$(0'',0'')$ & $\la 15''$(sm)        & 5$^{\rm p}$,6$^{\rm m}$\\
02071+3857 (NGC\,828, VI\,Zw\,177)  & 02 10 09.43 & $+$39 11 26.3  & 0.0178 (76.2)  &
$(0'',0'')$ & $14''$(co)            & 7$^{\rm p,m}$,8$^{\rm m}$\\
02080+3725$^*$ (NGC\,834)           & 02 11 01.55 & $+$37 40 01.3  & 0.0154 (65.8)  &
$(-2.5'',-2.0'')$ & $13''$(cm)      & 9$^{\rm p}$,10$^{\rm m}$\\
02114+0456$^*$ (Mrk\,1027)          & 02 14 05.60 & $+$05 10 27.7  & 0.0297 (128.3) &
$(0'',0'')$ & $17''$(sm,x)          & 5$^{\rm p}$,6$^{\rm m}$\\
02321--0900 (NGC\,985, Mrk\,1048)   & 02 34 37.74 & $-$08 47 14.7  & 0.0430 (187.7) & 
$(0'',0'')$ & $22''$(co)            & 11$^{\rm p,m}$\\
02401--0013$^*$ (NGC\,1068)         & 02 42 40.74 & $-$00 00 47.6  & 0.0037 (13.3)  &
$(0'',0'')$ & $40''$(co,sm,x)       & 12$^{p,m}$\\
02483+4302                          & 02 51 36.01 & $+$43 15 10.8  & 0.0514 (225.8) & 
$(0'',0'')$ & $1.75''$(co)          & 1$^{\rm p,m}$\\
02512+1446$^*$ (UGC\,2369)          & 02 54 01.80 & $+$14 58 14.0  & 0.0312 (135.0) & 
$(0'',0'')$ & $30''$(sm,cm,x)       & 5$^{\rm p}$,6$^{\rm m}$,13$^{\rm m}$\\
03359+1523                          & 03 38 46.90 & $+$15 32 55.0  & 0.0353 (153.2) &
$(0'',0'')$ & $4.5''$(cm)           & 5$^{\rm p}$,13$^{\rm m}$\\
04232+1436                          & 04 26 04.94 & $+$14 43 37.9  & 0.0796 (356.4) & 
$(-1.13'', +0.4'')$& $\la 8''$(cm,sm,x) & 14$^{\rm p}$,6$^{\rm m}$,13$^{\rm m}$ \ \ \\
05083+7936 (VII\,Zw\,031)           & 05 16 46.51 & $+$79 40 12.5  & 0.0543 (239.0) & 
$(0'',0'')$       & $2.3''$(co)     & 1$^{\rm p,m}$\\
05189--2524                         & 05 21 01.11 & $-$25 21 45.9  & 0.0427 (186.4) &
$(+4.0'',+1.0'')$ & $\la 3''$(cm,ir)& 9$^{\rm p}$,15$^{\rm m}$,13$^{\rm m}$\\
08030+5243                          & 08 06 50.10 & $+$52 35 05.4 & 0.0835 (375.4)  &
$(0'',0'')$       & $\la 0.8''$(ir) & 14$^{\rm p}$,16$^{\rm m}$\\
08354+2555$^{*}$ (NGC\,2623, Arp\,243)    & 08 38 24.10 & $+$25 45 16.5 & 0.0185 (79.3)   &
$(0'',0'')$       & $1.65''$(co)    & 17$^{\rm p,m}$\\
08572+3915$^{g}$         & 09 00 25.41 & $+$39 03 54.1 & 0.0582 (256.9)  & 
$(0'',0'')$       & $\la 2.1''$(co) & 18$^{\rm p,m}$\\
09126+4432 (Arp\,55)                & 09 15 54.90 & $+$44 19 54.4 & 0.0399 (173.3)  &
$(0'',0'')$      & $1.9'', 12''$(co,cm,dbl)& 5$^{p}$, 13$^{\rm m}$,19$^{\rm m}$\\
09320+6134 (UGC 05101)              & 09 35 51.53 & $+$61 21 11.6 & 0.0393 (171.1) &
$(0'',0'')$      & $1''$-$2''$(ir,co)    & 14$^{\rm p}$,15$^{\rm m}$,20$^{\rm m}$\\
09586+1600$^*$ (NGC\,3094)          & 10 01 26.00 & $+$15 46 14.0 & 0.0080 (34.0)  & 
$(-2.2'',-1.0'')$ & $11''$(ir,cm)   & 21$^{\rm p}$,22$^{\rm m}$,13$^{\rm m}$ \\
10039--3338$^*$ (IC\,2545)          & 10 06 04.50 & $-$33 53 03.0 & 0.0341 (147.9) &
$(0'',0'')$      & $\la 15''$(sm)     & 5$^{\rm p}$,6$^{\rm m}$\\
10035+4852                          & 10 06 45.83 & $+$48 37 46.1 & 0.0648 (287.5) &
$(+5.2'',+2.2'')$& $\la 15''$(sm)       & 14$^{\rm p}$,6$^{\rm m}$\\
10173+0828                          & 10 20 00.19 & $+$08 13 34.5 & 0.0489 (214.4) &
$(0'',0'')$ & $\la 3''$(co,cm)      & 23$^{\rm p,m}$,13$^{\rm m}$\\
10190+1322                          & 10 21 42.60 & $+$13 06 54.4 & 0.0765 (342.2) &
$(0'',0'')$ & $1.2'', 4''$(co,dbl)  & 24$^{\rm p,m}$\\
10356+5345$^*$ (NGC\,3310)          & 10 38 45.90 & $+$53 30 11.7 & 0.0033 (14.0)  &
$(0'',0'')$ &   $50''$(co,sm,x)     & 2$^{\rm p,m}$\\
10565+2448                          & 10 59 18.15 & $+$24 32 34.4 & 0.0428 (188.2) &
$(0'',0'')$ & $1.5''$(co)           & 1$^{\rm p,m}$\\ 
11191+1200 (PG\,1119+120)           & 11 21 47.12 & $+$11 44 18.3 & 0.0500 (219.4) &
$(0'',0'')$ & $\la 5'' $(co)        & 25$^{\rm p,m}$\\
11231+1456$^*$ (IC\,2810, UGC\,6436)& 11 25 45.00 & $+$14 40 36.0 & 0.0341 (147.9) &
$(+1.2'',0'')$ & $ 6''$(cm,sm)      & 21$^{\rm p}$,13$^{\rm m}$,26$^{\rm m}$\\
11257+5850 (Arp\,299)               & 11 28 32.45 & $+$58 33 45.8 & 0.0103 (43.8)  &
$(0'',0'')$   & $35''$(co,sm,x)     & 27$^{\rm p,m}$,6$^{\rm m}$\\
12001+0215$^*$ (NGC\,4045)          & 12 02 42.30 & $+$01 58 38.0 & 0.0066 (28.0)  &
$(-0.9'',-1.2'')$ & $7''$(cm)       & 21$^{\rm p}$,13$^{\rm m}$\\
12112+0305                          & 12 13 45.77 & $+$02 48 39.3 & 0.0727 (324.4) &
$(+3.9'',+2.1'')$ &$\la 2''\,(2.9'')$(co,dbl)& 9$^{\rm p}$,18$^{\rm m}$\\
12224--0624$^*$                     & 12 25 03.90 & $-$06 40 53.0 & 0.0263 (113.4) &
$(0'',0'')$      & $\la 2''$(cm)    & 21$^{\rm p}$,13$^{\rm m}$\\
12243-0036$^*$ (NGC\,4418)          & 12 26 54.70 & $-$00 52 39.0 & 0.0073 (31.0)  &
$(0'',0'')$ & $\la 3''$(cm)         & 21$^{\rm p}$,13$^{\rm m}$\\
12540+5708 (Mrk\,231)               & 12 56 14.21 & $+$56 52 25.1 & 0.0422 (184.1) &
$(0'',0'')$ & $0.85''$(co)          & 1$^{\rm p,m}$\\
13001--2339$^*$                     & 13 02 52.10 & $-$23 55 19.0 & 0.0215 (92.3)  &
$(0'',0'')$ & $2''$(ir)             & 5$^{\rm p}$,22$^{\rm m}$\\
13102+1251$^*$ (NGC\,5020)          & 13 12 39.90 & $+$12 35 59.0 & 0.0112 (47.7)  &
$(-1.0'',-0.9'')$& $12''$(cm)       & 21$^{\rm p}$,13$^{\rm m}$\\
Arp\,238$^*$ (UGC\,08335)           & 13 15 30.20 & $+$62 07 45.0 & 0.0315 (136.3) &
$(0'',0'')$ & $5''\,(35'')$(cm,dbl) & 5$^{\rm p}$, 10$^{\rm m}$\\
13183+3423 (Arp\,193)               & 13 20 35.32 & $+$34 08 22.2 & 0.0233 (100.2) &
$(0'',0'')$ & $1.5''$(co)           & 1$^{\rm p,m}$\\
13188+0036$^*$ (NGC\,5104)          & 13 21 23.10 & $+$00 20 32.0 & 0.0186 (79.7)  &
$(0'',0'')$ & $2.5''$(cm)           & 21$^{\rm p}$,13$^{\rm m}$\\
13229--2934 (NGC\,5135)             & 13 25 43.97 & $-$29 50 01.3 & 0.0136 (58.0)  &
$(0'',0'')$ & $6''$(cm)             & 10$^{\rm p,m}$\\
13362+4831 (NGC\,5256)              & 13 38 17.90 & $+$48 16 41.0  & 0.0278 (120.0) &
$(0'',0'')$ & $8''$(cm, sm)         & 5$^{\rm p}$,13$^{\rm m}$,6$^{\rm m}$\\
13428+5608 (Mrk\,273)               & 13 44 42.12 & $+$55 53 13.5 & 0.0378 (164.4) &
$(0'',0'')$ & $3''$(co)             & 1$^{\rm p,m}$\\
13470+3530$^*$ (UGC\,8739)          & 13 49 14.20 & $+$35 15 23.0 & 0.0168 (71.9)  &
$(-2.5'',+1.5'')$ & $11''$(cm,x)    & 21$^{\rm p}$,13$^{\rm m}$\\
F13500+3141 (3C\,293)               & 13 52 17.82 & $+$31 26 46.4 & 0.0446 (194.9) &
$(0'',0'')$ & $7''$(co)             & 28$^{\rm p,m}$\\
F13564+3741$^*$ (NGC\,5394)         & 13 58 33.60 & $+$37 27 13.0 & 0.0125 (53.3)  &
$(0'',0'')$ & $5''$(cm)             & 21$^{\rm p}$,13$^{\rm m}$\\
14003+3245$^*$ (NGC\,5433)          & 14 02 36.00 & $+$32 30 38.0 & 0.0145 (61.9)  &
$(0'',0'')$ & $8.5''$(cm)           & 21$^{\rm p}$,13$^{\rm m}$\\
14151+2705$^*$ (Mrk\,673)           & 14 17 21.00 & $+$26 51 28.0 & 0.0366 (159.0) &
$(0'',0'')$ & $\la 10''$(opt)       & 5$^{\rm p,m}$\\
14178+4927$^*$ (Zw\,247.020, Mrk\,1490)& 14 19 43.20 & $+$49 14 12.0 &0.0256 (110.3)&
$(0'',0'')$ & $2''$(cm)             & 21$^{\rm p}$,13$^{\rm m}$\\
14280+3126$^*$ (NGC\,5653)          & 14 30 10.40  & $+$31 12 54.0 & 0.0119 (50.7) &
$(-2.0'',+1.6'')$ & $17''$(cm)      & 21$^{\rm p}$, 13$^{\rm m}$\\
14348--1447                         & 14 37 38.32 & $-$15 00 22.7 & 0.0825 (370.7) &
$(0'',0'')$ & $\la 2'',3.5''$(co,dbl)& 29$^{\rm p,m}$,13$^{\rm m}$\\
15107+0724 (Zw\,049.057)            & 15 13 13.07 & $+$07 13 32.0 & 0.0129 (55.0)  &
$(0'',0'')$ & $5''$(co)             & 23$^{\rm p,m}$\\
15163+4255$^*$ (Mrk\,848, Zw\,107)  & 15 18 06.20 & $+$42 44 42.0 & 0.0402 (175.1) &
$(0'',0'')$ & $7''$(sm,cm)       & 5$^{\rm p}$, 6$^{\rm m}$, 13$^{\rm m}$\\  
15243+4150$^*$ (NGC\,5930, Arp\,090)& 15 26 07.90 & $+$41 40 34.0 & 0.0089 (37.8)  &
$(0'',0'')$ & $2.5''$(cm)           & 21$^{\rm p}$,13$^{\rm m}$\\
15322+1521$^*$ (NGC\,5953p, Arp\,091)& 15 34 32.30 & $+$15 11 38.0 & 0.0065 (27.6) &
$(0'',0'')$ & $10''$(cm)            & 21$^{\rm p}$,13$^{\rm m}$\\ 
15327+2340 (Arp\,220)               & 15 34 57.24 & $+$23 30 11.2 & 0.0182 (78.0)  &
$(0'',0'')$ & $1.8''$(co)           & 1$^{\rm p,m}$\\
15437+0234$^*$ (NGC\,5990)          & 15 46 16.50 & $+$02 24 56.0 & 0.0128 (54.6)  &
$(-2.1'', -0.8'')$ & $11''$(cm)     & 21$^{\rm p}$,13$^{\rm m}$\\
16104+5235$^*$ (NGC\,6090, Mrk\,496)& 16 11 40.70 & $+$52 27 25.0 & 0.0292 (126.1) &
$(0'',0'')$ & $6''$(cm)             & 5$^{\rm p}$,13$^{\rm m}$\\
16284+0411$^*$ (MCG\,+01-42-008)    & 16 30 56.50 & $+$04 04 59.0 & 0.0245 (105.5) &
$(0'',0'')$ & $3.9''$(cm)           & 21$^{\rm p}$,13$^{\rm m}$\\
16504+0228 (NGC\,6240)              & 16 52 59.05 & $+$02 24 05.8 & 0.0243 (104.6) &
$(0'',0'')$ &  $3''$(co)            & 7$^{\rm p}$,30$^{\rm m}$\\
17132+5313                          & 17 14 20.48 & $+$53 10 31.4 & 0.0507 (222.6) &
$(0'',+1.0'')$& $2.4''$(cm)         & 31$^{\rm p}$,13$^{\rm m}$\\
17208--0014                         & 17 23 21.92 & $-$00 17 00.7 & 0.0428 (186.8) &
$(0'',0'')$ & $1.7''$(co)           & 1$^{\rm p,m}$\\
\hline
\end{tabular}
\end{minipage}
\end{table*}

\begin{table*}
\centering
\begin{minipage}{152mm}
\contcaption{Observational parameters of the combined LIRG sample}
\begin{tabular}{@{}lcclcll@{}}
\hline
Name$^{a}$ & RA (J2000)$^{b}$ & Dec (J2000)$^{b}$ & 
$z$ ($\rm D_L$)$^{c}$ & $(\Delta \theta_{\alpha}, \Delta \theta_{\delta})^{d}$ &
$\rm \langle \theta_{s}\rangle^{e}$ & Refs$^{f}$ \\  
\hline
18425+6036$^*$ (NGC\,6701)          & 18 43 12.27 & $+$60 39 10.5 & 0.0132 (56.3)  &
$(+1.5'',+2.0'')$ &  $15''$(cm)     & 9$^{\rm p}$,10$^{\rm m}$\\
19458+0944                          & 19 48 15.47 & $+$09 52 01.3 & 0.1000 (454.8) & 
$(0'',0'')$ & $\la 0.8''$(ir)       & 14$^{\rm p}$,16$^{\rm m}$\\
20550+1656$^*$ (II\,Zw\,96)         & 20 57 23.70 & $+$17 07 44.0 & 0.0363 (157.7) &
$(0'',0'')$ & $20''$(sm,x)          & 5$^{\rm p}$,6$^{\rm m}$\\
22491--1808                         & 22 51 49.86 & $-$17 52 24.4 & 0.0773 (346.0) &
$(-7.3'',0'')$ & $2.5''$(ir,cm,x)   & 9$^{\rm p}$,15$^{\rm m}$,13$^{\rm m}$\\
23007+0836 (NGC\,7469)              & 23 03 15.60 & $+$08 52 26.3 & 0.0163 (69.7)  &
$(0'',0'')$ & $8''$(co)             & 32$^{\rm p,m}$\\
23365+3604                          & 23 39 01.25 & $+$36 21 08.4 & 0.0644 (285.6) &
$(0'',0'')$ & $0.95''$(co)          & \ \ \ \ \ 1$^{\rm p,m}$ \ \ \ \ \ \\
\hline
\end{tabular}
$^{a}$IRAS  name and the most common alternative(s), the
asterisk marks sources for which  CO line fluxes were obtained from an extensive 
literature search and line flux rectification process (see sections 2, 4.2).\\
$^{b}$Source coordinates (${\bf R_{beam}}$) used for CO observations.\\
$^{c}$The redshift used for receiver tuning (or the  $\rm z_{co}$  
reported in the literature for LIRGs not in the original sample), and the 
corresponding luminosity distance in Mpc.\\
$^{d}$Offsets between  expected CO source center ${\bf R_{cm}}$ (assumed
coincident with peak cm continuum) and  observed position: ${\bf \Delta R}$=${\bf R_{cm}-R_{beam}}$
(${\bf  \Delta R}$={\bf 0} when available CO or sub-mm images  defined the CO source 
center,  see 4.2). Thus ${\bf R_{cm}}$ marks the true CO source center
whenever ${\bf \Delta R}$$\neq ${\bf 0}.\\
$^{e}$CO region angular size $\rm\langle \theta _{s}\rangle$=$
 \rm (\theta _{min}\theta _{maj})^{0.5}$, from interferometric maps. If these were not
 available then cm, near-IR, or sub-mm continuum images were used to set  upper limits on
 $\rm\langle \theta _{s}\rangle$. 
The qualifiers for the images used  are: (co), (cm), (ir), (sm) for CO, cm, IR, and
 sub-mm images, (x)=complex source morphology ($\rm \langle \theta_s \rangle$ then
 denotes the overall source size),  (dbl)=double CO-bright nuclei. 
In the latter case  $\rm \langle \theta _{s} \rangle$ refers to the largest one, and the
second number denotes their~separation.\\
$^{f}$References used for source position (=p) and morphology/size (=m)
 information:
 1=Downes \& Solomon 1998; 2=Zhu et al.\ 2009; 3=Schinnerer et al.\ 1998;
 4=Eckart et al.\ 1994; 5=Leech et al.\ 2010; 6=Mortier et al.\ 2009; 7=Wang et al.\ 1991;
 8=Casoli et al.\ 1992; 9=Sanders et al.\ 1991; 10=Condon et al.\ 1996; 11=Appleton et al.\ 2002; 
 12=Papadopoulos \& Seaquist 1998a; 13=Condon et al.\ 1990; 14=Solomon et al.\ 1997;
 15=Scoville et al.\ 2000; 16=Murphy et al.\ 1996;  17=Bryant \& Scoville 1999;  18=Evans et al.\ 2002;
 19=Sanders et al.\ 1988; 20=Wilson et al.\ 2008; 21=Yao et al.\ 2003; 22=Zenner \& Lenzen 1993;
 23=Planesas et al.\ 1991; 24=Gracia-Carpio et al.\ 2007; 25=Evans et al.\ 2001; 
 26=Lisenfeld et al.\ 2000; 27=Aalto et al.\ 1997; 28=Evans et al.\ 1999; 
 29=Evans et al.\ 2000; 30=Tacconi et al.\ 1999; 31=Young et al.\ 1995; 32=Davies et al.\ 2004;33=Crawford
 et al.\ 1996\\
$^{g}$Double nuclei in near-IR  (Scoville et al.\ 2000), but only one is
 CO-bright (Evans et al.\ 2002).
\end{minipage}
\end{table*}

\begin{table*}
\centering
\begin{minipage}{122mm}
\caption{$ ^{12}$CO J=1--0, 2--1, 3--2 data for the combined LIRG sample}
\begin{tabular}{@{}lcccl@{}}
\hline
Name & CO J=1--0$^{a}$ & CO J=2--1$^{a}$ & CO J=3--2$^{a}$ & Refs.$^{b}$\\  
\hline
00057+4021     & $46\pm 7$ & $178\pm 30 $ & $267\pm 45$(1.087, G)& x,1\\
00322--0840$^*$&$500\pm 125^{c}$&$2000\pm 500^{c}$&$2610\pm 650^{c}$& 2\\
00509+1225    &$34\pm 7$(1.11, K$_c$)&$114\pm 23$(1.11, K$_c$)& $356\pm 103$(1.27, K$_c$) &  x,3\\
01053--1746$^*$& $691\pm 138$&    & $3324\pm 660$       & 4,5\\
01077--1707$^*$& $178\pm 36$&     & $507\pm 100 $       & 4,5\\
01418+1651$^*$& $75\pm 15$  &     & $417\pm 83$         & 4,5\\
02071+3857    & $408\pm 33$ & $1238\pm 248^{d}$ &$2560\pm 515 $& x,6\\
02080+3725$^*$& $147\pm 29$ &    & $586\pm 115$(1.052,K$_{\rm sh}$) & 6,7\\
02114+0456$^*$& $159\pm 32$ &   & $1196\pm 240 $ & 4,5\\
02321--0900 &$49\pm 6$ &$215\pm 35$(1.387, K$_c$)& $245\pm 55$(2.087, K$_c$)$^{e}$& x,8,9\\
02401--0013$^*$& $2830\pm 424$& $(1.13\pm0.22)\times 10^4$ & $(1.71\pm 0.34)\times 10^4$ & 10$^{f}$\\ 
02483+4302     & $30\pm 4$    & $116\pm 21$  & $119\pm 17$(1.087, G) &  x,1,4,11\\
02512+1446$^*$ &  $194\pm 39$ &        & $387\pm 77$ & 4, 5\\
03359+1523     & $155\pm 23^{g}$&    & $256\pm 51$ & 4,7\\
04232+1436     & $34\pm7$      &       & $335\pm 76$ &  x,11\\
05083+7936     & $87\pm 9$     & $252\pm 33$ & $340\pm 145$(1.087, G) & x,1,11,12\\
05189--2524    & $48\pm 7$& $130\pm22$(1.10, K$_{\rm sh}$) & $256\pm 36$(1.27, K$_{\rm sh}$)& x\\
08030+5243     & $30\pm 6$     & $77\pm 12$  & $98\pm 17 $(1.087, G) &  x,11\\
08354+2555$^*$ & $162\pm 20$   & $267\pm40^{i}$ & $614\pm 92$ &  7,13,14,15\\
08572+3915     & $10.5\pm 1.5 $& $41\pm 12$  & $\la 240^{h}$   &  x,11,16\\
09126+4432     & $162\pm 23$   & $779\pm 126$(3.08, K$_{c}$)$^{j}$& $775\pm 130$ & x,4,7,14,17\\
``  `` (SW nucleus) &    &       & $540\pm 61$                  & x, 4\\
``  `` (NE nucleus) &    &       & $235\pm 40$                  & x, 4\\
09320+6134     & $70\pm 14$    & $345\pm 62$  & $589\pm 95$(1.087, G) &  x,11\\
09586+1600$^*$ & $126\pm 26$(1.086,K$_{\rm sh}$) &   & $837\pm 126$(1.086,K$_{\rm sh}$) &  x,6,18\\
10039--3338$^*$& $65\pm 14$   &        & $316\pm 65$ & 4, 5\\
10035+4852     & $48\pm 10$(1.20, K$_{\rm sh}$) & $168\pm 29$(1.20, K$_{\rm sh}$) &   & x,11\\
10173+0828     & $56\pm 10$   &         & $127\pm 24$ (1.087, G)         &  x,7,18\\
10190+1322     & $37\pm 7$    & $100\pm 20$(1.06, K$_c$) & $225\pm 37$         &  x,4,19\\
10356+5345$^*$ & $140\pm 28^{c}$    & $822\pm 166$   & $1525\pm 300^{c}$ & 2\\
10565+2448     & $77\pm 8$    & $327\pm 35$   & $560\pm 56$ & x,1,4,7,11,18,20\\
11191+1200     & $4.5\pm 0.8$ &         & $22\pm 5$(1.045, K$_c$) & x,21\\
11231+1456$^*$ & $145\pm 20$  &         & $413\pm 87$(1.07, K$_c$) & 7,18\\
11257+5850     & $586\pm 115$ &         & $4360\pm 655$ &  x,4,5,15,22\\
12001+0215$^*$ & $138\pm 28 $ &         & $311\pm 65$    & 18\\
12112+0305     & $42\pm 5$    & $152\pm 30$(1.45, K$_{\rm sh}$) & $598\pm 113$(1.26, K$_{\rm sh}$) & x,7,16\\
12224--0624$^*$& $22\pm 9$    &         & $235\pm 52$    & 18\\
12243--0036$^*$& $132\pm 28$  &         & $995\pm 204$   & 18\\
12540+5708     & $88\pm 9$    & $315\pm 30$   & $568\pm 80$    & 23\\
13001--2339$^*$& $152\pm 30$  &         & $864\pm 173$   & 4,5\\ 
13102+1251$^*$ & $94\pm 19$   &         & $424\pm 90$    & 18\\
Arp\,238$^*$   & $110\pm 22$  &         & $300\pm 60$    & 4, 5\\
13183+3423     & $194\pm 16$  & $850\pm 130$  & $1294\pm 171$(1.087, G) & x,1,7,11,13,18,24,25\\
13188+0036$^*$ & $150\pm 30$  &         & $706\pm 143 $  & 18\\
13229--2934    & $382\pm 48$  & $1236\pm 120 $& $1960\pm 295^{k}$ & x,26\\
13362+4831     & $175\pm 20$  &         & $912\pm 90^{k}$ & x,4,5,14,17\\
13428+5608     & $82\pm 9 $   & $270\pm 35$   & $482\pm 82$(1.087, G) & x,1,7,11\\
13470+3530$^*$ & $118\pm 24$(1.11, K$_{\rm sh}$)&    & $1126\pm 231$(1.11,K$_{\rm sh}$)& 18\\
F13500+3141    & $52\pm 8$    & $155\pm 28$(1.035, K$_c$) & $208\pm 55$(1.09,K$_c$) &  x,27\\
F13564+3741$^*$& $244\pm 51$  &         & $756\pm 150$ & 18\\
14003+3245$^*$ & $58\pm 17 $  &         & $853\pm 181$ & 18\\
14151+2705$^*$ & $78\pm 16 $  &         & $340\pm 70$  & 4\\
14178+4927$^*$ & $68\pm 15 $  &         & $544\pm 115$ & 18\\
14280+3126$^*$ & $187\pm 27$  &         & $806\pm 162$ & 7,18\\
14348--1447    & $56\pm 8$    & $212\pm 35$   & $360\pm 67$  & x,7,28,29\\
15107+0724     & $120\pm 11$  & $605\pm 92$   & $710\pm 100$(1.045, K$_c$) & x,7,14,18,30\\
15163+4255$^*$ & $114\pm 23$  &         & $364\pm 73$  & 4,5\\
15243+4150$^*$ & $52\pm 14$   &         & $246\pm 55$  & 18\\
NGC\,5953p$^*$& $110\pm 24$   &         & $813\pm 166$ & 18\\ 
15327+2340    & $419\pm 36$   & $1127\pm 69$  & $3674\pm 405$& 31\\
15437+0234$^*$& $224\pm 52$(1.077, K$_{\rm sh}$)&  & $636\pm 133 $(1.077, K$_{\rm sh}$)& 18\\ 
16104+5235$^*$& $155\pm 31$   &         & $1062\pm 212$& 4,5\\
16284+0411$^*$& $63\pm 15$    &         & $580\pm 120$ & 18\\
\hline
\end{tabular}
\end{minipage}
\end{table*}

\begin{table*}
\centering
\begin{minipage}{122mm}
\contcaption{$ ^{12}$CO J=1--0, 2--1, 3--2 data for the combined LIRG sample}
\begin{tabular}{@{}lcccl@{}}
\hline
Name & CO J=1--0$^{a}$ & CO J=2--1$^{a}$ & CO J=3--2$^{a}$ & Refs.$^{b}$\\  
\hline
16504+0228    & $322\pm 29$   & $1492\pm 253 $& $3205\pm 642$& 31\\
17132+5313    & $127\pm 18$   & $168\pm 29$   & $264\pm 48$  & x,14,17\\
17208--0014   & $160\pm 16$   & $688\pm 109$  & $1198\pm 190$(1.087, G) &  x,1,11,29,30\\
18425+6036$^*$& $234\pm47 $   &         & $1440\pm 288$          & 6,7\\
19458+0944    & $29\pm 6$     &         & $261\pm 65 $(1.087, G)  &  x,11\\
20550+1656$^*$& $121\pm 24$   &         & $527\pm 105$           & 4,5\\
22491--1808   &$33\pm 6$&$145\pm 34$(1.36, K$_{\rm sh}$)&$558\pm 190$(2.1, K$_{\rm sh}$)$^{l}$&  x,7\\
23007+0836&$298\pm 27$  & $890\pm 103$  &$1600\pm 240$(1.12, K$_c$) & x,7,13,14,26,32,33,34\\
23365+3604\ \ \ \  & \ \ \  $39\pm 6 $ \ \ \ & \ \ \ $117\pm 20$ \ \ \ & \ \ \ $288\pm 71$(1.087, G) \ \ \  
&  x,1,11 \ \ \ \ \ \ \ \ \ \ \\ 
\hline
\end{tabular}
$^{a}$Velocity-integrated line flux densities in Jy\,km\,s$^{-1}$, with
the value and type of any applied corrections reported in the parentheses (see 4.1),
 G: the pointing error bias, $\rm K_{sh}$=$\rm K_{sh}(\Delta \theta)$
 (with $\rm \Delta \theta$=$\rm \left[(\Delta \theta _{\alpha})^2+(\Delta
 \theta _{\delta})^2\right]^{1/2}$): position offset correction, $\rm K_c$: beam-source geometric
 coupling correction for sources with {\it known} CO or sub-mm dust emission source size 
$\rm \theta _{s}$.\\
$^{b}$x=this work,  1=Downes \& Solomon 1998; 2=Zhu et al.\ 2009; 3=Barvainis et al.\ 1989; 
4=Leech et al.\ 2010; 5=Gao et al.\ 1999; 6=Narayanan et al.\ 2005; 7=Sanders et al.\ 1991; 8=Alloin et al.\ 1992; 9=Appleton et al.\ 2002;
10=Papadopoulos \& Seaquist 1998a; 11=Solomon et al.\ 1997;
 12=Scoville et al.\ 1989; 13=Bryant \& Scoville 1999; 14=Young et al.\ 1995;
15=Wilson et al.\ 2008;
 16=Evans et al.\ 2002; 17=Sanders et al.\ 1986; 18=Yao et al.\ 2003; 19=Gracia-Carpio et al.\ 2007;
 20=Bayet et al.\ 2006; 21=Evans et al.\ 2001; 22=Casoli et al.\ 1992; 23=Papadopoulos et al.\ 2007;
 24=Evans et al.\ 2005; 25=Mazzarela et al.\ 1993; 26=Papadopoulos \& Seaquist 1998b; 27=Evans et al.\ 1999;
 28=Evans et al.\ 2000; 29=Mirabel et al.\ 1990; 30=Planesas et al.\ 1991; 31=Greve et al.\ 2009;
 32=Davies et al.\ 2004; 33=Meixner et al.\ 1990; 34=Papadopoulos \& Allen~2002\\
$^{c}$J=1--0, 3--2 for the central 40$''$, J=2--1 for the inner 20$''$ (from Zhu et al.\ 2009).\\
$^{d}$From CO J=1--0 and the  $\langle$(2--1)/(1--0)$\rangle$ ratio over
                  the inner 22$''$  (Casoli et al. 1992).\\
$^{e}$Probably an underestimate (source significantly larger than the JCMT~beam).\\
$^{f}$Fluxes for the inner 20$''$ of the galaxy.\\
$^{g}$Shift factor $\rm K_{sh}$=1.25 applied for the Sanders et al.\ 1991 value.\\
$^{h}$$3\sigma$ upper limit.\\
$^{i}$Flux from an SMA image most likely collecting all of it given
                 its single dish CO 3--2 flux is also recovered by the interferometer.\\
$^{j}$Uncertain value, large correction of a single CO 2-1 observation of
                  a widely separated pair, pointing $\sim $6$''$ away from SW nucleus.\\
$^{k}$A 3x3 grid map at $7''$ grid size was made to obtain total line flux.\\
$^{l}$A large correction factor makes this value uncertain.
\end{minipage}
\end{table*}

\newpage

\section{The molecular gas in LIRGs}

The  ratios of  velocity/area-averaged brightness  temperatures  of CO
lines provide  the excitation indicators  of the average state  of the
molecular gas, allowing comparisons among LIRGs and well-studied local
ISM environments in the  Galaxy where individual excitation mechanisms
(e.g. SNR shocks,  far-UV/IR photons, cosmic rays) can  be more easily
identified.   The luminosity  of a  line  (x) used  to compute  such
brightness temperature line ratios is

\begin{eqnarray}
\rm L_x ^{'} &=& \int  _{\Delta  V}\int  _{A_s} T_{b,x}\,da\,dV \nonumber \\
 &=& 
\frac{c^2}{2k_B\nu^2 _{x,rest}}\left(\frac{D^2 _L}{1+z}\right) \int _{\Delta V} S_{\nu}\,dV,
\end{eqnarray}

\noindent
where $\rm  T_{b,x}$ is the  rest-frame brightness temperature  of the
line, $\rm  \Delta V$, $\rm  A_s$ are the  line FWZI and  source area,
$\rm D_{L}$ the  luminosity distance, and $\rm S_{\nu}$  the line flux
density. After substituting astrophysical units

\begin{equation}
\rm L^{'} _x=
 3.25\times 10^3 \left[\frac{D^2 _{L}(Mpc)}{1+z}\right]\left(\frac{\nu_{x,rest}}{100\,GHz}\right)^{-2} 
\left[\frac{\int _{\Delta V} S_{\nu }\,dV}{Jy\,km\,s^{-1}}\right],
\end{equation}

\noindent
where  $\rm  L^{'}_x$  is  in  K\,km\,s$^{-1}$\,pc$^2$  and  $\rm  \nu
_{x,rest}$ is the  rest frame frequency of line  (x).  This expression
is used to compute the CO line ratios with $\rm r_{J+1\,J}$=$\rm L^{'}
_{J+1\,,J}/L^{'} _{10}$.  The  conversion to ordinary luminosity units
($\rm L_{\odot}$),  used to express the total  line luminosities ($\rm
L_{x}$=$\rm \int L_{\nu}\,d\nu$) in CO SLEDs is

\begin{eqnarray}
\rm L_x &=& \frac{8\pi k_B\nu ^3 _{x,rest}}{c^3}\,L^{'} _x \nonumber \\
 &=& 3.18\times 10^4\left(\frac{\nu _{x,rest}}{100\,GHz}\right)^3
\left[\frac{L^{'} _x}{10^{9}L_l}\right]L_{\odot}.
\end{eqnarray}

\noindent
where $\rm L_l=K\,km\,s^{-1}\,pc^2$.

In Tables 7  and 8 we give the  CO line ratios for our  sample, and in
Figures  5,   6,  7   their  frequency  distributions.   These  reveal
well-excited  lines ($\rm \langle  r_{21}\rangle $=0.91,  $\rm \langle
r_{32} \rangle$=0.67)  but also  a significant excitation  range, from
low  ($\rm  r_{21}$$\sim  $0.6--0.7,  $\rm  r_{32}$$\sim  $0.3)  to  a
high-excitation  phase ($\rm r_{21,32}$$\ga  $1).  Luminous  CO J=4--3
lines clearly  mark the emergence  of a second highly  excitated phase
with $\rm r_{43}$$\ga
$$\rm  r_{32}$ {\it  and}  $\rm r_{32}$$\la  $$\rm  r_{21}$$\la $1  in
several LIRGs (Table  7).  This could not happen  if the average state
was dominated  by one phase with  optically thick CO  lines since then
$\rm  1\ga r_{21}\ga  r_{32}\ga r_{43}\ga...$  (higher-J  lines become
fainter as subthermal excitation  progressively sets in and/or as $\rm
E_{J+1,J}/(k_BT_{k})$   becomes   $>$1).    A  high-excitation   phase
overtaking the  lower-excitation one at the transition  where the line
ratios  of  the  latter  fall  below unity  can  indeed  reverse  this
inequality series, provided that  this new phase has well-excited {\it
  and} optically  thin CO  SLEDs.  In some  cases this  occurs already
from J=3--2  with $\rm r_{32}$$\ga $$\rm r_{21}$  and $\rm r_{21}$$\la
$0.6-1  (e.g.  IRAS\,10190+1322,  Arp\,220).  We  note that  while the
emergence  of  a  second  highly-excitated  phase  in  global  CO  and
$^{13}$CO line emission  has been known for some  time (e.g.  Aalto et
al.  1995  and references  therein), indications for  well-excited and
partially optically thin global CO SLEDs are rare.  Indicatively, even
for the ULIRG/QSO Mrk\,231  where ground and Herschel/FTS observations
completed its CO SLED from J=1--0  up to J=13--12 (van der Werf et al.
2010) it is always $\rm r_{J+1\,J}$$<$$\rm r_{J\,J-1}$.

\begin{table*}
\centering
\begin{minipage}{95mm}
\caption{The CO J=4--3, 6--5 data}
\begin{tabular}{@{}lcc@{}}
\hline
Name & CO J=4--3$^{a}$ & CO J=6--5$^{a}$ \\
\hline
00057+4021 & $1049\pm 370 $ (1.15, G)    & $\la 380^{b}$  (1.29, G) \\
00509+1225 &                       & $430\pm 155$ (1.97, G)$^{c}$\\
02483+4302 &                       & $401\pm 131$ (1.27, G)\\
04232+1436 &                       & $\la 345^{b}$ (1.21,G)\\
05083+7936 & $2033\pm 593$ (1.15, G)     & $688\pm 207$ (1.22, G) \\
05189--2524& $\la 689^{b}$ (1.47, K$_{\rm sh}$) & $638 \pm 220$(1.21,G)\\
08572+3915 (NW nucleus)&           & $405 \pm 158$(1.21,G)\\
09126+4432 (NE nucleus) &          & $705\pm 230$ (1.90, $\rm K_{sh}$)$^{d}$\\
09320+6134 &                       & $\la 422^{b}$ (1.35, G)\\
10173+0828 &                       & $\la 156^{b}$ (1.27, G)\\
10565+2448 &                       & $506\pm 148$ (1.35, G)\\
11257+5850 & $6340\pm 1585^{e}$ &   \\
11191+1200 &                       & $261\pm 82$ (1.14, K$_c$)\\
12112+0305 (NE nucleus)&           & $327\pm 143^{f}$ (1.22, G)\\
12112+0305 (SW nucleus)&           & $\la 123^{b}$ (1.21,G)\\
12540+5708 & $1127\pm 265$ (1.15, G)     & $1320\pm 400$(1.29, G)\\
13183+3423 &                       & $\la 630^{b}$ (1.28, G) \\
13428+5608 &                       & $488\pm 156$(1.45,G)\\
F13500+3141& $842\pm 245$ (1.15, K$_c$)  & $1086\pm 347$ (1.29, K$_c$)\\
15107+0724 & $1122\pm 262$ (1.073, K$_c$)& $904\pm 241$ (1.14, K$_c$)\\
15237+2340 &                       & $3130\pm 810$ (1.21, G)\\
16504+0228 &                       & $3321\pm 860$ (1.21,G)\\
17208--0014& $2312\pm 654$ (1.15, G)     & $340\pm 140^{g}$ (1.17, G)\\
23007+0836 & $3970\pm 820$ (1.19, K$_c$) & $2355\pm 590$\\
23365+3604 & $\la 489^{b}$ (1.15, G) & $295\pm 110$(1.37,G)$^{g}$\\
\hline
\end{tabular}
$^{a}$Velocity-integrated line flux densities in Jy\,km\,s$^{-1}$, with 
                 the value and type of any corrections applied (section 4.1) 
                 reported in the parentheses. A few CO J=6--5 fluxes differ slightly from
                 those in P10b, a result of additional data.\\
$^{b}$3$\sigma $ upper limit.\\
$^{c}$The high velocity part of a double horn line tentatively detected, large
    correction applied because of extended source, uncertain total line flux.\\
$^{d}$Possible detection, highly uncertain because a pointing offset of $\sim $3.85$''$ 
                  results to a beam-shift correction $\rm K_{sh}$=1.90.\\
$^{e}$From a partially completed CO J=4--3 map, and thus a likely underestimate.\\
$^{f}$Tentative detection of the NE nucleus of a double source.\\
$^{g}$Very uncertain value.\\
\end{minipage}
\end{table*}

\begin{figure*}
\includegraphics[width=\columnwidth]{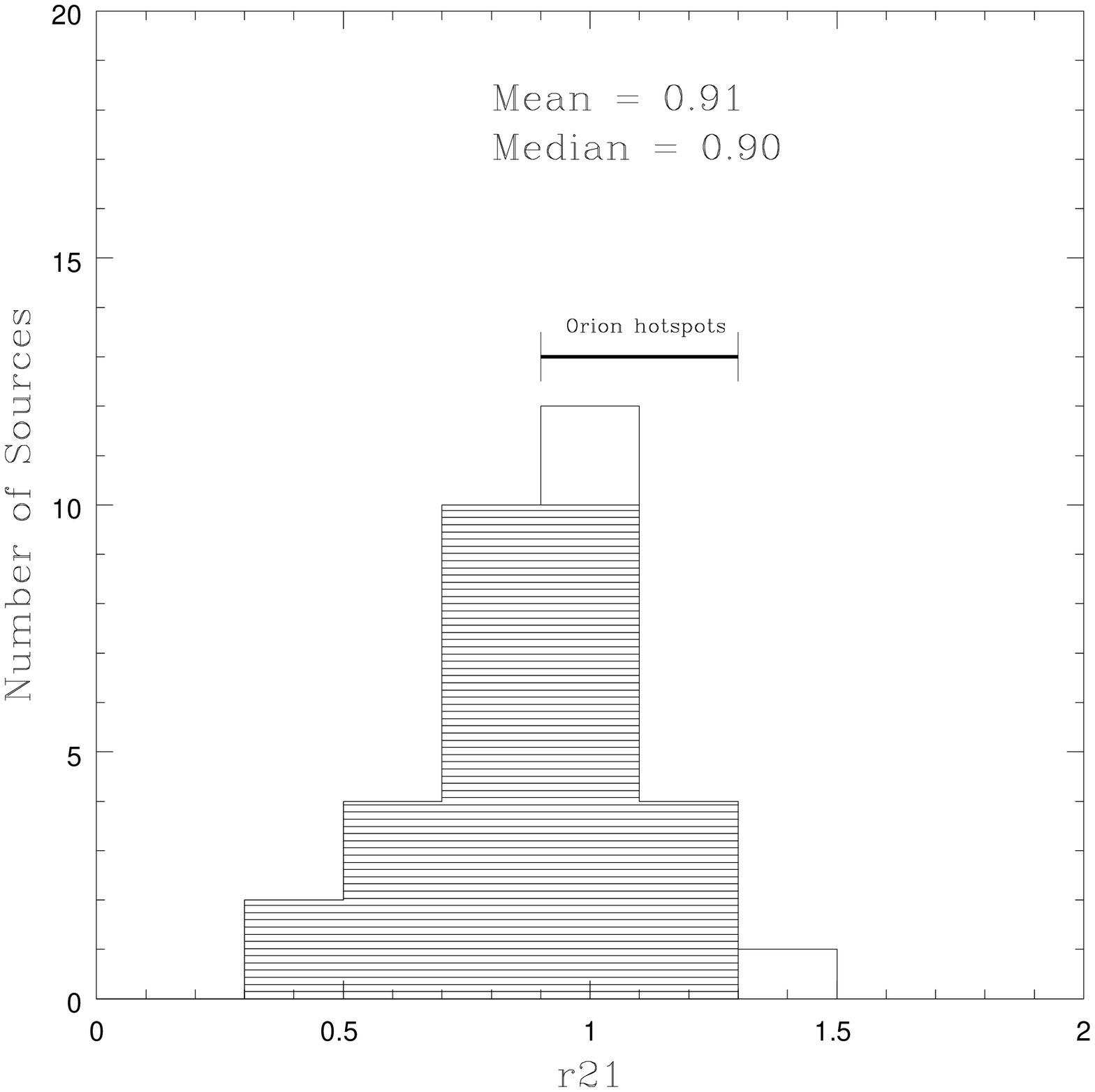}
\includegraphics[width=\columnwidth]{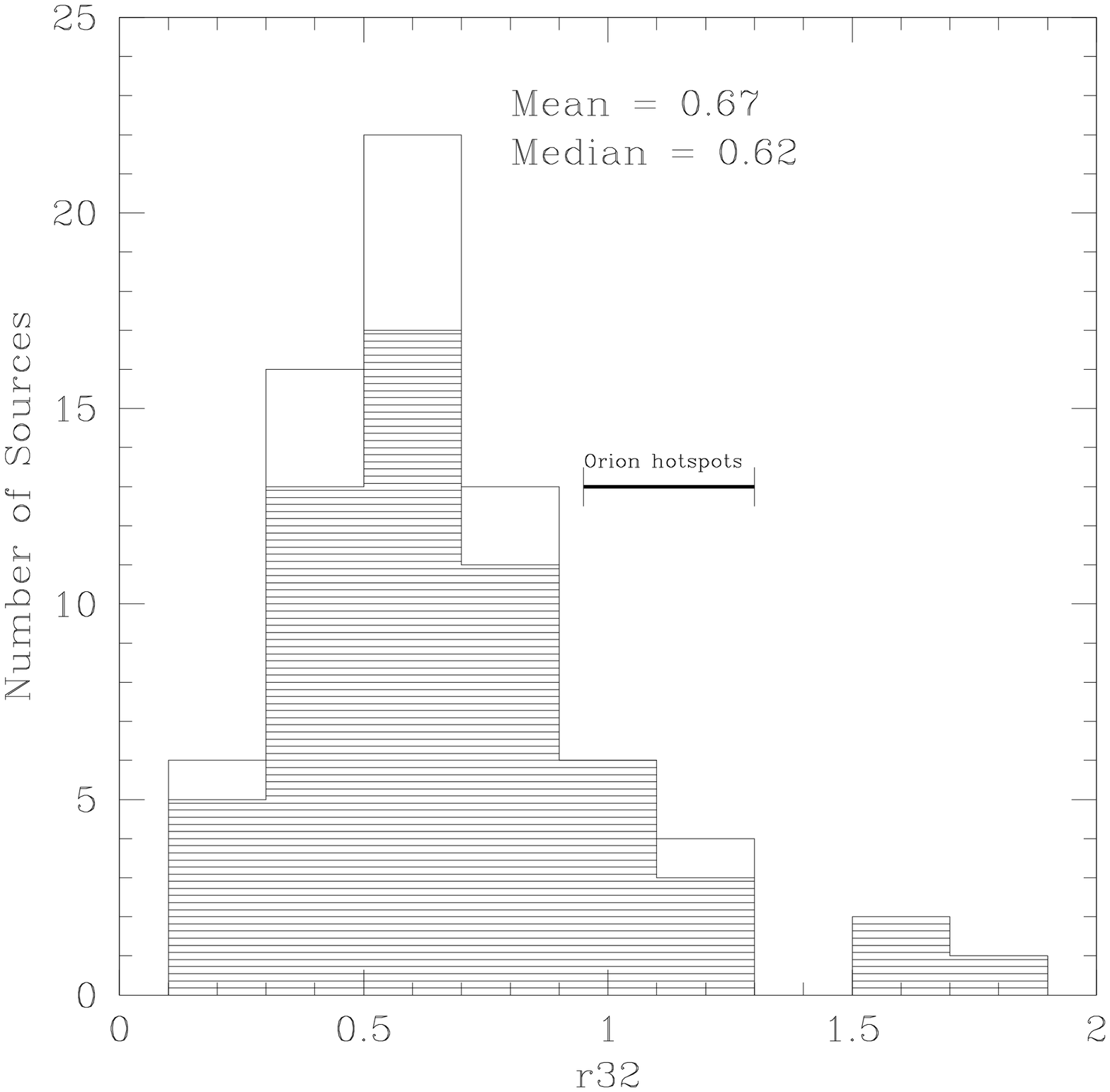}
\caption{The   distributions  of   the  CO   (2--1)/(1-0)   (left)  and
(3--2)/(1--0)  (right) global brightness  temperature ratios  for the
LIRGs  in  our   sample.   The  shaded  area  marks   the  line  ratio
distributions  for sources with  CO emission  region angular  sizes of
$\leq $15$''$ (the HPBW of  the JCMT at 345\,GHz).  The measured range
of the (2--1)/(1--0) line ratio in the hot spots near H\,II regions or
PDRs  around  O, B  stars  in  the Orion  A,  and  B molecular  clouds
(Sakamoto  et   al.   1994),  and   the  expected  range  of   the  CO
(3--2)/(1--0) ratio  in these high-excitation regions  are indicated by
the horizontal bar.}
\end{figure*}

\subsection{Very high excitation gas: ``out-exciting'' Orion A and B}

  The  qualitatively  different  state  of the  molecular  gas  in
  (U)LIRGs than in spiral disks, has been noted in the past by several
  groups (e.g.   Aalto et al.  1995;  Solomon et al.   1997; Downes \&
  Solomon   1998;  Papadopoulos   \&   Seaquist  1998;   Yao  et   al.
  2003). Moreover ISM conditions similar  to those in ULIRGs have been
  revealed in  the nuclei of  nearby spirals but involving  only small
  fractions  of  their total  molecular  gas  (e.g.   Bradford et  al.
  2003).  Nevertheless most such studies lacked J=3--2 and higher-J CO
  lines,  and did  not benchmak  their very  limited CO  SLEDs  to any
  regions  where excitation mechanisms  are well-determined  (e.g.  SF
  ´hot´-spots in Galactic~GMCs). 

For the highly excited CO SLEDs  of some ULIRGs a superposition of two
optically thick CO SLEDs,  a high-excitation and a low-excitation one,
can at  most yield a ``flattening''  or a slower decline  of high-J CO
line  ratios with  respect  to those  expected  from a  low-excitation
phase.  Only dense and warm gas with at least partially optically thin
CO  lines   can  produce  $\rm  r_{43}$$>$$\rm   r_{32}$  and/or  $\rm
r_{32}$$>$$\rm r_{21}$ and the high  excitation tails seen in Figure 5
($\rm r_{21,32}$$>$1).  Such CO line  ratios are rare, and reported in
the past only  for the starburst nuclei of  M\,82 and NGC\,3310 (Knapp
et  al.  1980; Olofsson  \& Rydbeck  1984; Braine  et al.   1993), and
although confirmed at lower values,  they are still $>$1 (Weiss et al.
2001; Zhu  et al.  2009), implying extraordinary  ISM conditions.  Our
study recovers this  as a statistically robust result  for our sample.
Indeed, while  measurement errors  still allow $\rm  r_{J+1,J}$$\la $1
for individual objects with measured $\rm r_{J+1,J}$$>$1 (J=1,2), this
cannot be so for all the  LIRGs in the high excitation tails in Figure
5  as  the  probability  for  this  is:  $\rm  P_{21}$$\sim  $5$\times
$10$^{-7}$  and  $\rm  P_{32}$$\sim  $4.4$\times  $10$^{-7}$  for  the
(2-1)/(1-0)  and   (3--2)/(1-0)  line  ratios   (assuming  independent
gaussian probability  distributions for each  measurement).  Thus some
extraordinary physical  conditions are possible for the  {\it bulk} of
the molecular gas of some (U)LIRGs.

  In  the Galaxy  such high  CO line  ratios are  found {\it  only} in
  isolated  ``hot''  spots in  GMCs  as  results  of strong  and  very
  localized excitation mechanisms.  These are intense far-UV radiation
  from O,B  stars inducing Photon-dominated  Regions (PDRs) containing
  warm gas near  H\,II regions (Sakamoto et al.   1994), or supernovae
  remant (SNR)-induced shocks (e.g.  Seta et al.  1998; Arikawa et al.
  1999;  Bolatto   et  al.   2003)  which  yield   warm  and  strongly
  kinematically stirred gas ($\rm K_{vir}$$\gg $1).  In the Orion A, B
  molecular clouds, the site of active  formation of O, B stars in the
  Galaxy, the  cloud-average $\rm \langle  r_{21}\rangle _{A,B} $$\sim
  $0.62--0.75,  with $\rm  r_{21}$$\sim  $0.90-1.3 found  only in  few
  isolated PDRs near HII regions  and O, B star associations (Sakamoto
  et  al.  1994).   Thus $\rm  r_{21}$$\ga $0.8,  observed  in several
  LIRGs, already  places their entire molecular  gas reservoirs higher
  than the Orion A, and B clouds in terms of average excitation, while
  $\rm r_{21}$$>$1  puts them on  par with their ``hottest''  SF spots
  and  the excitation state  of molecular  gas in  SNR-GMC interfaces!
  This  is  indeed extraordinary  given  that  the extreme  conditions
  needed  to produce optically  thin CO  lines in  LIRGs must  then be
  maintained    over   large    molecular   gas    reservoirs   ($\sim
  $(10$^{9}$-10$^{10}$)\,M$_{\odot}$).  The  CO/$^{13}$CO J=1--0, 2--1
  line ratios (Figure 7), with $\rm \langle R_{10}\rangle $$\sim $$\rm
  \langle R_{21}\rangle $=$18$,  are significantly larger than typical
  GMCs in the Galactic disk ($\rm R_{10}$$\sim $4--6, e.g.  Solomon et
  al.   1979;  Polk et  al.   1988),  and  independently indicate  low
  optical depths.   Using again the Orion  A and B  clouds to indicate
  how  extraordinary  such  large   $\rm  R_{10,21}$  ratios  are  for
  galaxy-sized molecular  gas reservoirs we note  that $\rm R_{21}$=10
  is observed only in ``hot''  isolated interfaces of molecular gas to
  O,B stellar associations while for the bulk of the gas in Orion $\rm
  R_{21}$$\sim  $4--6 (Sakamoto  et al.   1994).  Even  in  the highly
  turbulent and  warm molecular clouds  in the Galactic Center  the CO
  line emission remains mostly  optically thick with $\rm R_{10}$$\sim
  $5 and $\rm \tau _{10}$$\sim $5 (Oka et al.~1998). In other galactic
  nuclei  such  a dense  and  warm gas  phase  amounts  to only  $\sim
  $(1-2)\% of the total molecular gas mass of a typical spiral disk.

The extreme thermal and/or kinematic conditions needed to achieve even
partially optically thin CO SLEDs can be simply demonstrated using the
CO J=1--0 optical depth:

\begin{equation}
\rm \tau _{10}=\frac{c^3 g_1 A_{10}}{8\pi \nu^3 _{10}}\left(\frac{1-e^{-E_{10}/k_B T_k}}{Z_{LTE}}\right) 
\left[\frac{CO}{H_2}\right] \frac{n(H_2)}{(dV/dr)},
\end{equation}

\noindent
(assuming LTE),  where $\rm Z_{LTE}$ is the  partition function. After
substituting the various physical constants this becomes

\begin{equation}
\rm \tau _{10} = 1.29\times 10^4 \left(\frac{1-e^{-5.5/T_k}}{T_k}\right) \left[\frac{CO}{H_2}\right] \frac{n(H_2)}{(dV/dr)},
\end{equation}

\noindent
where $\rm  dV/dr$ is in  units of km\,s$^{-1}$\,pc$^{-1}$.   For $\rm
[CO/H_2]$$\sim  $$10^{-4}$ (for Solar  metallicities), and  setting the
average  velocity gradient  of  the gas  phase  as $\rm  (dV/dr)$=$\rm
K_{vir}\times (dV/dr)_{vir}$ where

\begin{equation}
\rm \left(\frac{dV}{dr}\right)_{vir}\approx 0.65\sqrt{\alpha } \left(\frac{n(H_2)}{10^3\,cm^{-3}}\right)^{1/2}
km\,s^{-1}\,pc^{-1},
\end{equation}

\noindent 
is the average  velocity gradient expected for a  virialized gas phase
(with $\alpha  $$\sim $0.5-2.5 depending on the  assumed cloud density
profile, Greve et al. 2009; Bryant \& Scoville 1996), we obtain

\begin{equation}
\rm \tau _{10} = 63 \alpha^{-1/2}  \left(\frac{1-e^{-5.5/T_k}}{T_k}\right) \frac{\sqrt{n(H_2)}}{K_{vir}}. 
\end{equation}

\noindent
For  most gas in  typical GMCs:  $\rm T_{kin}$$\sim  $(15-30)\,K, $\rm
n(H_2)$$\sim  $(500-10$^3$)\,cm$^{-3}$,   and  $\rm  K_{vir}$$\sim  $1
(virial  gas  motions),  yielding  $\rm \tau_{10}$$\sim  $6.5-33  (for
$\alpha$=1.5).   Only high  temperatures  ($\rm T_{kin}$$\ga  $100\,K)
and/or  highly  non-virial gas  motions  ($\rm  K_{vir}$$\ga $10)  can
produce $\rm \tau_{10}$$\la $1.  Moreover the conditions necessary for
$\rm r_{J+1,J}$$>$1 become even more extreme for higher-J lines since,
along   with   the   high    temperatures   needed   (so   that   $\rm
E_{J+1,J}/(k_BT_k)$$\ll$1), such ratios  need densities high enough to
thermalize  the J+1$\rightarrow $J  line while  retaining moderate/low
optical depths.   The latter becomes progressively  more difficult for
higher-J lines since $\rm \tau _{J+1,J}$$\sim $$\rm (J+1)^2\tau _{10}$
(for   LTE  and   $\rm  E_{J+1,J}/(k_BT_k)$$\ll$1).    Thus   to  keep
$\tau_{10}$$\ll  $1  (so that  e.g.   $\rm  \tau _{43}$$\sim  $16$\tau
_{10}$ will  remain $\la  $1 yielding an  optically thin  J=4--3 line)
both high  $\rm T_{kin}$ {\it  and} highly non-virial gas  motions are
necessary.   Indeed for  $\rm T_{k}$=100\,K,  Equation 11  yields $\rm
\tau    _{10}$$\sim    $2.8$\times    $$\rm    10^{-2}K^{-1}    _{vir}
\sqrt{n(H_2)}$,     which      for     $\rm     n(H_2)$$>$(1-2)$\times
$10$^4$\,cm$^{-3}$ (the critical densities  of the J=3--2, 4--3 lines)
gives  $\rm \tau_{10}$$\ga  $(3-4)$\rm  K^{-1} _{vir}$,  which can  be
brought into the optically thin regime provided that $\rm K_{vir}$$\ga
$10, which corresponds to highly unbound dynamical states for the gas.

   As  the star-forming Orion  A and  B molecular  clouds demonstrate,
   with  low CO line  excitation and  large optical  depths prevailing
   when  averaged over  their entire  volume, such  extreme conditions
   cannot easily  dominate the  CO SLEDs of  individual SF  clouds let
   alone  the   global  CO  SLEDs  of  entire   galaxies.   Thus  $\rm
   r_{J+1,J}$$\ga  $1  (J=1,2,3)  ratios  are not  expected,  even  in
   vigorously star-forming  LIRGs, and  are indicative of  strong {\it
     galaxy-wide} effects maintaining high gas temperatures as well as
   highly    non-virial    velocity    fields    up    to    densities
   n$>$10$^4$\,cm$^{-3}$   throughout   their   large  molecular   gas
   reservoirs.   As  we  will  argue  in  more  detail  later  such  a
   high-excitation gas phase is  irreducible to an ensemble-average of
   individual  SF  molecular  clouds.   For  the rest  of  LIRGs  more
   ordinary excitation  states of SF  ($\rm r_{21, 32}$$\la  $0.9) and
   non-SF  ($\rm r_{21}$$\sim  $0.5-0.7, $\rm  r_{32}$$\sim  $0.3) gas
   adequately account for their observed CO SLEDs.

  High resolution submm imaging with  ALMA will reveal extreme CO line
  excitation   and  thus  extraordinary   ISM  conditions   much  more
  frequently  as   the  high  CO  ratios  of   SF-powered  regions  or
  AGN-excited   gas   will    no   longer   be   ``watered-down''   by
  lower-excitation  cloud  ensembles.   Line  ratio imaging  can  then
  reveal deeply dust-enshrouded AGN  in galactic centers where neither
  optical nor  even IR lines are  capable of doing so.   An early such
  demonstration is the imaging of  the central $\sim $100\,pc of grand
  design spiral galaxy M\,51 which  finds high CO line excitation with
  $\rm r_{32}$$\sim  $1.9, and possibly  caused by an  AGN-powered jet
  (Matsushita~et~al.~2004).

\begin{table}
\centering
\begin{minipage}{74mm}
\caption{The $ ^{13}$CO line data}
\begin{tabular}{@{}lccc@{}}
\hline
Name & $ ^{13}$CO J=1--0$^{a}$ & $ ^{13}$CO J=2--1$^{a}$ & 
References$^{b}$ \\  
\hline
00057+4021     &                      & $\la 24^{c}$  & this work\\
00509+1225     & $4.0\pm 0.8$               & (R=$6\pm 2$)$^{d}$ & 1, this work\\
01053--1746$^*$& (R$\ga$$17$)$^{c,d}$&                     & 2\\
02071+3857     & $37\pm 6$                  & (R=$12\pm 3$)$^{d}$& this work\\
02401--0013$^*$& $185\pm 28$                & $1034\pm 200$               & 5\\
02483+4302     & $\la 2.1^{c}$ & $\la 21^{c}$   & this work\\
03359+1523     &                      & (R=$12 \pm 3$)$^{e}$ & this work\\
05083+7936     & $7.9\pm 2.3$               & $23.0\pm 4.5$               & this work\\
05189-2524     &                      & $21\pm 7$                   & this work\\
08354+2555     &                      & (R$\ga $18)$^{c,d}$ & 6\\
09126+4432     &                      & $\la 55^{c}$   & this work\\
09320+6134     &                      & $18\pm 3$                   & this work\\
10035+4852     &                      & $9.9\pm 2.5^{f}$ & this work\\
10173+0828     &                      & $5.2\pm 1.0$                & this work\\
10565+2448     & $4.7\pm 1.3$               & (R$\ga 18$)$^{c}$& this work, 6\\
11257+5850     & (R=$27\pm 5$)$^{d}$ &                   & 7\\
12540+5708     &                      & $6.1\pm 2.1 $               & this work\\
13183+3423     &                      & $26\pm 4$                   & this work\\
13229--2934    & $13.5\pm 2.0$              & $87\pm 13 $                 & this work, 8 \\
13428+5608     &                      & $36\pm 9$                   & this work\\
F13500+3141    & $\la 11.8^{c}$ &                       & this work\\  
14348--1447    &                      & $16.7\pm 5.6$               & this work\\
15107+0724     & $6.8\pm 1.8$               & $23\pm 7$                   & this work\\
15327+2340     & $9\pm 2 $                  & $59\pm 7 $                  & this work, 9\\
16504+0228     & $6.5\pm 1.9$               & $26\pm 4$                   & this work, 9\\
17208--0014    &                      & $\la 18^{c}$   & this work\\
23007+0836     &                      & $54\pm 13$                  & 8\\
23365+3604     &                      & $6.2\pm 1.4$                & this work\\
\hline
\end{tabular}
$^{a}$The velocity-integrated line flux densities in Jy\,km\,s$^{-1}$.\\
$^{b}$1=Eckart et al.\ 1994; 2=Aalto et al.\ 1995;
3=Aalto et al.\ 1991; 4=Casoli et al.\ 1992; 5=Papadopoulos \& Seaquist 1998a; 6=Wilson et al.\ 2008;
7=Aalto et al.\ 1997; 8=Papadopoulos \& Seaquist 1998b; 9=Greve et al.\ 2009.\\
$^{c}3\sigma $ upper limit.\\
$^{d}$Only  R=$^{12}$CO/$^{13}$CO  is available (reported in parenthesis) for areas
 smaller than the total CO-emitting region (see 4.2).\\
$^{e}$Large pointing offset for both CO and $^{13}$CO line
measurements makes absolute fluxes uncertain but leaves line ratio invariant.\\
$^{f}$Uncertain value, large $\rm K_{sh}$ correction applied ($\rm K_{sh}$=2).\\
\end{minipage}
\end{table}

\subsection{High-excitation gas in LIRGs: ``bottom''-stirred by SF feedback? }

 A  warm ($\rm  T_{kin}$$\ga $100\,K)  unbound ($\rm  K_{vir}$$\gg $1)
 phase  concomitant with a  less excited  cooler and  denser one  is a
 feature of standard two-phase models  in (U)LIRGs (e.g.  Aalto et al.
 1995; Papadopoulos \& Seaquist  1999), and could in principle explain
 the very high-excitation CO SLEDs  in our sample.  In these systems a
 frequent  non-convergence  of  global  CO  line  ratios  (mostly  CO,
 $^{13}$CO 1--0,  2--1) to  an average ISM  state is attributed  to an
 envelope-like  diffuse   (n$\sim  $(10$^2$-10$^{3}$)\,cm$^{-3}$)  non
 self-gravitating warm  gas with  high turbulent linewidths  (and thus
 small/moderate    $\rm   \tau_{10}$),    surrounding    much   denser
 self-gravitating gas  where CO and even $^{13}$CO lines have large
 optical  depths (Aalto  et al.   1995).  However,  the  difficulty of
 maintaining low optical depths {\it and} well-excited CO J=3--2, 4--3
 lines (so that $\rm r_{43,32}$$\ga $1) makes this two-phase ISM model
 problematic  as   the  ``envelope''  phase  should   then  have  $\rm
 n(H_2)$$>$10$^{4}$\,cm$^{-3}$      ($\rm      n_{crit}$(3-2,4-3)$\sim
 $(1-2)$\times      $$10^4$\,cm$^{-3}$)     rather      than     n$\la
 $10$^{3}$\,cm$^{-3}$.

 SF-feedback effects capable of producing highly-excited CO SLEDs with
 moderate/low optical depths up to  J=3--2, 4--3 do occur in dense gas
 regions deep inside  molecular clouds, and are driven  by the massive
 stars. The warm  PDRs and SNR-shocked gas regions,  along with strong
 radiation pressure onto the dust  grains mixed with the molecular gas
 can produce a warm {\it and} non self-gravitating ($\rm K_{vir}$$>$1)
 dense phase near SF sites. It is worth noting that radiation pressure
 from  massive  stars   is  powerful  enough  to  even   be  the  main
 SF-regulator in  the starbursts found in ULIRGs  (Andrews \& Thompson
 2011).  Such  a ``bottom''-stirred ISM  would yield a  very different
 two-phase differentiation,  with a warm, dense,  and non-virial phase
 lying deeply embedded in molecular clouds, surrounded by a less dense
 cooler one extending further from  the vicinity of SF sites.  A warm,
 dense, and  non-virial phase  is indeed compatible  with the  CO line
 ratios of  the SF  ``hot-spots'' of Orion  A,B molecular  clouds (see
 7.2.1), with  CO SLEDs  expected to be  highly excited  and partially
 optically thin up to high-J  lines, a result of the high temperatures
 and $\rm K_{vir}$ values.

\subsubsection{A ``bottom''-stirred ISM,  some predictions}

In the context of the  old 2-phase model, the high CO/$^{13}$CO J=1--0
ratios in  ULIRGs ($\rm  R_{10}$$\ga $20) are  attributed to  the much
more turbulent  ($\rm K_{vir}$$\gg$1) diffuse/warm  ``envelopes'' than
those of  molecular clouds in less extreme  star-forming systems ($\rm
R_{10}$$\sim  $10-15) or SF-quiescent  ones like  the Milky  Way ($\rm
R_{10}$$\sim  $4-6) (Aalto  et al.   1995).  A  much larger  amount of
turbulent  energy per gas  mass is  available in  ULIRGs, a  result of
strong  mergers, and its  dissipation can  readily warm  molecular gas
(Pan   \&   Padoan   2009)   and   increase  $\rm   R_{10}$.    In   a
``bottom''-stirred  ISM  however, the  dense  phase  is  the one  most
affected  by   the  SF  feedback   and  thus  {\it  high   ($\ga  $15)
  $^{12}CO/^{13}CO$  $ R_{J+1,J}$  J+1=3, 4,  5, 6  and HCN/H$^{13}$CN
  J=1--0  ratios   are  expected,}  as   dense  gas  can   now  attain
low/moderate molecular line  optical depths.  These transitions (which
would  be   very  optically  thick   for  dense  gas  that   has  $\rm
K_{vir}$$\sim        $1),        with       $\rm        n_{crit}$$\sim
$(10$^{4}$-10$^{5}$)\,cm$^{-3}$, probe  the dense gas  with negligible
contributions  from diffuse  cloud  envelopes.  High  C/$^{13}$C-based
isotopologue line  ratios at ever increasing  critical densities (e.g.
HCN/H$^{13}$CN J=3--2, $\rm n_{crit}$$\sim $$4\times 10^6$\,cm$^{-3}$)
would  then   provide  an   ever  sharpening  distinction   between  a
``bottom''-stirred and a standard two-phase ISM model in~(U)LIRGs.

\subsection{The quest for a new ISM excitation mechanism in ULIRGs}
 
The small amount  of gas mass typically found  in the SF ``hot-spots''
of ordinary  star-forming GMCs  renders even a  ``bottom''-stirred ISM
model problematic  in producing the highest excitation  CO SLEDs found
in   our    sample.    Indeed,   even    if   all   the    dense   gas
($>$10$^{4}$\,cm$^{-3}$) in GMCs  was feedback-affected it would still
amount only to $\sim $few\% of their total mass.  This cannot ``skew''
their CO  SLED anywhere close  to the high excitation  levels observed
for  some ULIRGs  (e.g.   IRAS\,12112+0305, IRAS\,17208--0014).   {\it
  Much larger mass  fractions of GMCs must be  dense and SF-active for
  such  SLEDs to  emerge.}  High  cloud boundary  pressures  and tidal
striping of  outer GMC envelopes  in a merger environment  can produce
much  denser clouds, a  situation exemplified  by the  Galactic Center
(G\"usten  \& Phillip 1994).   Large amounts  of molecular  gas ($\sim
$10$^{10}$\,M$_{\odot}$)          at          densities          $\sim
$(10$^5$-10$^6$)\,cm$^{-3}$, and a possible steepening of the (average
density)-size power law $\langle n \rangle
$$\propto $$\rm L^{-k}$ to k$>$1  (k=1 for Galactic GMCs, Larson 1981)
certainly argue  for a much denser  molecular ISM in  ULIRGs (Greve et
al.  2009).  At  the same time this hinders  far-UV photon propagation
even more (i.e.  the PDRs will be smaller) while SNR-induced shocks in
such dense ISM will dissipate fast and produce shock-excited molecular
gas regions  that are even more  confined around the  SNR sites.  Thus
{\it a quest  for mechanisms that can globally  heat the large amounts
  of dense gas found in ULIRGs to high temperatures opens up.}

\subsection{Strong AGN feedback: a new global driver of ISM excitation}

Two  intriguing objects,  the optically/IR-luminous  QSO PG\,1119+120,
and  the powerful  FR\,II radio  galaxy  3C\,293, widen  the range  of
possibilities  regarding globally  operating molecular  gas excitation
mechanisms in galaxies.  This is particularly clear in 3C\,293 where a
strong jet-ISM interaction injects  large amounts of mechanical energy
into  its highly  turbulent ISM  and powers  luminous CO  J=4--3, 6--5
lines in an  otherwise SF-quiescent ISM with subthermal  low-J CO line
ratios  and cold  dust emission  (Papadopoulos et  al.   2008, 2010b).
Strong,  penetrating   irradiation  by  an  X-ray   luminous  AGN  can
volumetrically heat molecular gas by creating giant X-ray Disossiation
Regions  (XDRs),  and is  another  powerful  and  possibly global  ISM
excitation  mechanism in  AGN-hosting LIRGs  (Meijerink et  al.  2007;
Schleicher  et   al.   2010).    Unfortunately  there  are   no  X-ray
observations  of PG\,1119+120 to  assess the  X-ray luminosity  of its
AGN, while we find no particular predominance of AGN amongst the LIRGs
populating  the high-excitation  tails in  Figure 5.   Very  high-J CO
SLEDs (J$\geq$8-10) will provide  the tools for distinguishing between
AGN-induced XDRs, SF-induced PDRs as  the cause of high ISM excitation
(Meijerink et al.  2006; van der Werf et al. 2010). Finally the recent
discovery of a massive molecular gas outflow, likely driven by the AGN
in Mrk\,231, by Aalto et al. (2012) inserts an exciting new element in
this   debate  by   demonstrating   that  much   higher  density   gas
($>$10$^4$\,cm$^{-3}$) can also be affected by AGN feedback.

\begin{table*}
\centering
\begin{minipage}{140mm}
\caption{$^{12}$CO line ratios and SF-powered IR luminosities}
\begin{tabular}{@{}lcccccc@{}}
\hline
Name & $\rm r_{21}^{a}$ & $\rm r_{32}^{a}$ &$\rm r_{43}^{a}$ &
$\rm r_{65}^{a}$ & $\rm L_{CO(1-0)}^{'\,b}$ & $\rm L^{(*)}_{IR} (T_{dust})^{c}$\\
 & & & & & $\rm (\times 10^9\,L_{l})$ & $\rm (\times 10^{11}\,L_{\odot}) (K)$\\  
\hline
00057+4021       & $0.97\pm0.22$& $0.64\pm0.14$ & $1.43\pm0.54$     & $\la 0.23^{d}$ & $4.08\pm0.62$ 
& 2.9 (37)\\ 
00322--0840$^{*}$& $1.00\pm0.35$& $0.58\pm0.20$ &   &   & $0.66\pm0.17$ & 0.55 (28) \\ 
00509+1225       &$0.84\pm0.24$ & $1.16\pm0.40$ &   & $0.35\pm0.15 $& $5.73\pm1.18$ & 0.96 (44) \\
01053--1746$^{*}$&        & $0.53\pm0.14$ &   &   & $12.2\pm2.4$ & 3.00 (33) \\
01077--1707$^{*}$&        & $0.32\pm0.09$ &   &   & $9.8\pm2.0$  & 3.00 (33) \\
01418+1651$^{*}$ &        & $0.62\pm0.17$ &   &   & $2.5\pm0.5 $ & 3.10 (45) \\
02071+3857       & $0.76\pm0.16$& $0.70\pm0.15$ &   &   & $5.7\pm0.46$ & 0.78 (46) \\
0208+3725$^{*}$  &        & $0.44\pm0.12$ &   &   & $1.5\pm0.3$  & 0.39 (50) \\
02114+0456$^{*}$ &        & $0.84\pm0.24$ &   &   & $6.2\pm1.2$  & 1.64 (32) \\
02321--0900      & $1.10\pm0.22$& $0.56\pm0.14$ &   &   & $4.0\pm0.5$  & 0.98 (61) \\
02401--0013$^{*}$& $1.00\pm0.25$& $0.67\pm0.16$ &   &   & $1.22\pm0.18$& 0.66 (74) \\
02483+4302       & $0.97\pm0.22$& $0.44\pm0.09$ &   & $0.37\pm0.13$ & $3.6\pm0.5$& 3.50 (36) \\
02512+1446$^{*}$ &        & $0.22\pm0.06$ &   &   & $8.4\pm1.7$  &  1.85 (69) \\
03359+1523       &        & $0.18\pm0.05$ &   &   & $8.6\pm1.3$  &  2.05 (52) \\ 
04232+1436       &        & $1.09\pm0.33$ &   & $\la 0.28^{d}$ & $9.8\pm2.0$& 7.45 (35) \\
05083+7936       & $0.72\pm0.12$& $0.43\pm0.19$ & $1.46\pm0.45$ & $0.22\pm0.07$ & $11.53\pm1.20$ & 4.90 (52) \\
05189--2524      & $0.67\pm0.15$& $0.59\pm0.12$ & $\la 0.90$ & $0.37\pm 0.14$& $3.91\pm0.57$ & 8.88 (60) \\ 
08030+5243       & $0.64\pm0.16$& $0.36\pm 0.10$&   &   & $9.54\pm1.90$& 5.96 (44) \\ 
08354+2555$^{*}$ & $0.41\pm0.08$& $0.42\pm0.08$ &   &   & $2.45\pm0.30$&  2.01 (51) \\
08572+3915       & $0.98\pm0.31$&      &   & $1.07\pm0.44^{e}$& $1.60\pm0.23$    & 9.76 (52) \\
09126+4432       & $1.20\pm0.26$& $0.53\pm0.12$ &   &   & $11.44\pm1.62$& 2.22 (53) \\
09320+6134       & $1.23\pm0.33$& $0.93\pm0.24$ &   & $\la 0.17^{d}$ & $4.82\pm 0.95$& 3.88 (40) \\
09586+1600$^{*}$ &        & $0.74\pm0.19$ &   &   & $0.35\pm0.07$& 0.25 (58) \\
10039--3338$^{*}$&        & $0.54\pm0.16$ &   &   & $3.36\pm0.72$& 3.91 (52) \\
10035+4852       & $0.88\pm0.24$&         &   &   & $9.11\pm1.90$& 6.17 (48) \\
10173+0828       &        & $0.25\pm 0.07$&   & $\la 0.08^{d}$ & $6.0\pm 1.1$ & 4.16 (48) \\
10190+1322       & $0.68\pm 0.18$& $0.68\pm0.17$ &   &   & $9.84\pm1.85$   & 5.81 (51) \\
10356+5345$^{*}$ & $1.47\pm 0.42$& $1.21\pm0.34$ &   &   & $0.067\pm0.013$ & 0.27 (34) \\
10565+2448       & $1.06\pm 0.16$& $0.80\pm0.12$ &   & $0.18\pm 0.055$& $6.40\pm 0.65$&  6.45 (41) \\
11191+1200       &         & $0.54\pm0.16$ &   & $1.61\pm 0.58$ & $0.50\pm0.09$ &  0.29 (41) \\
11231+1456$^{*}$ &         & $0.32\pm0.08$ &   &   & $7.50\pm1.03$ &  2.00 (51) \\
11257+5850       &         & $0.83\pm0.20$ & $0.68\pm0.21^{f}$ &   & $2.72\pm0.53$&4.67 (47) \\
12001+0215$^{*}$ &         & $0.25\pm 0.07$&   &   & $0.26\pm 0.053$  &  0.070 (54) \\
12112+0305       & $0.90\pm0.21$ & $1.58\pm0.35$ &   & $0.29\pm0.13^{g}$ & $10.1\pm1.2$& 12.1 (44) \\
12224--0624$^{*}$&     & $1.19\pm0.55$     &   &   & $0.67\pm0.27$ &  1.21 (37)\\
12243--0036$^{*}$&     & $0.84\pm 0.25$    &   &   & $0.31\pm0.07$ &  0.66 (43)\\
12540+5708       & $0.89\pm 0.12$ &  $0.72\pm 0.13$ & $0.80\pm0.20$  & $0.42\pm0.13$ & $7.0\pm0.72$ & 13.8 (44) \\
13001--2339$^{*}$&    & $0.63\pm0.18$      &   &   & $3.10\pm 0.62$& 1.81 (47)\\
13102+1251$^{*}$ &    & $0.50\pm0.15$      &   &   & $0.52\pm 0.10$& 0.20 (50)\\
Arp\,238$^{*}$   &    & $0.30\pm 0.08$     &   &   & $4.85\pm0.97$ & 4.00 ( )\\
13183+3423       & $1.10\pm0.19$ & $0.74\pm0.12$ &   & $\la 0.09^{d}$ & $4.65\pm0.40$& 2.21 (42)\\
13188+0036$^{*}$ &   & $0.52\pm0.15$       &   &   & $2.28\pm0.46$& 0.54 (34)\\
13229--2934      & $0.81\pm0.13$ & $0.57\pm0.11$ &   &   & $3.10\pm0.39$& 0.68 (37)\\
13362+4831       &         & $0.58\pm0.09$ &   &   & $6.0\pm0.70$ & 2.09 (33)\\
13428+5608       & $0.82\pm0.14$ & $0.65\pm0.13$ &   & $0.17\pm0.06$ & $5.20\pm0.60$ & 8.37 (52)\\  
13470+3530$^{*}$ &         & $1.06\pm0.30$ &   &   & $1.47\pm0.30$ & 0.50 (31)\\
F13500+3141      & $0.75\pm0.18$ & $0.44\pm0.13$ & $1.01\pm0.33$ & $0.58\pm0.20$ & $4.63\pm0.71$ & 0.12 (57)\\ 
F13564+3741$^{*}$&         & $0.34\pm0.10$ &   &   & $1.67\pm0.33$ & 0.48 (31)\\
14003+3245$^{*}$ &         & $1.63\pm0.59$ &   &   & $0.54\pm0.16$ & 0.50 (30)\\
14151+2705$^{*}$ &         & $0.48\pm0.14$ &   &   & $4.65\pm0.93$ & 0.88 (52)\\
14178+4927$^{*}$ &         & $0.88\pm0.27$ &   &   & $1.97\pm0.43$ & 0.48 (71)\\
14280+3126$^{*}$ &         & $0.48\pm0.12$ &   &   & $1.16\pm0.17$ & 0.31 (52)\\
14348--1447      & $0.95\pm0.21$ & $0.71\pm0.16$ &   &   & $17.40\pm2.5$ & 14.1 (50)\\
15107+0724       & $1.26\pm0.22$ & $0.66\pm0.11$ & $0.58\pm0.15$ & $0.21\pm0.06$ & $0.87\pm0.08$ & 1.18 (33)\\
15163+4255$^{*}$ &         & $0.35\pm0.10$ &   &   & $8.22\pm1.65$  & 4.10 (46)\\
15243+4150$^{*}$ &         & $0.53\pm0.18$ &   &   & $0.18\pm0.05$  & 0.21 (57)\\
15322+1521$^{*}$ &         & $0.82\pm0.24$ &   &   & $0.20\pm0.04$  & 0.13 (54)\\
15327+2340       & $0.67\pm0.07$ & $0.97\pm0.14$ &   & $0.21\pm0.06$ & $6.12\pm0.53$ & 10.0 (46)\\
15437+0234$^{*}$ &         & $0.32\pm0.10$ &   &   & $1.61\pm0.37$  & 0.51 (52)\\
16104+5235$^{*}$ &         & $0.76\pm0.21$ &   &   & $5.85\pm1.17$  & 1.87 (58)\\
16284+0411$^{*}$ &         & $1.02\pm0.32$ &   &   & $1.67\pm0.40$  & 1.08 (40)\\
\hline
\end{tabular}
\end{minipage}
\end{table*}

\begin{table*}
\centering
\begin{minipage}{140mm}
\contcaption{$^{12}$CO line ratios and SF-powered IR luminosities}
\begin{tabular}{@{}lcccccc@{}}
\hline
Name & $\rm r_{21}^{a}$ & $\rm r_{32}^{a}$ &$\rm r_{43}^{a}$ &
$\rm r_{65}^{a}$ & $\rm L_{CO(1-0)}^{'\,b}$ & $\rm L^{(*)}_{IR} (T_{dust})^{c}$\\
 & & & & & $\rm (\times 10^9\,L_{l})$ & $\rm (\times 10^{11}\,L_{\odot}) (K)$\\  
\hline
16504+0228       & $1.16\pm0.22$ & $1.10\pm0.24$ &   & $0.29\pm0.08$ & $8.41\pm0.8$ & 3.78 (53)\\
17132+5313       & $0.33\pm0.07$ & $0.23\pm0.05$ &   &   & $14.65\pm2.1$  & 1.26 (58)\\
17208--0014& $1.08\pm0.20$ & $0.83\pm0.16$ & $0.90\pm0.27$ & $0.06\pm0.025^{h}$ &$13.10\pm1.30$ &15.6 (46)\\
18425+6036$^{*}$ &         & $0.68\pm0.19$ &   &   & $1.79\pm0.36$  & 0.44 (52)\\
19458+0944       &         & $1.00\pm0.32$ &   &   & $13.33\pm 2.80$& 10.0 (41)\\ 
20550+1656$^{*}$ &         & $0.48\pm0.13$ &   &   & $7.10\pm1.40$  & 4.71 (42)\\
22491--1808      & $1.10\pm0.32$ & $1.88\pm0.72^{h}$ &   &   & $8.97\pm1.63$ & 13.0 (49)\\
23007+0836       & $0.75\pm0.11$ & $0.60\pm0.10$ & $0.83\pm0.19$ & $0.22\pm0.06$ & $3.48\pm0.32$ & 1.61 (42)\\
23365+3604       & $0.75\pm0.17$ & $0.82\pm0.24$ & $\la 0.78^{d}$ & $0.21\pm0.09^{h}$ &
$7.30\pm1.12$    & 6.19 (61)\\
\hline
\end{tabular}
$^{a}$The CO line ratios $\rm r_{J+1 J}$=$\rm L^{'} _{CO(J+1-J)}/L^{'} _{CO(1-0)}$\\
$^{b}$The CO 1--0 luminosity in units of $\rm L_{l}$=K\,km\,s$^{-1}$\,pc$^2$ (see Equation 6).\\
$^{c}$The SF-powered part of  $\rm L_{IR}$ and the associated dust temperature (see P10a for details).\\
$^{d}$3-$\sigma $ upper limit.\\
$^{e}$CO J=6--5 flux from the NW nucleus where most of the CO emission is (Evans et al.\ 2002).\\
$^{f}$Likely higher as CO 4--3 is obtained from a partially completed map.\\
$^{g}$Estimated for the NE nucleus where 75\% of CO(1--0) emission arises (Evans et al.\ 2002).\\
$^{h}$Uncertain value, see footnotes of Tables 4,5.\\
\end{minipage}
\end{table*}

\begin{table}
\centering
\begin{minipage}{43mm}
\caption{The CO/$^{13}$CO line ratios}
\begin{tabular}{@{}lcc@{}}
\hline
Name & J=1--0$^{a}$ & J=2--1$^{a}$\\  
\hline
00057+4021     &                        & $\ga 7^{b}$ \\
00509+1225     &  $8\pm 2$                    & $6\pm 2$\\
01053--1746$^*$&  $\ga 17^{b,c}$ &   \\
02071+3857     &  $10\pm 2$                   & $12\pm 3$ \\
02401--0013$^*$&  $14\pm 3$                   & $10\pm 3$ \\
02483+4302     &  $\ga 13^{b}$   & $\ga 5^{b}$ \\
03359+1523     &                        & $12 \pm 3$\\
05083+7936     &  $10\pm 3$                   & $10\pm 2$\\
05189--2524    &                        & $6\pm 2$\\
08354+2555     &                        & $\ga 18^{b,c}$\\
09126+4432     &                        & $\ga 13^{b}$\\
09320+6134     &                        & $18\pm 4$\\
10035+4852     &                        & $16\pm 5$\\
10565+2448     &   $15\pm 4$                  & $\ga 18^{b}$\\
11257+5850     &   $27\pm 5^{c}$ &   \\               
12540+5708     &                        & $47\pm 16$\\
13183+3423     &                        & $30\pm 7$\\
13229--2934    &  $26\pm 5$                   & $13\pm 2$ \\
13428+5608     &                        & $7\pm 2$\\
F13500+3141    &  $\ga 4^{b}$    &  \\
14348--1447    &                        & $12\pm 4$\\
15107+0724     & $16\pm 4$                    & $24\pm 8$\\
15327+2340     & $43\pm 10$                   & $17\pm 2$\\ 
16504+0228     & $45\pm 14$                   & $53\pm 11$\\
17208--0014    &                        & $\ga 35^{b}$\\
23007+0836     &                        & $15\pm 4$\\
23365+3604     &                        & $17\pm 5$\\
\hline
\end{tabular}
$^{a}${$\rm L^{'}(CO)/L^{'}(^{13}CO)$ line ratios.}\\
$^{b}$Ratio that corresponds to 3-$\sigma$ upper limit on $ ^{13}$CO flux.\\
$^{c}$Ratio measured over a smaller area than that of total CO J=1--0 emitting region.\\
\end{minipage}
\end{table}

\begin{figure}
\includegraphics[width=\columnwidth]{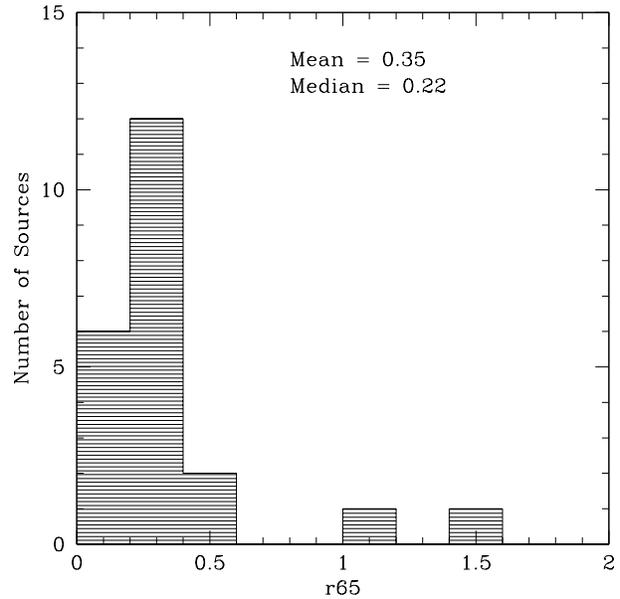}
\caption{The   distribution    of   the   CO(6--5)/(1--0)   brightness
temperature  ratios  for  the  LIRGs  in our  sample.   The  optically
luminous  QSO  PG\,1119+120  has  the  highest  excitation  with  $\rm
r_{65}$$\sim $1.6. Some  of the low ratios ($\leq  $0.3) may have been
surpressed by a strong dust continuum becoming optically thick even at
short submm wavelengths (see P10a).}
\end{figure}

\begin{figure}
\includegraphics[width=\columnwidth]{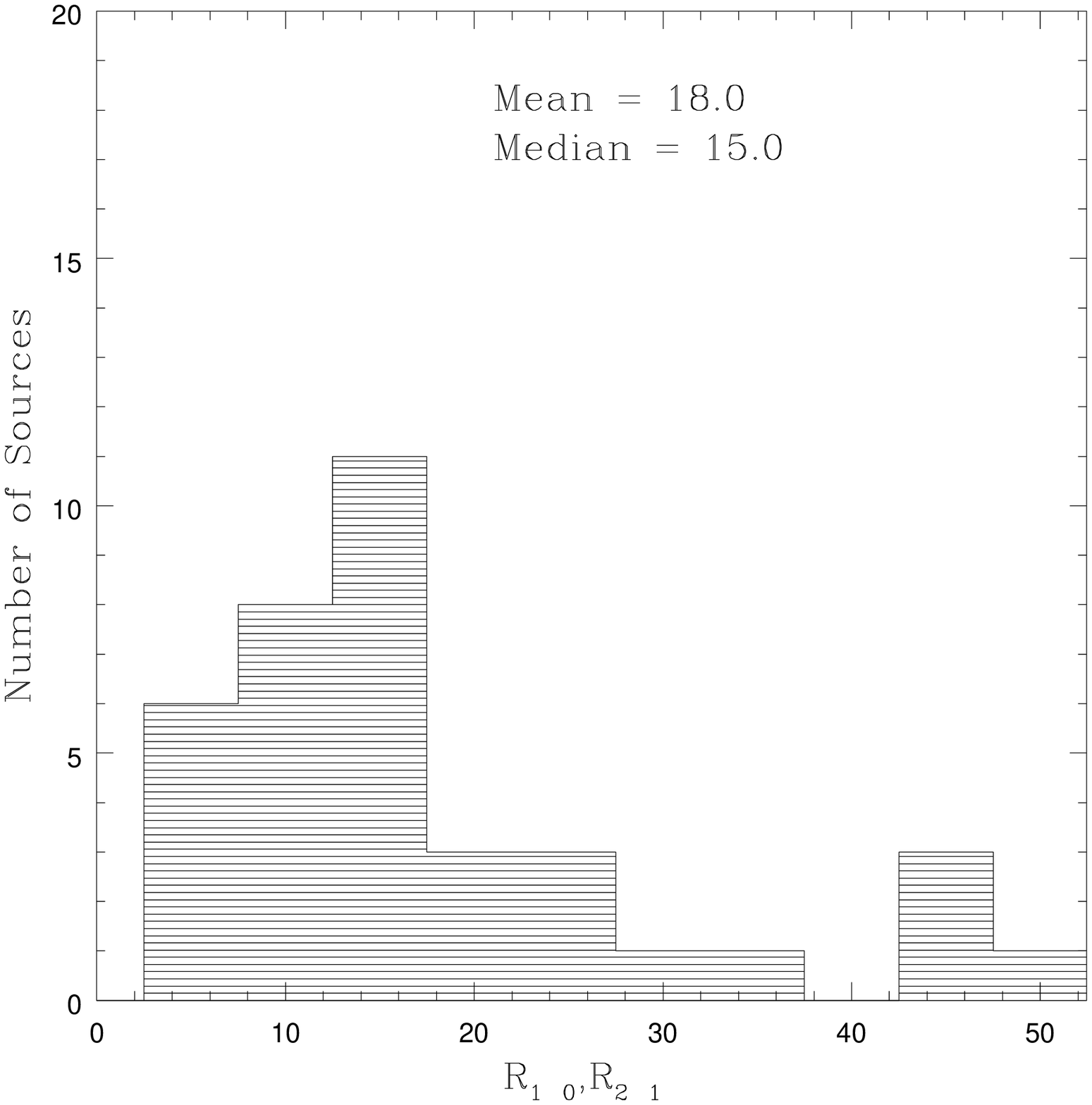}
\caption{The joint distribution of the CO/$^{13}$CO J=1--0, 2--1 line
ratios for the LIRGs in our sample.}
\end{figure}

\section{CO line radiative transfer models}

 Interpreting  global  CO line  ratios  in  terms  of underlying  {\it
   average}  gas physical  conditions is  the task  of  molecular line
 radiative  transfer  models.  In  the  present  analysis  of the  CO,
 $^{13}$CO  lines we  will use  a  Large Velocity  Gradient code  (see
 Papadopoulos \&  Sequist 1998 and  references therein) that  has been
 shown adept at discerning the average thermal and kinematic states of
 the molecular gas when constrained by CO, $^{13}$CO line ratios (e.g.
 Mao et al. 2000;  Weiss et al. 2001; Yao et al.  2003).  Our LVG code
 uses the observed CO ratios $\rm R^{(obs)} _k$ to find the regions in
 the  $\rm \left[n(H_2),  T_k, \Lambda_{co}\right]  $  parameter space
 where $\rm \chi ^2$=$\rm \Sigma _k (1/\sigma_k)^2[R^{(obs)}_k-R_k]^2$
 (where  $\rm R_k$  is a  model  line ratio  and $\rm  \sigma _k$  the
 measurement   error)   is  minimum.    $\rm   \Lambda  _{co}   $=$\rm
 [CO/H_2]/(dV/dR)$  (in $\rm (km\,s^{-1}\,pc^{-1})^{-1}$),  where $\rm
 [CO/H_2]$ is  the CO abundance and  $\rm dV/dR$ is  the average local
 velocity  gradient  of  the  turbulent  CO line  emitting  gas.   The
 temperature  range we  considered  is: 15\,K$\leq$$\rm  T_{kin}$$\leq
 $150\,K, with the minimum value  typical of cold SF-quiescent gas and
 the  maximum  of the  Cold  Neutral  Medium  HI.  Moreover  only  LVG
 solutions with $\rm K_{vir}\ga 1$ where

\begin{equation}
\rm K_{vir}=\frac{\left(dV/dR\right)}{\left(dV/dR\right)_{virial}}\sim 
1.54\frac{[CO/H_2]}{\sqrt{\alpha}\Lambda _{co}}\left(\frac{n(H_2)}{10^3\,
 cm^{-3}}\right)^{-1/2}
\end{equation}

\noindent
(using equation  10) are accepted.  The latter  determines the average
dynamic state of the molecular  gas, with $\rm K_{vir}$$\sim $1--3 for
the  (mostly) self-gravitating  GMCs in  the Galactic  disk  (and $\rm
K_{vir}$$\ll$1 corresponding to dynamically unattainable gas motions).
There  is no  theoretical  upper  limit for  $\rm  K_{vir}$, and  $\rm
K_{vir}$$\gg $1 in regions  where strong injection of turbulent energy
occurs  (e.g.   in  GMC/SNR  interfaces).  Indicatively,  for  average
densities  of  $\sim  $(10$^{2}$--10$^4$)\,cm$^{-3}$ $\rm  K_{vir}$=20
corresponds  to  $\rm (dV/dR)$$\sim  $(5--50)\,km\,s$^{-1}$\,pc$^{-1}$
(for $\alpha  =1.5$), encompassing the  values observed in  the highly
turbulent gas of the Galactic Center  (Dahmen et al.  1998; Oka et al.
1999), in  pre-star-forming molecular clouds in  the Galaxy (Falgarone
et   al.   1998)  and   galactic  SF   nuclei  where   typically  $\rm
(dV/dR)$$\sim  $(2--6)\,km\,s$^{-1}$\,pc$^{-1}$ (Weiss et  al.  2001).
Furthermore  we adopt  $\rm  [CO/^{13}CO]$=50 as  appropriate for  the
environments in  LIRGs (e.g.  Henkel  et al.  1993) and  measured also
for the Galactic disk in  the Solar vicinity (Wilson \& Penzias 1993),
while  we  set $\rm  [CO/H_2]$=$10^{-4}$  (needed  for computing  $\rm
K_{vir}$ from Equation 12).  Finally  we note that some CO J=6--5 line
luminosities may be diminished by significant dust optical depths even
at  short submm  wavelengths  (P10a,b),  while in  a  few cases  (e.g.
Arp\,193) they  may be underestimated because  of substantial pointing
offsets.  Thus we  do not use them as constraints  in our LVG modeling
of the entire sample.  Finally, we only made LVG models when N$\geq $3
lines  were  available,  except   in  the  special  cases  where  $\rm
r_{J+1,J}$$>$1 that  indicate quite unique  high-excitation regions of
the LVG parameter space.

   In Figure 8 we show  the $\rm T_{kin}$, $\rm T_{r}$(1-0) (CO J=1--0
   brightness   temperature),   $\rm   n(H_2)$,  and   $\rm   K_{vir}$
   distributions as  derived from  the LVG radiative  transfer models,
   including   the   degenerate   or   poor  solutions.    A   general
   n(H$_2$)-$\rm T_{kin}$  degeneracy exists when  $\rm r_{J+1,J}$$<$1
   and  $^{13}$CO  lines are  unavailable  as  constraints, with  such
   ratios  compatible  with dense/cold  and  with diffuse/warm  phases
   (e.g.     for    IRAS\,14348-144:    10$^4$\,cm$^{-3}$/20\,K    and
   300\,cm$^{-3}$/150\,K).    These    nevertheless   typically   have
   different $\rm K_{vir}$, which  can be used to discriminate between
   such  disparate LVG  solution  ranges once  $^{13}$CO lines  become
   available.  The few  states with $\rm n(H_2)$$\ga$10$^4$\,cm$^{-3}$
   are associated  with LIRGs where $\rm r_{J+1,J}$$\geq  $1 (the high
   excitation  tails in Figure  5), they  also have  $\rm T_{kin}$$\ga
   $100\,K,  and   involve  large   molecular  gas  masses   of  $\sim
   $(4-6)$\times  $10$^{9}$\,M$_{\odot}$.  These  extreme  systems are
   studied  in  more  detail  in  our  final  paper  (Papadopoulos  et
   al.~2012).

  The wide $\rm  K_{vir}$ distribution is a result  of the large range
  of $\rm R_{10,21}$ ratios  (Figure 7), with average dynamical states
  having  $\rm K_{vir}$$\sim  $0.7--2 ($\sim  $virial values)  to $\rm
  K_{vir}$$\ga  $20  (strongly  unbound  states).  This  diversity  is
  expected  given that  the galaxies  in our  sample  include isolated
  disk-dominated systems (where  GMCs are mostly self-gravitating) and
  strongly  evolving mergers  (where GMCs  can be  fully  disrupted by
  tidal fields),  and this has been  noted from early  studies of such
  systems (e.g.  Aalto et al.  1995).  The distribution of the average
  brightness  temperature of the  CO J=1--0  line $\rm  T_{r}$(1-0) is
  rather   narrow,  with  an   average  of   $\sim  $16\,K   and  $\rm
  \sigma(T_r)$$\sim   $4\,K  (for  the   distribution  of   the  good,
  non-degenerate    LVG    solutions).     The   $\rm    \chi_{g-d}$=$\rm
  T_{kin}/T_{dust} $ and  $\rm \chi_{th}$=$\rm T_{ex}$(3-2)/$\rm T_{kin}$
  distributions   (Figure  9)   indicate  how   much   ``warmer''  (or
  ``colder'') a CO SLED is with  respect to the SED of the concomitant
  dust  reservoir,   and  the  thermalization  level   of  the  J=3--2
  transition.

 For  most  LVG  solutions  $\rm T_{kin}/T_{dust}$$\ga  $1,  which  is
 expected  since  photoelectric, and  even  more  so  turbulent or  CR
 heating of  the ISM  affect the  dust much less.   Only at  very high
 densities ($\ga  $10$^{5}$\,cm$^{-3}$ for Galactic  ISM environments)
 thermal  equillibrium   between  gas  and  dust   is  attained  ($\rm
 T_{kin}$$\sim $$\rm T_{dust}$).  Some large thermal decouplings ($\rm
 \chi_{g-d}$$\ga$3)  exist,  with one  of  the  highest  found in  the
 powerful  radio galaxy  3C\,293  where a  strong jet-ISM  interaction
 drives a ``hot'' CO SLED  while its dust SED remains cold, reflecting
 Milky-Way levels  of star  formation (P10b).  Thus  turbulent heating
 (in this  case AGN-driven) can heat {\it  galaxy-sized} molecular gas
 reservoirs  to  high temperatures,  even  in  the  absence of  strong
 SF-powered radiation fields, an  important issue discussed further in
 section 7.1.   The $\rm \chi_{th}$ values range  from very subthermal
 ($\sim   $0.1-0.2),  to   fully   thermalized  ($\sim   $1)  in   the
 high-excitation ISM  of some ULIRGs.   Low $\rm \chi_{th}$  values on
 the  other hand  are obtained  mostly  for a  high temperature  ($\rm
 T_{kin}$$\sim    $(60--150)\,K),     diffuse    ($\rm    n(H_2)$$\sim
 $(10$^{2}$-10$^{3}$)\,cm$^{-3}$)  phase.   This  can  be  typical  in
 turbulent GMC  envelopes, and may  even dominate the CO  J=1--0, 2--1
 line emission  in ULIRGs (DS98), yet  {\it it cannot  contain much of
   their molecular gas mass.}  Indeed in such galaxies with SF-powered
 IR luminosities  of $\rm L^{(*)}  _{IR}$$\sim $10$^{12}$\,L$_{\odot}$
 (see P10b) the  dense gas phase associated with SF  sites will have a
 minimum mass of $\rm M_{SF}$(H$_2$)$\sim $$\rm L^{(*)} _{IR}/\epsilon
 _{g,*}$$\sim $4$\times $10$^9$\,M$_{\odot}$ (for an Eddington-limited
 SF      efficiency       of      $\rm      \epsilon      _{g,*}$$\sim
 $250\,L$_{\odot}$/M$_{\odot}$).   This  is  already  $\sim  $2$\times
 $M(HI+H$_2$) of gas-rich spirals  (e.g.  Young \& Scoville 1991), the
 typical progenitors of ULIRGs  (Sanders \& Ishida 2004), thus leaving
 little room  for any significant  additional molecular gas mass  in a
 diffuse phase in  such systems.  A very dense  dense star-forming gas
 phase containing  most of the  molecular gas mass is  indeed strongly
 suggested  by  HCN J=1--0  observations  (Gao  \&  Solomon 2004)  and
 conclusively  shown for  individual  ULIRGs using  heavy rotor  (e.g.
 HCN, CS) and $^{13}$CO lines (Papadopoulos et al.  2007; Greve et al.
 2009).

\begin{figure*}

\includegraphics[width=\textwidth]{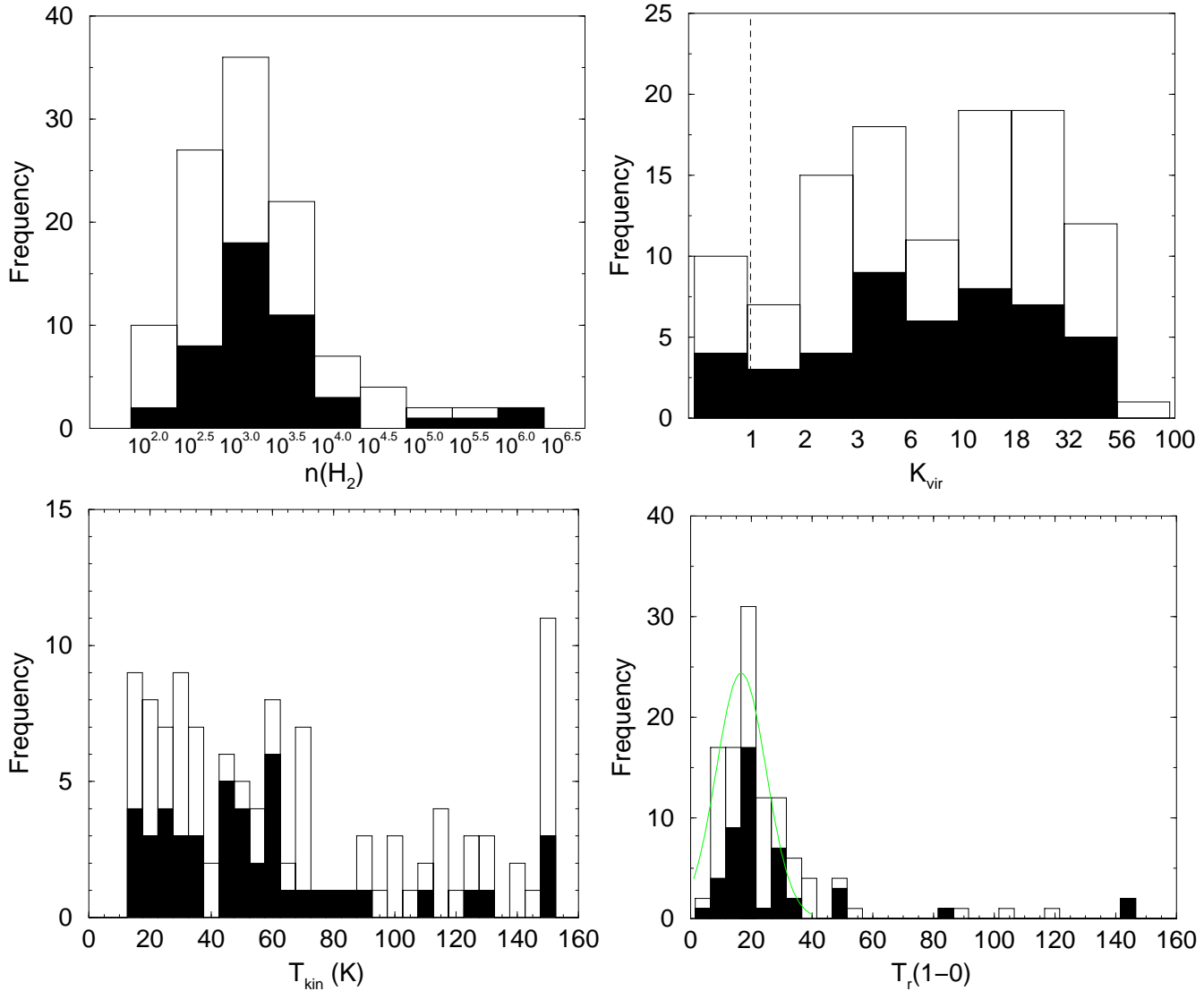}

\caption{The  $\rm n(H_2)$,  $\rm  K_{vir}$, $\rm  T_{kin}$, and  $\rm
  T_{r}$(1-0)  distributions  obtained  from one-phase  LVG  radiative
  transfer models of the CO lines of the LIRG sample. White bars: from
  all solutions,  black bars: using only the  best, non-degenerate LVG
  solutions obtained for each system (see section 6).}
\end{figure*}

\subsection{Low-excitation  CO SLEDs and the average ISM states}
 
 Only few ``cold'' CO SLEDs are  found in our sample,  indicated by the
 low excitation ``tail'' of CO line ratios with $\rm r_{21}$$\leq $0.6
 and/or $\rm r_{32}$$\leq $0.30 (Figure 5).  This is expected for what
 at the end  is an IR-selected (and thus  SFR-selected) galaxy sample,
 with the  limitations of such  samples in selecting large  numbers of
 (cold-ISM)-dominated  systems discussed  thoroughly by  Dunne  et al.
 2000.  Furthermore, unlike the high excitation CO SLEDs that uniquely
 correspond   to  warm,   dense  (and   often  unbound)   gas  states,
 sub-thermally  excited  ``cold''  low-J  CO SLEDs  cannot  be  uniquely
 attributed to a  quiescent ISM state.  Indeed they  could mark a cold
 ($\sim $(10-20)\,K),  self-gravitating ($\rm K_{vir}$$\sim  $1) phase
 of  moderate  densities   ($\sim  $10$^{3}$\,cm$^{-3}$),  typical  of
 SF-quiescent GMCs, but also a warmer ($\ga $30\,K), low-density ($\la
 $10$^3$\,cm$^{-3}$),   and    highly   unbound   gas    phase   ($\rm
 K_{vir}$$\gg$1)  more  typical of  turbulent  cloud ``envelopes''  or
 intercloud  gas in  ULIRGs (Aalto  et  al.  1995;  Downes \&  Solomon
 1998).  This  renders low-excitation CO  J+1$\rightarrow $J, J+1$\leq
 $3 SLEDs  {\it highly degenerate to  the average state of  the ISM in
   galaxies.}  This  is obvious in individual systems  such as 3C\,293
 and Arp\,220 where subthermal low-J CO lines give no hint about their
 massive  and  highly-excited  gas  reservoirs revealed  only  via  CO
 J+1$\rightarrow$J, J+1$\geq$4  or heavy  rotor (e.g.  HCN,  CS) lines
 (Papadopoulos et al.  2008; Greve et al.  2009).

 This degeneracy  casts doubts on recent claims  for Galactic-type ISM
 in  distant gas-rich  star-forming disks  (Dannerbauer et  al.  2009;
 Daddi et al.   2010) as current observations do  not extend beyond CO
 J=3--2 where  such degeneracies  diminish (see section  7.2.1). Among
 the galaxies  with low-excitation CO  line ratios in our  sample this
 was   possible    for   only   two    LIRGs   (IRAS\,03359+1523   and
 IRAS\,05189--2524) where  available $^{13}$CO line  observations (see
 Table 8) played  a key role in determining  Galaxy-type ISM states as
 the  most likely  ones. Observations  of $^{13}$CO  lines  of distant
 systems like the BzK galaxies will soon become possible with ALMA and
 will  be decisive  in ´breaking´  the aforementioned  degeneracies of
 low-J  CO SLEDs  with  $\rm R_{10,21}$$\sim  $5--10 corresponding  to
 SF-quiescent Galactic-type  ISM conditions while  $\rm R_{10,21}$$\ga
 $15 indicating vigorous star-forming and/or merger ISM environments.

\section{CO SLEDs of LIRGs: their excitation range, and power sources}

  The    range    of    normalized   $\rm    L^{(n)}    _{J+1,J}$=$\rm
  L_{J+1,J}/L_{IR}$ CO SLEDs as determined  from our study is shown in
  Figure 10.  From this it becomes obvious that: a) beyond J=3--2 very
  large variations  become possible, b)  some global SLEDs are  on par
  with those  of the Orion  SF ``hot-spots''.  The  highest excitation
  SLED  is   AGN-driven  (3C\,293)  followed  by   those  for  extreme
  starbursts  (e.g.   IRAS\,17208--0014,  IRAS\,12112+0305).  We  must
  note  that the large  $\rm L^{(n)}  _{J+1,J}$ range,  spanning $\sim
  $3-5 orders of magnitude beyond J=4--3, reflects both the degeneracy
  of the one-phase radiative transfer models as well as the wide range
  of  ISM  excitation  found   in  our  sample.   Recent  Herschel/FTS
  observations of  some these  (U)LIRGs fully determine  this higher-J
  part of their  CO SLEDs confirm this picture,  and will be presented
  in a forthcoming paper.  Finally from Figure 10 it is obvious that a
  positive K-correction  for redshifted CO line  emission from distant
  galaxies due to a rising  $\rm L^{(n)} _{J+1,J}$ with J-level (which
  makes possible the detection of very distant systems in high-J lines
  by  nearly countering luminosity-diminishing  with distance)  can be
  guaranteed only up to the J=3--2, 4--3 transitions.

\subsection{Cosmic rays and  turbulence: the dominant molecular gas heating mechanisms in~ULIRGs? }

What is to be made of those global CO SLEDs that are so highly excited
that they are  on par or even surpass those expected  for the Orion SF
``hot-spots''?    Moreover,   maintaining   high  temperatures   ($\rm
T_{kin}$$\ga $100\,K)  for metal-rich gas with  average densities $\rm
n(H_2)$$\ga $10$^5$\,cm$^{-3}$, as implied  by some of these SLEDs, is
difficult  since  cooling   $\rm  \Lambda  $$\propto$$\rm  [n(H_2)]^2$
(Lequeux 2005),  and is thus $\sim $10$^4$-10$^6$  times stronger than
for     average    densities     of     molecular    clouds     ($\sim
$10$^2$-10$^3$\,cm$^{-3}$).   The galaxy-sized  gas  reservoirs ($\sim
$(2-5)$\times $10$^9$\,M$_{\odot}$)  involved make such highly-excited
CO SLEDs even  harder to maintain energetically since,  for their high
average densities, far-UV radiation  fields will be reduced by factors
of $\sim$10$^4$  over distances  of $\la $0.1\,pc.   The corresponding
SF-powered  PDRs will  thus be  very localized,  involving  only small
fractions of a GMC's mass,  as demonstrated by the SF ``hot-spots'' in
the  Orion A, B  clouds, while  the same  is expected  for SNR-shocked
regions (i.e.  the ``mechanical'' counterpart of SF feedback).

Cosmic rays  (CRs) and/or  turbulence, already implicated  as powerful
heating sources  of the molecular gas in  Galactic Center (Yusef-Zadeh
et  al.   2007), and  powering  the  high-excitation  CO SLED  of  the
starburst nucleus of NGC\,253 (Bradford et al.  2003; Hailey-Dunsheath
et  al.  2008),  can provide  strong  volumetric heating  on the  mass
scales necessary to power the high-excitation CO SLEDs of ULIRGs.  For
CRs  this has  been recently  demonstrated, with  the large  CR energy
densities  of   $\rm  U_{CR}$$\sim$[(few)$\times$10$^{3}$]U$_{\rm  CR,
  Gal}$ expected  in such galaxies penetrating much  deeper into dense
molecular regions than far-UV, optical and even IR photons and heating
their  mass  up  to  $\sim  $100\,K  (Papadopoulos  2010).   The  very
turbulent gas disks found in such galaxies (DS98) ($\rm \sigma_z$$\sim
$(40-140)\,km\,s$^{-1}$      versus      $\rm     \sigma      _z$$\sim
$(5-6)\,km\,s$^{-1}$ for ordinary face-on spirals) and the dissipation
of  their  supersonic  velocity   fields  can  considerably  warm  the
molecular gas throughout their  volume.  Following Pan \& Padoan 2009,
the average turbulent heating rate~is


\begin{equation}
\rm \Gamma _{turb}=\mu m_{H} \langle \epsilon (\vec{x}, t) \rangle, 
\end{equation}

\noindent
where $\mu $=2.35 is the mean molecular weight for molecular gas, n is
the gas density, and  $\langle \epsilon\rangle $=$\rm \langle \epsilon
(\vec{x},t)\rangle $=1/2$\rm (\sqrt{3}  \sigma _v)^3/L$ is the average
turbulence  dissipation  rate per  unit  mass  (assuming an  isotropic
velocity    field).    Using    the   $\rm    \sigma_v$-L    (velocity
dispersion)-(cloud size) relation

\begin{equation}
\rm \sigma _v = 
\sigma _{\circ} \left(\frac{P_e}{P_{\circ}}\right)^{1/4}\left(\frac{L}{pc}\right)^{\alpha}
\end{equation}

\noindent
(Larson  1981), which  is  now well-established  for  the Galaxy  with
nearly constant $\rm  \sigma _{\circ}$=1.2\,km\,s$^{-1}$ and power-law
index $\alpha $=1/2 (Heyer \& Brunt 2004).  The pressure-dependence of
the normalization,  with $\rm P_{\circ}/k_B$=$10^4$\,K\,cm$^{-3}$, can
be derived  from the virial  theorem applied for clouds  with boundary
pressure   $\rm   P_e$   (Chi\'eze   1987;  Elmegreen   1989).    This
pressure-dependence  is  very  important   as  it  predicts  a  higher
normalization  of the  $\rm  \sigma _v$-L  relation for  high-pressure
environments.  This has actually  been observed in the Galactic Center
(G\"usten \& Phillip 2004) and  astonishingly also in a molecular gas
disk in a  distant ULIRG at z$\sim $2.3 (Swinbank  et al. 2011).  From
Equations  13  and 14  after  replacing  the  expression for  $\langle
\epsilon  \rangle$,  setting  $\alpha  $=1/2, and  substituting  units
we~obtain

\begin{equation}
\rm \Gamma _{turb}=3.3\times 10^{-27}n\, \sigma_{\circ, n} \left(\frac{P_e}{P_{\circ}}\right)^{3/4}
L^{1/2} _{pc}\, erg\,cm^{-3}\,s^{-1}
\end{equation}

\noindent
where $\rm  \sigma _{\circ, n}$=$\rm  \sigma_{\circ}/(km\,s^{-1})$ and
$\rm L_{pc}$=$\rm  L/(pc)$.  On the other hand  the (mainly) molecular
line  cooling for  self-gravitating  gas is  well-approximated by  the
expression:

\begin{equation}
\rm \Lambda _{line}=1.9\times 10^{-27}n\,\left(\frac{T_{kin}}{10K}\right)^3\, erg\,cm^{-3}\,s^{-1}
\end{equation}

\noindent
(Papadopoulos 2010 and references therein), with most of it due to the
CO  SLEDs  though  other  molecules  (e.g.   $^{13}$CO)  significantly
contribute  when n$\ga $$10^4$\,cm$^{-3}$  (Goldsmith 2001).  Thus for
thermal equillibrium: $\rm \Gamma_{turb}$=$\rm \Lambda _{line}$, we
obtain

\begin{equation}
\rm T_{kin}\sim 
12\sigma^{1/3} _{\circ,n}\left[\left(\frac{P_e/k_B}{10^4\,K\,cm^{-3}}\right)^{1/4}L^{1/6}_{pc}\right]\,K.
\end{equation}

\noindent
Hence,  for the  typical  pressures  expected in  the  Galaxy and  the
midplane of ordinary  spirals ($\sim $10$^4$\,K\,cm$^{-3}$), turbulent
heating can easily account for the temperatures of UV-shielded regions
($\sim  $(10-15)\,K), while  insensitive to  the spatial  scales ($\rm
L_{pc}$) involved, and with a  rather weak dependance on the pressure.
However, for the dense gas-rich disks in ULIRGs the expected turbulent
pressures can reach  $\sim $(1-3)$\times $10$^7$\,K\,cm$^{-3}$.  Thus,
for $\sigma  _{\circ}$=1.2\,km\,s$^{-1}$ and turbulent-driving  at the
largest scales $\rm L_{pc}$=25-50 (disk scale-heights in local ULIRGs,
DS98)  it  would be  $\rm  T_{kin}$$\sim  $(120-140)\,K.  Atomic  line
cooling and gas-dust thermal  interaction can lower these temperatures
somewhat (e.g.   Papadopoulos et al.  2011)  which nevertheless remain
high enough to  power the extreme-excitation global CO  SLEDs is these
systems.

In reality  both CR  and turbulent  heating will occur  in the  ISM of
ULIRGs, powering  high-excitation global CO SLEDs which  would then be
irreducible  to ensembles  of PDRs  and/or SNR-shocked  regions.  This
``irreducibility''  has  indeed  been  noted  in the  past  for  small
molecular gas reservoirs ($\sim $(10$^6$--10$^7$)\,M$_{\odot}$) in the
Galactic Center (Yusef-Zadeh et al.  2007) and the nucleus of NGC\,253
(Bradford et al.  2003; Hailey-Dunsheath  et al.  2008), yet in ULIRGs
it will involve  3-4 orders of magnitude larger  molecular gas masses.
Unique observational  signatures of such  CR-dominated regions (CRDRs)
or turbulently-heated  regions (THRs) have been  recently discussed in
great  detail (Papadopoulos  2010;  Meijerink et  al.  2011; Bayet  et
al.  2011)  and most  will  be  accessible  to ALMA.   Strong  thermal
decoupling of  gas and dust  with $\rm T_{kin}$$\gg$$\rm  T_{dust}$, a
common  aspect  of  both  heating  mechanisms,  can  be  revealed  via
high-resolution mm/submm imaging of dust continuum and line imaging of
dense  gas tracers dense  gas (e.g.   CS, HCN).   High-J CO  {\it and}
$^{13}$CO  line  (e.g.  J=6--5,  7--6)  imaging  will be  particularly
valuable  for distinguishing between  the volumetrically  heated CRDRs
and  THRs versus  the  surface-heated PDRs  as  those responsible  for
powering the  observed CO SLEDs in (U)LIRGs  (e.g. Hailey-Dunsheath et
al.~2008).


\begin{figure*}
\includegraphics[width=\textwidth]{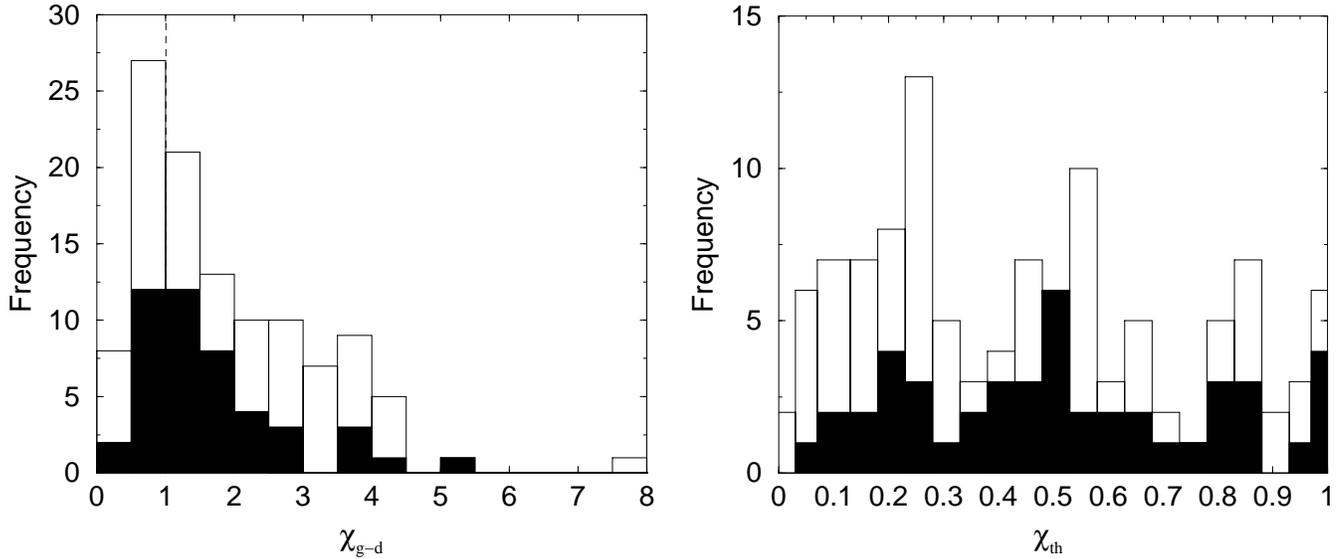}
\caption{The  $\bf{ \chi}_{g-d}$=$\rm  T_{kin}$/$\rm  T_{dust}$  and  $\rm
 \chi_{th}$=$\rm T_{ex}$(3--2)/$\rm  T_{kin}$ distributions obtained from
 the one-phase LVG radiative transfer models (see section 6)}
\end{figure*}

\subsection{CO SLEDs and X$_{\bf co}$ values as  SF mode indicators in galaxies}

     The  low-efficiency (i.e.   low $\rm  L^{(*)}  _{IR}/M(H_2)$) and
  spatially distributed star formation in isolated gas-rich disks, and
  the  high-efficiency  one  occuring  in  the  compact  merger-driven
  starbursts in ULIRGs  can be thought as two  distinct SF modes.  The
  latter accounts for  only 1\% of the total SFR  density in the local
  Universe,  but  could be  dominant  in  the  distant Universe  (e.g.
  Hughes et al.  1998).  However  the discovery of very gas-rich disks
  ($\rm   f_{gas}$$\sim  $0.5-0.6)   with   extended  low   efficiency
  star-formation, and  a space density  10-30 times higher  than dusty
  mergers at high  redshifts (Daddi et al.  2007),  suggests this mode
  as the  prevelant one  in the distant  Universe.  Recent  studies of
  such disks  reported a low  CO(3-2)/(2-1) ratio and a  near Galactic
  $\rm X_{co}$ factor, which were considered independent indicators of
  a disk-like  low SFE  mode (Dannerbauer et  al.  2009; Daddi  et al.
  2010).  Given  the vast potential of molecular  line observations in
  the  upcoming era  of ALMA,  it  is important  to determine  whether
  global CO SLEDs (by far the  easiest to obtain for galaxies) and the
  corresponding  $\rm  X_{co}$  values  are indeed  reliable  SF  mode
  indicators.

   The $\rm  r_{32/21}$ distribution  obtained for our  sample (Figure
   11) makes obvious  that the $\rm r_{32/21}$ recently  measured in a
   distant  gas-rich disk  (Dannerbauer et  al.  2009)  is  within the
   range found for the vigorously star-forming LIRGs.  Furthermore, as
   discussed previously,  uniquely attributing MW-type  ISM conditions
   to  subthermal low-J  CO ratios  is problematic  without additional
   constraints (e.g.  $^{13}$CO~lines).  Finally, a near-Galactic $\rm
   X_{co}$  factor  {\it  does  not necessarily  imply  low-excitation
     SF-quiescent ISM.}   Indeed Galactic  $\rm X_{co}$ values  can be
   associated    with   both    low-excitation    SF-quiescent   ($\rm
   T_{b,1-0}$$\sim  $10\,K, $\rm n$$\sim  $500\,cm$^{-3}$) as  well as
   dense   warm  gas  ($\rm   T_{b,1-0}$$\sim  $60\,K,   $\rm  n$$\sim
   $10$^{4}$\,cm$^{-3}$)    in   star-forming    sites    since   $\rm
   X_{co}$$\propto $$\rm \sqrt{n}/T_{b,1-0}$.  This is actually one of
   the main reasons  for the robustness of $\rm  X_{co}$ in estimating
   the  mass of  large molecular  cloud ensembles  which inadvertendly
   encompass  SF-quiescent as  well as  SF gas  in spiral  disks (e.g.
   Young \&  Scoville 1991).  Finally,  any AGN-powering of  high-J CO
   lines  in  otherwise  SF-quiescent  galaxies, CR  and/or  turbulent
   heating,  inject further  uncertainties into  any estimates  of the
   dense/SF          gas          mass          fraction          $\rm
   f_{d}$=M(n$>$10$^5$\,cm$^{-3}$)/$\rm    M_{tot}$(H$_2$)    (another
   potent measure of the SF mode) using CO SLEDs.

 For  a SF  efficiency  per  dense molecular  gas  mass $\rm  \epsilon
 _{*,g}$=$\rm   L^{(*)}   _{IR}$/$\rm   M_{SF}$(H$_2$)   that   is   a
 near-constant ($\sim $(250-500)\,L$_{\odot}$/M$_{\odot}$), and set by
 the physics underlying SF feedback  onto the ISM (e.g. Scoville 2004;
 Andrews \& Thompson 2011), the  $\rm f_{d}$ becomes an equivalent and
 much better  measure of SF  modes with $f_d$(mergers/starbursts)$\sim
 $(5-10)$\times $$ f_d$(disks/low-SF-efficiency) (e.g.  Solomon et al.
 1992; Gao \&  Solomon 2004; Wu et al.  2005).   The latter asumes the
 HCN/CO J=1--0 ratio  $\rm r_{HCN/CO}$ as a good  proxy, which is also
 more practical to obtain  (low frequency observations) than high-J CO
 SLEDs.  An extensive  body of data in the  local Universe allows both
 comparative  studies  of  SF  modes using  ``raw''  $\rm  r_{HCN/CO}$
 ratios, and their ``calibration'' in terms of actual $\rm f_d$ values
 (e.g.  Papadopoulos et al.  2007;  Gao \& Solomon 2004). Sensitive CO
 and HCN  J=1--0 and other  types of molecular line  observations that
 can discern SF  modes of galaxies at high  redshifts will be possible
 in the future with MeerKAT,  the SKA, and ALMA (Geach \& Papadopoulos
 2012).


\subsubsection{BzK galaxies as SF disks: well-excited CO SLEDs beyond J=3--2?}

  For the gas-rich  disks  of  BzK galaxies  at  high redshifts:  $\rm
L^{(*)}   _{IR}$$\sim  $(1-4)$\times   $10$^{12}$\,L$_{\odot}$,  which
corresponds to $\rm M_{SF}$(H$_2$)$\sim $(0.4-1.6)$\times
$$10^{10}$\,M$_{\odot}$  of  high-excitation  gas (for  $\rm  \epsilon
_{*,g}$=250\,M$_{\odot}$/L$_{\odot}$).  This is $\sim $(5-13)\% of the
$\rm M_{tot}$(H$_2$)  reported for these systems (Daddi  et al.  2010)
with $\sim $10\% being average.   Local spirals on the other hand have
$\rm M_{SF}(H_2)/M_{tot}(H_2)$$\sim  $(3-10)\% (e.g.  I\,Zw\,1  in our
sample),   while  in   ULIRGs   $\rm  M_{SF}(H_2)/M_{total}(H_2)$$\sim
$(20$\rightarrow $50)\%  (e.g.  Solomon et  al.  1992; Gao  \& Solomon
2004; Greve et al.  2009).  For a two-phase decomposition of molecular
line emission  (a SF-quiescent  and a SF-active  phase) the  global CO
$\rm  \langle  T_{b}$(J+1-J)$\rangle$/$\langle \rm  T_b$(1-0)$\rangle$
line ratio  ($\langle..\rangle$ denote velocity/area  averages) can be
easily shown to be

\begin{equation}
\rm r_{J+1,J}=r^{(h)} _{J+1,J}\left[\frac{1+\frac{r^{(l)}_{J+1,J}}{r^{(h)}_{J+1,J}}
\frac{X^{(h)} _{co}}{X^{(l)} _{co}}\left(\frac{1-f_d}{f_d}\right)}
{1+\frac{X^{(h)} _{co}}{X^{(l)} _{co}}\left(\frac{1-f_d}{f_d}\right)}\right].
\end{equation}

\noindent
where  (l) and  (h) denote  the quantities  for the  low and  the high
excitation  phase.  From COBE  measurements (Fixsen  et al.   1999) we
obtain  $\rm r^{(l)}  _{J+1,J}$=0.6, 0.28,  0.10, 0.046,  $\la $0.028,
$\la $0.017  for J+1=2,3,4,5,6,7 (upper limits are  3$\sigma$) for the
the inner Galaxy which we  use, along with $\rm X^{(l)} _{co}$=5\,$\rm
X_l$, to characterize the low-excitation, SF-quiescent phase.  For the
high-excitation phase  we use  the CO (2--1)/(1--0)  ($\sim $1.2--1.3)
and  CO/$^{13}$CO  J=2--1  ($\sim  $10)  ratios obtained  for  the  SF
``hot-spots'' of Orion A and  B (Sakamoto et al.  1994) as constraints
on a  LVG one-phase model.   The densest and warmest  phase compatible
with       them       ($\rm       T_{kin}$=(125--150)\,K,       n$\sim
$3$\times$10$^5$\,cm$^{-3}$,  and $\rm  K_{vir}$=7) yields  a  CO SLED
template  with $\rm  r^{(h)} _{J+1,J}$=1.35,  1.33, 1.30,  1.27, 1.25,
1.22   for   J+1=2,3,4,5,6,7,  and   a   corresponding  $\rm   X^{(h)}
_{co}$=2.2\,$\rm X_l$.  Then  Equation 18 yields $\rm r_{J+1,J}$=0.75,
0.49, 0.34,  0.29, 0.27, 0.26  for J+1=2,3,4,5,6,7.  The  $\rm r_{65}$
and  $\rm r_{76}$  values assume  the  upper limits  for $\rm  r^{(l)}
_{65}$  and $\rm  r^{(l)} _{76}$,  while  setting the  latter to  zero
yields $\rm r^{(l)}_{65}$=0.25 and $\rm r^{(l)} _{76}$=0.24 as minimum
values.

 Figure 12 shows the CO SLED for a fiducial gas-rich disk with 10\% of
 its molecular  gas mass in  the SF phase  along with the one  for the
 inner  Galaxy.    For  typical  measurement   uncertainties  of  $\rm
 \sigma(r)/r$$\sim  $0.25-0.30,  only   ratios  $\rm  r_{J+1,J}$  with
 J+1$\geq $4  markedly deviate from  a Galactic CO SLED  while lower-J
 lines will  remain compatible with it.   It is thus  obvious that for
 the average SF gas mass fraction  in BzK galaxies {\it their CO SLEDs
   will be significantly  more excited from that of  the Galaxy beyond
   J=3--2.}  Future observations of  higher-J CO lines in such systems
 are necessary to verify~this.











\begin{figure*}
\includegraphics[width=\textwidth]{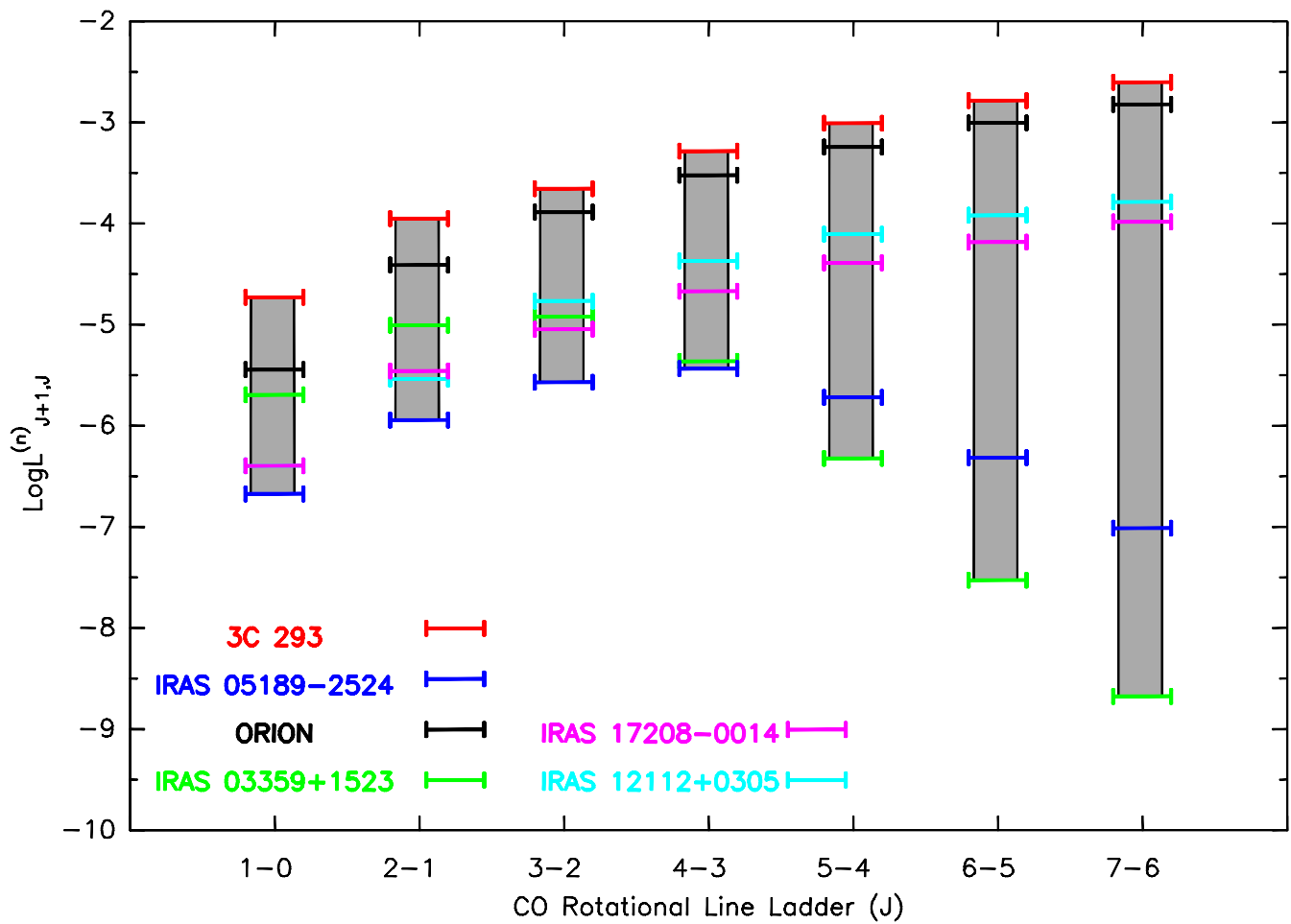}
\caption{The range of the  normalized $\rm L^{(n)} _{J+1,J}$=$\rm L _{J+1,J}/L^{(*)} _{IR}$ CO 
SLEDs of the sample, along with a few key systems (lines beyond J=4--3 are interpolated
using one-phase LVG models).  }
\end{figure*}

\begin{figure}
\includegraphics[width=\columnwidth]{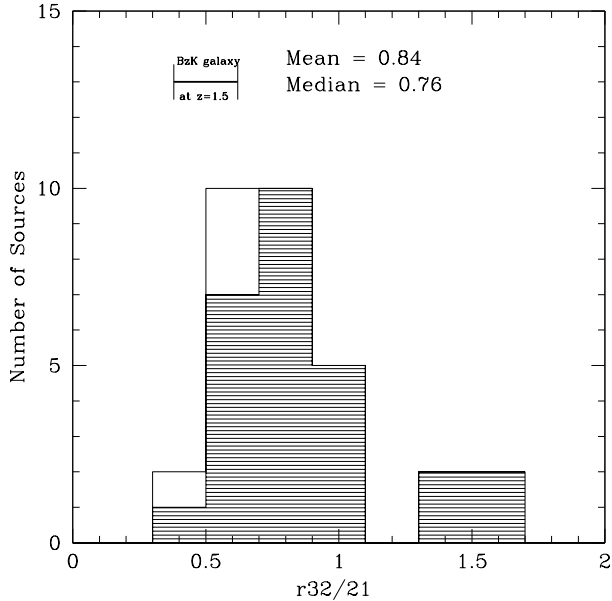}
\caption{The CO (3-2)/(2-1) brightness temperature distribution for our sample (shaded area:
 sources with CO source size $\leq $14$''$), with the range corresponding
to a measured value for a gas-rich disk in a BzK galaxy at z$\sim $1.5 
(Dannerbauer et  al. 2009). }
\end{figure}

\begin{figure}
\includegraphics[width=\columnwidth]{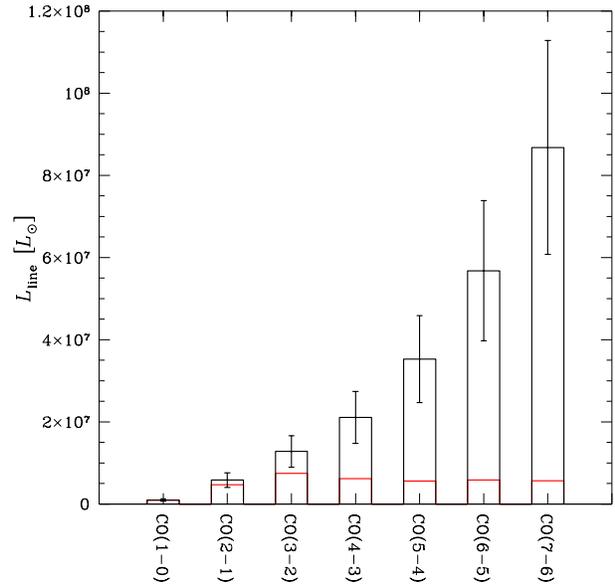}
\caption{The CO SLED of a gas-rich disk with 10\% of its molecular gas
mass in  a SF gas  phase (typical for  BzK galaxies) with  the typical
30\% dispersion expected from measurement errors (black line). The
CO SLED of the inner Galaxy is marked by the red lines (see 7.2.1).}
\end{figure}

\section{Conclusions, and  some important open questions}

We present  multi-J $^{12}$CO  and $^{13}$CO line  data of  total line
luminosities  for  a  sample  of  IR-luminous  galaxies  (N=36)  which
includes the most  intense starbursts known in the  local Universe. We
then conducted a detailed  literature search for all other IR-luminous
galaxies  with available  total CO  line luminosities  up to  at least
J=3--2, the transition where the dense and warm star-forming gas phase
starts dominating the CO SLEDs.  The final sample contains 70 galaxies
with IR luminosities  of $\rm L^{(*)} _{IR}$$\sim $(10$^{10}$-2$\times
$10$^{12}$)\,$\rm L_{\odot}$, morphologies ranging from isolated disks
to  strong  mergers,  and  is  an  excellent  resource  for  follow-up
molecular  line observations  (e.g.   HCN, CS,  HCO$^+$) and  mm/submm
imaging of  the CO or other  molecular SLEDs with ALMA  ($\sim $2/3 of
the sample is accessible from  Llano Chajnantor). In the present study
it was used to obtain a  better overal picture of the average physical
conditions of the molecular gas in star-forming galaxies, our findings
are as follows:
\begin{enumerate}
\item
There  is  a surprisingly  wide  range  of  CO line  excitation,  from
subthermally-excitated  low-J transitions, to  thermalized ones  up to
J=4--3,  6--5.  A  positive K-correction  is  assured only  up to  the
J=3--2,  4--3  lines,  beyond  which  large  variations  of  the
emergent   IR-normalized   ($\rm   L_{J+1\,J}/L^{(*)}   _{IR}$)   line
luminosities are expected.

\item
The highly  excited global CO  SLEDs are found solely  in starburst
   galaxies  with   $\rm  L_{IR}$$>$$10^{11}$\,L$_{\odot}$,  with  the
   exception  of two  found  in  galaxies with  low  or moderate  star
   formation but  hosting powerful AGN.  Such  SLEDs imply tremendeous
   amounts    of    molecular    gas   mass    ($\sim    $(few)$\times
   $10$^{9}$\,M$_{\odot}$)  in a  dense ($\rm  \ga $10$^4$\,cm$^{-3}$)
   and very warm ($\rm T_{kin}$$\ga $100\,K) state.

\item
The high densities  and temperatures, as well as  the strongly unbound
motions  ($\rm K_{vir}$$\gg$1) often  found for  the molecular  gas of
ULIRGs  underlie global  CO SLEDs  which can  remain  well-excited and
become partly optically thin up  to high-J transitions.  In the Galaxy
such  SLEDs are  found  only  in strongly  irradiated  gas near  H\,II
regions  and  shocked  gas   in  SNR-molecular  cloud  interfaces  and
involving  only  $\sim  $(1-5)\%  of  the  mass  of  individual  Giant
Molecular Clouds.  In some ULIRGs of our sample the warm and dense gas
phase can contain $\ga $50\% of their total molecular gas mass.

\item
Highly  supersonic turbulence  and high  CR energy  densities, both
  permeating the ISM of  ULIRGs, can volumetrically heat large amounts
  of  dense gas  to high  temperatures, and  easily account  for their
  observed  highly  excited  SLEDs.    In  the  compact,  and  heavily
  dust-enshrouded ISM  environments of ULIRGs  such heating mechanisms
  may  actually   dominate  over  the  classical   mechanism  via  the
  far-UV/optical radiation  fields and the  photoelectric effect.  The
  resulting  highly excited  global CO  SLEDs will  be  irreducible to
  ensemble  averages of Photon-dominated  Regions (PDRs),  an exciting
  possibility that will be further explored with Herschel observations
  of  the complete  J-ladder above  J=4--3 for  several ULIRGs  in our
  sample (Key project HerCULES).

\item
As  expected  for  a   sample  of  IR-selected  (and  thus  vigorously
star-forming) galaxies,  only few  low-excitation CO SLEDs  are found.
For  only two  of them  a cold  and gravitationally  bound  phase with
moderate/low densities (typical of SF-quiescent molecular gas) emerges
as the most likely ISM state,  while the rest of the ``cold'' CO SLEDs
may  still belong  to  starburst systems.   Only  $^{13}$CO and/or  CO
J+1$\rightarrow   $J,  J+1$>$3   lines  can  reduce   this  well-known
degeneracy, and as a result the low-J part of global CO SLEDs (J=1--0,
2--1,  3--2)  cannot  be a  unique  indicator  of  the SF  mode  (i.e.
merger-driven  compact starburst  versus  isolated-disk low-efficiency
star formation) in galaxies.

\item
The gas-rich disks recently  discovered at high redshifts are examples
of galaxies whose  ``cold'' low-J CO SLEDs have  been used to indicate
Galactic-type low-excitation ISM, yet  higher-J (J=4--3 and higher) CO
transitions are  expected to be highly-excited and  indicative of much
more vigorous star formation environments than those found in local
spirals.

\end{enumerate}

\subsection{Important open questions}

If strong  turbulence and/or high CR energy  densities are responsible
for the large amounts of very warm and dense gas found in some ULIRGs,
they will also {\it change the initial conditions of star-formation in
  such galaxies,} set deep  inside UV-shielded dense gas regions.  The
new  conditions can  lead  towards a  top-heavy  stellar Initial  Mass
Function (IMF) (Hocuk \& Spaans  2010; Papadopoulos et al.  2011) with
ground-breaking implications for the interpretation of IR/optical SEDs
and H\,II  region optical-IR  lines of such  systems in terms  of star
formation  rates.   Nevertheless  the  current  data,  while  strongly
suggestive  of different  dominant heating  mechanisms  operating deep
inside dense  molecular clouds in ULIRGs, they  cannot determine their
state in  detail.  Recent work (Bayet  et al.  2011;  Meijerink et al.
2011)  demonstrates  this  to  be possible  using  the  high-frequency
coverage provided  by Herschel, and  the increased sensitivity  in the
mm/submm regime soon to be provided by ALMA.

Finally,  the   wide  range  of   average  ISM  conditions   found  in
star-forming IR-luminous  galaxies will strongly  impact the so-called
$\rm X_{co}$ factor.  Indeed it  becomes rather hard to argue in favor
of one  convenient, ULIRG-appropriate,  $\rm X_{co}$ factor  given the
diversity of CO  SLEDs found even within the  ULIRG class.  The strong
supersonic  turbulence  alone  will  ``re-settle'' large  portions  of
molecular gas mass in such systems towards high density phases.  These
would be traceable only via high-J  CO and heavy rotor (e.g.  HCN, CS)
transitions while low-J CO lines  will be dominated by a warm, diffuse
and unbound gas  phase that contains little mass  and has misleadingly
``cold'' low CO  ratios.  These issues are now  addressed in detail in
Paper\,II (Papadopoulos et al. 2012).

\section*{Acknowledgments}

YG's research is partially supported  by China NSF frants No 11173059,
10833006  and 10621303.   The  project  was funded  also  by the  John
S. Latsis Public Benefit  Foundation.  The sole responsibility for the
content  lies with its  authors. We  would like  to thank  the referee
Jonathan  Braine  for  his  thorough  reading of  a  rather  extensive
manuscript and comments  that improved the clarity of  this work.  PPP
would like  to also thank  Zhi-Yu Zhang for  help with Figure  10, and
Yiping Ao for his diligent reading and commenting on the plots and our
results.  Finally  a decade-long  work of this  scope would  have been
impossible without the continuous and expert support provided from all
the people  at the Joint Astronomy  Center in Hilo,  Hawaii, that made
the JCMT the success story it has been over two two and a half decades
of its operation.

\bsp

\label{lastpage}

\end{document}